\documentclass{article}

\usepackage{amsmath,amssymb,amsthm,amsfonts,mathrsfs,tikz-cd}
\usepackage[margin=1in]{geometry}
\usepackage{hyperref}
\usepackage{xcolor}
\usepackage{authblk}

\usepackage{tikz}
\usetikzlibrary{fit,backgrounds,calc}

\newtheorem{definition}{Definition}
\newtheorem{proposition}{Proposition}
\newtheorem{remark}{Remark}
\newtheorem{example}{Example}
\newtheorem{theorem}{Theorem}
\newtheorem{lemma}{Lemma}
\newtheorem{corollary}{Corollary}
\newtheorem*{theoremA}{Theorem A}
\newtheorem*{theoremB}{Theorem B}
\newtheorem*{theoremC}{Theorem C}
\newtheorem*{theoremD}{Theorem D}
\newtheorem*{theoremE}{Theorem E}
\newtheorem*{theoremF}{Theorem F}
\newtheorem*{theoremG}{Theorem G}
\newtheorem*{theoremH}{Theorem H}

\newcommand{\ra}{\rightarrow}
\newcommand{\CF}{{\mathcal F}}
\newcommand{\CC}{\mathcal C}
\newcommand{\CA}{{\mathcal A}}
\newcommand{\CQ}{{\mathcal Q}}
\newcommand{\CG}{{\mathcal G}}
\newcommand{\CH}{{\mathcal H}}
\newcommand{\CI}{{\mathcal I}}
\newcommand{\CK}{{\mathcal K}}

\newcommand{\CE}{\mathcal E}

\newcommand{\CU}{{\mathcal U}}

\newcommand{\Uloc}{\CU_{\mathrm{loc}}}
\newcommand{\Aloc}{\CA_{\mathrm{loc}}}
\newcommand{\ZZ}{\mathbb Z}
\newcommand{\RR}{\mathbb R}
\newcommand{\eps}{\epsilon}

\newcommand{\Ad}{\operatorname{Ad}}

\newcommand{\Cat}{\mathbf{Cat}}
\newcommand{\Circ}{\operatorname{Circ}}
\newcommand{\Part}{\operatorname{Part}}
\newcommand{\Diag}{\operatorname{diag}}

\newcommand{\Hom}{\operatorname{Hom}}
\newcommand{\id}{\operatorname{id}}

\newcommand{\Map}{\operatorname{Map}}

\newcommand{\Fun}{\operatorname{Fun}}
\newcommand{\fd}{\mathsf{d}}   
\newcommand{\fs}{\mathsf{s}}   
\newcommand{\Moore}{\operatorname{Moore}}
\newcommand{\Ho}{\operatorname{Ho}}
\newcommand{\Pro}{\operatorname{Pro}}
\newcommand{\Mat}{{\rm Mat}}
\newcommand{\bott}{{\boldsymbol{\bot}}}
\newcommand{\BL}{B_{\mathrm L}}
\newcommand{\BDelta}{\mathcal B_{\mathrm L}}

\newcommand{\supp}{\operatorname{supp}}
\newcommand{\dist}{{\rm dist}}
\newcommand{\Fin}{\mathbf{Fin}}
\newcommand{\Cech}{\check{\mathsf C}}
\newcommand{\SL}{{\mathsf L}}
\newcommand{\XCube}{\mathbf{XCube}}

\newcommand{\cat}{\mathrm{cat}}
\newcommand{\Grp}{\mathbf{Grp}}
\newcommand{\im}{\operatorname{im}}
\newcommand{\Set}{\mathbf{Set}}
\newcommand{\SSet}{\mathbf{SSet_*}}

\newcommand{\fU}{{\mathfrak U}}
\newcommand{\frl}{{\mathfrak l}}

\renewcommand{\RR}{{\mathbb R}}
\renewcommand{\ZZ}{{\mathbb Z}}
\newcommand{\NN}{{\mathbb N}}
\newcommand{\bbC}{{\mathbb C}}

\newcommand{\chB}{\check{B}}
\newcommand{\chbfB}{\check{\mathbf{B}}}
\newcommand{\bW}{\bar{W}}
\newcommand{\sGrp}{\mathbf{sGrp}}
\newcommand{\CSets}{\mathrm{CS}}
\newcommand{\SCSets}{\mathrm{SCS}}

\DeclareMathOperator{\Codiag}{Codiag}
\DeclareMathOperator{\Disc}{Disc}
\DeclareMathOperator{\Cov}{Cov}
\DeclareMathOperator{\hofib}{hofib}
\DeclareMathOperator{\Fan}{Fan}
\DeclareMathOperator{\ev}{ev}
\newcommand{\tgt}{\mathrm{tgt}}
\newcommand{\src}{\mathrm{src}}
\newcommand{\idt}{\mathrm{idt}}

\newcommand{\scrA}{\mathscr{A}}
\newcommand{\scrB}{\mathscr{B}}
\newcommand{\scrC}{\mathscr{C}}
\newcommand{\scrD}{\mathscr{D}}
\newcommand{\scrH}{\mathscr{H}}
\newcommand{\scrP}{\mathscr{P}}
\newcommand{\fV}{\mathfrak{V}}
\newcommand{\bPi}{\overline{\Pi}}
\newcommand{\bfk}{\mathbf{k}}
\newcommand{\bfm}{\mathbf{m}}
\newcommand{\bfell}{\boldsymbol{\ell}}

\newcommand{\CHpath}{\CH^{\mathrm{path}}}
\newcommand{\CGpc}{\CG_{\mathrm{pc}}}
\newcommand{\CGos}{\CG_{\mathrm{os}}}
\newcommand{\hA}{\hat{A}}
\newcommand{\hB}{\hat{B}}
\newcommand{\hR}{\hat{R}}
\newcommand{\hP}{\hat{P}}
\newcommand{\bx}{\bar{x}}
\newcommand{\by}{\bar{y}}
\newcommand{\ba}{\bar{a}}
\newcommand{\bdH}{\partial H}
\newcommand{\talpha}{\tilde{\alpha}}

\title{Anomalies, higher symmetries, and homotopy theory I:\\
crossed precosheaves and a nonabelian \v{C}ech construction}
\author[1,2]{Anton Kapustin}
\author[1]{Shixiong Xu}
\affil[1]{California Institute of Technology}
\affil[2]{Institut des Hautes \'{E}tudes Scientifiques}

\begin{document}

\maketitle

\begin{abstract}
Symmetries of a quantum field theory or a quantum lattice model are expected to be described not by a group but by a homotopy type, whose homotopy groups are the higher symmetry groups and whose Postnikov invariants encode 't~Hooft anomalies. We propose a homotopy-theoretic definition of 't Hooft anomalies based on the structure of localized symmetries in the operator-algebraic approach. The input is a crossed precosheaf: a collection of groups attached to regions of space, equipped with pairings recording that the commutator of two transformations localized in two regions is localized in their intersection. To a crossed precosheaf and a finite family of regions we attach a pointed connected homotopy type, the \v{C}ech-Loday classifying space. As the family of regions ranges over finite covers of space, the classifying spaces form a pro-object in the homotopy category. There is a canonical map from the ordinary classifying space of the symmetry group to this pro-type. Composing this map with a group action gives an anomaly whose nontriviality obstructs gauging. In favorable cases, the \v{C}ech-Loday space is essentially independent of the cover. In such cases, the homotopy fiber of the canonical map is a candidate for the classifying space of the higher group of symmetries.
We give examples of crossed precosheaves arising from gauge transformations in continuum field theory and from quantum cellular automata and finite-depth circuits. For quantum spin systems in one and two dimensions, our construction recovers the Else-Nayak index and its generalizations to two dimensions.
\end{abstract}

\tableofcontents

\section{Introduction}\label{sec:intro}

This introduction has two parts. The first states the mathematical problem, the construction that solves it, and its scope, and assumes no physics. The second delves into the physical meaning of the problem and of the construction. 

Let $G$ be a group and $X$ a space (for present purposes, one can take $X=\RR^d$ or a more general smooth manifold). In the simplest situation, the group
of gauge transformations associated to $G$ is the group $\CG(X)$ of
maps $X\to G$ of some regularity class: continuous, smooth, piecewise constant with respect to a stratification of $X$, etc. The group $\CG(X)$ has additional structure which is essential for its physical applications.  First, for every region $U\subseteq X$ there
is the subgroup $\CG(U)\subseteq\CG(X)$ of maps supported in $U$, and
$\CG(U)\subseteq\CG(V)$ whenever $U\subseteq V$; that is, $U\mapsto\CG(U)$
is a precosheaf of groups on the regions of $X$. Second,
$[\CG(U),\CG(V)]\subseteq\CG(U\cap V)$, because a commutator of two maps
is trivial wherever either of them is; in particular, transformations
supported on disjoint regions commute, and every $\CG(U)$ is normal in
$\CG(X)$. Third, for any two regions $U,V$ we have $\CG(U)\cap\CG(V)=\CG(U\cap V)$. Fourth, in some cases (for example, for piecewise-constant gauge transformations on $\RR^d$), a multiplicative partition of unity holds: for any two regions $U,V$ one has $\CG(U\cup V)=\CG(U)\,\CG(V)$ (a transformation
supported on a union is a product of transformations supported on the
pieces). 
These four properties are the axioms of what we call a
\emph{gauge precosheaf}.
The global symmetry group $G$ is embedded in $\CG(X)$ as the constant
maps.

Now let $F$ be the symmetry group of some physical system on $X$, and suppose
that $F$ likewise comes with subgroups $\CF(U)\subseteq F$ of symmetries
localized on $U$, satisfying the first three of the above properties but not necessarily the last one.
To \emph{gauge} an action $\rho\colon G\to F$ is to extend it to a
morphism of precosheaves $\varrho\colon\CG\to\CF$ with
$\varrho(X)\circ\iota=\rho$, where $\iota\colon G\to\CG(X)$ is the
embedding as constant maps: a gauge transformation supported on $U$
should act by a symmetry localized on $U$. Whether such an extension
exists is a question about $\rho$, $\CF$, and the regions of $X$, and
nothing else. One may test the problem on a finite cover $U_1,\ldots,U_n$ of $X$.
Write $U_A=\bigcap_{i\in A}U_i$ for $A\subseteq\{1,\ldots,n\}$, with
$U_\varnothing=X$. A morphism $\varrho$ has components
$\varrho_A\colon\CG(U_A)\to\CF(U_A)$, compatible with the inclusions
among the $U_A$ and with $\varrho_\varnothing\circ\iota=\rho$ on the
constant maps, and one asks whether such a compatible family exists.
Since a constant map $g\in\CG(X)$ is a product $f_1\cdots f_n$ of maps
$f_i\in\CG(U_i)$, compatibility gives
$\rho(g)=\varrho_{\{1\}}(f_1)\cdots\varrho_{\{n\}}(f_n)$; so, in the
first place, each $\rho(g)$ must be a product of elements
$\rho_i(g)\in\CF(U_i)$. Such a factorization, if it  exists, is not unique. Part of
this ambiguity is explained by the pairwise intersections, since an
element of $\CF(U_i\cap U_j)$ may be moved from the $i$th factor to the
$j$th; the rest is constrained on triple intersections, and so on.

If the groups $\CF(U)$ were abelian, constructing obstructions would be a problem in homological algebra. In the abelian case, the groups $\CF(U_A)$ assemble into the augmented
\v{C}ech complex
\begin{equation}\label{eq:abelian-cech-intro}
\CF(X)\longleftarrow\bigoplus_i\CF(U_i)\longleftarrow
\bigoplus_{i<j}\CF(U_i\cap U_j)\longleftarrow\cdots,
\end{equation}
where $\CF(X)$ sits in degree $-1$. Let   $H^{\mathrm{aug}}_k$, $k\geq -1$, be its homology groups. A homomorphism
$\rho\colon G\to\CF(X)$ determines a sequence of classes in
$H^{k+2}(G;H^{\mathrm{aug}}_k)$, $k\geq -1$, each defined when the
previous ones vanish, all of which vanish if $\rho$ admits a gauging.
The leading one is the composite $G\to\CF(X)\to H^{\mathrm{aug}}_{-1}$,
which obstructs the factorization of each $\rho(g)$ into elements of
$\CF(U_i)$; the next lies in $H^2(G;H^{\mathrm{aug}}_0)$, where
$H^{\mathrm{aug}}_0$ is the group of ambiguities of factorization modulo
those explained by pairwise intersections, and obstructs a choice of
factorizations compatible with the group law up to such explained
ambiguities; and so on.\footnote{One might expect the stronger requirement that the
factorizations be compatible with the group law on the nose, with the
corresponding obstruction taking values in the cycles of
\eqref{eq:abelian-cech-intro} rather than in its homology. But a
gauging does not in general produce such factorizations: the
decomposition of a constant map into maps supported on the $U_i$ need
not be multiplicative, and the axioms of a gauge precosheaf only
guarantee that its failure to be multiplicative is explained by
pairwise intersections. Homology, not cycles, is therefore the right
coefficient group.}

For local quantum systems, the groups $\CF(U)$ are not abelian, and one needs a nonabelian version of this \v{C}ech theory. This is what we develop in this paper. To a precosheaf
of groups $\CF$ with commutator locality and a finite cover
$\xi=(U_1,\ldots,U_n\,\vert\,X)$ of $X$ we attach a pointed connected simplicial
set, the \emph{\v{C}ech-Loday classifying complex}
$\chB_{\CF}(\xi)\in\SSet$,
together with a canonical map $\kappa_\CF(\xi)\colon B\CF(X)\to\chB_{\CF}(\xi)$ from the classifying complex of the global symmetry group. The
construction uses the crossed $n$-cubes of Ellis and Steiner
and the $\cat^n$-groups of Loday: the groups $\CF(U_{i_1}\cap\cdots\cap
U_{i_k})$ and their inclusions form a crossed $n$-cube, and Loday's
classifying functor turns a crossed $n$-cube into an $(n+1)$-truncated
homotopy type. For abelian $\CF$ one can show that one recovers
\eqref{eq:abelian-cech-intro}; in general, the homotopy class of
$\kappa_\CF(\xi)\circ B\rho\colon BG\to\chB_\CF(\xi)$ replaces the
sequence of classes above and obstructs gauging in the sense defined
above. Borrowing a term from physics, we call it the \emph{anomaly}
of $\rho$ at the cover $\xi$. 
We also study the behavior of $\chB_{\CF}(\xi)$ and $\kappa_\CF(\xi)$ under refinements of covers. 

Apart from the nonabelian nature of the problem, two additional complications occur in applications to quantum systems. First,
the localized symmetries of a quantum lattice system do not form
subgroups of a single group. A symmetry localized in an unbounded region
is an automorphism of the algebra of observables, but a symmetry
localized in a bounded region is most naturally a unitary observable
$u$, acting by $a\mapsto uau^{-1}$, and the map $u\mapsto\Ad_u$ has
kernel $U(1)$. The corestriction maps $\CF(U)\to\CF(V)$ therefore need not be
injective (the precosheaf $\CF$ need not be \emph{coflasque}) and commutator locality
can no longer be a property of subgroups: it becomes a datum, a
\emph{pairing} $\CF(U)\times\CF(V)\to\CF(U\cap V)$ lifting the
commutator and subject to identities. In the examples studied below, it is this kernel that carries the
final obstruction. The condition $\CF(U\cap V)=\CF(U)\cap\CF(V)$ also needs to be either reformulated or dropped. Second, localization in a lattice
system is defined only up to a bounded thickening of regions, so the
regions of $X$ form a partially ordered set with meets and joins which
is not a set of subsets. We therefore work over an arbitrary
distributive lattice $\SL$ of regions. A precosheaf of groups on $\SL$
with pairings satisfying the identities of a crossed cube is what we
call a \emph{crossed precosheaf}
(Definition~\ref{def:crossed-precosheaf-structure}); a gauge precosheaf
is a coflasque, meet-preserving crossed precosheaf with the
partition-of-unity property. 

The general theory applies to crossed precosheaves in the continuum
and on the lattice alike; the applications in this paper are to
quantum spin systems. In the continuum, the precosheaves of
piecewise-constant and of continuous gauge transformations are the
motivating examples (Section~\ref{sec:gaugingnew}); what should play the
role of $\CF$ for a Quantum Field Theory is not addressed here. On the
lattice, quantum cellular automata, finite-depth circuits, and a
unitary-lifted version of the latter carry natural crossed-precosheaf
structures, and it is for these that we compute
\v{C}ech-Loday types and identify the anomaly with known invariants.

The problem of gauging a symmetry comes from physics, and we now explain the
physics. A quantum field theory or a quantum
lattice model with a symmetry $G$ may or may not admit a coupling to
background gauge fields for $G$; the obstructions are called 't Hooft anomalies
\cite{tHooft1980}. For symmetries which act trivially on space-time coordinates, it is believed that 't Hooft anomalies can always be canceled by coupling the original theory to a classical topological gauge theory in one dimension higher. In the case of a finite symmetry $G$, bosonic topological gauge theories are Dijkgraaf-Witten theories \cite{DW}, leading to the conjecture that for a bosonic theory in $d+1$
space-time dimensions the anomalies of a finite symmetry group $G$ are classified
by $H^{d+2}(BG,U(1))$ \cite{Wenanomalies,KapustinThorngren2014}. 

More
recently it was proposed that a local quantum theory in $d+1$ space-time dimensions (a QFT or a quantum lattice model) has, besides the symmetry group
$G=G_0$, \emph{higher symmetry groups} $G_p$, $p=1,\ldots,d$
\cite{GKSW}, whose elements label invertible topological defects of space-time
codimension $p+1$; these groups are abelian and $G_0$ acts on them by
automorphisms. It is natural to conjecture that all of them are homotopy
groups of a single pointed connected $(d+1)$-truncated homotopy type
$Y$, $G_p=\pi_{p+1}(Y)$, the classifying homotopy type of a
$(d+1)$-group \cite{KapustinThorngren2013,Sharpe,CDI,BeniniCordovaHsin}.
Since every theory has scalar operators, $G_d$ always contains $U(1)$;
if there is nothing else, $Y$ has exactly two nontrivial homotopy groups,
$\pi_1(Y)=G$ and $\pi_{d+1}(Y)=U(1)$, and is classified by a
$k$-invariant in $H^{d+2}(BG,U(1))$. It was suggested in
\cite{Sharpe,KapustinXu} that the 't Hooft anomaly should be
reinterpreted as the failure of $Y$ to split, that is, as a mixing
between $G$ and the higher symmetries. There is at present no general
procedure for constructing $Y$ from a QFT, but the question has been studied for
quantum lattice models, whose operator-algebraic description is well
understood \cite{NayakElse,KapSop,YuAnetal,4auth,KapustinXu,KawShirley,KS,ShirleyZhangJiLevin,CGT,JiYangLinearization}.
In particular, in \cite{KS}, L.~Spodyneiko and one of the authors proposed to construct
$Y$ from transformation groups attached to regions of space and the
locality of their commutators. The present paper carries this out.

It is useful to distinguish the group of kinematic symmetries $F=\CF(X)$ of a concrete system (typically, a group of automorphisms of the algebra of observables) and an abstract symmetry group $G$ which acts on the system via a homomorphism $\rho:G\to F$. The classifying space construction being functorial, this homomorphism gives a continuous map $B\rho:BG\to BF$ well-defined up to homotopy. We define the anomaly as the homotopy class of the composite map $\kappa_\CF(\xi)\circ B\rho:BG\to \chB_\CF(\xi)$. The anomaly is trivial when this map is homotopic to a constant map. The higher homotopy groups of the \v{C}ech-Loday homotopy 
type $\chB_\CF(\xi)$ encode higher symmetry groups of the system. The conjectural classification of bosonic anomalies in $d+1$ space-time dimensions is recovered if the target of the map happens to be the Eilenberg-MacLane space $K(U(1),d+2)$. 

The \v{C}ech-Loday complex of a cover
$\xi$ detects anomalies and higher symmetries at the ``scale'' of $\xi$;
refining the cover gives finer information. In favorable cases, all anomalies and higher symmetries can be computed from a single sufficiently
fine cover. For trivial reasons, this holds for one-dimensional quantum spin systems, and we also prove that this is so for two-dimensional spin systems.
Conjecturally, it also holds for crossed precosheaves arising from
well-behaved QFT models.

Even then, the \v{C}ech-Loday classifying complex should not be identified with the classifying space of higher symmetries $Y$.
For a one-dimensional spin system and a natural cover, for example, the
former is a simply-connected 3-type $K(U(1),3)$, whereas we expect $Y$ to be a $2$-type with
$\pi_1(Y)=G_0$ and $\pi_2(Y)=G_1=U(1)$. If the homotopy type of $\chB_\CF(\xi)$ is independent of $\xi$ for a sufficiently fine cover $\xi$, a natural candidate for the classifying space of the higher symmetry group is a space  $Y_\CF$ defined by the homotopy fibration sequence
$$
Y_\CF\longrightarrow B F\longrightarrow\chB_{\CF}(\xi),
\qquad F=\CF(X).
$$
Then the long exact sequence of homotopy groups gives $G_p:=\pi_{p+1}(Y_\CF)=\pi_{p+2}(\chB_{\CF}(\xi))$
for $p>0$. The information about anomalies and higher symmetries
contained in $Y_\CF\to BF$ is the same as that contained in
$BF\to\chB_{\CF}(\xi)$. While $Y_\CF$ is a natural candidate for $Y$, we do not know if this is a reasonable definition for all systems of interest. For example, the Whitehead bracket $\pi_k(Y_\CF)\times \pi_l(Y_\CF)\to\pi_{k+l-1}(Y_\CF)$ vanishes for $k,l>1$, a property which is not expected to be universally true for the higher symmetry type $Y$. 

The standard account of higher symmetries and anomalies is formulated
in terms of the Euclidean path integral: higher symmetries are represented by 
invertible topological defects, while anomalies are detected by the response
to background gauge fields. This approach is heuristic and semi-classical and offers no prescription for
computing the higher symmetry groups of a theory which is not defined
by a classical Lagrangian. In the Euclidean approach, it is also far from obvious how to extract the anomalies from the properties of topological defects. The approach of this paper differs in three respects.
It applies to any theory with the notion of a localized symmetry and does not need a classical Lagrangian or a Euclidean continuation. It
produces anomalies and higher symmetries from a single object, the
\v{C}ech-Loday type and its canonical map, so that the relation
between the two, which in the Euclidean approach is a conjecture, is
built into the construction: the obstructions to triviality of the map $\kappa_\CF(\xi)\circ B\rho$ are read off the Postnikov tower of $\chB_\CF(\xi)$ whose fibers are the Eilenberg-MacLane spaces of higher symmetry groups.
And in low dimensions this object is computable: for spin chains and for
two-dimensional spin systems the \v{C}ech-Loday type is determined
explicitly, the anomaly reduces to the invariants of Else and Nayak
\cite{NayakElse} and of \cite{KapustinXu,4auth,KawShirley}, and we
prove that these invariants are the whole anomaly, for every cover. On the other hand, there is some  overlap between our operator-algebraic approach and the Euclidean one. For example, the group of piecewise-constant gauge transformations includes elements which can be interpreted as networks of invertible topological defects (Section \ref{sec:gaugingnew}).

\medskip\noindent
\textbf{Acknowledgments.} In preparing this paper the authors used ChatGPT-6 and Claude 5.1 to assist in proving technical results and to edit the paper. All mathematical arguments were independently checked by the authors, who take full responsibility for the content of the paper. This work was supported in part by the U.S.\ Department of Energy, Office of Science, Office of High Energy Physics, under Award Number DE-SC0011632 and by the Simons Investigator Award. A.K. would like to thank Institut des Hautes \'{E}tudes Scientifiques for hospitality during the preparation of this work.

\section{Summary of main results}\label{sec:summary}

Here we summarize the main results, with pointers to where they are proved.
Throughout, $\SL$ is a meet-semilattice (a poset with finite nonempty meets),
$\CF$ is a crossed precosheaf of groups on $\SL$ (Definition~\ref{def:crossed-precosheaf-structure}), and a
\v{C}ech configuration of arity $n$ is an $n$-tuple $U_1,\dots,U_n\in\SL$
together with a common upper bound $W$, written $\xi=(U_1,\dots,U_n\,|\,W)$. $U_1,\ldots,U_n$ need not form a cover of $W$. 

\medskip\noindent
\textbf{The abelian model.}
It is helpful to keep in mind the case of a precosheaf $\CA$ of abelian
groups. To a configuration $\xi=(U_1,\dots,U_n\,|\,W)$ one attaches the
augmented \v{C}ech complex
$$
\CA(W)\longleftarrow\bigoplus_i \CA(U_i)\longleftarrow\bigoplus_{i<j}\CA(U_i\wedge U_j)
\longleftarrow\cdots\longleftarrow \CA(U_1\wedge\dots\wedge U_n),
$$
whose homology $\check H^{\mathrm{aug}}_*(\xi;\CA)$ is the augmented \v{C}ech
homology of the family $(U_i)$ relative to $W$. This complex is natural
with respect to shrinking the $U_i$ or $W$, permuting the $U_i$, and
adjoining or repeating an entry; adjoining an entry $V$ produces the
mapping cone of the map of complexes induced by $U_i\wedge V\le U_i$; and
an inverse system over refinements supplies a cover-independent pro-object. Our construction is the nonabelian
version of this picture: precosheaves of groups replace precosheaves of
abelian groups, commutator pairings encode non-commutativity of elements of groups on different subsets, and
homotopy types replace chain complexes. When $\CA$ is abelian with trivial
pairings, one can show that the construction below returns a space with
$\pi_{k+2}\chB_{\CA}(\xi)\cong\check H^{\mathrm{aug}}_k(\xi;\CA)$, and each of
Theorems A--C specializes to the corresponding statement about the
\v{C}ech complex.

\medskip\noindent
\textbf{The \v{C}ech-Loday classifying complex.}
To each $\xi$ we attach a crossed $n$-cube of groups $\Cech_{\CF}(\xi)$,
the nonabelian counterpart of the \v{C}ech complex above: its vertices are
the groups $\CF(W)$ and $\CF(U_{i_1}\wedge\dots\wedge U_{i_k})$, its edges are
the corestriction maps, and the data of the crossed precosheaf supply
commutator pairings between the vertices. Composing with the
Ellis-Steiner equivalence between crossed cubes and Loday's
$\cat^n$-groups and then with Loday's classifying space functor, we obtain a
pointed simplicial set $\chB_{\CF}(\xi)$.

Configurations are compared along the same operations as \v{C}ech
complexes. We write $\Cech(\SL)$ for the category whose objects are
configurations of all arities and in which a morphism
$(V_1,\dots,V_m\,|\,W')\to(U_1,\dots,U_n\,|\,W)$ is a function
$r\colon\{1,\dots,m\}\to\{1,\dots,n\}$ with $V_j\le U_{r(j)}$ for all $j$
and $W'\le W$ (Definition~\ref{def:total-cech-indexing}). Thus a morphism enlarges the members and the ambient
object, and simultaneously reindexes; in particular permuting, repeating,
or adjoining entries are morphisms.

\begin{theoremA}[Corollary~\ref{cor:loday-connected-truncated}, Theorem~\ref{thm:total-loday-functor}, Definition~\ref{def:les-indexed-type}]
$\chB_{\CF}(\xi)$ is a pointed connected $(n+1)$-truncated homotopy type, and
$\xi\mapsto\chB_{\CF}(\xi)$ is a functor from $\Cech(\SL)$ to the
homotopy category of pointed simplicial sets.
\end{theoremA}

For $n=0$ this is the ordinary classifying complex $B\CF(W)$, and for every
$\xi=(\fU\,|\,W)$ the morphism $(\,|\,W)\to\xi$ gives a canonical map
$\kappa_{\CF}(\xi)\colon B\CF(W)\to\chB_{\CF}(\xi)$, the analog of the
augmentation $\CA(W)\to$ (\v{C}ech complex).

\medskip\noindent
\textbf{The basic fibration.}
Objects of the category $\Cech(\SL)$ are called \v{C}ech configurations. One can build arbitrary configurations by starting with a \v{C}ech configuration $(\,|W)$ and successively adding $U_i\leq W$, $i=1,2,\ldots,n$. The following result describes how this affects the \v{C}ech-Loday complex.

\begin{theoremB}[Theorem~\ref{thm:one-coordinate-fibration}]
Let $\xi=(U_1,\dots,U_n\,|\,W)$, let $V\le W$, and write
$\xi^+=(U_1,\dots,U_n,V\,|\,W)$ and $\xi|_V=(U_1\wedge V,\dots,U_n\wedge V\,|\,V)$.
Then
$$
\chB_{\CF}(\xi|_V)\longrightarrow\chB_{\CF}(\xi)\longrightarrow\chB_{\CF}(\xi^+)
$$
is a homotopy fibration sequence. It is natural with respect to morphisms of crossed precosheaves and morphisms of $\xi^+$ which preserve the distinguished added coordinate and the remaining coordinates.
\end{theoremB}

\medskip\noindent
\textbf{Dependence on the cover.}
When $\SL$ is a distributive lattice and $X\in\SL$, finite covers of $X$ form a
cofiltered refinement preorder.

\begin{theoremC}[Lemma~\ref{lem:refinement-independence}, Corollary~\ref{cor:nonabelianprotype}, Definition~\ref{def:nonabelian-cech-obstruction-diagram}]
The homotopy types $\chB_{\CF}(\fU\,|\,X)$, as $\fU$ ranges over
finite covers of $X$, depend on refinements only through the refinement
relation and assemble into a pro-object
$$
\chbfB_{\CF}(X)\in\Pro\bigl(\Ho(\SSet)\bigr),
$$
functorial in $\CF$ and receiving a canonical map $\kappa_\CF\colon B\CF(X)\to\chbfB_{\CF}(X)$.
\end{theoremC}

If $\CF$ is coflasque (i.e. all corestriction maps are injective), preserves meets, and satisfies a multiplicative
partition-of-unity property, then $\chbfB_{\CF}(X)$ is constant and
contractible (Proposition~\ref{prop:meetpreserving}, Corollary~\ref{cor:contractible-protype}). Thus nontrivial invariants arise
only from crossed precosheaves that fail one of these conditions. 

\medskip\noindent
\textbf{Anomalies and gauging.}
Section~\ref{sec:tHooft} connects the \v{C}ech-Loday construction to the discussion of
gauging in Section~\ref{sec:gaugingnew}. For a crossed precosheaf $\CF$ and a
homomorphism $\rho\colon G\to\CF(X)$, the composites
$BG\to B\CF(X)\to\chB_{\CF}(\fU|X)$ over all finite covers $\fU$ form the
\emph{anomaly} $\alpha_{\CF,\rho}\in\Hom_{\Pro(\Ho(\SSet))}(BG,\chbfB_{\CF}(X))$
(Definition~\ref{def:anomaly}); it is compatible with refinement of covers
(Proposition~\ref{prop:monotone}), and in favorable cases it can be
computed from a single cover (Proposition~\ref{prop:represented-by-cover}).
A \emph{gauge precosheaf} $(\CG,\iota)$ for $G$ is as in the
Introduction and Section~\ref{sec:gaugingnew}
(Definition~\ref{def:gauge-precosheaf}). A gauging of a $G$-action $\rho:G\to\CF(X)$ is defined as a triple $(\CG,\iota,\varrho)$ where $(\CG,\iota)$ is a gauge precosheaf and $\varrho:\CG\to\CF$ is a morphism of crossed precosheaves which reduces to $\rho$ on constant gauge symmetries. Thus, according to our definition, gauging involves a choice of an abstract gauge symmetry and its local action on a concrete system. 

\begin{theoremD}[Theorem~\ref{thm:gauging-obstruction}]
If a gauging of a $G$-action $(\CF,\rho)$ exists, then $(\CF,\rho)$ is anomaly-free.
\end{theoremD}

For a quantum spin system the abstract on-site transformations form a gauge precosheaf for every group $G$ (Lemma  \ref{lem:fuzzification}). For a QFT on $\RR^d$ and a discrete $G$, there is a natural candidate gauge precosheaf as well, the precosheaf of piecewise constant gauge transformations (Section \ref{sec:gaugingnew}). Here we focus on lattice systems and  leave QFT applications for future work.

\medskip\noindent
\textbf{Examples from quantum spin systems.}
Let $X$ be a metric space and $q$ be a locally finite function $q:X\ra\NN$. Such a pair $(X,q)$ defines a quantum spin system \cite{JiYang}. Let $\SL$ be a meet-subsemilattice of regions
of $X$ up to bounded thickening (Section~\ref{sec:meetsubsemi}). We assume that $\SL$ contains the equivalence class of bounded regions which we denote $\bott$.
For $X=\RR^d$ one may take $\SL$ to be the set of conical closed semilinear subsets of $\RR^d$; this gives a natural notion of localization for quantum spin systems on $\RR^d$.

For quantum spin systems, there are several natural choices of crossed precosheaves of symmetries. 
\begin{theoremE}[Theorems~\ref{thm:crossed-precosheaf-QCA}, \ref{thm:crossed-precosheaf-circuits}, and~\ref{thm:E}]
The assignments
$U\mapsto \CQ(U)$, $U\mapsto \CC(U)$, $U\mapsto \CE(U)$,
sending a region to the group of quantum cellular automata localized on $U$,
to the group of finite-depth circuits localized on $U$, and to the
central extension of the latter obtained by replacing $\CC(\bott)$ by the
group of local unitaries, are crossed precosheaves on $\SL$. For $\CQ$ and $\CC$
the commutator pairings are ordinary commutators; for $\CE$ the pairings
between regions with bounded intersection are given by the 
lift $\eta$ of \cite{KapustinXu}. 
\end{theoremE}
The precosheaf $\CE$ is not coflasque: for an unbounded region $V$ its corestriction map $\CE(\bott)=\Uloc\to \CE(V)=\CC(V)$ has kernel $U(1)$. In the 1d case, this is the source of all the
anomaly information. 

Once a particular choice of a crossed precosheaf of symmetries $\CF$ has been made, one can evaluate its anomaly with respect to a $G$-action. If the anomaly is nontrivial, then there is no morphism from the crossed precosheaf of abstract on-site symmetries to $\CF$ which gauges the $G$-action. Corollary~\ref{cor:onsite-up-to-conjugacy} makes this concrete: nontrivial $\CC$-, $\CQ$-, and $\CE$-anomalies obstruct, respectively, conjugacy by a finite-depth circuit to a projectively on-site action, conjugacy by a QCA to a projectively on-site action, and conjugacy by a finite-depth circuit to an on-site action.

\medskip\noindent
\textbf{Anomalies in low dimensions.}
For $X=\RR$ the lattice $\SL$ has four elements $\bott,L,R,X$, where $L$ and $R$ are the left and right half-lines. The cover $(L,R)$ is initial under refinement pre-order, so the
pro-object of Theorem C is represented by the homotopy type corresponding to the cover of $\RR$ by $L$ and $R$.

\begin{theoremF}[Section~\ref{sec:1d}; Theorem~\ref{thm:1d-anomaly}]
$\chB_{\CE}(L,R\,|\, X)\simeq K(U(1),3)$, and the canonical map
$B\CC(X)\to\chB_{\CE}(L,R\,|\,X)$ classifies, up to sign, the
pullback of the universal Else-Nayak class along the truncation
$\CC(X)\to \CC(R)/\im\Ad$. Consequently, for any group
action $\rho\colon G\to \CC(X)$, the Else-Nayak anomaly of $\rho$ is the
obstruction to null-homotoping the composite
$BG\to B\CC(X)\to\chB_{\CE}(L,R\,|\, X)$.
\end{theoremF}

The homotopy fiber $Y$ of $B\CC(X)\to K(U(1),3)$ is a $2$-type with
$\pi_1(Y)=\CC(X)$, $\pi_2(Y)=U(1)$, and $k$-invariant the anomaly class,
which is the higher-symmetry homotopy type one expects for a one-dimensional
quantum spin system \cite{KapustinXu}. Theorem F is the model for the general proposal, made
above, that the higher-symmetry type be the homotopy fiber of $\kappa_\CF$. It also showcases the mechanism by which the classification of 't Hooft anomalies can become model-independent.

Similarly, we show that the 2d anomaly indices defined in \cite{KapustinXu,4auth,KawShirley} are obstructions to null-homotoping maps from $BG$ to the \v{C}ech-Loday classifying complex of the crossed precosheaf $\CE$ and a particular cover of $\RR^2$. 

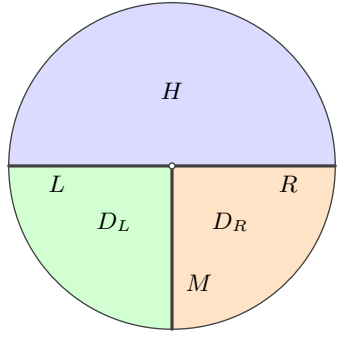
\begin{figure}[t]
\centering
\begin{tikzpicture}[scale=0.6, line join=round, line cap=round, font=\small]
  \colorlet{Hcol}{blue!14}
  \colorlet{DLcol}{green!18}
  \colorlet{DRcol}{orange!22}
  \colorlet{BoundaryCol}{black!75}

  \def\R{3.6}

  \fill[Hcol]
    (0,0) -- (\R,0)
    arc[start angle=0,end angle=180,radius=\R] -- cycle;
  \fill[DLcol]
    (0,0) -- (-\R,0)
    arc[start angle=180,end angle=270,radius=\R] -- cycle;
  \fill[DRcol]
    (0,0) -- (0,-\R)
    arc[start angle=270,end angle=360,radius=\R] -- cycle;

  \draw[BoundaryCol, line width=0.5pt] (0,0) circle (\R);
  \draw[BoundaryCol, line width=1.2pt] (0,0) -- (\R,0);
  \draw[BoundaryCol, line width=1.2pt] (0,0) -- (-\R,0);
  \draw[BoundaryCol, line width=1.2pt] (0,0) -- (0,-\R);

  \fill[white] (0,0) circle (0.065);
  \draw[BoundaryCol, line width=0.5pt] (0,0) circle (0.065);

  \node at (0,1.65) {$H$};
  \node at (-1.28,-1.23) {$D_L$};
  \node at (1.28,-1.23) {$D_R$};

  \node[anchor=north east] at (-2.15,0) {$L$};
  \node[anchor=north west] at (2.15,0) {$R$};
  \node[anchor=north west] at (0.10,-2.18) {$M$};
\end{tikzpicture}
\caption{A cover of $\RR^2$ by three cones.}
\label{fig:three-cone-cover}
\end{figure}

Let $X=\RR^2$, let
$\fU=(D_L,D_R,H)$
be the three-cone cover shown in Fig.~\ref{fig:three-cone-cover}. Let $D=D_L\cup D_R$ and put
$$
\CI:=\CC(H)\cap\CC(D)/\CC(\bdH).
$$
One can show that $\CI$ is a group which naturally embeds into $ {\mathbb Q}^{\times}_{>0}$, the target group of the GNVW index \cite{GNVW}. In particular, $\CI$ is abelian. 

\begin{theoremG}[Proposition~\ref{prop:two-dimensional-homotopy-types}
and Theorem~\ref{thm:2d-anomaly}]
The homotopy type $\chB_{\CE}(\fU\vert X)$ has nontrivial homotopy groups only in degrees $2$ and $4$. The corresponding  Postnikov fibration is 
$$
K(U(1),4)
\longrightarrow
\chB_{\CE}(\fU\vert X)
\longrightarrow
K(\CI,2).
$$
Consequently, the anomaly at $\fU$ of an action
$\rho:G\to\CC(X)$ has a primary obstruction
$o_2(\rho)\in H^2(G;\CI)$.
If this obstruction vanishes, each choice $\nu$ of its trivialization
determines a secondary obstruction 
$o_4(\rho,\nu)\in H^4(G,U(1))$.
These are the obstruction to on-site-ability identified in \cite{KapustinXu,4auth,KawShirley}. The anomaly at $\fU$ is trivial if and only if
$o_2(\rho)=0$ and there is a trivialization $\nu$ for which
$o_4(\rho,\nu)=0$.
\end{theoremG}

The three-cone cover $\fU$ is not initial among covers of $\RR^2$, so
Theorem~G by itself says nothing about the anomaly at finer covers. It
does, however, determine the whole pro-homotopy type.

\begin{theoremH}[Theorem~\ref{thm:planar-protype-constant},
Corollary~\ref{cor:universal-planar-anomaly}]
For $X=\RR^2$ and $\SL=\SCSets_2$, the pro-object
$\chbfB_{\CE}(X)$ is essentially constant, isomorphic in
$\Pro(\Ho(\SSet))$ to the constant pro-object at
$\chB_{\CE}(\fU\vert X)$, and under this isomorphism $\kappa_{\CE}$
is $\kappa_{\CE}(\fU\vert X)$. Consequently, for every action
$\rho\colon G\to\CC(X)$, the anomaly $\alpha_{\CE,\rho}$ is trivial if
and only if its component at $\fU$ is, that is, if and only if
$o_2(\rho)=0$ and $o_4(\rho,\nu)=0$ for some trivialization $\nu$.
\end{theoremH}

The proof exhibits a cofinal family of covers, the planar fans, along
which all refinement maps are homotopy equivalences
(Section~\ref{sec:2d-cofinal}). Together with the corresponding statement in
one dimension, where the cover $(L,R)$ is initial, this is the sense in
which the anomaly invariants of \cite{NayakElse,KapustinXu,4auth,KawShirley}
are exhaustive.

\subsection*{Relation to prior work}

The approach to lattice anomalies through the
restriction of symmetries to regions goes back to Else and Nayak
\cite{NayakElse} who considered symmetries of quantum spin systems in one spatial dimension. It has been generalized to quantum spin systems in two dimensions in \cite{KapustinXu,4auth,KawShirley}. The interpretation of some of the
resulting anomaly indices as Postnikov invariants of a higher group (a crossed
module in one dimension, a crossed square in two) is explained in 
\cite{KapustinXu}, and the proposal that crossed $n$-cubes arise from
covers of space by a nonabelian \v{C}ech construction was made in
\cite{KS}. Relative to these papers, the present one supplies the definition of a crossed precosheaf, the construction of the homotopy type for arbitrary covers, its
functoriality under refinement, a cover-independent definition of the
anomaly, the theorem that anomalies obstruct gauging, and, in one and two
dimensions, the proof that a single cover suffices.

A second line of work, largely independent of ours, studies the homotopy theory
of quantum cellular automata directly and interprets it in terms of anomalies. Czajka, Geiko and Thorngren
\cite{CGT} construct a classifying space of QCA on $\ZZ^d$ as a globular
$\infty$-groupoid whose higher cells are ``blends'' along the coordinate
axes, define the anomaly of a QCA action of $G$ as the homotopy class of
the induced map from $BG$, and obtain, after stabilization by ancillas,
an $\Omega$-spectrum. The globular model is weak, and checking the coherence axioms of the $\infty$-groupoid is deferred to a sequel. Ji and Yang \cite{JiYang} construct a
space $\mathbf Q(X)$ of QCA on a general metric space $X$ as the loop
space of the algebraic $K$-theory of a category of spin systems, and
show that $\mathbf Q(\ZZ^{n-1})\simeq\Omega\mathbf Q(\ZZ^n)$; Jones,
Sopenko and Thorngren \cite{JST} propose a $K$-theoretic
$\Omega$-spectrum of invertible phases and construct nontrivial
invertible QCA on $\ZZ^2$ from conformal nets. Closest to the present
paper is \cite{JiYangLinearization}, which uses $\mathbf Q(X)$ to develop
an obstruction theory for the \emph{linearization} of a homomorphism
$G\to\CQ(X)$, that is, for its conjugacy to an on-site action, and
computes the relevant homotopy types for $X$ a point, a line and a
plane. This is the question addressed by our 
Corollary~\ref{cor:onsite-up-to-conjugacy} and Theorem~G, and we expect
the obstruction classes of \cite{JiYangLinearization} in the plane to
coincide with $o_2$ and $o_4$; we have not verified this.

The approaches differ in their input and their generality. The spaces
of \cite{CGT,JiYang,JiYangLinearization} are built from QCA on $\ZZ^d$ or on a
metric space, with stabilization by ancillas or by $K$-theory built in,
and are infinite loop spaces; this gives anomalies an abelian group
structure and places them in a generalized cohomology theory. The
\v{C}ech-Loday type is built from an arbitrary crossed precosheaf on an
arbitrary distributive lattice of regions, without stabilization, and
is in general not an infinite loop space; in exchange it applies
uniformly to the continuum and the lattice, it is a strict model whose
functoriality under refinement of covers is proved, and it carries the
gauging theorem (Theorem~D). The precise relation between the
\v{C}ech-Loday type of $\CQ$ or $\CE$ for a cover of $\ZZ^d$ by
coordinate half-spaces and the spaces of \cite{CGT,JiYang} is
an interesting question which we do not pursue here.

Our construction of the \v{C}ech-Loday classifying complex  also has a mathematical antecedent in the higher van Kampen theorem of Brown and Loday \cite{BrownLoday}, which
attaches to a space with $n$ subspaces a fundamental crossed $n$-cube
whose vertex groups are fundamental groups of iterated homotopy fibers
and whose pairings are generalized Whitehead products
\cite{EllisSteiner}. The two constructions run in opposite directions:
Brown and Loday extract a crossed cube from the topology of a cover of a
space, whereas here the input is a precosheaf of groups on an abstract
lattice of regions and the output is a homotopy type.

\subsection*{Reading guide}

The rest of the paper has three parts, which can be read largely
independently.

Section~\ref{sec:gaugingnew} discusses symmetries and gauging from an
abstract viewpoint. It motivates the definition of a crossed precosheaf
and the definition of gauging used in Section~\ref{sec:tHooft}, but it
is not needed for the formal development and assumes no homotopy
theory.

Sections~\ref{sec:crossed-cubes}--\ref{sec:cech-loday} contain the
general theory. Section~\ref{sec:crossed-cubes} recalls crossed cubes
and $\cat^n$-groups from \cite{Loday1982,EllisSteiner};
Section~\ref{sec:crossed-precosheaves} defines crossed precosheaves and
gives the continuum examples; Section~\ref{sec:internal-cats} recalls
internal categories and nerves and describes the multinerve of a
crossed cube; Section~\ref{sec:loday} constructs the Loday classifying
complex and proves the fibration theorems;
Section~\ref{sec:cech-loday} assembles the \v{C}ech-Loday functor and
the pro-homotopy type. Long or routine proofs are collected in the
appendices and referenced where they are used.

Sections~\ref{sec:tHooft}--\ref{sec:computations} are the applications.
Section~\ref{sec:tHooft} defines anomalies as homotopy-theoretic
obstructions to gauging; Section~\ref{sec:qspin} constructs three
crossed precosheaves attached to a quantum spin system (quantum
cellular automata, finite-depth circuits, and a central extension of
the latter) and expresses obstructions to being on-site up to
conjugacy in these terms; Section~\ref{sec:computations} computes the
\v{C}ech-Loday types for one- and two-dimensional spin systems,
identifies the anomalies with the Else-Nayak index and its
two-dimensional generalizations, and proves essential constancy in two
dimensions. From the general theory these sections use only the
definitions of a crossed precosheaf and of the \v{C}ech-Loday type,
Theorems~B and~C, and the contractibility criterion of
Corollary~\ref{cor:contractible-protype}; a reader interested mainly
in the applications can take these as black boxes and start at
Section~\ref{sec:tHooft}.

\section{Symmetry and gauging}\label{sec:gaugingnew}

Let $G$ be an abstract group acting by global symmetries on a quantum system occupying a ``space'' $X$. For a QFT on the space-time $X\times\RR$, $X$ is the spatial manifold; for a quantum lattice model, $X$ is a discrete set of sites, such as $\ZZ^d\subset\RR^d$. In either case the action is a homomorphism $\rho\colon G\to F$ from $G$ to the group $F$ of symmetries of a concrete system.\footnote{Without coupling to gauge fields, an extended action is a symmetry only at the kinematic level: it preserves the algebraic relations between observables but need not commute with time evolution. Throughout, $F$ denotes the group of kinematic symmetries.} Typically $G$ is a finite or Lie group, while $F$ may be very large.

Informally, to gauge $G$ is to let the parameters of symmetry transformations depend on position, that is, to extend $\rho$ to a much larger group of \emph{gauge transformations}, in a way that respects locality. In the Introduction, localized symmetries were subgroups of a single group; in general, symmetries localized on $U$ form a group $\CF(U)$, with $\CF(X)=F$, and whenever $U\subseteq V$ there is a homomorphism $c^{\CF}_{UV}\colon\CF(U)\to\CF(V)$, not necessarily injective, with $c^{\CF}_{VW}\circ c^{\CF}_{UV}=c^{\CF}_{UW}$ for $U\subseteq V\subseteq W$. That is, $\CF$ is a \emph{precosheaf of groups} on the poset of regions of $X$: a covariant functor from that poset, viewed as a category, to the category of groups. Abstract gauge transformations likewise form a precosheaf $\CG$, and a gauging of $\rho$ is a morphism of precosheaves $\varrho\colon\CG\to\CF$, a family of homomorphisms $\varrho(U)\colon\CG(U)\to\CF(U)$ commuting with the corestrictions, with $\varrho(X)\circ\iota=\rho$ for a specified embedding $\iota\colon G\to\CG(X)$.\footnote{Later, when we introduce crossed structure on precosheaves, we will also demand $\varrho$ to preserve this structure.} 

The precosheaf $\CG$ is not determined by $G$ alone. One option is to fix $\CG$ and an embedding $\iota:G\to\CG(X)$ on physical grounds. Given such a choice, a morphism of precosheaves $\varrho$ may or may not exist. Obstructions to the existence of $\varrho$ may be called anomalies relative to a choice of $(\CG,\iota)$. If $\varrho$ does exist, it is usually non-unique; these are gauging ambiguities. Another option is to list the properties which $\CG$ must have in order to reflect our intuition about gauging and regard a choice of $\CG$ within this class as part of the gauging ambiguity. Obstructions to the existence of triples $(\CG,\iota,\varrho)$ satisfying all the requirements may be called absolute anomalies.  

If one takes the latter approach, some constraints on $\CG$ are necessary, otherwise gauging is always possible and all obstructions vanish. If one were free to choose $\CG$ after seeing the system, one could take $\CG=\CF$ and $\varrho=\id$. In the absence of any constraints, another universal choice is $\CG(X)=G$, $\CG(U)=1$ for $U\neq X$, and $\varrho(X)=\rho$. To exclude such trivial solutions to the gauging problem, we make the following definition.

\begin{definition}[Gauge precosheaf]\label{def:gauge-precosheaf-concrete}
Let $G$ be a group and let $\SL$ be a set of regions of $X$, partially ordered and closed under finite intersections and unions.\footnote{In the body of the paper, $\SL$ is a poset, the order need not be inclusion of subsets (see the discussion of fuzzy localization below), and intersection and union are replaced by meet and join.} A \emph{gauge precosheaf} for $G$ over $X$ consists of a group $\CG(X)$, a subgroup $\CG(U)\subseteq\CG(X)$ for every region $U\in \SL$, and an embedding $\iota\colon G\hookrightarrow\CG(X)$, such that for all regions $U,V$ we have 
\begin{enumerate}
\item \emph{monotonicity:} $\CG(U)\subseteq\CG(V)$ whenever $U\subseteq  V$, and $\CG(X)$ is the value at the top region;
\item \emph{commutator locality:} $[\CG(U),\CG(V)]\subseteq\CG(U\cap V)$;
\item \emph{meet preservation:} $\CG(U)\cap\CG(V)=\CG(U\cap V)$.
\item \emph{multiplicative partition of unity:} $\CG(U\cup V)=\CG(U)\,\CG(V)$.
\end{enumerate}
\end{definition}

Condition 1 ensures that the collection of subgroups $\CG(U)$ for all $U$ can be regarded as a precosheaf of groups on $\SL$. This precosheaf has a special feature that corestriction maps are injective; such precosheaves are usually called coflasque. Coflasqueness means that ``being localized on $U$'' is a property, not a datum. Condition 2 implies that each $\CG(U)$ is a normal subgroup of $\CG(X)$, so the product in condition 4 is a subgroup. If $\varnothing\in\SL$ and $\CG(\varnothing)=1$, Condition 2 also implies that symmetries localized on disjoint regions commute, another desirable property. 
Condition 3 says that localization regions behave as supports: being localized on $U\cap V$ is exactly the same as being localized on $U$ and $V$. Condition 4 reflects the essence of gauge symmetry; it fails for the degenerate choice $\CG(U)=1$, $U\neq X$. 


Let us describe the two main examples of gauge precosheaves, one in the continuum and one on the lattice. Both work for an \emph{arbitrary} group $G$ carrying no topology; this is the situation in which the theory developed in this paper is unconditional. The case when $G$ is a Lie group with its topology is discussed at the end of the section.

\medskip\noindent\textbf{Piecewise-constant gauge transformations.}

Let $X=\RR^d$. A subset of $\RR^d$ is \emph{semilinear} if it is a finite union of sets
each cut out by finitely many affine equalities and affine inequalities,
strict or non-strict. Let $X=\RR^d$ and let $\SL$ consist of all closed semilinear subsets of $\RR^d$. One can easily see that $\SL$ partially ordered by inclusion has pairwise meets and joins given by intersection and union, respectively, and thus is a distributive lattice.

Call a
function $f\colon\RR^d\to G$ \emph{piecewise constant} if it takes
finitely many values and every level set $f^{-1}(g)$, $g\in G$, is
semilinear. Equivalently, $f$ is constant on each cell of some finite
partition of $\RR^d$ into relatively open semilinear sets, namely the
partition generated by the affine hyperplanes appearing in the
description of the level sets. Note that the cells of such a partition
are not open subsets of $\RR^d$: the walls across which $f$ jumps are
themselves cells, carrying their own values of $f$.

\begin{definition}\label{def:pc-precosheaf}
For a region $U\in \SL$, let
$$
\CGpc(U):=\bigl\{\,f\colon\RR^d\to G\ \text{piecewise
constant}\ \bigm|\ \supp f\subseteq U\,\bigr\},
\qquad
\supp f:=\overline{\{f\neq e\}},
$$
and let $\iota\colon G\to\CGpc(\RR^d)$ send $g$ to the
constant function $g$.
\end{definition}

The support of a piecewise-constant function is a closed semilinear set,
so the condition $\supp f\subseteq U$ is compatible with
the lattice $\SL$.

\begin{proposition}\label{prop:pc-gauge}
$\CGpc$, with corestrictions the inclusions, is a gauge
precosheaf for $G$ over $\RR^d$ in the sense of
Definition~\ref{def:gauge-precosheaf-concrete}.
\end{proposition}

\begin{proof}
Since semilinear sets are closed under finite Boolean operations,
piecewise-constant functions $\RR^d\to G$ form a group under pointwise multiplication.
The relations
$$
\supp(fg)\subseteq\supp f\cup\supp g,\qquad
\supp(f^{-1})=\supp f,\qquad
\supp[f,g]\subseteq\supp f\cap\supp g
$$
imply monotonicity, commutator locality, and meet preservation.

For $f\in\CGpc(U\cup V)$, define
$$
f_U(x)=
\begin{cases}
f(x),&x\in U,\\
e,&x\notin U,
\end{cases}
\qquad
f_V=f_U^{-1}f.
$$
Both functions are piecewise constant, and, since $U$ and $V$ are closed,
$\supp f_U\subseteq U$, $\supp f_V\subseteq V$.
Thus $f=f_Uf_V$ with
$f_U\in\CGpc(U)$ and
$f_V\in\CGpc(V)$, proving the multiplicative
partition-of-unity property.
\end{proof}

The group $\CGpc(\RR^d)$ is generated by the functions equal
to $g\in G$ on a polyhedron $P$ cut out by finitely many affine
inequalities, strict or not, and to $e$ elsewhere. A gauging
$\varrho\colon\CGpc\to\CF$ therefore amounts to a choice,
for every $g$ and every such $P$, of a \emph{truncated} symmetry
operator localized on $P$ and acting as $g$ in its interior, subject
to the multiplicative relations that hold in $\CGpc(X)$. For
$P$ of full dimension this is, in QFT language, an invertible
topological defect of codimension one along $\partial P$; for $P$ of
lower dimension it is a symmetry operator supported on $P$, such as
the ratio $g\cdot1_{\{y\ge0\}}\,(g\cdot1_{\{y>0\}})^{-1}$ of two
truncations differing only on their common wall. Obstructions to
gauging in the sense of this paper are obstructions to such consistent
systems of operators.

\medskip\noindent\textbf{On-site gauge transformations.}
Let now $X$ be a countable discrete metric space of sites, for instance
$\ZZ^d\subset\RR^d$. A \emph{quantum spin system} on $X$ attaches
to each site $x$ a finite-dimensional Hilbert space, to each subset
$U\subseteq X$ the algebra of observables supported on $U$, and localized
symmetries act on observables accordingly; the precise definitions are
given in Section~\ref{sec:qca1}. Here we only need the sites. For a
subset $U\subseteq X$ put
$$
\CGos(U):=\Bigl\{\,(g_x)_{x\in X}\in\textstyle\prod_{x\in X}G
\ \Bigm|\ g_x=e\ \text{for all}\ x\notin U\,\Bigr\},
$$
with $\iota\colon G\to\CGos(X)$ the diagonal. This is a gauge
precosheaf on the lattice of all subsets of $X$, with ordinary
intersection and union: monotonicity is clear; commutators and
intersections of subgroups are computed sitewise, which gives commutator
locality and meet preservation; and a family $g$ supported on $U\cup V$
is the product $g^Ug^V$, where $g^U_x=g_x$ for $x\in U$ and $g^U_x=e$
otherwise, and $g^V:=(g^U)^{-1}g$ is supported on $V$. A gauging
$\varrho\colon\CGos\to\CF$ assigns to every family $(g_x)$
supported on $U$ a symmetry of the spin system localized on $U$,
multiplicatively in the family.

\medskip\noindent\textbf{Fuzzy localization.}
Ordinary inclusion of subsets is, however, not the right notion of
locality for the concrete side. On the lattice, the natural symmetries
of a spin system (quantum cellular automata, Section~\ref{sec:qca1}) may 
displace observables by a bounded distance, so an observable localized on
$U$ is, after composition with such an automorphism, localized only on a
bounded thickening $U^r=\{x:\dist(x,U)\leq r\}$ of $U$. In relativistic
QFT, evolution for a finite time moves observables within the light
cone, so an observable localized in $U$ at one time is localized in a
bounded thickening of $U$ at another. A notion of localized symmetry
that is meaningful independently of such choices must be stable under
bounded thickening. Accordingly, regions should be ordered by
\emph{fuzzy inclusion}, $U\leq V$ iff $U\subseteq V^r$ for some $r\geq0$,
and regarded up to the equivalence $U\sim V$ iff $U\leq V$ and $V\leq U$.
We write $P_X$ for the resulting partially ordered set of fuzzy classes
of nonempty subsets of $X$. All bounded regions collapse to a single bottom
element $\bott$ of $P_X$; the union $U\cup V$ represents the join of the
classes of $U$ and $V$, since $(U\cup V)^r=U^r\cup V^r$; meets, on the
other hand, need not exist in $P_X$ in general, and the body of the
paper works with sublattices $\SL\subseteq P_X$ in which they do, meets
and joins being computed in $P_X$ (Section~\ref{sec:meetsubsemi}). For
$X=\RR^d$ the fuzzy classes of closed semilinear sets form such a
sublattice, identified in Section~\ref{sec:meetsubsemi} with the lattice
$\SCSets_d$ of spherical closed semilinear subsets of the sphere at
infinity $S^{d-1}$.

Passing to fuzzy classes is compatible with gauge precosheaves, by a
general construction.

\begin{definition}[Fuzzification]\label{def:fuzzification}
Let $\CG$ be a precosheaf of subgroups of a group $\CG(X)$ on the power
set of $X$, with corestrictions the inclusions. For $U\subseteq X$ put
$$
\CG^{\mathrm f}(U):=\bigcup_{r\geq0}\CG(U^{r}).
$$
This depends only on the class of $U$ in $P_X$ and is monotone for fuzzy
inclusion, so it defines a precosheaf of subgroups of $\CG(X)$ on $P_X$,
the \emph{fuzzification} of $\CG$.
\end{definition}

\begin{lemma}\label{lem:fuzzification}
Let $\CG$ be as above, with an embedding $\iota\colon G\to\CG(X)$, and
suppose that for all subsets $S,T\subseteq X$
$$
[\CG(S),\CG(T)]\subseteq\CG(S\cap T),\qquad
\CG(S)\cap\CG(T)=\CG(S\cap T).
$$
Let $\SL\subseteq P_X$ be a sublattice of $P_X$ containing the class of $X$, and suppose that for all
$U,V$ representing elements of $\SL$ and every $r\geq0$ there is
$r'\geq r$ with
$$
\CG\bigl((U\cup V)^r\bigr)\subseteq\CG(U^{r'})\,\CG(V^{r'}).
$$
Then $(\CG^{\mathrm f},\iota)$ is a gauge precosheaf for $G$ on $\SL$,
where in Definition~\ref{def:gauge-precosheaf-concrete} intersections
and unions are replaced by meets and joins in $P_X$.
\end{lemma}

\begin{proof}
Monotonicity holds by construction. Let $U,V\in\SL$ and $r,s\geq0$ sufficiently large so that $U^r\cap V^s$ is nonempty. Then $U^r\cap V^s$ is a lower bound of $U$ and $V$ in $P_X$, hence
$U^r\cap V^s\subseteq(U\wedge V)^t$ for some $t\geq0$, and therefore
$$
[\CG(U^r),\CG(V^s)]\subseteq\CG(U^r\cap V^s)\subseteq\CG^{\mathrm f}(U\wedge V),
\qquad
\CG(U^r)\cap\CG(V^s)=\CG(U^r\cap V^s)\subseteq\CG^{\mathrm f}(U\wedge V).
$$
Taking the union over $r,s$ gives commutator locality and
$\CG^{\mathrm f}(U)\cap\CG^{\mathrm f}(V)\subseteq\CG^{\mathrm f}(U\wedge V)$;
the reverse inclusion follows from monotonicity. Finally $U\vee V$ is
represented by $U\cup V$, and the last hypothesis gives
$\CG^{\mathrm f}(U\vee V)\subseteq\CG^{\mathrm f}(U)\,\CG^{\mathrm f}(V)$;
the reverse inclusion again follows from monotonicity.
\end{proof}

The first two hypotheses hold for any precosheaf of the form
$\CG(S)=\{g\in\CG(X):\supp g\subseteq S\}$ with a support
satisfying $\supp(gh)\subseteq\supp g\cup
\supp h$ and
$\supp[g,h]\subseteq\supp g\cap\supp h$,
and both examples above are of this form. The third hypothesis is the
one to check. For
$\CGos$ this is cutting sitewise, and
Lemma~\ref{lem:fuzzification} applies to every sublattice
$\SL\subseteq P_X$. For $\CGpc$, extended to arbitrary
$S\subseteq\RR^d$ by the same formula, we use the $\ell^\infty$-metric on
$\RR^d$; the $r$-thickening of a closed semilinear set is then again
closed semilinear, and the hypothesis holds for $\SL=\SCSets_d$: for
closed semilinear $U,V$ and $f$ supported in $(U\cup V)^r=U^r\cup V^r$,
the functions $f_1:=f\cdot1_{U^r}$ and $f_2:=f_1^{-1}f$ are piecewise
constant, with $\supp f_1\subseteq U^r$ and
$\supp f_2\subseteq\overline{V^r\setminus U^r}\subseteq V^r$.
The choice of metric on $\RR^d$ is immaterial for everything else: any
two norms on $\RR^d$ are equivalent, thickenings with respect to one are
sandwiched between thickenings with respect to the other, and so the
fuzzy preorder, the poset $P_{\RR^d}$, and the lattice $\SCSets_d$
do not depend on it.

\emph{From now on, $\CGos$ and $\CGpc$ denote the
fuzzified precosheaves $\CGos^{\mathrm f}$ and
$\CGpc^{\mathrm f}$, regarded as gauge precosheaves on a
sublattice of $P_X$.} In particular, a family $(g_x)$ counts as
localized on $U$ if it vanishes outside \emph{some} bounded thickening
of $U$; this is the convention used in Sections~\ref{sec:qspin}
and~\ref{sec:computations}.

\medskip\noindent\textbf{Lie groups.}
For this discussion, we return to $\SL={\mathrm Open}(X)$ with the usual partial order. When $G$ is a connected Lie group, piecewise-constant functions are unnatural, and one is led to consider the precosheaf $\CH$ of all continuous maps $f\colon\RR^d\to G$ with closed support in $U$, or the precosheaf $\CHpath$ of \emph{small} gauge transformations (it assigns to $U$ the group of continuous functions $f:\RR^d\to G$ whose closed support is in $U$ and which can be continuously deformed to the identity element within $\CH(U)$). Both satisfy conditions 1-2, but in general fail conditions 3-4. The precosheaf $\CH$ preserves meets but need not satisfy multiplicative partition of unity. Conversely, $\CHpath$ satisfies multiplicative partition of unity (see  Appendix \ref{app:path-epi}), but need not preserve meets. As a consequence, for connected Lie group symmetries of a continuum QFT the present formalism yields a notion of anomaly \emph{relative} to a chosen precosheaf of gauge transformations, rather than an absolute one. We expect that the absolute theory requires coupling to background gauge fields, that is, replacing the precosheaf of groups $\CG$ by a precosheaf of \emph{groupoids}, with connections as objects and gauge transformations as morphisms. This refinement, and the perturbative anomalies it should capture, is beyond the scope of this paper.

\medskip
The examples above concern the ``abstract'' side of gauging. On the
``concrete'' side, the precosheaf $\CF$ of localized symmetries of a
quantum system is typically not coflasque, for the reason explained in
the Introduction, and commutator locality must then be supplied as a
datum. This leads to the notion of a crossed precosheaf, defined in
Section~\ref{sec:crossed-precosheaves}, to which we now turn.

\section{Crossed cubes and categorical groups}\label{sec:crossed-cubes}

To define crossed precosheaves we need a group-theoretic gadget introduced by Ellis and Steiner \cite{EllisSteiner}. We begin with the motivating example.

\begin{example}[Normal-subgroup crossed cubes]\label{ex:normal-subgroup-crossed-cube}
Let $G$ be a group, and let $N_1,\ldots,N_n\trianglelefteq G$ be normal subgroups. For $A\subseteq\langle n\rangle:=\{1,\ldots,n\}$, set
$$
M_A=
\begin{cases}
G, & A=\emptyset,\\[4pt]
\displaystyle\bigcap_{i\in A}N_i, & A\neq\emptyset.
\end{cases}
$$
These groups are arranged as the vertices of an $n$-dimensional cube: for every $i\in A$ there is an inclusion $\mu_i:M_A\hookrightarrow M_{A\setminus\{i\}}$, and the $\mu_i$ commute. Since every $N_i$ is normal,
$$
[M_A,M_B]\subseteq M_{A\cup B},
$$
so the commutator defines a pairing $h_{A,B}:M_A\times M_B\to M_{A\cup B}$, $h_{A,B}(a,b)=[a,b]$. If $A\subseteq B$, then $M_A$ acts on $M_B$ by automorphisms via $a:b\mapsto {}^ab:=h_{A,B}(a,b)b$. 
The pairings satisfy the usual commutator identities in $G$: skew-symmetry, the identities $[aa',b]={}^a[a',b]\,[a,b]$ and $[a,bb']=[a,b]\,{}^b[a,b']$ expressing multiplicativity up to conjugation, and the Hall-Witt identity:
\begin{equation}
{}^{a}
\!\left[\left[a^{-1},b\right],c\right]
\cdot
{}^{c}
\!\left[\left[c^{-1},a\right],b\right]
\cdot
{}^{b}
\!\left[\left[b^{-1},c\right],a\right]
=1.
\end{equation}
\end{example}

Many examples arising from physics produce cubes of normal subgroups of a common transformation group, with structure maps given by inclusions and pairings given by ordinary commutators. Example 1 is the special case in which the higher vertices are intersections. However, the precosheaves relevant for anomalies (Section~\ref{sec:E}) produce cubical diagrams of groups with non-injective structure maps. Commutators in the original transformation group do not determine their lifts to the vertex groups. These lifts must instead be supplied as additional data. A crossed $n$-cube is what remains of Example~\ref{ex:normal-subgroup-crossed-cube} in that situation: the inclusions $\mu_i$ are replaced by arbitrary homomorphisms, the commutators by pairings $h_{A,B}$ which are part of the data, and the commutator identities of $G$ by axioms. The following definition is due to Ellis and Steiner \cite{EllisSteiner}.

\begin{definition}[Crossed $n$-cube of groups \cite{EllisSteiner}]
 \label{def:crossed-n-cube}
A \emph{crossed $n$-cube of groups} consists of the following input
data.

\begin{itemize}
\item \textbf{Vertex groups.}
For every subset $A\subseteq\langle n\rangle$, a group $M_A$.

\item \textbf{Structure maps.}
For every $i\in\langle n\rangle$ and
$A\subseteq\langle n\rangle$, a group homomorphism
$$
\mu_i:M_A\longrightarrow M_{A\setminus\{i\}},
$$
called the \emph{structure map in the $i$-th direction}. It is
required as part of the input that
$$
\mu_i=\id_{M_A}
\qquad\text{whenever }i\notin A.
$$

\item \textbf{Pairings.}
For every pair of subsets $A,B\subseteq\langle n\rangle$, a function
$$
h_{A,B}:M_A\times M_B\longrightarrow M_{A\cup B},
$$
called the \emph{commutator pairing associated with $A$ and $B$}.
The functions $h_{A,B}$ are not required to be group
homomorphisms.
\end{itemize}

Whenever $A\subseteq B$, let 
$$
{}^a b:=h_{A,B}(a,b)b,
\qquad
a\in M_A,\quad b\in M_B.
$$
The axioms below imply that this defines an action of $M_A$ on $M_B$ by automorphisms.

The following conditions are required to hold for all
$A,B,C\subseteq\langle n\rangle$, $a,a'\in M_A$, $b,b'\in M_B$, $c\in M_C$,
and all $i,j\in\langle n\rangle$.

\begin{enumerate}
\item \textbf{Commutativity of the cube.}
$$
\mu_i\mu_j=\mu_j\mu_i.
$$

\item \textbf{Compatibility of the pairings with the structure maps.}
$$
\mu_i\bigl(h_{A,B}(a,b)\bigr)
=
h_{A\setminus\{i\},\,B\setminus\{i\}}
\bigl(\mu_i(a),\mu_i(b)\bigr).
$$

\item \textbf{Independence of a common coordinate.}
If $i\in A\cap B$
$$
h_{A,B}(a,b)
=
h_{A\setminus\{i\},B}\bigl(\mu_i(a),b\bigr)
=
h_{A,B\setminus\{i\}}\bigl(a,\mu_i(b)\bigr).
$$

\item \textbf{Diagonal commutator identity.}
$$
h_{A,A}(a,a')=[a,a'].
$$

\item \textbf{Skew-symmetry.}
$$
h_{A,B}(a,b)=h_{B,A}(b,a)^{-1}.
$$

\item \textbf{Normalization.}
$$
h_{A,B}(1,b)=1,
\qquad
h_{A,B}(a,1)=1.
$$

\item \textbf{Crossed multiplicativity.}
$$
h_{A,B}(aa',b)
=
{}^a h_{A,B}(a',b)\,
h_{A,B}(a,b)
$$
and
$$
h_{A,B}(a,bb')
=
h_{A,B}(a,b)\,
{}^b h_{A,B}(a,b').
$$

\item \textbf{Hall-Witt identity.}
$$
\begin{aligned}
&{}^{a}
h_{A\cup B,C}\!\left(h_{A,B}(a^{-1},b),c\right)
\\
&\qquad\cdot
{}^{c}
h_{C\cup A,B}\!\left(h_{C,A}(c^{-1},a),b\right)
\\
&\qquad\cdot
{}^{b}
h_{B\cup C,A}\!\left(h_{B,C}(b^{-1},c),a\right)
=1.
\end{aligned}
$$

\item \textbf{Equivariance of the pairings.}
Whenever $A\subseteq B$ and $A\subseteq C$
$$
{}^a h_{B,C}(b,c)
=
h_{B,C}({}^a b,{}^a c).
$$
\end{enumerate}
\end{definition}

\begin{definition}[Morphism of crossed $n$-cubes]
\label{def:crossed-n-cube-morphism}
Let $M_\bullet$ and $N_\bullet$ be crossed $n$-cubes. A
\emph{morphism of crossed $n$-cubes}
$f_\bullet:M_\bullet\to N_\bullet$
is a family of group homomorphisms
$f_A:M_A\to N_A$, $A\subseteq\langle n\rangle$,
satisfying the following two conditions.

\begin{enumerate}
\item \textbf{Compatibility with the structure maps.}
For every $A\subseteq\langle n\rangle$ and
$i\in\langle n\rangle$,
$$
f_{A\setminus\{i\}}\mu_i^M
=
\mu_i^N f_A.
$$

\item \textbf{Compatibility with the commutator pairings.}
For every $A,B\subseteq\langle n\rangle$,
$a\in M_A$, and $b\in M_B$,
$$
f_{A\cup B}\bigl(h^M_{A,B}(a,b)\bigr)
=
h^N_{A,B}\bigl(f_A(a),f_B(b)\bigr).
$$
\end{enumerate}

We denote the category of crossed $n$-cubes and their morphisms by
$\XCube_n$.
\end{definition}

The axioms fall into three groups. Axioms 1--3 concern the structure maps: they form a commutative cube, the pairings are compatible with them, and the pairing of $a$ with $b$ does not see a coordinate common to $A$ and $B$; in Example~\ref{ex:normal-subgroup-crossed-cube} all three hold trivially, since the $\mu_i$ are inclusions. Axioms 4--7 say that $h_{A,B}$ behaves like a commutator: it is the commutator on the diagonal, is skew-symmetric, is normalized, and is multiplicative up to the action. Axioms 8 and 9 are the Hall-Witt identity and the equivariance of the pairings under the action, which for honest commutators are the identities $[{}^ab,{}^ac]={}^a[b,c]$. In Example~\ref{ex:normal-subgroup-crossed-cube} every axiom is an identity in $G$, so that example is indeed a crossed $n$-cube.

\begin{remark}\label{rem:crossed-modules-and-squares}
For $n=0$, a crossed $0$-cube is just a group $M_{\emptyset}$.

For $n=1$, put
$P:=M_{\emptyset}$, $M:=M_{\{1\}}$, $\partial:=\mu_1:M\to P$.
The pairing $h_{\emptyset,\{1\}}$ defines an action of $P$ on $M$ by
$$
{}^p m
:=
h_{\emptyset,\{1\}}(p,m)\,m.
$$
With this action, $\partial:M\to P$ is a crossed module. Compatibility of
the pairings with the structure map, Axiom~2, gives
$$
\partial({}^p m)
=
p\,\partial(m)\,p^{-1}.
$$
The Peiffer identity
$$
{}^{\partial(m)}m'
=
m\,m'\,m^{-1}
$$
is Axiom~3, independence of a common coordinate, applied with
$A=B=\{1\}$, together with the diagonal commutator identity
$h_{\{1\},\{1\}}(m,m')=[m,m']$. Thus a crossed $1$-cube is precisely a
crossed module.

For $n=2$, write
$P:=M_{\emptyset}$, $M:=M_{\{1\}}$, $N:=M_{\{2\}}$, $L:=M_{\{1,2\}}$.
The structure maps form a commutative square
$$
\begin{tikzcd}[column sep=large,row sep=large]
L
  \arrow[r,"\lambda'=\mu_1"]
  \arrow[d,"\lambda=\mu_2"']
&
N
  \arrow[d,"\nu=\mu_2"]
\\
M
  \arrow[r,"\mu=\mu_1"']
&
P .
\end{tikzcd}
$$
Its commutativity is Axiom~1. Each of the four sides is a crossed module,
because every one-dimensional face of a crossed $2$-cube is a crossed
$1$-cube. More explicitly, $\mu:M\to P$ and $\nu:N\to P$ are crossed
modules for the actions of $P$ determined by
$h_{\emptyset,\{1\}}$ and $h_{\emptyset,\{2\}}$, while
$\lambda:L\to M$ and $\lambda':L\to N$ are crossed modules for the
actions of $M$ and $N$ determined by
$h_{\{1\},\{1,2\}}$ and $h_{\{2\},\{1,2\}}$. The group $P$ also acts on
$L$, $M$, and $N$ through the pairings $h_{\emptyset,A}$, and the
structure maps are compatible with these actions.

Most of the $h$-pairings merely encode these actions and the ordinary
commutators. Indeed, whenever $A\subseteq B$,
$$
h_{A,B}(a,b)
=
{}^a b\,b^{-1},
$$
the pairing with the arguments reversed is determined by skew-symmetry,
and the diagonal pairing is the ordinary commutator. The only additional
pairing between incomparable vertices is
$h:=h_{\{1\},\{2\}}: M\times N\to L$.
Its opposite is given by
$h_{\{2\},\{1\}}(n,m)=h(m,n)^{-1}$,
and compatibility with the structure maps gives
$$
\lambda'\bigl(h(m,n)\bigr)
=
{}^{\mu(m)}n\,n^{-1},
\qquad
\lambda\bigl(h(m,n)\bigr)
=
m\,{}^{\nu(n)}m^{-1}.
$$
In particular, the common image of $h(m,n)$ in $P$ is
$[\mu(m),\nu(n)]$.
Thus $h$ is a specified lift to $L$ of the commutator of the images of
$m$ and $n$ in $P$. The remaining crossed-cube axioms become the usual
crossed-square axioms. Hence a crossed $2$-cube is precisely a crossed
square \cite{Loday1982}.
\end{remark}

The notion of a crossed $n$-cube of groups is a reformulation of the notion of a $\cat^n$-group introduced by Loday \cite{Loday1982}. 

\begin{definition}[$\cat^n$-group \cite{Loday1982}]
Let $n\geq 0$. For $n=0$, a $\cat^0$-group is just a
group.

For $n\geq 1$, a $\cat^n$-group is a group $G$ equipped
with endomorphisms
$s_i,t_i:G\to G$, $i=1,\ldots,n$,
called the source and target maps in the $i$-th categorical direction,
satisfying the following identities.

For every $i$,
$$
t_i s_i=s_i,
\qquad
s_i t_i=t_i,
\qquad
[\ker s_i,\ker t_i]=1.
$$
For $i\neq j$,
$$
s_i s_j=s_j s_i,
\qquad
t_i t_j=t_j t_i,
\qquad
s_i t_j=t_j s_i.
$$

A morphism of $\cat^n$-groups
$f:(G,s_i,t_i)\to (G',s'_i,t'_i)$
is a group homomorphism $f:G\longrightarrow G'$ such that
$f s_i=s'_i f$, $f t_i=t'_i f$
for every $i=1,\ldots,n$. The category of $\cat^n$-groups will be denoted $\cat^n\Grp$. 
\end{definition}

\begin{remark}
The relations
$t_i s_i=s_i$, $s_i t_i=t_i$
imply that both $s_i$ and $t_i$ are idempotent and that
$\im\, s_i=\im\, t_i$.
\end{remark}

\begin{remark}
The maps $s_i$ and $t_i$ are to be thought of as source and target maps: a $\cat^1$-group is the same thing as an internal category in the category of groups, with object group $\im\, s=\im\, t$, arrow group $G$, and composition forced by $s$ and $t$, and a $\cat^n$-group is an $n$-fold internal category in groups. This point of view is developed in Section~\ref{sec:internal-cats}, where it is used to define the nerve of a $\cat^n$-group.
\end{remark}

\begin{definition}[The crossed $n$-cube associated to a $\cat^n$-group]\label{def:catn-to-xcube}
Let
$(G,s_1,t_1,\ldots,s_n,t_n)$
be a $\cat^n$-group. For $A\subseteq\{1,\ldots,n\},$ define
$$
M_A
:=
\left(\bigcap_{i\in A}\ker s_i\right)
\cap
\left(\bigcap_{i\notin A}\im\,s_i\right).
$$
Equivalently, an element $x\in G$ lies in $M_A$ precisely when
$$
s_i(x)=1
\quad\text{for }i\in A,
\qquad
s_i(x)=x
\quad\text{for }i\notin A.
$$

For $i\in\{1,\ldots,n\}$, define $\mu_i:M_A\to M_{A\setminus\{i\}}$ to be the restriction of $t_i$.
If $i\notin A$, then $x\in M_A\subseteq\im\,s_i$, and hence
$t_i(x)=x,$
so in this case $\mu_i$ is the identity map on $M_A$.

For $A,B\subseteq\{1,\ldots,n\}$, define the pairing $h_{A,B}:M_A\times M_B\to M_{A\cup B}$ by $h_{A,B}(a,b):=[a,b]$, using the commutator in $G$. Then $M_\bullet=(M_A,\mu_i,h_{A,B})$ is the crossed $n$-cube associated to $G$.
\end{definition}

\begin{proposition}[\cite{EllisSteiner}]\label{prop:catn-to-xcube}
The construction of Definition~\ref{def:catn-to-xcube}, with morphisms
sent to their restrictions on the vertex subgroups, defines a functor
$$
\cat^n\Grp
\longrightarrow
\XCube_n.
$$
\end{proposition}

\begin{definition}[Recovering the $\cat^n$-group from a crossed $n$-cube]\label{def:catrecover}
Let
$M_\bullet=(M_A,\mu_i,h_{A,B})$, $A,B\subseteq \{1,\ldots,n\}$,
be a crossed $n$-cube of groups. We write $\langle n\rangle:=\{1,\ldots,n\}.$
The associated $\cat^n$-group $\Gamma_n(M_\bullet)$ is constructed as follows.

First, let
$$
\Gamma_n(M_\bullet)
:=
\left(
\underset{A\subseteq\langle n\rangle}{\ast} M_A
\right)
\Big/
\left\langle\!\left\langle
[j_A(a),j_B(b)]
j_{A\cup B}\bigl(h_{A,B}(a,b)\bigr)^{-1}
\right\rangle\!\right\rangle ,
$$
where $j_A:M_A\to \underset{A}{\ast}\, M_A$ denotes the free-product inclusion. Let $\iota_A=q\circ j_A$ where $q:\underset{A}{\ast}\, M_A\to \Gamma_n(M_\bullet)$ is the canonical projection.

For each $i\in\langle n\rangle$, define endomorphisms
$s_i,t_i:\Gamma_n(M_\bullet)\to \Gamma_n(M_\bullet)$
on generators by
$$
s_i(\iota_A(a))
=
\begin{cases}
1, & i\in A,\\[4pt]
\iota_A(a), & i\notin A,
\end{cases}
$$
and
$$
t_i(\iota_A(a))
=
\begin{cases}
\iota_{A\setminus\{i\}}\bigl(\mu_i(a)\bigr), & i\in A,\\[4pt]
\iota_A(a), & i\notin A.
\end{cases}
$$
The Ellis-Steiner crossed $n$-cube identities imply that these formulas
respect the defining relations of $\Gamma_n(M_\bullet)$. Hence they extend
to well-defined endomorphisms of $\Gamma_n(M_\bullet)$.

The group $\Gamma_n(M_\bullet)$ together with the endomorphisms $s_i,t_i$, $1\leq i\leq n$, is the
$\cat^n$-group associated to the crossed $n$-cube $M_\bullet$.
\end{definition}

\begin{proposition}[\cite{EllisSteiner}]\label{prop:catrecover}
The endomorphisms $s_i,t_i$ defined in Definition~\ref{def:catrecover}
make $\Gamma_n(M_\bullet)$ into a $\cat^n$-group.
The assignment $M_\bullet\mapsto \Gamma_n(M_\bullet)$ extends to a functor
$$
\Gamma_n:\XCube_n\longrightarrow \cat^n\Grp.
$$
For a morphism $f_\bullet:M_\bullet\to N_\bullet$ of crossed $n$-cubes,
this functor is given on generators by
$$
\Gamma_n(f_\bullet)(\iota_A^M(a))
=
\iota_A^N(f_A(a)),
\qquad a\in M_A,
$$
where $\iota_A^M$ and $\iota_A^N$ denote the canonical maps into
$\Gamma_n(M_\bullet)$ and $\Gamma_n(N_\bullet)$, respectively. Together
with the functor of Proposition~\ref{prop:catn-to-xcube}, $\Gamma_n$ gives
an equivalence between crossed $n$-cubes and $\cat^n$-groups.
\end{proposition}


\begin{theorem}[Ellis-Steiner normal form \cite{EllisSteiner}]\label{thm:normalform}
Let $M_\bullet=(M_A,\mu_i,h_{A,B})$ be a crossed $n$-cube of groups, and
let $\Gamma_n(M_\bullet)$ be the $\cat^n$-group of
Definition~\ref{def:catrecover}. Fix any total ordering of the set of
subsets of $\{1,\ldots,n\}$. Then the multiplication map
$$
\prod_{A\subseteq\{1,\ldots,n\}} M_A
\longrightarrow
\Gamma_n(M_\bullet),
\qquad
(m_A)_A\longmapsto \prod_A \iota_A(m_A),
$$
where the product is taken in the chosen total order, is a bijection of
sets.

In particular, every element of $\Gamma_n(M_\bullet)$ has a unique normal
form
$\prod_A \iota_A(m_A)$, $m_A\in M_A$,
with the product taken in the chosen order.
\end{theorem}

It is often convenient to label the directions by an arbitrary finite
set $I$ rather than by $\{1,\ldots,n\}$. We then speak of crossed
$I$-cubes and $\cat^I$-groups, write $\Gamma_I$ for the Ellis-Steiner
functor with this labeling, and put $\Gamma_\varnothing(M_\bullet)=M_\varnothing$.
For $B\subseteq A\subseteq I$ we write
$\mu_{A,B}:M_A\to M_B$
for the composite of the structure maps $\mu_i$, $i\in A\setminus B$;
by Axiom~1 the order in which they are applied is immaterial.

In particular $\mu_{A,A}=\id$ and $\mu_{B,C}\,\mu_{A,B}=\mu_{A,C}$ for
$C\subseteq B\subseteq A$. Axioms~2 and~3 combine into a single
compatibility of the pairings with these composites.

\begin{lemma}\label{lem:pairing-restriction}
Let $M_\bullet$ be a crossed $I$-cube, and let $A'\subseteq A\subseteq I$
and $B'\subseteq B\subseteq I$. Then for all $a\in M_A$ and $b\in M_B$,
\begin{equation}\label{eq:pairing-restriction}
\mu_{A\cup B,\,A'\cup B'}\,h_{A,B}(a,b)
=h_{A',B'}\bigl(\mu_{A,A'}a,\;\mu_{B,B'}b\bigr).
\end{equation}
\end{lemma}

\begin{proof}
We induct on $|A\setminus A'|+|B\setminus B'|$. If this number is zero,
there is nothing to prove. Otherwise pick $i\in A\setminus A'$; the case
$i\in B\setminus B'$ is symmetric.

If $i\notin B'$, then $A'\subseteq A\setminus\{i\}$,
$B'\subseteq B\setminus\{i\}$, and
$i\in(A\cup B)\setminus(A'\cup B')$, so that
$\mu_{A\cup B,A'\cup B'}=\mu_{(A\cup B)\setminus\{i\},\,A'\cup B'}\,\mu_i$.
By Axiom~2 and the induction hypothesis for the pair
$(A\setminus\{i\},B\setminus\{i\})$,
\begin{align*}
\mu_{A\cup B,A'\cup B'}\,h_{A,B}(a,b)
&=\mu_{(A\cup B)\setminus\{i\},\,A'\cup B'}\,
h_{A\setminus\{i\},B\setminus\{i\}}(\mu_ia,\mu_ib)\\
&=h_{A',B'}\bigl(\mu_{A\setminus\{i\},A'}\,\mu_ia,\;
\mu_{B\setminus\{i\},B'}\,\mu_ib\bigr),
\end{align*}
and $\mu_{A\setminus\{i\},A'}\,\mu_i=\mu_{A,A'}$ on $M_A$,
$\mu_{B\setminus\{i\},B'}\,\mu_i=\mu_{B,B'}$ on $M_B$ (recall that
$\mu_i$ is the identity of $M_B$ if $i\notin B$).

If $i\in B'$, then $i\in A\cap B$ and
$(A\setminus\{i\})\cup B=A\cup B$. By Axiom~3 and the induction
hypothesis for the pair $(A\setminus\{i\},B)$,
\begin{align*}
\mu_{A\cup B,A'\cup B'}\,h_{A,B}(a,b)
&=\mu_{A\cup B,A'\cup B'}\,h_{A\setminus\{i\},B}(\mu_ia,b)\\
&=h_{A',B'}\bigl(\mu_{A\setminus\{i\},A'}\,\mu_ia,\;\mu_{B,B'}b\bigr)
=h_{A',B'}\bigl(\mu_{A,A'}a,\;\mu_{B,B'}b\bigr).
\qedhere
\end{align*}
\end{proof}

It is sometimes useful to single out a single direction $i\in I$. The two
\emph{faces} of $M_\bullet$ in direction $i$ are the crossed
$(I\setminus\{i\})$-cubes $\partial^0_iM_\bullet$ and
$\partial^1_iM_\bullet$ with vertex groups
$$
(\partial^0_iM)_A=M_A,
\qquad
(\partial^1_iM)_A=M_{A\cup\{i\}},
\qquad
A\subseteq I\setminus\{i\},
$$
and the inherited structure maps and pairings; the structure maps in
direction $i$ form a morphism of crossed $(I\setminus\{i\})$-cubes
$\mu_i:\partial^1_iM_\bullet\to\partial^0_iM_\bullet$. Applying the Ellis-Steiner functor $\Gamma_{I\backslash\{i\}}$ to these faces and $\mu_i$ we get a morphism of $\cat^{I\backslash\{i\}}$-groups. The following lemma describes this morphism in terms of $\Gamma_I(M_\bullet)$.

\begin{lemma}\label{lem:faces}
Let $M_\bullet$ be a crossed $I$-cube and $i\in I$. In
$\Gamma_I(M_\bullet)$, the subgroups $\im s_i$ and
$\ker s_i$ are preserved by $s_j$ and $t_j$ for $j\neq i$, and the
assignments
$$
\iota_A(m)\longmapsto\iota_A(m),
\qquad
\iota_A(m)\longmapsto\iota_{A\cup\{i\}}(m),
\qquad
A\subseteq I\setminus\{i\},
$$
define isomorphisms of $\cat^{I\setminus\{i\}}$-groups, natural in
$M_\bullet$,
$$
\Gamma_{I\setminus\{i\}}(\partial^0_iM_\bullet)\xrightarrow{\ \cong\ }\im s_i,
\qquad
\Gamma_{I\setminus\{i\}}(\partial^1_iM_\bullet)\xrightarrow{\ \cong\ }\ker s_i,
$$
under which the restricted target map
$t_i|_{\ker s_i}:\ker s_i\to\im s_i$ becomes
$\Gamma_{I\setminus\{i\}}(\mu_i)$.
\end{lemma}

\begin{proof}
Since $s_i$ is a homomorphism fixing $\iota_A(m)$ for $i\notin A$ and
killing it for $i\in A$, it sends a normal form
$\prod_A\iota_A(m_A)$ to $\prod_{A\not\ni i}\iota_A(m_A)$. By the
uniqueness of normal forms, $\im s_i$ consists of the
elements whose normal form has $m_A=1$ for all $A\ni i$, and $\ker s_i$
of those with $m_A=1$ for all $A\not\ni i$; these are the subgroups
generated by the $\iota_A(M_A)$ with $i\notin A$, respectively with
$i\in A$. The operators $s_j,t_j$, $j\neq i$, send generators of each
kind to generators of the same kind, so both subgroups are
$\cat^{I\setminus\{i\}}$-groups. The generators of
$\im s_i$, respectively $\ker s_i$, are copies of the vertex
groups of $\partial^0_iM_\bullet$, respectively $\partial^1_iM_\bullet$,
and they satisfy the defining relations of
$\Gamma_{I\setminus\{i\}}$ of that face, because the pairings of the face
are the pairings of $M_\bullet$ restricted to the relevant subsets. The
displayed assignments therefore extend to surjective homomorphisms
compatible with $s_j,t_j$, $j\neq i$, and they are injective because
they match the normal forms of Theorem~\ref{thm:normalform} on both
sides. Finally
$t_i\bigl(\iota_{A\cup\{i\}}(m)\bigr)=\iota_A\bigl(\mu_i(m)\bigr)$ by
Definition~\ref{def:catrecover}, which is the last claim. Naturality
is clear.
\end{proof}

\begin{remark}[The abelian case]\label{rem:abelian-gamma}
Suppose that all vertex groups $M_A$ are abelian and all pairings are
trivial, $h_{A,B}\equiv1$. Then Axioms 2--9 of
Definition~\ref{def:crossed-n-cube} hold automatically, and a crossed
$n$-cube of this kind is just a commuting $n$-cube of abelian groups. The
defining relations of $\Gamma_n(M_\bullet)$ say that the images of the
generators commute, so
$$
\Gamma_n(M_\bullet)=\bigoplus_{A\subseteq\langle n\rangle}M_A ,
$$
$s_i$ is the projection killing the summands with $i\in A$, and $t_i$ is
$\mu_i$ on those summands and the identity on the others. The normal form
of Theorem~\ref{thm:normalform} is the direct-sum decomposition. In
general the theorem says that $\Gamma_n(M_\bullet)$ still ``looks like''
this direct sum once an order of the factors is fixed, the failure of the
factors to commute being recorded by the pairings. We return to this
picture in Section~\ref{sec:grid}.
\end{remark}

\section{Crossed precosheaves}\label{sec:crossed-precosheaves}
\subsection{Definition of a crossed precosheaf}
We are going to define a {\it crossed precosheaf} over a ``space'' as a precosheaf of groups with additional structure. Crossed precosheaves are gadgets which axiomatize the notion of a symmetry localized on a region of ``space''. Examples arise from QFT and quantum lattice models. For most of the formal theory, a ``space'' only needs to be a meet-semilattice $\SL$: a poset with finite nonempty meets. An example to keep in mind is the poset $Open(X)$ of open subsets of a topological space $X$. For quantum lattice systems on $\RR^d$, $\SL$ can be taken to be the poset of closed spherical semilinear subsets of $S^{d-1}$ (i.e. the poset whose elements are finite unions of closed spherical polyhedra in $S^{d-1}$), ordered by inclusion \cite{AKY}.
\begin{remark}
    Finite joins are not needed in the definition of a crossed precosheaf. They enter later, when one wants to speak about finite covers and refinements of covers. 
\end{remark}

\begin{definition}[Meet-semilattice]
A \emph{meet-semilattice} is a partially ordered set $\SL$ in which
every pair $x,y\in \SL$ has a greatest lower bound, denoted
$x\wedge y$. Thus
$x\wedge y\leq x$, $x\wedge y\leq y$,
and if $z\leq x$ and $z\leq y$, then
$z\leq x\wedge y$.
Equivalently, $\SL$ has all finite nonempty meets.
\end{definition}

\begin{definition}[Join-semilattice]
A \emph{join-semilattice} is a partially ordered set $\SL$ in which
every pair $x,y\in \SL$ has a least upper bound, denoted $x\vee y$.
Thus
$x\leq x\vee y$, $y\leq x\vee y$,
and if $x\leq z$ and $y\leq z$, then
$x\vee y\leq z$.
Equivalently, $\SL$ has all finite nonempty joins.
\end{definition}

Roughly speaking, a crossed structure on a precosheaf of groups $\CF$ attaches to every $W\in \SL$ and every finite family $U_1,\ldots,U_n\in \SL$ with $U_i\leq W$ a crossed $n$-cube with vertex groups $\CF(W)$ and $\CF(U_{i_1}\wedge\cdots\wedge U_{i_k})$, compatibly with shrinking the $U_i$ and with relabeling, repeating, and deleting members of the family. Such families are the objects of the following definition.

\begin{definition}[\v{C}ech configurations]
\label{def:Cech-config-presheaf}
Let $\SL$ be a meet-semilattice, and let $I$ be a finite set. An $I$-indexed \emph{\v{C}ech configuration} in $\SL$
is a pair
$\xi=(\fU\vert W)$, $\fU=(U_i)_{i\in I}$,
where $U_i,W\in \SL$ and
$U_i\leq W$ for every $i\in I$.
Thus $W$ is a common upper bound for the indexed family $\fU$.
Let $\Cech_I(\SL)$ denote the poset of $I$-indexed \v{C}ech
configurations in $\SL$.

The order is given by $(\fU\vert W)\leq(\fU'\vert W')$ if and only if $U_i\leq U_i'$ for every $i\in I$ and $W\leq W'$. For $I=\varnothing$, configurations are simply elements of $\SL$, so $\Cech_\varnothing(\SL)=\SL$.

For a function $r:J\to I$ of finite sets, define the pullback, or
reindexing, of $(\fU\vert W)$ along $r$ by
$$
r^*(\fU\vert W)
:=
\bigl((U_{r(j)})_{j\in J}\vert W\bigr).
$$
Thus pulling back along an injection deletes members of the family, pulling back along a surjection repeats members, and pulling back along a bijection relabels them. If $\xi\le\xi'$ in $\Cech_I(\SL)$ then $r^*\xi\le r^*\xi'$ in $\Cech_J(\SL)$, since $U_{r(j)}\le U'_{r(j)}$ for every $j\in J$. Moreover $(r\circ q)^*=q^*r^*$ for composable $q:K\to J$, $r:J\to I$, and $\id_I^*=\id$.
If $\SL$ also has finite joins and $I\neq\varnothing$, the conditions $U_i\leq W$ for all
$i\in I$ are equivalent to
$\bigvee_{i\in I}U_i\leq W$.
\end{definition}

We can now give the main definition. It is phrased directly in terms of a precosheaf of groups together with commutator pairings; a more concise, but equivalent, formulation is recorded in Appendix~\ref{app:categorical}.

\begin{definition}[Crossed precosheaf]
\label{def:crossed-precosheaf-structure}
Let $\SL$ be a meet-semilattice. A \emph{crossed precosheaf}
$\CF$ consists of a precosheaf
$\CF_{\varnothing}:\SL\to\Grp$,
called the \emph{underlying precosheaf} of $\CF$, together with the
data described below. The subscript $\varnothing$ refers to the empty family: $\CF_\varnothing(W)$ will be the crossed $0$-cube attached to $(\,\vert W)$.

For every finite set $I$, every \v{C}ech configuration
$\xi=((U_i)_{i\in I}\vert W)\in\Cech_I(\SL)$,
and every $A,B\subseteq I$, the crossed precosheaf includes a function
$$
h^\xi_{A,B}:
\CF_{\varnothing}(U_A^\xi)
\times
\CF_{\varnothing}(U_B^\xi)
\longrightarrow
\CF_{\varnothing}(U_{A\cup B}^\xi),
$$
where
$$
U_A^\xi=
\begin{cases}
W, & A=\varnothing,\\[2mm]
\displaystyle\bigwedge_{i\in A}U_i, & A\neq\varnothing.
\end{cases}
$$

These data are required to satisfy the following conditions.

\begin{enumerate}
\item
\textbf{Crossed-cube condition.}
For every $I$-indexed \v{C}ech
configuration $\xi$, define the vertex groups by
$$
M_A=\CF_{\varnothing}(U_A^\xi),
\qquad A\subseteq I.
$$
For every $i\in I$ and $A\subseteq I$, the structure map
$\mu_i:M_A\to M_{A\setminus\{i\}}$
is the corestriction map of the precosheaf $\CF_{\varnothing}$
associated with the relation
$U_A^\xi\leq U_{A\setminus\{i\}}^\xi$.
 The vertex groups and structure maps,
together with the pairings $h^\xi_{A,B}$ form a crossed $I$-cube, as defined in \ref{def:crossed-n-cube}, which we denote by $\Cech_{\CF}(\xi)$.

\item
\textbf{Fixed-fiber functoriality.}
Suppose
$\xi=((V_i)_{i\in I}\vert W) \leq \xi'=((U_i)_{i\in I}\vert W')$.
For every $A\subseteq I$, let
$\rho_A: \CF_{\varnothing}(U_A^\xi) \to \CF_{\varnothing}(U_A^{\xi'})$
be the corresponding corestriction map.

The family
$(\rho_A)_{A\subseteq I}$
must be a morphism of crossed $I$-cubes. By
Definition~\ref{def:crossed-n-cube-morphism}, a morphism of crossed
$I$-cubes must be compatible with both the structure maps and the
pairings. Here the maps $\rho_A$ and all the structure maps are
corestriction maps of the precosheaf $\CF_{\varnothing}$, so
compatibility with the structure maps follows automatically from the
precosheaf property. Consequently, the family
$(\rho_A)_{A\subseteq I}$ is a morphism of crossed $I$-cubes if
and only if, for every $A,B\subseteq I$,
$x\in\CF_{\varnothing}(U_A^\xi)$, and
$y\in\CF_{\varnothing}(U_B^\xi)$,
$$
\rho_{A\cup B}\bigl(h^\xi_{A,B}(x,y)\bigr)
=
h^{\xi'}_{A,B}\bigl(\rho_A(x),\rho_B(y)\bigr).
$$

\item
\textbf{Reindexing compatibility.}
For every function $r:J\to I$ of finite sets, every $\xi\in\Cech_I(\SL)$, and every $A,B\subseteq J$,
$$
h^{r^*\xi}_{A,B}
=
h^\xi_{r(A),r(B)}.
$$
This makes sense because $U^{r^*\xi}_A=U^\xi_{r(A)}$ for every $A\subseteq J$, so that both sides are functions $\CF_\varnothing(U^\xi_{r(A)})\times\CF_\varnothing(U^\xi_{r(B)})\to\CF_\varnothing(U^\xi_{r(A)\cup r(B)})$.
\end{enumerate}
The third condition is automatic whenever the pairings are given by a single formula in terms of the vertex groups, which is the case in all examples in this paper. It says that the crossed cube attached to a relabeled, repeated, or shortened family is determined by the one attached to the original family; see Section~\ref{sec:cech-loday}, where this is made precise.
\end{definition}

For example, for a two-element \v{C}ech configuration
$\xi=(U,V\vert W)$,
the crossed cube $\Cech_{\CF}(\xi)$ is the crossed square
$$
\begin{tikzcd}[column sep=large,row sep=large]
\CF(U\wedge V)
  \arrow[r,"c_{U\wedge V,V}"]
  \arrow[d,"c_{U\wedge V,U}"']
& \CF(V)
  \arrow[d,"c_{V,W}"]
\\
\CF(U)
  \arrow[r,"c_{U,W}"']
& \CF(W).
\end{tikzcd}
$$
In addition to the four corestriction maps, it carries the pairing
$$
h^\xi_{\{1\},\{2\}}:
\CF(U)\times\CF(V)\longrightarrow\CF(U\wedge V).
$$
This pairing is a specified lift to the intersection group of the
commutator of the images of its two arguments in $\CF(W)$.  When the
corestriction maps are inclusions into a common ambient group, it is simply
the ordinary commutator.

\begin{definition}[Morphism of crossed precosheaves]
\label{def:morphism-crossed-precosheaves-precosheaf}
Let $\SL$ be a meet-semilattice, and let $\CF$ and $\CG$ be crossed
precosheaves on $\SL$. A \emph{morphism of crossed precosheaves} $\phi:\CF\longrightarrow\CG$
consists of a natural transformation between their underlying
precosheaves,
$\phi_{\varnothing}: \CF_{\varnothing} \to \CG_{\varnothing}$,
satisfying the following conditions. 

For every $I$-indexed \v{C}ech
configuration $\xi\in\Cech_I(\SL),$ the homomorphisms
$\phi_{U_A^\xi}: \CF_{\varnothing}(U_A^\xi) \to \CG_{\varnothing}(U_A^\xi)$,
must form a morphism of crossed $I$-cubes
$\phi_\xi: \Cech_{\CF}(\xi) \to \Cech_{\CG}(\xi)$.

These crossed cube morphisms are automatically natural in $\xi$: for every 
$\xi\leq\xi'$ in $\Cech_I(\SL)$ the diagram
$$
\begin{tikzcd}[column sep=large]
\Cech_{\CF}(\xi)
    \arrow[r,"\phi_\xi"]
    \arrow[d]
&
\Cech_{\CG}(\xi)
    \arrow[d]
\\
\Cech_{\CF}(\xi')
    \arrow[r,"\phi_{\xi'}"']
&
\Cech_{\CG}(\xi')
\end{tikzcd}
$$
commutes, where the vertical arrows are induced by fixed-fiber
functoriality. Compatibility with reindexing is also automatic, since the vertex component of $\phi_{r^*\xi}$ at $A\subseteq J$ is $\phi_{U^\xi_{r(A)}}$, the vertex component of $\phi_\xi$ at $r(A)$.

Compatibility with the structure maps follows from the naturality of
the underlying precosheaf morphism $\phi_{\varnothing}$, since the
structure maps of $\Cech_{\CF}(\xi)$ and $\Cech_{\CG}(\xi)$ are corestriction maps.
Consequently, for a fixed $\xi$, the condition that $\phi_\xi$ be a
morphism of crossed $I$-cubes is equivalent to compatibility with the
pairings: for every $A,B\subseteq I$,
$x\in\CF_{\varnothing}(U_A^\xi)$, and
$y\in\CF_{\varnothing}(U_B^\xi)$, one has
$$
\phi_{U_{A\cup B}^\xi}
\bigl(h^{\CF,\xi}_{A,B}(x,y)\bigr)
=
h^{\CG,\xi}_{A,B}
\bigl(
\phi_{U_A^\xi}(x),
\phi_{U_B^\xi}(y)
\bigr).
$$

Crossed precosheaves on $\SL$, together with these morphisms, form a
category denoted $\mathbf{XPreCoSh}(\SL)$.
\end{definition}

\begin{remark}[Categorical reformulation]\label{rem:categorical}
Definitions~\ref{def:crossed-precosheaf-structure} and~\ref{def:morphism-crossed-precosheaves-precosheaf} can be rephrased more concisely in terms of functors between categories of crossed cubes and of \v{C}ech configurations indexed by finite sets, with reindexing along functions as additional structure. This reformulation is recorded in Appendix~\ref{app:categorical}; it is not needed in the body of the paper.
\end{remark}

In the remainder of the paper, we suppress the distinction between a crossed precosheaf
$\CF$ and its underlying precosheaf $\CF_{\varnothing}$, denoting
both by $\CF$. In particular, $\CF(U)$ denotes
$\CF_{\varnothing}(U)$ for $U\in \SL$.

\begin{definition}
\label{def:colasqueness}
A precosheaf of groups is called coflasque if all corestriction maps are injective. A crossed precosheaf is called coflasque if the corresponding precosheaf of groups is coflasque.
\end{definition}

\subsection{Examples of crossed precosheaves}

The following is a prototypical example of a crossed precosheaf.

\begin{example}[Crossed precosheaf of subgroups]
\label{ex:crossed-precosheaf-subgroups}
Let $\SL$ be a meet-semilattice and $G_0$ a group. Denote by
$\operatorname{Sub}(G_0)$ the lattice of subgroups of $G_0$, ordered
by inclusion. An order-preserving map
$\CF:\SL\to\operatorname{Sub}(G_0)$
defines a precosheaf of subgroups in the common ambient group $G_0$:
whenever $U\leq V$, the corestriction map is the subgroup inclusion
$\CF(U)\hookrightarrow\CF(V)\subseteq G_0$.
In particular, $\CF$ is coflasque.
Assume additionally that $\CF$ satisfies the
\textbf{commutator containment} condition
$$
[\CF(U),\CF(V)]\subseteq\CF(U\wedge V)
\qquad (\forall
U,V\in \SL).
$$

For every $I$-indexed \v{C}ech configuration $\xi$ and every
$A,B\subseteq I$, define
$$
h^\xi_{A,B}:
\CF(U_A^\xi)\times\CF(U_B^\xi)
\longrightarrow
\CF(U_{A\cup B}^\xi),
$$
by the standard group commutator
$h^\xi_{A,B}(a,b):=[a,b]$.
This is well defined because of the commutator containment condition
$$
[\CF(U_A^\xi),\CF(U_B^\xi)]
\subseteq
\CF(U_A^\xi\wedge U_B^\xi)
=
\CF(U_{A\cup B}^\xi).
$$
We claim that these pairings make $\CF$ into a coflasque crossed precosheaf.

Coflasqueness is immediate since all corestriction maps are subgroup inclusions. 

\textbf{Crossed-cube condition.}
For any \v{C}ech configuration $\xi=((U_i)_{i\in I}\vert W)$, the definition of a crossed
precosheaf prescribes the vertex groups
$M_A=\CF(U_A^\xi)$ and the structure maps
$\mu_i:M_A\hookrightarrow M_{A\setminus\{i\}}$, which are the
subgroup inclusions. All $M_A$ are subgroups of the common
ambient group $\CF(W)$. The structure maps therefore commute, and
their compatibility with the pairings follows because subgroup
inclusions preserve commutators.

If $A\subseteq B$, then $U_B^\xi\leq U_A^\xi$, and the action of $M_A$ on $M_B$ is given by ordinary group conjugation
${}^a b=[a,b]b=aba^{-1}$.
 The remaining crossed-cube axioms thus reduce to the standard
commutator identities in this ambient group. Hence the prescribed data form a crossed $I$-cube.

\textbf{Fixed-fiber functoriality.}
For $\xi\leq\xi'$, all induced vertex maps are subgroup inclusions.
Since inclusions preserve commutators, they form a morphism of crossed
$I$-cubes.

\textbf{Reindexing compatibility.}
For every map $r:J\to I$,
$$
U_A^{r^*\xi}=U_{r(A)}^\xi,
\qquad
h^{r^*\xi}_{A,B}=h^\xi_{r(A),r(B)}.
$$
Indeed, both pairings are restrictions of the same commutator in
$G_0$. 
\end{example}

\begin{example}[Functions with prescribed support]
\label{ex:support-functions}
Let $X$ be a topological space, and take $\SL=Open(X)$, the lattice
of open subsets of $X$, ordered by inclusion, with meet and join given
by intersection and union. Let $G$ be a topological group with identity
$e$, and let $C(X,G)$ denote the group of continuous maps
$X\to G$ under pointwise multiplication.

For $f\in C(X,G)$, define
$$
Z(f):=\{x\in X\mid f(x)\neq e\},
\qquad
\supp f:=\overline{Z(f)}.
$$
For $U\in \SL$, let
$$
\CH(U):=\{f\in C(X,G)\mid \supp f\subseteq U\}.
$$
The identities $Z(fg)\subseteq Z(f)\cup Z(g)$ and
$Z(f^{-1})=Z(f)$ show that each $\CH(U)$ is a subgroup of
$C(X,G)$. Moreover, $U\subseteq V$ implies $\CH(U)\subseteq \CH(V)$.
Thus $\CH$ is a coflasque precosheaf of subgroups of $C(X,G)$.

If $f\in \CH(U)$ and $g\in \CH(V)$, then $[f,g](x)=e$ whenever
$f(x)=e$ or $g(x)=e$. Consequently,
$$
\supp[f,g]
\subseteq
\supp f\cap\supp g
\subseteq
U\cap V.
$$
Therefore,
$$
[\CH(U),\CH(V)]
\subseteq
\CH(U\cap V)
=
\CH(U\wedge V).
$$
This is precisely the commutator-containment condition. Therefore $\CH$,
equipped with the pointwise group commutator pairings, is an instance of
the coflasque crossed precosheaf of subgroups described in
Example~\ref{ex:crossed-precosheaf-subgroups}.
\end{example}

From the physics perspective, the group $\CH(U)$ is the group of gauge transformations which are localized on $U$. It is common to require gauge transformations to be "small", i.e. to belong to the connected component of the identity element of $\CH(U)$. More precisely, we equip $C(X,G)$ with the
compact-open topology; pointwise multiplication and inversion are then
continuous for arbitrary $X$ (via the homeomorphism
$C(X,G)\times C(X,G)\cong C(X,G\times G)$ and continuity of
post-composition with continuous maps), so $C(X,G)$ is a topological
group and every $\CH(U)$, with the subspace topology, is a topological
group whose inclusions $\CH(U)\hookrightarrow \CH(V)$ are continuous. Any
other group topology on $C(X,G)$ would do just as well in what follows.

\begin{lemma}\label{lem:identity-components}
Let $C$ be a topological group and let $H,K,M\le C$ be subgroups with
$[H,K]\subseteq M$. Write $H_0$ for the identity component of the
topological group $H$, and likewise $M_0$. Then
$$
[H_0,K]\subseteq M_0.
$$
\end{lemma}

\begin{proof}
Fix $g\in K$ and consider the map
$$
c_g\colon H\longrightarrow M,
\qquad
c_g(f):=[f,g]=fgf^{-1}g^{-1}.
$$
It takes values in $M$ by hypothesis, and it is continuous: it is a
composite of group operations of $C$, hence continuous into $C$, and
continuity into the subspace $M$ follows. Since $c_g(e)=e$, the image
$c_g(H_0)$ is a connected subset of $M$ containing $e$; connectedness of
a subset is intrinsic, so $c_g(H_0)$ is contained in the connected
component of $e$ in $M$, i.e.\ $[H_0,g]\subseteq M_0$. As $g\in K$ was
arbitrary, the claim follows.
\end{proof}

Note the asymmetry: only one of the two arguments needs to lie in the
identity component.

\begin{proposition}\label{prop:identity-component}
In the situation of Example~\ref{ex:support-functions}, let $\CH'(U)$ denote the identity component of the topological group $\CH(U)$ for every $U\in Open(X)$.
Then $U\mapsto \CH'(U)$, with corestrictions the inclusions and with all
$h$-pairings the ordinary commutators in $C(X,G)$, is a coflasque
crossed precosheaf on $Open(X)$.
\end{proposition}

\begin{proof}
Let $\SL=Open(X)$ . For every $U\in \SL$, the
identity component $\CH'(U)$ is a subgroup of $\CH(U)$, and hence of
$C(X,G)$.

If $U\subseteq V$, the inclusion $\CH(U)\hookrightarrow \CH(V)$ is
continuous and preserves the identity component. It therefore maps the connected
set $\CH'(U)$ into the identity component $\CH'(V)$. Thus
$\CH'(U)\subseteq \CH'(V)$, so
$\CH'$
is a coflasque precosheaf of subgroups of $C(X,G)$.

By Example~\ref{ex:support-functions},
$$
[\CH(U),\CH(V)]\subseteq \CH(U\cap V).
$$
Applying Lemma~\ref{lem:identity-components} with
$H=\CH(U)$, $K=\CH(V)$, and $M=\CH(U\cap V)$ gives 
$[\CH'(U),\CH(V)]\subseteq \CH'(U\cap V),$
and therefore 
$$
[\CH'(U),\CH'(V)]
\subseteq
\CH'(U\cap V)
=
\CH'(U\wedge V).
$$
Thus $\CH'$ satisfies the commutator-containment condition. Therefore
$\CH'$, equipped with the ordinary pointwise commutator pairings, is an
instance of the coflasque crossed precosheaf of subgroups described in
Example~\ref{ex:crossed-precosheaf-subgroups}. 
\end{proof}

\begin{remark}\label{rem:Fpath}
The same argument, with ``connected'' replaced by ``path-connected''
throughout, shows that
$U\mapsto \CHpath(U)$ (path component of the identity in $\CH(U)$) is likewise a
coflasque crossed precosheaf with commutator pairings.
\end{remark}

As mentioned in Section \ref{sec:gaugingnew}, the coflasque crossed precosheaves of Example  \ref{ex:support-functions} and Proposition~\ref{prop:identity-component} can be used to model gauge symmetries in QFT. Each of them has special properties. By construction, the precosheaf $\CH$ of Example \ref{ex:support-functions} satisfies $\CH(U\cap V)=\CH(U)\cap \CH(V)$. On the other hand, for $X$ a manifold and $G$ a Lie group, the precosheaf $\CHpath$ of Remark \ref{rem:Fpath} obeys the multiplicative partition of unity property  which was mentioned in Section \ref{sec:gaugingnew}.

\begin{definition}\label{def:epi}
    A precosheaf of groups $\CF$ on a join-semilattice $\SL$ is called an epiprecosheaf if for any $U,V\in \SL$ the images of $\CF(U)$ and $\CF(V)$ in $\CF(U\vee V)$ under corestriction maps generate $\CF(U\vee V)$. In the case of a coflasque crossed precosheaf on a lattice $\SL$, this is equivalent to $\CF(U\vee V)=\CF(U)\CF(V).$
\end{definition}

Let $X$ be a paracompact manifold and $G$ a Lie group. Then $\CHpath$ is a coflasque crossed epiprecosheaf on $\mathrm{Open}(X)$: for every finite open cover $W=U_1\cup\cdots\cup U_m$ one has $\CHpath(W)=\CHpath(U_1)\cdots \CHpath(U_m)$. A proof is sketched in Appendix~\ref{app:path-epi}. This fact is not used in the rest of the paper.

\section{Internal categories and the nerve construction}\label{sec:internal-cats}\label{sec:nerve}

In this section we recall internal categories in groups and their relation to
$\cat^1$-groups. The relation is classical; see
\cite{BrownSpencer,Loday1982}. We also collect  routine properties of nerves, multinerves, and diagonals which are used repeatedly, and with their strict forms, in Section~\ref{sec:loday}; the expert reader may skim the statements and skip the proofs. The last subsection, Section~\ref{sec:grid}, is not routine: it gives an explicit description of the multinerve of the $\cat^n$-group attached to a crossed cube. Its results are used only in the proof of Theorem~\ref{thm:total-loday-functor} in Section~\ref{sec:cech-loday}.

\subsection{Internal categories}

\begin{definition}\label{def:internal-cat}
An \emph{internal category} in $\Grp$ consists of groups $C_0$ and $C_1$
together with group homomorphisms
$$
\sigma,\tau:C_1\longrightarrow C_0,
\qquad
\varepsilon:C_0\longrightarrow C_1,
\qquad
\mu:C_1\times_{C_0}C_1\longrightarrow C_1,
$$
where
$$
C_1\times_{C_0}C_1
=
\{(g,f)\in C_1\times C_1\mid \sigma g=\tau f\},
$$
such that $\sigma\varepsilon=\tau\varepsilon=\id_{C_0}$, the source and
target of $\mu(g,f)$ are respectively $\sigma f$ and $\tau g$, and the
usual unit and associativity identities hold. We write $\mu(g,f)=g\circ f$.
If $\scrC$ and $\scrD$ are internal categories in $\Grp$, with object and arrow groups $C_0,C_1$ and $D_0,D_1$, respectively, an \emph{internal functor}
$F=(F_0,F_1):\scrC\to\scrD$
consists of homomorphisms
$F_0:C_0\to D_0$, $F_1:C_1\to D_1$
such that
$$
\sigma_{\scrD}F_1
=
F_0\sigma_{\scrC},
\qquad
\tau_{\scrD}F_1
=
F_0\tau_{\scrC},
\qquad
F_1\varepsilon_{\scrC}
=
\varepsilon_{\scrD}F_0,
$$
and
$$
F_1(g\circ f)=F_1(g)\circ F_1(f)
$$
for every composable pair $(g,f)$ in $\scrC$.

Let
$F,G:\scrC\to\scrD$
be internal functors. An \emph{internal natural transformation}
$\eta:F\Longrightarrow G$
is a homomorphism
$\eta:C_0\to D_1$
such that
$\sigma_{\scrD}\eta=F_0$, $\tau_{\scrD}\eta=G_0$,
and, for every $a\in C_1$,
$$
G_1(a)\circ\eta(\sigma_{\scrC}a)
=
\eta(\tau_{\scrC}a)\circ F_1(a).
$$
For internal categories in simplicial groups, the same definitions are
used with all these maps required to be homomorphisms of
simplicial groups.
\end{definition}

One can define internal categories in any category with pull-backs. However, internal categories in $\Grp$ have some special properties, as the following result shows.

\begin{proposition}[\cite{BrownSpencer}]\label{prop:internal-groupoid}
In an internal category $(C_0,C_1,\sigma,\tau,\eps,\mu)$ in $\Grp$ the composition law $\mu$ is not an independent datum: it is given by
$$
\mu(g,f)=g\cdot (\varepsilon \sigma g)^{-1}\cdot f.
$$
Every internal category in $\Grp$ is an internal groupoid. The categorical inverse of $a\in C_1$ is
$$
a^{\circ-1}=\varepsilon(\sigma a)\,a^{-1}\,\varepsilon(\tau a),
$$
and $a\mapsto a^{\circ-1}$ is a group automorphism of $C_1$ which exchanges
$\sigma$ and $\tau$. 
\end{proposition}

Below we will encounter internal categories in other categories with pull-backs, such as simplicial groups, multisimplicial groups, and $\cat^n$-groups. The following almost obvious lemma allows one to pass between them. 

\begin{lemma}\label{lem:change-of-base}
Let $\Phi\colon\scrA\to\scrB$ be a functor between finitely
complete categories which preserves finite limits. Then $\Phi$ induces a
functor from internal categories in $\scrA$ to internal categories
in $\scrB$: the internal category
$\scrC=(C_0,C_1,\sigma,\tau,\varepsilon,\mu)$ is sent to
$\Phi\scrC=(\Phi C_0,\Phi C_1,\Phi\sigma,\Phi\tau,\Phi\varepsilon,
\Phi\mu)$, with the composition defined on
$\Phi C_1\times_{\Phi C_0}\Phi C_1$ via the canonical isomorphism
$\Phi(C_1\times_{C_0}C_1)\cong\Phi C_1\times_{\Phi C_0}\Phi C_1$, and an
internal functor $(T_0,T_1)$ is sent to $(\Phi T_0,\Phi T_1)$. The same
isomorphisms identify $\Phi$ applied to the object of composable
$r$-strings of $\scrC$ with the object of composable $r$-strings of
$\Phi\scrC$, for every $r\geq0$. See
\cite[Proposition~1.8]{Ghiorzi}; the change-of-base functor is treated
in the general setting of internal category theory in
\cite[\S B2.3]{JohnstoneElephant}.
\end{lemma}

Next we recall the relation between internal categories in $\Grp$ and $\cat^1$-groups. Recall that a $\cat^1$-group is a group $G$ equipped with endomorphisms $s,t:G\to G$
satisfying
$$
ts=s,
\qquad
st=t,
\qquad
[\ker s,\ker t]=1.
$$
A morphism of $\cat^1$-groups is a homomorphism commuting with $s$ and $t$.

\begin{theorem}[\cite{BrownSpencer}, \cite{Loday1982}]\label{thm:cat1-internal}
The category of $\cat^1$-groups is equivalent to the category of internal
categories in $\Grp$. Under the normalization
in which the object group is a subgroup of the arrow group and the identity
map is the inclusion, this equivalence is an isomorphism of categories. Concretely, an internal category
$(C_0,C_1,\sigma,\tau,\varepsilon,\mu)$ determines the $\cat^1$-group
$(C_1,\varepsilon\sigma,\varepsilon\tau)$,
while a $\cat^1$-group $(G,s,t)$ determines the normalized internal category
with arrow group $G$, object group $\im\, s=\im\, t$, source and target $s$ and
$t$ with their codomains restricted, and composition
$$
\mu(g,f)=g(sg)^{-1}f
\qquad (sg=tf).
$$
Consequently, a $\cat^n$-group may be viewed as an $n$-fold internal
category in groups, one internal-category direction for each pair
$(s_i,t_i)$.
\end{theorem}

\begin{remark}\label{rem:internal-transfer}
Proposition~\ref{prop:internal-groupoid} and
Theorem~\ref{thm:cat1-internal} remain true with $\Grp$ replaced by
simplicial groups or by $\cat^I$-groups. Indeed, pull-backs in either of
these categories are pull-backs of the underlying groups, with the
additional structure maps acting componentwise. Hence an internal
category in simplicial groups or in $\cat^I$-groups is an internal
category in $\Grp$ whose structure maps are morphisms of the category in
question, and similarly for internal functors and for $\cat^1$-objects.
Since the formulas for $\mu$, $a^{\circ-1}$, $\varepsilon\sigma$ and
$\varepsilon\tau$ involve only the structure maps and the group
operations, they define morphisms of the category in question whenever
the structure maps do.
\end{remark}

\subsection{The nerve and the diagonal}
\label{sec:multinerve}

\begin{definition}[Nerve and multinerve]\label{def:multinerve}
Let
$\scrC=(C_0,C_1,\sigma,\tau,\varepsilon,\mu)$
be an internal category in groups. We write strings in the order in
which the arrows are traversed:
$$
x_0\xrightarrow{g_1}x_1\xrightarrow{g_2}\cdots
\xrightarrow{g_m}x_m.
$$
Composition retains the convention
$\mu(g,f)=g\circ f$ of Definition~\ref{def:internal-cat}.

The \emph{nerve} of $\scrC$ is the simplicial group
$N\scrC:\Delta^{\mathrm{op}}\to\Grp$ with
$(N\scrC)_0=C_0$, $(N\scrC)_1=C_1$,
and, for $m\geq2$,
$$
(N\scrC)_m=
\left\{(g_1,\ldots,g_m)\in C_1^m
\ \middle|\
\tau(g_i)=\sigma(g_{i+1})\text{ for }1\leq i<m
\right\},
$$
with componentwise group multiplication. We identify a one-term string
with its arrow. Thus a two-simplex is written $(f,g)$, while its
composite is $\mu(g,f)=g\circ f$.

For $m\geq1$, the vertices of a string are
$$
x_0=\sigma(g_1),\qquad x_i=\tau(g_i)\quad(1\leq i\leq m).
$$
Throughout, face and degeneracy maps of simplicial objects are written in sans-serif, $\fd_i$ and $\fs_i$, to distinguish them from the source and target endomorphisms $s_i,t_i$ of a $\cat^n$-group. In degree one the faces are
$\fd_0(g)=\tau(g)$, $\fd_1(g)=\sigma(g)$.
For $m\geq2$, they are
$$
\begin{aligned}
\fd_0(g_1,\ldots,g_m)&=(g_2,\ldots,g_m),\\
\fd_i(g_1,\ldots,g_m)
&=(g_1,\ldots,g_{i-1},\mu(g_{i+1},g_i),g_{i+2},\ldots,g_m)
&& (0<i<m),\\
\fd_m(g_1,\ldots,g_m)&=(g_1,\ldots,g_{m-1}).
\end{aligned}
$$
The degeneracies are $\fs_0(x)=\varepsilon(x)$ in degree zero and
$$
\fs_i(g_1,\ldots,g_m)
=(g_1,\ldots,g_i,\varepsilon(x_i),g_{i+1},\ldots,g_m)
\qquad(0\leq i\leq m)
$$
for $m\geq1$, with empty portions of a string omitted.
The internal-category axioms give the simplicial identities, and all
these maps are group homomorphisms. An internal functor
$T=(T_0,T_1):\scrC\to\scrH$ induces $NT$ by applying $T_0$
in degree zero and $T_1$ to each arrow in every positive degree.
The same construction applies levelwise to internal categories in
simplicial or multisimplicial groups.

For a $\cat^n$-group $G$, regarded as an $n$-fold internal category in
groups by Theorem~\ref{thm:cat1-internal}, take the nerve successively
in the categorical directions $1,\ldots,n$, retaining these labels
on the resulting simplicial directions. This defines the
\emph{multinerve functor}
$$
N^n:\cat^n\Grp\longrightarrow\mathbf{s}^n\Grp,
\qquad
N^nG:(\Delta^{\mathrm{op}})^n\longrightarrow\Grp.
$$
Its value in multidegree $(m_1,\ldots,m_n)$ uses the $m_i$-simplices
in direction $i$. The order in which the nerves are taken is
immaterial up to canonical natural isomorphism
(Lemma~\ref{lem:nerve-limits}(2)). For $I\subseteq\{1,\ldots,n\}$,
we write $N^I$ when only the categorical directions indexed by $I$
are nerved, and set $N^0G=G$.
\end{definition}

\begin{lemma}[Source-kernel normal form]
\label{lem:source-kernel-normal-form}
Let $\scrC$ be an internal category in groups, and put
$$
K=\ker(\sigma:C_1\to C_0),\qquad
\partial=\tau|_K:K\longrightarrow C_0.
$$
Use the $C_0$-action on $K$ given by
$$
{}^xk=\varepsilon(x)k\varepsilon(x)^{-1}.
$$
For $\bfk=(k_1,\ldots,k_m)\in K^m$, write
$$
p_0(\bfk)=1,\qquad
p_i(\bfk)=k_i\cdots k_1\quad(1\leq i\leq m).
$$
Set $K^{[0]}=1$. For $m\geq1$, let $K^{[m]}$ have underlying set
$K^m$ and multiplication
$$
(\bfk\star\bfell)_i
=
k_i\,{}^{\partial(p_{i-1}(\bfk))}\ell_i
\qquad(1\leq i\leq m).
$$
This makes $K^{[m]}$ a group. It is an iterated semidirect product:
$$
K^{[m]}\cong K\rtimes K^{[m-1]},
$$
where the leftmost $K$ is recorded as the last coordinate, and
$\bfk\in K^{[m-1]}$ acts on it through
$\partial(p_{m-1}(\bfk))$. The group $C_0$ acts on $K^{[m]}$ by coordinatewise automorphisms.
In particular, multiplication in $K^{[m]}\rtimes C_0$ is explicitly
$$
(\bfk;x)(\bfell;y)=(\mathbf r;xy),
\qquad
r_i=k_i\,{}^{\partial(p_{i-1}(\bfk))x}\ell_i.
$$

There is a group isomorphism, natural in internal functors,
$$
\Theta_m:K^{[m]}\rtimes C_0\xrightarrow{\ \cong\ }(N\scrC)_m.
$$
For $m\geq1$, it is given by
$$
\Theta_m(\bfk;x)=(g_1,\ldots,g_m),\qquad
x_0=x,\quad g_i=k_i\varepsilon(x_{i-1}),\quad
x_i=\partial(k_i)x_{i-1}.
$$
Its inverse is
$$
x=\sigma(g_1),\qquad
k_i=g_i\varepsilon(\sigma(g_i))^{-1}
\quad(1\leq i\leq m).
$$
For $m=0$, the isomorphism sends $(();x)$ to $x$.
If $T=(T_0,T_1):\scrC\to\scrH$ is an internal functor and
$T_K=T_1|_K$ is its map on source kernels, then $NT$ is expressed in
these coordinates by
$$
(k_1,\ldots,k_m;x)
\longmapsto
(T_K(k_1),\ldots,T_K(k_m);T_0(x)).
$$
\end{lemma}

\begin{proof}
The case $m=0$ is immediate. For $m\geq1$, the reconstruction formulas
give
$\sigma(g_i)=x_{i-1}$, $\tau(g_i)=x_i$,
so the arrows are composable. Conversely, for a composable string,
$g_i\varepsilon(\sigma(g_i))^{-1}$ lies in $K$, and the displayed
inverse recovers the unique tuple $(\bfk;x)$. Thus the formulas
define mutually inverse bijections.

To determine the multiplication, let $(g_i)$ and $(h_i)$ have vertices
$x_i$ and $y_i$, and source-kernel coordinates $k_i$ and $\ell_i$.
Their componentwise product has vertices $x_i y_i$, and its $i$-th
source-kernel coordinate is
$$
\begin{aligned}
g_i h_i\varepsilon(x_{i-1}y_{i-1})^{-1}
&=k_i\varepsilon(x_{i-1})\ell_i\varepsilon(x_{i-1})^{-1}\\
&=k_i\,{}^{x_{i-1}}\ell_i.
\end{aligned}
$$
Since $x_{i-1}=\partial(p_{i-1}(\bfk))x_0$, this is precisely
the multiplication stated in the lemma.

The tuples with $x_0=1$ correspond to the kernel of the initial-vertex
homomorphism $(N\scrC)_m\to C_0$. Hence $\star$ is a group law
on $K^{[m]}$. The initial-vertex homomorphism is split by the string
of identity arrows at $x\in C_0$, and conjugation by this string acts
coordinatewise on $K^{[m]}$. This proves the asserted semidirect-product
decomposition and the group-isomorphism property of $\Theta_m$.
Forgetting the last coordinate of $K^{[m]}$ is a homomorphism with
kernel $K$, split by appending $1$; its conjugation action is the one
displayed above. This proves the iterated semidirect-product description.
Finally, naturality follows because an internal functor commutes with
$\sigma$, $\tau$, and $\varepsilon$.
\end{proof}

In words: a composable string is determined by its initial vertex
together with the ``step'' $k_i\in K$ taken at each stage, and the
group law on strings is the product of the steps twisted by the
vertices reached so far. For the $\cat^1$-group attached to a crossed
module, this is the one-direction case of the normal form of
Section~\ref{sec:grid}, which extends it to the multinerve of a crossed
cube.

\begin{lemma}\label{lem:nerve-limits}
Let $\scrC$ be a finitely complete category. Finite limits in
$\Cat(\scrC)$ are computed on objects and arrows
\cite[Proposition~1.7]{Ghiorzi}. Consequently:
\begin{enumerate}
\item
Each nerve level $N_m$, and hence the nerve $N$, preserves finite
limits. For every finite diagram there is a canonical natural
isomorphism
$$
N\!\left(\lim_\alpha\scrC_\alpha\right)
\xrightarrow{\ \cong\ }
\lim_\alpha N\scrC_\alpha.
$$
In particular, the nerve preserves fiber products and, when
$\scrC$ is pointed, kernels.

\item
Writing $N_i$ for the nerve in categorical direction $i$, there are
canonical natural isomorphisms
$$
N_j(N_iG)\cong N_i(N_jG)
\qquad(i\neq j)
$$
for every $\cat^n$-group $G$, with the simplicial direction labels
fixed. These respect all structure maps and are compatible with
successive changes of order. Thus the multinerve is independent of
the order of taking nerves up to canonical natural isomorphism.

\item
The multinerve $N^I$ preserves finite limits. Applied in directions
distinct from an internal-category direction, it sends internal
categories and internal functors to internal categories and internal
functors, as in Lemma~\ref{lem:change-of-base}. For an internal
category with object and arrow objects $H_0,H_1$, composition is
transported along the canonical isomorphism
$$
N^I(H_1\times_{H_0}H_1)
\xrightarrow{\ \cong\ }
N^I H_1\times_{N^I H_0}N^I H_1.
$$
\end{enumerate}
\end{lemma}

\begin{remark}
    In subsequent formulas, we suppress these canonical comparison
isomorphisms.
\end{remark}

\begin{proof}
(1) The evaluations $N_0,N_1$ preserve finite limits. Each $N_m$ for
$m\geq2$ is a finite limit of these evaluations, so the result follows
because finite limits commute. The comparisons commute with the
simplicial operators by naturality.

(2) In each multidegree, the two orders give iterated finite limits
of the same diagram of composable cells. Their universal properties
give the comparison isomorphisms, compatibility with all structure
maps, and compatibility under successive changes of order.

(3) A composite of finite-limit-preserving functors preserves finite
limits. Apply Lemma~\ref{lem:change-of-base} to this composite.
\end{proof}

The nerve construction applies to internal categories in any category with pullbacks, in particular in $\Set$. A group $G$ is an internal category in $\Set$ with one object and the arrow set $G$. We use the convention where the composition of morphisms is opposite to the composition of group elements: $g\circ g'=g'g$.  The nerve of this category is the classifying complex of $G$.

\begin{definition} [Degreewise classifying complex]
\label{def:degree-wise-classifying-complex}
For a group $G$, the \emph{classifying complex} $B_\bullet G$ is the nerve of the one-object category just described: the simplicial set with
$$
(B_\bullet G)_m=G^m\quad(m\geq0),\qquad (B_\bullet G)_0=*,
$$
whose faces and degeneracies are those of the nerve,
$$
\fd_i(g_1,\ldots,g_m)=
\begin{cases}
(g_2,\ldots,g_m),& i=0,\\
(g_1,\ldots,g_ig_{i+1},\ldots,g_m),& 0<i<m,\\
(g_1,\ldots,g_{m-1}),& i=m,
\end{cases}
\qquad
\fs_i(g_1,\ldots,g_m)=(g_1,\ldots,g_i,1,g_{i+1},\ldots,g_m).
$$
A homomorphism $G\to G'$ induces a simplicial map $B_\bullet G\to B_\bullet G'$ by applying it in every entry, so $B_\bullet$ is a functor $\Grp\to\SSet$; we also write $BG:=B_\bullet G$.

More generally, if
$X:(\Delta^{\mathrm{op}})^n\to\Grp$
is an $n$-simplicial group, applying $B_\bullet$ levelwise gives the
$(n+1)$-simplicial set
$$
B_\bullet X:(\Delta^{\mathrm{op}})^{n+1}\longrightarrow\Set,
\qquad
(B_\bullet X)_{m,k_1,\ldots,k_n}
=
\bigl(X_{k_1,\ldots,k_n}\bigr)^m,
$$
with the faces and degeneracies above in the first index and those of $X$ in the remaining ones. We use the same notation $B_\bullet$ for all of these degreewise
extensions.
\end{definition}

In what follows we will make an extensive use of the functor $\Diag$ from $n$-simplicial sets or   groups to simplicial sets or groups. We finish this section by recording some of its properties.

\begin{lemma}\label{lem:diag-basic}
Let $\scrC$ be a complete category, $\scrD$ a small category, $J$ a
finite set, and let
$$
\Diag_J\colon
\Fun\bigl((\Delta^{\mathrm{op}})^{J}\times\scrD,\ \scrC\bigr)
\longrightarrow
\Fun\bigl(\Delta^{\mathrm{op}}\times\scrD,\ \scrC\bigr)
$$
be precomposition with the functor $\delta_J:([m],d)\mapsto\bigl(([m])_{j\in J},d\bigr)$.
\begin{enumerate}
\item $\Diag_J$ preserves limits, which in both functor
categories are computed levelwise (=objectwise); it also sends levelwise surjections of
(multi)simplicial groups to levelwise surjections.
\item Diagonals compose under grouping: 
for a partition $J=J_1\sqcup J_2$, first taking the partial diagonal in each block and then taking the diagonal of the two resulting simplicial directions gives $\Diag_J$. The two blockwise operations may be performed in either order.
\end{enumerate}
\end{lemma}

\begin{proof}
All statements hold because $\Diag_J$ is precomposition with
a functor. Limits of diagram categories are computed
levelwise, and $\Diag_J$ evaluates a diagram at the levels
$\bigl(([m])_{j\in J},d\bigr)$, so it carries the levelwise limit to the
levelwise limit, giving the first part of (1); and each component of
$\Diag_Jf$ is a component of $f$, so levelwise surjectivity
is inherited, giving the second. Statement (2) follows from $\delta_J=(\delta_{J_1}\times \delta_{J_2})\circ\delta_{\{1,2\}}$.
\end{proof}

\begin{lemma}\label{lem:levelwise-diag}
Let $J$ be a finite set of simplicial directions. Applied levelwise in the
$J$-directions, the following constructions commute with
$\Diag_J$, as identities of functors:
\begin{enumerate}
\item the classifying complex construction: for a multisimplicial group $G$,
$B_\bullet(\Diag_JG)=\Diag_J(B_\bullet G)$;
\item the nerve of an internal category: if $\scrH$ is an internal
category in $J$-simplicial groups, then
 $N(\Diag_J\scrH)=\Diag_J(N\scrH)$,
where $\Diag_J\scrH$ is the internal category in
simplicial groups obtained by applying $\Diag_J$ to the
objects, arrows, and structure maps of $\scrH$;
\item kernels of morphisms of multisimplicial groups:
$\ker(\Diag_Jf)=\Diag_J(\ker f)$.
\end{enumerate}
\end{lemma}
\begin{proof}
In each case both sides have the same value at every level: $B_pG$ is the
$p$-fold product, the nerve level $N_r$ is the $r$-fold fiber product of
the arrow object over the object object, and kernels are
equalizers; all are finite limits, computed levelwise by
Lemma~\ref{lem:diag-basic}(1), and $\Diag_J$ is
precomposition, so the comparison maps are identities. The face and
degeneracy maps agree because they are induced maps between these limits
(multiplication, units, projections; source, target, composition,
identity section) and $\Diag_J$ is functorial.
\end{proof}

\subsection{The multinerve of a crossed cube}\label{sec:grid}

In the next two sections we will extensively use the multinerve of internal $n$-categories in groups arising from crossed $n$-cubes. In this subsection we describe $N^I\Gamma_I(M_\bullet)$ for a crossed cube $M_\bullet$ explicitly, in
every multidegree, in terms of the vertex groups $M_A$, the structure
maps and the pairings. The description contains the Ellis-Steiner
normal form (Theorem~\ref{thm:normalform}) as the case of multidegree
$(1,\ldots,1)$ and the source-kernel normal form
(Lemma~\ref{lem:source-kernel-normal-form}) as the case $|I|=1$. This explicit description is
used only once, to construct the reindexing maps in the proof of
Theorem~\ref{thm:total-loday-functor} in Section~\ref{sec:cech-loday};
a reader willing to take that theorem on trust may skip this subsection.
We begin with the abelian case as a motivation.

Let all $M_A$ be abelian and all pairings trivial, so that $M_\bullet$
is a commuting cube of abelian groups and
$\Gamma_I(M_\bullet)=\bigoplus_{A\subseteq I}M_A$, with $s_i$ and $t_i$
as in Remark~\ref{rem:abelian-gamma}. For brevity write $M_1$, $M_2$,
$M_{12}$ for $M_{\{1\}}$, $M_{\{2\}}$, $M_{\{1,2\}}$.

\emph{Abelian cube, one direction.} For $I=\{1\}$ the cube is a homomorphism
$\partial=\mu_1:M_1\to M_\varnothing$, and $\Gamma_1(M_\bullet)$ is the
internal category in abelian groups with object group $M_\varnothing$,
arrow group $M_1\oplus M_\varnothing$, source $\sigma(c,x)=x$, target
$\tau(c,x)=x+\partial c$, and identities $\varepsilon(x)=(0,x)$. A
string of $m$ composable arrows is therefore a sequence of vertices
$x_0,\ldots,x_m\in M_\varnothing$ together with elements
$c_1,\ldots,c_m\in M_1$ such that $x_j=x_{j-1}+\partial c_j$. The
vertices are determined by $x_0$ and the $c_j$:
\begin{equation}\label{eq:heaviside-1d}
x_u=x_0+\sum_{k=1}^{m}(\partial c_k)\,H_k(u),
\qquad
H_k(u)=
\begin{cases}1,&u\geq k,\\ 0,&u<k,\end{cases}
\end{equation}
which is the elementary fact that a function on $\{0,\ldots,m\}$ is
the sum of its initial value and of its increments, each increment
switched on from the point where it occurs. Thus
$$
(N\Gamma_1(M_\bullet))_m\cong M_\varnothing\oplus M_1^{\,m},
$$
the summand $M_\varnothing$ consisting of the constant strings and the
$k$-th copy of $M_1$ of the strings with a single non-identity arrow
$c$, at step $k$, whose vertex function is $(\partial c)H_k$. We call
$k$ the \emph{threshold} of $H_k$, and of this string: it is the value
of $u$ at which $H_k$ jumps from $0$ to $1$, and the edge from $u=k-1$
to $u=k$ is the one that \emph{crosses} the threshold.

\begin{figure}[htb]
\centering
\begin{tikzpicture}[scale=0.55]
\tikzset{grid/.style={gray!60,thin},
         on/.style={circle,fill=black,inner sep=0pt,minimum size=4.2pt},
         off/.style={circle,draw=gray!70,fill=white,inner sep=0pt,minimum size=3.2pt},
         arr/.style={line width=2pt,blue!70!black}}
\begin{scope}
\draw[step=1,grid] (0,0) grid (4,4);
\foreach \x in {0,...,4} \foreach \y in {0,...,4} \node[on] at (\x,\y) {};
\foreach \x in {0,...,4} \node[below] at (\x,-0.1) {\scriptsize $\x$};
\foreach \y in {0,...,4} \node[left] at (-0.1,\y) {\scriptsize $\y$};
\node at (2,-1.1) {\small (a) $\bfk=(0,0)$, $x\in M_\varnothing$};
\end{scope}
\begin{scope}[xshift=6.3cm]
\draw[step=1,grid] (0,0) grid (4,4);
\foreach \y in {0,...,4} \draw[arr] (1,\y)--(2,\y);
\foreach \x in {0,...,4} \foreach \y in {0,...,4} \node[off] at (\x,\y) {};
\foreach \x in {2,...,4} \foreach \y in {0,...,4} \node[on] at (\x,\y) {};
\node[above] at (1.5,4.05) {\scriptsize $a$};
\foreach \x in {0,...,4} \node[below] at (\x,-0.1) {\scriptsize $\x$};
\node at (2,-1.1) {\small (b) $\bfk=(2,0)$, $a\in M_{1}$};
\end{scope}
\begin{scope}[xshift=12.6cm]
\draw[step=1,grid] (0,0) grid (4,4);
\foreach \x in {0,...,4} \draw[arr] (\x,2)--(\x,3);
\foreach \x in {0,...,4} \foreach \y in {0,...,4} \node[off] at (\x,\y) {};
\foreach \x in {0,...,4} \foreach \y in {3,4} \node[on] at (\x,\y) {};
\node[right] at (4.05,2.5) {\scriptsize $b$};
\foreach \x in {0,...,4} \node[below] at (\x,-0.1) {\scriptsize $\x$};
\node at (2,-1.1) {\small (c) $\bfk=(0,3)$, $b\in M_{2}$};
\end{scope}
\begin{scope}[xshift=18.9cm]
\fill[blue!25] (0,2) rectangle (1,3);
\draw[step=1,grid] (0,0) grid (4,4);
\foreach \y in {3,4} \draw[arr] (0,\y)--(1,\y);
\foreach \x in {1,...,4} \draw[arr] (\x,2)--(\x,3);
\foreach \x in {0,...,4} \foreach \y in {0,...,4} \node[off] at (\x,\y) {};
\foreach \x in {1,...,4} \foreach \y in {3,4} \node[on] at (\x,\y) {};
\node at (0.5,2.5) {\scriptsize $c$};
\foreach \x in {0,...,4} \node[below] at (\x,-0.1) {\scriptsize $\x$};
\node at (2,-1.1) {\small (d) $\bfk=(1,3)$, $c\in M_{12}$};
\end{scope}
\end{tikzpicture}
\caption{The four kinds of orthant steps in bidegree $(4,4)$; the
horizontal coordinate is direction $1$, the vertical one direction $2$.
A filled dot marks a vertex carrying the element of $M_\varnothing$
indicated below, a thick edge an arrow carrying an element of $M_1$ or
$M_2$, and the shaded cell the one square carrying the element of
$M_{12}$; open dots, thin edges, and unshaded cells carry $0$. All
filled vertices in one panel carry the same element: $x$ in (a),
$\mu_1a$ in (b), $\mu_2b$ in (c), $\mu_1\mu_2c$ in (d); likewise all
thick horizontal edges in (d) carry $\mu_2c$ and all thick vertical
edges carry $\mu_1c$. The nonzero elements are supported on the orthant
$\{u\ge k_1,\ v\ge k_2\}$ together with the edges and the cell
crossing into it.}
\label{fig:orthant-steps}
\end{figure}
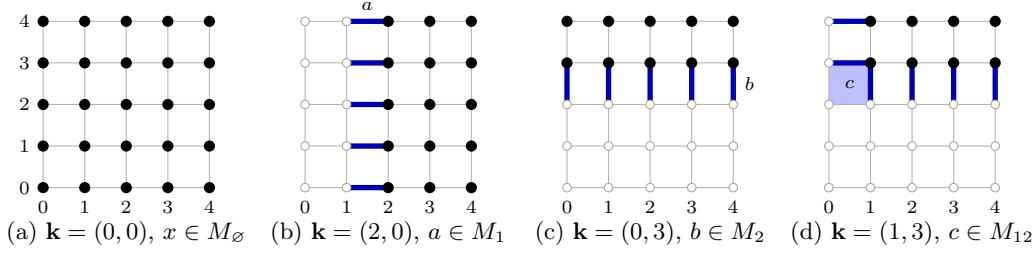

\emph{Abelian cube, two directions.} For $I=\{1,2\}$ the cube is a commuting square
$$
\begin{tikzcd}
M_{12}\ar[r,"\mu_2"]\ar[d,"\mu_1"'] & M_1\ar[d,"\mu_1"]\\
M_2\ar[r,"\mu_2"'] & M_\varnothing ,
\end{tikzcd}
$$
and an element of the binerve in bidegree $(m_1,m_2)$ is a decorated two-dimensional grid:
vertices $x_{u,v}\in M_\varnothing$ for $0\le u\le m_1$, $0\le v\le
m_2$; horizontal arrows $a_{u,v}\in M_1$ on the edges from $(u-1,v)$ to
$(u,v)$; vertical arrows $b_{u,v}\in M_2$ on the edges from $(u,v-1)$
to $(u,v)$; and squares $c_{u,v}\in M_{12}$ on the cells
$[u-1,u]\times[v-1,v]$, $u,v\geq1$. They are subject to constraints 
$$
x_{u,v}-x_{u-1,v}=\mu_1 a_{u,v},\quad
x_{u,v}-x_{u,v-1}=\mu_2 b_{u,v},\quad
a_{u,v}-a_{u,v-1}=\mu_2 c_{u,v},\quad
b_{u,v}-b_{u-1,v}=\mu_1 c_{u,v}.
$$
The free data are the vertex $x_{0,0}$, the arrows $a_{u,0}$ and
$b_{0,v}$ along the two axes, and all squares $c_{u,v}$. Everything
else is recovered by summation:
\begin{equation}\label{eq:heaviside-2d}
x_{u,v}=x_{0,0}
+\sum_{k_1}\mu_1a_{k_1,0}\,H_{k_1}(u)
+\sum_{k_2}\mu_2b_{0,k_2}\,H_{k_2}(v)
+\sum_{k_1,k_2}\mu_1\mu_2c_{k_1,k_2}\,H_{k_1}(u)H_{k_2}(v).
\end{equation}
This is a two-dimensional version of \eqref{eq:heaviside-1d}: a function on
a grid is determined by its value at the origin, its increments along
the two axes, and its mixed second differences. Thus the binerve in
bidegree $(m_1,m_2)$ is canonically isomorphic to 
$$
M_\varnothing\oplus M_1^{\,m_1}\oplus M_2^{\,m_2}\oplus M_{12}^{\,m_1m_2},
$$
with one summand $M_{A(\bfk)}$ for each pair $\bfk=(k_1,k_2)$
with $0\le k_i\le m_i$, where $A(\bfk)=\{i:k_i>0\}$. The summand
indexed by $\bfk$ consists of the decorated grids supported on the orthant
$\{(u,v):u\ge k_1,\ v\ge k_2\}$: for $a\in M_{A(\bfk)}$ the element
$a$ itself sits on the cells that span the interval $[k_i-1,k_i]$ in
each direction $i\in A(\bfk)$ and are reduced to a point in the
other directions. Thus, if $k_1,k_2>0$, $a\in M_{12}$ decorates a single square $[k_1-1,k_1]\times[k_2-1,k_2]$; if $\bfk=(k_1,0)$, $a\in M_1$ decorates a  whole column of horizontal edges
$[k_1-1,k_1]\times\{v\}$, $0\le v\le m_2$; if $\bfk=(0,k_2)$, $a\in M_2$ decorates 
a whole row of vertical edges $\{u\}\times[k_2-1,k_2]$,
$0\le u\le m_1$; if $k_1=k_2=0$, $a\in M_\varnothing$ decorates all vertices of the grid. Its images under the structure
maps fill the rest of the orthant, and everything outside the orthant
is zero. We call such a grid the \emph{orthant step} of $a$ with corner
$\bfk$; the positive coordinates $k_i$, $i\in A(\bfk)$, are its
\emph{thresholds}, and the cells carrying $a$ are those crossing all
of them. The four kinds of orthant steps, according to which of
$k_1,k_2$ are positive, are shown in Figure~\ref{fig:orthant-steps}.
Note that the vertices of the orthant all carry the same element
$\mu_{A(\bfk),\varnothing}a$, which may be nonzero, while the cells of positive dimension
inside the orthant carry $0$ unless they cross a threshold: for
$\bfk=(2,0)$, say, every vertex with $u\ge2$ carries $\mu_1a$,
the edges from $u=1$ to $u=2$ carry $a$, and all other edges and
all squares carry $0$.

\emph{The nonabelian case.}
In general, the vertex groups are not abelian and the summands do not
commute. The lemma below says that the above description of the multinerve survives, with sums
replaced by ordered products: every element of the multinerve is
uniquely an ordered product of orthant steps, one for each corner
$\bfk$, and the commutator of two orthant steps is an orthant step
supported on the intersection of the two orthants, with the pairing
$h_{A,B}$ supplying the element at its corner.

For a multidegree $\bfm=(m_i)_{i\in I}\in\NN^I$ put
$$
T_{\bfm}=\prod_{i\in I}\{0,1,\ldots,m_i\},
\qquad
A(\bfk)=\{i\in I:k_i>0\}\quad(\bfk\in T_{\bfm}),
$$
and for $0\le k\le m$ regard the Heaviside function $H_k$ of
\eqref{eq:heaviside-1d}, $H_k(u)=1$ for $u\ge k$ and $0$ for $u<k$,
as a monotone map $H_k:[m]\to[1]$; thus $H_0$ is the constant map with
value $1$.

Let $\scrC=(C_0,C_1,\sigma,\tau,\varepsilon,\mu)$ be an internal
category in groups. The nerve $N\scrC$ is a functor
$\Delta^{\mathrm{op}}\to\Grp$, so every morphism $\beta:[m']\to[m]$ of
$\Delta$ induces a homomorphism
$\beta^*=(N\scrC)(\beta):(N\scrC)_m\to(N\scrC)_{m'}$, the
\emph{simplicial operator} induced by $\beta$, and
$(\beta\circ\beta')^*=\beta'^*\beta^*$. We need only the operators
induced by the monotone maps $[m]\to[1]$, which are the Heaviside maps
$H_k$, $0\le k\le m$, and the constant map $c_0$ with value $0$. Since
an $m$-simplex of the nerve is a functor $[m]\to\scrC$, the operator
induced by such a map sends an arrow $g:x_0\to x_1$ to the string
obtained by composing it with the map: for $1\le k\le m$,
$$
H_k^*(g)=
\bigl(\varepsilon\sigma g,\ldots,\varepsilon\sigma g,\ g,\
\varepsilon\tau g,\ldots,\varepsilon\tau g\bigr)
\quad\text{with $g$ in position $k$},
$$
while
$H_0^*(g)=(\varepsilon\tau g,\ldots,\varepsilon\tau g)$ and
$c_0^*(g)=(\varepsilon\sigma g,\ldots,\varepsilon\sigma g)$. 
For a $\cat^I$-group $\Gamma$ and $i\in I$, we write $H_k^{*}$ and
$c_0^*$ also for the same operators applied in direction $i$ of the
multinerve, i.e.\ for the nerve of the internal category
$(\Gamma,s_i,t_i)$ applied levelwise; operators in distinct
directions commute (Lemma~\ref{lem:nerve-limits}). Recall that in each
direction the nerve in degree one is the arrow object, so that
$(N^I\Gamma_IM_\bullet)_{(1,\ldots,1)}=\Gamma_I(M_\bullet)$.

\begin{definition}[Orthant steps]\label{def:orthant-step}
For $\bfk\in T_{\bfm}$ define
$$
u_{\bfk}=u^{M}_{\bfk}:
M_{A(\bfk)}\longrightarrow (N^I\Gamma_I M_\bullet)_{\bfm},
\qquad
u_{\bfk}(a)=\Bigl(\prod_{i\in I}H_{k_i}^{*}\Bigr)\,\iota_{A(\bfk)}(a),
$$
where $H_{k_i}^*$ is the operator induced by $H_{k_i}:[m_i]\to[1]$ in
direction $i$ and the operators are applied in any order. For
$I=\varnothing$, $u_{()}$ is the identity of $M_\varnothing$. We omit the
superscript $M$ when no confusion can arise.
\end{definition}

Each $u_{\bfk}$ is a homomorphism, being a composite of
$\iota_{A(\bfk)}$ with simplicial operators. In a direction
$i\notin A(\bfk)$, the element $\iota_{A(\bfk)}(a)$ lies in
$\im s_i$, where $s_i=t_i$, so $H_0^*$ (or equally $c_0^*$)
makes $u_{\bfk}(a)$ constant in that direction. In a direction
$i\in A(\bfk)$, $H_{k_i}^*$ places $\iota_{A(\bfk)}(a)$ at step
$k_i$, with identity arrows at $s_i\iota_{A(\bfk)}(a)=1$ before that
step and at
$t_i\iota_{A(\bfk)}(a)=\iota_{A(\bfk)\setminus\{i\}}(\mu_ia)$
after it. Thus $u_{\bfk}(a)$ is the orthant step of $a$ with corner
$\bfk$ of the abelian discussion, with sums replaced by products:
the positive $k_i$ are its thresholds and it is constant in the
directions with $k_i=0$.

\begin{lemma}[Grid normal form]\label{lem:grid-normal-form}
Let $M_\bullet$ be a crossed $I$-cube and $\bfm\in\NN^I$. Fix a
total order on $I$ and order $T_{\bfm}$ lexicographically,
comparing coordinates in the order of $I$.
\begin{enumerate}
\item Every element of $(N^I\Gamma_I M_\bullet)_{\bfm}$ has a unique
expression
\begin{equation}\label{eq:grid-normal-form}
\prod_{\bfk\in T_{\bfm}}u_{\bfk}(a_{\bfk}),
\qquad a_{\bfk}\in M_{A(\bfk)},
\end{equation}
with the product taken in decreasing order, so that the last factor
is $u_{\mathbf 0}(a_{\mathbf 0})$ with $a_{\mathbf 0}\in M_\varnothing$.
\item For $\bfk,\bfell\in T_{\bfm}$, $a\in M_{A(\bfk)}$
and $b\in M_{A(\bfell)}$,
\begin{equation}\label{eq:grid-presentation}
[u_{\bfk}(a),u_{\bfell}(b)]
=u_{\bfk\vee\bfell}\bigl(h_{A(\bfk),A(\bfell)}(a,b)\bigr),
\end{equation}
where $(\bfk\vee\bfell)_i=\max(k_i,\ell_i)$, so that
$A(\bfk\vee\bfell)=A(\bfk)\cup A(\bfell)$.
Moreover, these relations present the group: the homomorphism
$$
\ast_{\bfk\in T_{\bfm}}M_{A(\bfk)}\longrightarrow
(N^I\Gamma_I M_\bullet)_{\bfm}
$$
from the free product which is $u_{\bfk}$ on the factor
$M_{A(\bfk)}$ is surjective, and its kernel is the normal subgroup
generated by the elements
$[j_{\bfk}(a),j_{\bfell}(b)]\,
j_{\bfk\vee\bfell}\bigl(h_{A(\bfk),A(\bfell)}(a,b)\bigr)^{-1}$,
where $j_{\bfk}$ denotes the free-product inclusion of
$M_{A(\bfk)}$.
\item A morphism $f_\bullet:M_\bullet\to M'_\bullet$ of crossed $I$-cubes
acts on generators by
$N^I\Gamma_I(f_\bullet)\,u^{M}_{\bfk}(a)=u^{M'}_{\bfk}(f_{A(\bfk)}a)$.
\end{enumerate}
\end{lemma}

The proof is not complicated, in principle, but is rather long, so we relegate it to Appendix \ref{app:internal-categories}.

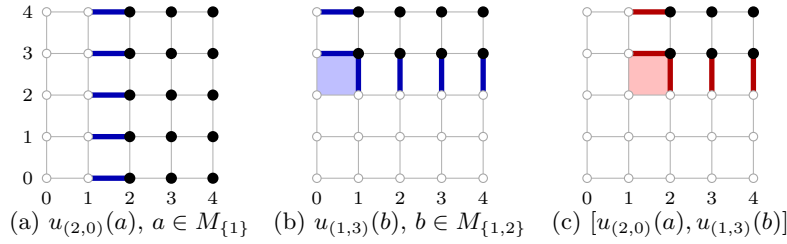
\begin{figure}[htb]
\centering
\begin{tikzpicture}[scale=0.55]
\tikzset{grid/.style={gray!60,thin},
         on/.style={circle,fill=black,inner sep=0pt,minimum size=4.2pt},
         off/.style={circle,draw=gray!70,fill=white,inner sep=0pt,minimum size=3.2pt},
         arr/.style={line width=2pt,blue!70!black}}
\begin{scope}
\draw[step=1,grid] (0,0) grid (4,4);
\foreach \y in {0,...,4} \draw[arr] (1,\y)--(2,\y);
\foreach \x in {0,...,4} \foreach \y in {0,...,4} \node[off] at (\x,\y) {};
\foreach \x in {2,...,4} \foreach \y in {0,...,4} \node[on] at (\x,\y) {};
\foreach \x in {0,...,4} \node[below] at (\x,-0.1) {\scriptsize $\x$};
\foreach \y in {0,...,4} \node[left] at (-0.1,\y) {\scriptsize $\y$};
\node at (2,-1.1) {\small (a) $u_{(2,0)}(a)$, $a\in M_{\{1\}}$};
\end{scope}
\begin{scope}[xshift=6.5cm]
\fill[blue!25] (0,2) rectangle (1,3);
\draw[step=1,grid] (0,0) grid (4,4);
\foreach \y in {3,4} \draw[arr] (0,\y)--(1,\y);
\foreach \x in {1,...,4} \draw[arr] (\x,2)--(\x,3);
\foreach \x in {0,...,4} \foreach \y in {0,...,4} \node[off] at (\x,\y) {};
\foreach \x in {1,...,4} \foreach \y in {3,4} \node[on] at (\x,\y) {};
\foreach \x in {0,...,4} \node[below] at (\x,-0.1) {\scriptsize $\x$};
\node at (2,-1.1) {\small (b) $u_{(1,3)}(b)$, $b\in M_{\{1,2\}}$};
\end{scope}
\begin{scope}[xshift=13cm]
\fill[red!25] (1,2) rectangle (2,3);
\draw[step=1,grid] (0,0) grid (4,4);
\foreach \y in {3,4} \draw[line width=2pt,red!70!black] (1,\y)--(2,\y);
\foreach \x in {2,...,4} \draw[line width=2pt,red!70!black] (\x,2)--(\x,3);
\foreach \x in {0,...,4} \foreach \y in {0,...,4} \node[off] at (\x,\y) {};
\foreach \x in {2,...,4} \foreach \y in {3,4} \node[on] at (\x,\y) {};
\foreach \x in {0,...,4} \node[below] at (\x,-0.1) {\scriptsize $\x$};
\node at (2,-1.1) {\small (c) $[u_{(2,0)}(a),u_{(1,3)}(b)]$};
\end{scope}
\end{tikzpicture}
\caption{The commutator of two orthant steps, drawn with the
conventions of Figure~\ref{fig:orthant-steps}. The commutator (c) is the
orthant step $u_{(2,3)}\bigl(h(a,b)\bigr)$ with corner
$(2,3)=(2,0)\vee(1,3)$: its filled vertices are exactly the vertices
filled in both (a) and (b), and its cell is the one crossing both
thresholds.}
\label{fig:grid}
\end{figure}

\begin{remark}[The grid picture]\label{rem:grid-picture}
For general $I$, an element of $(N^I\Gamma_IM_\bullet)_{\bfm}$ is a decorated 
grid $\prod_i[m_i]$ whose vertices, edges, and higher cells carry
elements of $\Gamma_I(M_\bullet)$, compatibly along faces, and
$u_{\bfk}(a)$ is the orthant step of $a$ with corner $\bfk$. Figure~\ref{fig:grid} illustrates the
relation \eqref{eq:grid-presentation} in the case $|I|=2$: the commutator of the orthant
steps with corners $\bfk$ and $\bfell$ is supported on the
intersection of the two orthants, whose corner is
$\bfk\vee\bfell$, and the pairing $h_{A,B}(a,b)$ is the
element placed there. This is the same pattern as in the definition of
a crossed precosheaf, where the commutator of transformations
localized in two regions is localized in their intersection, with
orthants of the grid in the role of regions.
\end{remark}

The simplicial operators act on orthant steps by moving thresholds.
For a morphism $\alpha:[m']\to[m]$ in $\Delta$ define
$$
\lambda_\alpha:\{0,\ldots,m\}\longrightarrow\{0,\ldots,m'\}\cup\{\infty\},
\qquad
\lambda_\alpha(k)=
\begin{cases}
0,&k=0,\\
\min\{v\in[m']:\alpha(v)\geq k\},
   &k>0\text{ and this set is nonempty},\\
\infty,&\text{otherwise}.
\end{cases}
$$

\begin{lemma}
\label{lem:grid-operators}
Let $\alpha=(\alpha_i)_{i\in I}$ with $\alpha_i:[m'_i]\to[m_i]$ in
$\Delta$, and let
$\alpha^*:(N^I\Gamma_IM_\bullet)_{\bfm}\to(N^I\Gamma_IM_\bullet)_{\bfm'}$
be the induced operator. Then for $\bfk\in T_{\bfm}$ and
$a\in M_{A(\bfk)}$,
\begin{equation}\label{eq:grid-operators}
\alpha^*u_{\bfk}(a)=
\begin{cases}
1,&\lambda_{\alpha_i}(k_i)=\infty\text{ for some }i\in A,\\
u_{\lambda_\alpha\bfk}\bigl(\mu_{A,A'}a\bigr),&\text{otherwise, where }
A'=\{i\in A:\lambda_{\alpha_i}(k_i)>0\},
\end{cases}
\end{equation}
where $A=A(\bfk)$ and $(\lambda_\alpha\bfk)_i=\lambda_{\alpha_i}(k_i)$,
so that $A(\lambda_\alpha\bfk)=A'$.
\end{lemma}
\begin{remark}
In words, for each direction $i\in A$ the threshold $k_i$ of the
orthant step is, relative to the new grid, in one of three positions:
\emph{unreached} if $\alpha_i(v)<k_i$ for all $v$
($\lambda_{\alpha_i}(k_i)=\infty$), so that the new grid lies entirely
in front of the threshold; \emph{passed} if $\alpha_i(0)\ge k_i$
($\lambda_{\alpha_i}(k_i)=0$), so that the new grid lies entirely beyond
it; and \emph{crossed} otherwise, between $\lambda-1$ and $\lambda$
where $\lambda=\lambda_{\alpha_i}(k_i)$, so that
$\alpha_i(\lambda-1)<k_i\le\alpha_i(\lambda)$. An unreached threshold
kills the orthant step. A passed threshold makes it constant in
direction $i$ and applies $\mu_i$ to $a$, so that $i$ leaves $A$; these
are the directions collected in $\mu_{A,A'}$. A crossed threshold
moves to its new position $\lambda$, with no structure map applied.
Directions with $k_i=0$ are unaffected. For example, in one direction
with $m=3$ and $c\in M_{\{1\}}$, the face $\fd_0$, induced by
$v\mapsto v+1$, sends $u_2(c)=(1,c,\varepsilon\tau c)$ to
$u_1(c)=(c,\varepsilon\tau c)$ (crossed) and $u_1(c)$ to
$u_0(\mu_1c)=(\varepsilon\tau c,\varepsilon\tau c)$ (passed), while the face $\fd_3$ sends
$u_3(c)$ to $1$ (unreached).
\end{remark}

\begin{proof}
Since $u_{\bfk}(a)=\bigl(\prod_iH_{k_i}^*\bigr)\iota_A(a)$,
operators in distinct directions commute, and
$\alpha_i^*H_{k_i}^*=(H_{k_i}\circ\alpha_i)^*$ in direction $i$, we have
$\alpha^*u_{\bfk}(a)=\bigl(\prod_i(H_{k_i}\circ\alpha_i)^*\bigr)\iota_A(a)$.
For $k_i=0$, $H_0\circ\alpha_i=H_0$.
Let $k_i>0$ and $\lambda=\lambda_{\alpha_i}(k_i)$. If $\lambda=\infty$,
then $\alpha_i(v)<k_i$ for all $v$, so $H_{k_i}\circ\alpha_i=c_0$, and
$c_0^*$ produces the string of identities at $s_i\iota_A(a)=1$, because
$i\in A$. If $0<\lambda<\infty$, then $H_{k_i}\circ\alpha_i=H_\lambda$
because $\alpha_i$ is monotone. If $\lambda=0$, then
$\alpha_i(0)\geq k_i$, so $H_{k_i}\circ\alpha_i=H_0$, and $H_0^*$
produces the string of identities at
$t_i\iota_A(a)=\iota_{A\setminus\{i\}}(\mu_ia)$. Combining the directions gives \eqref{eq:grid-operators}, with
$\mu_{A,A'}$ the composite of the $\mu_i$ over the directions $i$ with
$0=\lambda_{\alpha_i}(k_i)<k_i$.
\end{proof}

\section{The Loday classifying complex and fibration theorems}\label{sec:loday}

\subsection{The Loday classifying complex}

Recall how the classifying complex of a simplicial group \cite{EilenbergMacLane1953} is expressed through the Artin-Mazur codiagonal functor \cite{ArtinMazur1966}.

\begin{definition}[$\bW$-construction]
\label{def:W-construction}
Let $X:(\Delta^{\mathrm{op}})^2\to\Set$ be a bisimplicial set.
Write $\fd_i^h,\fs_i^h$ for the simplicial operators in the first index
and $\fd_i^v,\fs_i^v$ for those in the second. Its
\emph{Artin-Mazur codiagonal} is the simplicial set $\nabla X$ with
$$
(\nabla X)_n
=
\left\{
(x_0,\ldots,x_n)\in\prod_{p=0}^n X_{p,n-p}
\;\middle|\;
\fd_0^v x_p=\fd_{p+1}^h x_{p+1}\text{ for }0\leq p<n
\right\},
\qquad n\geq0.
$$
For $x=(x_0,\ldots,x_n)\in(\nabla X)_n$, the faces
$\fd_i:(\nabla X)_n\to(\nabla X)_{n-1}$, for $n\geq1$ and
$0\leq i\leq n$, are given by
$$
(\fd_i x)_p=
\begin{cases}
\fd_{i-p}^v x_p,&0\leq p<i,\\[2mm]
\fd_i^h x_{p+1},&i\leq p\leq n-1.
\end{cases}
$$
The degeneracies $\fs_i:(\nabla X)_n\to(\nabla X)_{n+1}$, for
$n\geq0$ and $0\leq i\leq n$, are given by
$$
(\fs_i x)_p=
\begin{cases}
\fs_{i-p}^v x_p,&0\leq p\leq i,\\[2mm]
\fs_i^h x_{p-1},&i<p\leq n+1.
\end{cases}
$$
These formulas preserve the matching conditions and satisfy the
simplicial identities; see
\cite[Section~6.2]{MutluPorterCrossedSquares}.
A morphism $f:X\to Y$ induces $\nabla f$ by applying $f_{p,n-p}$
to each coordinate $x_p$. If $X$ is a bisimplicial group, then
$(\nabla X)_n$ is a subgroup of the displayed product and all these
operators are homomorphisms, so $\nabla X$ is a simplicial group.

For a simplicial group $H$, the degreewise classifying complex of
Definition~\ref{def:degree-wise-classifying-complex} is the
bisimplicial set
$(B_\bullet H)_{p,q}=(H_q)^p$.
The \emph{reduced $W$-construction} is its codiagonal:
$$
\bW H:=\nabla(B_\bullet H).
$$
Thus $\bW:\sGrp\to\SSet$ is the reduced
classifying-complex functor.
\end{definition}

\begin{remark}
For any bisimplicial set $X_{\bullet,\bullet}$, $\nabla X$ is weakly equivalent to $\Diag X$ \cite{CegarraRemediosDiagonalBar}. Thus, for any simplicial group $H$, $\bW H\simeq \Diag(B_\bullet H)$. We use the codiagonal instead of the diagonal because $\bW H$ is a Kan complex for any $H$ \cite{MaySimplicialObjects}, and this simplifies some arguments. 
\end{remark}

\begin{definition}[Loday classifying functor]
\label{def:loday-classifying-complex}
For $n\geq1$, the \emph{Loday classifying functor} is the composite
$$
\BL\colon
\cat^n\Grp
\xrightarrow{\,N^n\,}
\mathbf{s}^n\Grp
\xrightarrow{\,\Diag\,}
\sGrp
\xrightarrow{\,\bW\,}
\SSet.
$$
Thus, for a $\cat^n$-group $G$, its \emph{Loday classifying
complex} is
$\BL G :=\bW\!\left(\Diag N^nG\right)$.

For a crossed $n$-cube $M_\bullet$, put $\BL M_\bullet:=\BL(\Gamma_n(M_\bullet))$, where $\Gamma_n$ is the Ellis-Steiner functor of Proposition~\ref{prop:catrecover}. The argument determines which construction $\BL$ denotes. Their images in $\Ho(\SSet)$ are called the corresponding \emph{Loday homotopy types}.
\end{definition}
\begin{remark}
    In \cite{Loday1982}, Loday defines the classifying space of a $\cat^n$-group $G$ as the geometric realization of $\Diag(B_\bullet N^nG)$. By the previous remark, this is weakly equivalent to the geometric realization of our $\BL G$.
\end{remark}

At each fixed arity these assignments are functorial. Thus
$\BL\colon\cat^n\Grp\to\SSet$, $\BL\colon\XCube_n\to\SSet$.
For $n=0$ both constructions reduce to the ordinary classifying complex
$BG$.

\begin{remark}
The category $\XCube_n$ carries a natural homotopy theory which, through
the Loday classifying functor $\BL$, is equivalent to the
homotopy theory of pointed connected $(n+1)$-types in simplicial sets.
The precise formulation and proof of this statement are given in
Appendix~\ref{app:model-structures-crossed-cubes}.
\end{remark}

\subsection{The basic fibration theorem}
Here we show that an internal category in simplicial groups gives rise to a homotopy fibration sequence. To formulate this result, it is convenient to introduce a classifying functor for $\cat^m$-objects in $\sGrp$, or equivalently in internal $m$-categories in $\sGrp$. We define
$$
\BDelta
:=\bW\circ\Diag\circ N^m:
\cat^m(\sGrp)\longrightarrow\SSet.
$$
For $\scrH \in \cat^m(\sGrp)$, $N^m\scrH$ is an $(m+1)$-simplicial group and $\Diag$ is its total diagonal. 
For $m=0$, this reduces to a classifying functor on simplicial groups
$\BDelta \scrH =\bW \scrH$ while for
$m=1$ it gives
$\BDelta\scrH=\bW\Diag N\scrH$ for an
internal category $\scrH$ in simplicial groups.

If $G$ is a $\cat^n$-group,
Lemma~\ref{lem:nerve-limits} shows that taking the nerve in one
internal-category direction preserves the remaining internal-category
structures and therefore gives
$$
NG
\in
\mathbf{s}(\cat^{n-1}\Grp)
\cong
\cat^{n-1}(\sGrp).
$$
Since the multinerve and the diagonal may be formed in stages, there is a
natural identification
$$
\BL G
\cong
\BDelta(NG).
$$

In this sense, $\BDelta$ is an analog of $\BL$ for internal multi-categories in simplicial groups. Thus the fibration theorem we prove here will extend naturally to $\BL$ in the next subsection. 

\begin{definition}[Homotopy fibration sequence]
\label{def:homotopy-fibration-sequence}
Let
$$
F\xrightarrow{i}E\xrightarrow{p}B
$$
be pointed maps between pointed Kan complexes. Use the homotopy-fiber
model
$$
\hofib_*(p)
:=E\times_{B\times B}B^{\Delta^1},
$$
where the maps to $B\times B$ are $(p,*)$ and
$(\ev_0,\ev_1)$, respectively, and
$B^{\Delta^1}$ is the simplicial path space.

The sequence is a \emph{homotopy fibration sequence} if there exists
a pointed simplicial nullhomotopy $H:p\circ i\simeq *$ such that
$$
(i,H^\#):F\longrightarrow\hofib_*(p)
$$
is a weak equivalence. Here $H^\#:F\to B^{\Delta^1}$ is adjoint to
$H:F\times\Delta^1\to B$, with
$H(-,0)=p\circ i$ and $H(-,1)=*$.
\end{definition}

This restricted formulation suffices here, since the classifying
complexes in our fibration sequences are of the form $\bW H$,
which is Kan for every simplicial group $H$.
For the general theory of homotopy fibers and homotopy fibration
sequences, see \cite[Ch.~II, \S9]{GoerssJardine} or \cite[\S\S13.3--13.4]{Hirschhorn}. A homotopy fibration sequence induces a long exact sequence of homotopy groups. 

\begin{theorem}[Crossed-module fibration]
\label{thm:crossed-module-fibration}
Let $\scrC=(C_0,C_1,\sigma,\tau,\varepsilon,\mu)$ be an internal
category in simplicial groups, and set
$$
K:=\ker(\sigma:C_1\to C_0),
\qquad
\partial:=\tau|_K:K\longrightarrow C_0.
$$
Let
$e_{\scrC}:C_0\to\Diag N\scrC$
be the homomorphism of simplicial groups which for $q>0$ sends $g\in C_{0,q}$ to the
string of $q$ identity arrows at $g$, $g\mapsto\left(\varepsilon(g),\ldots,\varepsilon(g)\right),$ and for $q=0$ is the identity map on $C_0$. Put
$p_{\scrC}:=\bW e_{\scrC}: \BDelta C_0\to\BDelta\scrC$.
Then
$$
\BDelta K\xrightarrow{\BDelta\partial}\BDelta C_0
\xrightarrow{p_{\scrC}}\BDelta\scrC
$$
is a homotopy fibration sequence. The displayed maps are natural with respect to internal functors $T:\scrC\to{\scrC}'$ where $K\to K'$ is the restriction of $T_1$.
\end{theorem}

The proof is rather long and is relegated to Appendix~\ref{app:crossed-module-fibration}.

\begin{remark}
    For internal categories in discrete groups (i.e. for discrete crossed modules), this result is proved by Brown and Higgins \cite{BrownHigginsClassifyingSpace}.
\end{remark}

\subsection{Face-fibration theorem for crossed cubes}

Let $M_\bullet$ be a crossed $(I\sqcup\{v\})$-cube, with faces
$\partial_v^0M$ and $\partial_v^1M$ in the $v$-direction and the
morphism $\mu_v:\partial_v^1M\to\partial_v^0M$ of Lemma~\ref{lem:faces}.

Apply the Ellis-Steiner functor to $M_\bullet$ to get a
$\cat^{I\sqcup\{v\}}$-group $\Gamma(M)$. Single out the $v$-direction and
regard $\Gamma(M)$ as a $\cat^1$-object in the category of
$\cat^I$-groups, with $s_v$ and $t_v$ denoting its source and target
morphisms. Theorem~\ref{thm:cat1-internal} applied to the category of $\cat^I$-groups (Remark \ref{rem:internal-transfer})
turns it into an internal category in $\cat^I$-groups with arrow object
$\Gamma(M)$ and object object $\im s_v$. Applying the
multinerve in the directions indexed by $I$, followed by the diagonal,
yields an internal category in simplicial groups
(Lemmas~\ref{lem:change-of-base} and~\ref{lem:nerve-limits}), which we
denote $\scrC_v(M)$; thus
$$
\scrC_v(M)_1=\Diag N^I\,\Gamma(M),
\qquad
\scrC_v(M)_0=\Diag N^I\,\im s_v.
$$

The following lemma expresses the ingredients of $\scrC_v(M)$ in terms of
the faces of $M_\bullet$.

\begin{lemma}\label{lem:Gv-identifications}
There are isomorphisms, natural in $M_\bullet$,
$$
\BDelta\ker\bigl(\sigma:\scrC_v(M)_1\to\scrC_v(M)_0\bigr)
\cong
\BL(\partial_v^1M),
\qquad
\BDelta\bigl(\scrC_v(M)_0\bigr)
\cong
\BL(\partial_v^0M),
$$
and a canonical natural isomorphism
$$
\BDelta\scrC_v(M)
\cong
\BL M,
$$
under which the map $\BDelta(\tau|_{\ker\sigma})$ becomes
$\BL(\mu_v)$.
\end{lemma}

\begin{proof}
Lemma~\ref{lem:faces} gives isomorphisms of $\cat^{I}$-groups, natural
in $M_\bullet$,
$$
\im(s_v)
\cong
\Gamma(\partial_v^0M),
\qquad
\ker(s_v)
\cong
\Gamma(\partial_v^1M),
$$
under which the restricted target map
$t_v|_{\ker(s_v)}:\ker(s_v)\to\im(s_v)$ is
$\Gamma(\mu_v)$.

Since the multinerve and the diagonal preserve kernels
(Lemmas~\ref{lem:nerve-limits} and~\ref{lem:levelwise-diag}),
applying $\Diag N^I$ gives
$$
\ker\sigma
=
\Diag N^I\ker(s_v)
\cong
\Diag N^I\,\Gamma(\partial_v^1M),
\qquad
\scrC_v(M)_0
\cong
\Diag N^I\,\Gamma(\partial_v^0M),
$$
with $\tau|_{\ker\sigma}$ becoming $\Diag N^I\,\Gamma(\mu_v)$. Applying
$\BDelta$ and computing the diagonal in stages
(Lemmas~\ref{lem:diag-basic} and~\ref{lem:levelwise-diag})
identifies $\BDelta\bigl(\Diag N^I\Gamma(-)\bigr)$ with
$\BL(-)$, which gives the first two isomorphisms and the
identification of $\BDelta(\tau|_{\ker\sigma})$ with
$\BL(\mu_v)$. For the last equality no normal form is needed:
by the same two lemmas,
$$
\BDelta\scrC_v(M)
=
\bW\Diag N\scrC_v(M)
\cong
\bW\Diag N^{I\sqcup\{v\}}\Gamma(M)
\cong
\BL M.
$$
\end{proof}

\begin{definition}\label{def:add-map}
The \emph{$v$-add-map}
$a_v(M): \BL(\partial_v^0M) \to \BL M$
is the composite
$$
\BL(\partial_v^0M)
\cong
\BDelta\bigl(\scrC_v(M)_0\bigr)
\xrightarrow{\,p_{\scrC_v(M)}\,}
\BDelta\scrC_v(M)
=
\BL M,
$$
where the first isomorphism is that of Lemma~\ref{lem:Gv-identifications} and
$p_{\scrC_v(M)}=\bW e_{\scrC_v(M)}$ is the map of
Theorem~\ref{thm:crossed-module-fibration} for the internal category
$\scrC_v(M)$. Thus $a_v(M)$ adds the $v$-nerve direction by filling that
direction with identity arrows. (In the grid picture of
Section~\ref{sec:grid}, $e_{\scrC_v(M)}$ is the inclusion of the
$I$-grids as the $(I\sqcup\{v\})$-grids that are constant in direction
$v$.)
\end{definition}

\begin{theorem}[Face fibration for crossed cubes]
\label{thm:face-fibration-crossed-cubes}
The maps in
$$
\BL(\partial_v^1M)
\xrightarrow{\,\BL(\mu_v)\,}
\BL(\partial_v^0M)
\xrightarrow{\,a_v(M)\,}
\BL M
$$
are natural in $M_\bullet$, and for every $M_\bullet$ this is a homotopy
fibration sequence.
\end{theorem}

\begin{proof}
Theorem~\ref{thm:crossed-module-fibration}, applied to the internal
category $\scrC_v(M)$ in simplicial groups, gives a homotopy fibration
sequence
$$
\BDelta\ker\sigma
\xrightarrow{\,\BDelta(\tau|_{\ker\sigma})\,}
\BDelta\bigl(\scrC_v(M)_0\bigr)
\xrightarrow{\,p_{\scrC_v(M)}\,}
\BDelta\scrC_v(M).
$$
Transporting
along the identifications of Lemma~\ref{lem:Gv-identifications} and
Definition~\ref{def:add-map} yields the asserted sequence. Naturality
follows from the functoriality of the Ellis-Steiner construction, the
naturality in Lemma~\ref{lem:Gv-identifications}, and the naturality
asserted in Theorem~\ref{thm:crossed-module-fibration}.
\end{proof}

\begin{corollary}
\label{cor:loday-connected-truncated}
Let $M_\bullet$ be a crossed $n$-cube, or $G$ a $\cat^n$-group. Then
$\BL M_\bullet$, respectively $\BL G$, is a connected
$(n+1)$-truncated homotopy type.
\end{corollary}

\begin{proof}
By Proposition~\ref{prop:catrecover}, every $\cat^n$-group is isomorphic
to $\Gamma_n(M_\bullet)$ for a crossed $n$-cube $M_\bullet$, so it
suffices to treat $\BL M_\bullet$. Connectedness holds because $\bW$
takes values in reduced simplicial sets. Truncation is proved by
induction on $n$. For $n=0$, $M_\bullet$ is a group $M$ and
$\BL M=BM$ is a $1$-type. For $n\geq1$, choose a direction $v$ and
apply Theorem~\ref{thm:face-fibration-crossed-cubes}: in the homotopy
fibration sequence
$\BL(\partial^1_vM)\to\BL(\partial^0_vM)\to\BL M$
the two faces are crossed $(n-1)$-cubes, so the outer terms are
$n$-truncated by the induction hypothesis, and the long exact sequence
of homotopy groups shows that $\BL M$ is $(n+1)$-truncated.
\end{proof}

\begin{remark}
This result is due to Loday \cite{Loday1982}; the proof of the key
Lemma~2.4 there is only sketched, and the above argument via
Theorem~\ref{thm:face-fibration-crossed-cubes} is independent of it.
\end{remark}

\subsection{Homotopy groups of crossed modules and crossed squares}
Here we apply the face-fibration theorem to crossed modules ($\XCube_1$) and crossed squares ($\XCube_2$). 

\begin{example}\label{ex:loday-crossed-module}
Let
$\partial:M\to P$
be a crossed module, regarded as a crossed $1$-cube. Under the
correspondence between crossed modules and internal categories in groups,
Theorem~\ref{thm:crossed-module-fibration} gives a homotopy fibration
sequence
$$
BM
\xrightarrow{\,B\partial\,}
BP
\longrightarrow
\BL(\partial).
$$
Since $BM$ and $BP$ are connected $1$-types, the associated long exact
homotopy sequence gives
$$
\pi_1 \BL(\partial)
\cong
P/\partial(M),
\qquad
\pi_2 \BL(\partial)
\cong
\ker\partial,
$$
and
$$
\pi_k \BL(\partial)=0
\qquad (k\geq3).
$$
Thus $\BL(\partial)$ is the connected $2$-type associated with
the crossed module. The action of $\pi_1$ on $\pi_2$ is induced by the
given action of $P$ on $M$; this descends to $P/\partial(M)$ because
$\partial(M)$ acts trivially on $\ker\partial$ by the Peiffer identity.
\end{example}

Before computing the homotopy groups of crossed square, we first need a few additional definitions and tools. 

Let
$U\colon\sGrp\to\SSet$
denote the forgetful functor. For a simplicial group $G$, we write
$\pi_n(G):=\pi_n(UG,e)$, where $e$ is the identity vertex. The underlying
simplicial set $UG$ is a Kan complex, and its set of components
$\pi_0(UG)$ inherits a natural group structure from $G$; see
\cite[Section~17]{MaySimplicialObjects}.

\begin{lemma}\label{lem:W-homotopy-shift}
For every simplicial group $G$, there are natural isomorphisms
$$
\pi_{n+1}(\bW G)\cong\pi_n(UG),
\qquad n\geq0.
$$
\end{lemma}

\begin{proof}
May's \cite{MaySimplicialObjects} universal principal twisted Cartesian product gives a simplicial
fibration sequence
$$
UG\longrightarrow WG\longrightarrow\bW G.
$$
Its total space $WG$ is contractible by
\cite[Proposition~21.5]{MaySimplicialObjects}. The result follows from the
associated homotopy exact sequence, including in degree zero.
\end{proof}

The homotopy groups of a simplicial group also admit a purely algebraic description in terms of the Moore complex. Recall that the Moore complex $\Moore(G)$ of a simplicial group $G$ is given by
$$
\Moore(G)_n=\bigcap_{j=0}^{n-1}\ker(\fd_j\colon G_n\to G_{n-1}),
\qquad
\partial_n=\fd_n|_{\Moore(G)_n}.
$$

\begin{lemma}\label{lem:moore-computes-homotopy}
For every simplicial group $G$, there are natural isomorphisms
$$
\pi_n(UG)\cong H_n(\Moore(G)),
\qquad n\geq0.
$$
In degree zero this reads
$$
\pi_0(UG)\cong H_0(\Moore(G))
=G_0\big/\fd_1(\ker \fd_0).
$$
\end{lemma}

\begin{proof}
This is \cite[Proposition~17.4]{MaySimplicialObjects}
\end{proof}

Now, we will compute the homotopy groups of a crossed square using Moore complex. 
\begin{example}[Homotopy groups of crossed square]
\label{ex:loday-crossed-square}
   Now let $S_\bullet$ be the crossed square 
$$
\begin{matrix}
L & \xrightarrow{\lambda'} & N\\
\lambda\big\downarrow && \big\downarrow\nu\\
M & \xrightarrow{\mu} & P,
\end{matrix}
\qquad
h:M\times N\longrightarrow L,
$$
Let
$X_{p,q}:=\left(N^2\Gamma_2(S_\bullet)\right)_{q,p}$
be the binerve of its associated $\cat^2$-group. The indices $p,q$ were exchanged to agree with the conventions of \cite{MutluPorterCrossedSquares}. By Lemma \ref{lem:diagonal-codiagonal} in Appendix \ref{app:model-structures-crossed-cubes}, taking the diagonal and codiagonal results in weakly equivalent simplicial groups:
$$
\Diag X\simeq \nabla X.
$$
Therefore $\pi_i(U \Diag X)\cong \pi_i(U\nabla X)$.

The Moore complex of $\nabla X$ is computed in \cite{MutluPorterCrossedSquares} as
$$
L
\xrightarrow{\,\partial_2\,}
M\rtimes N
\xrightarrow{\,\partial_1\,}
P,
$$
where $N$ acts on $M$ through $\nu\colon N\to P$ and the given
$P$-action. The maps are given by
$$
\partial_2(l)
=
\bigl(\lambda(l)^{-1},\lambda'(l)\bigr),
\qquad
\partial_1(m,n)
=
\mu(m)\nu(n).
$$
See \cite[Sections~6.3 and~7]{MutluPorterCrossedSquares}. By
Lemma~\ref{lem:moore-computes-homotopy}, it follows that
$$
\pi_i(U\nabla X)
\cong
H_i\!\left(
L\longrightarrow M\rtimes N\longrightarrow P
\right),
\qquad 0\leq i\leq2.
$$

Combining this calculation with Lemma~\ref{lem:W-homotopy-shift} gives
\begin{align*}
\pi_{i+1}\bigl(\BL S_\bullet\bigr)
&=
\pi_{i+1}\bigl(\bW\Diag X\bigr)
\\
&\cong
\pi_i\bigl(U\Diag X\bigr)
\\
&\cong
\pi_i(U\nabla X)
\\
&\cong
H_i\!\left(
L\longrightarrow M\rtimes N\longrightarrow P
\right),
\qquad 0\leq i\leq2.
\end{align*}
More explicitly,
\begin{align*}
\pi_1\bigl(\BL S\bigr)
&\cong
P\big/\bigl(\mu(M)\nu(N)\bigr),
\\
\pi_2\bigl(\BL S\bigr)
&\cong
\ker(\partial_1)\big/\im(\partial_2),
\\
\pi_3\bigl(\BL S\bigr)
&\cong
\ker(\partial_2).
\end{align*}
Moreover, $\pi_k(\BL S_\bullet)=0$ for $k\geq4$. 
\end{example}

\section{The \v{C}ech-Loday functor and pro-homotopy type}\label{sec:cech-loday}

\subsection{Reindexing crossed cubes}

The crossed-precosheaf axioms are stated fiberwise over finite indexing
sets.  We now combine fixed-arity morphisms and changes of arity in single
source and target categories.  This requires the operation of reindexing a
crossed cube along a function of indexing sets, which we define first.

\begin{definition}[Reindexing crossed cubes]\label{def:reindexing-crossed-cubes}
For a finite set $I$, let $\XCube_I$ denote the
category of crossed cubes of groups whose directions are labeled by the
elements of $I$. Thus a crossed $I$-cube is an ordinary crossed $|I|$-cube
together with the chosen labeling of its coordinates by $I$.

Let $r:J\to I$ be a function of finite sets, and let $M_\bullet$ be a crossed
$I$-cube. We define data indexed by subsets $A\subseteq J$ as follows. The
vertex at $A$ is
$$
(r^*M)_A:=M_{r(A)},
\qquad
A\subseteq J,
$$
where $r(A)\subseteq I$ denotes the image subset.

For clarity, write
$\mu^M_{i,B}:M_B\to M_{B\setminus\{i\}}$
for the $i$-direction structure map of $M_\bullet$ at the vertex $B\subseteq I$;
if $i\notin B$, this is the identity of $M_B$. For $j\in J$ and
$A\subseteq J$, the $j$-direction structure map of $r^*M$ at the vertex $A$ is
$$
\mu^{r^*M}_{j,A}:
(r^*M)_A=M_{r(A)}
\longrightarrow
(r^*M)_{A\setminus\{j\}}=M_{r(A\setminus\{j\})}.
$$
It is defined by
$$
\mu^{r^*M}_{j,A}:=
\begin{cases}
\mu^M_{r(j),r(A)},
& j\in A\text{ and }r(j)\notin r(A\setminus\{j\}),\\[4pt]
\id_{M_{r(A)}},
& j\notin A\text{ or }r(j)\in r(A\setminus\{j\}).
\end{cases}
$$
In the second case the displayed identity has the indicated codomain because
$r(A\setminus\{j\})=r(A)$. Thus the source and target are part of the notation:
$\mu^{r^*M}_{j,A}$ is a map from the vertex $A$ to the vertex
$A\setminus\{j\}$.
The pairings are
$$
h^{r^*M}_{A,B}:=h^M_{r(A),r(B)}:
M_{r(A)}\times M_{r(B)}\longrightarrow M_{r(A\cup B)}.
$$
If $f:M_\bullet\to N_\bullet$ is a morphism of crossed $I$-cubes, define
$r^*f:r^*M_\bullet\to r^*N_\bullet$ by
$(r^*f)_A:=f_{r(A)}$.
\end{definition}

\begin{lemma}[Reindexing as a functor]\label{lem:reindexing-crossed-cubes}
The data of Definition~\ref{def:reindexing-crossed-cubes} form a crossed
$J$-cube. Moreover, the assignments $M_\bullet\mapsto r^*M_\bullet$ and
$f\mapsto r^*f$ are functorial in $M_\bullet$, and reindexing is
contravariantly functorial in $r$.
\end{lemma}

The proof is a tedious verification of the definition of a crossed cube and is given in Appendix~\ref{app:categorical}.

\begin{lemma}\label{lem:reindex-cech-cube}
Let $\CF$ be a crossed precosheaf on $\SL$, let $r:J\to I$ be a function of finite sets, and let $\xi\in\Cech_I(\SL)$. Then
$$
\Cech_{\CF}(r^*\xi)=r^*\Cech_{\CF}(\xi)
$$
as crossed $J$-cubes.
\end{lemma}
\begin{proof}
Both sides have vertex group $\CF(U^\xi_{r(A)})$ at $A\subseteq J$. The structure map of the left side in the direction $j\in A$ is the corestriction along $U^\xi_{r(A)}\leq U^\xi_{r(A\setminus\{j\})}$, which is the identity when $r(j)\in r(A\setminus\{j\})$ and the structure map $\mu_{r(j)}$ of $\Cech_{\CF}(\xi)$ otherwise; this is the definition of the structure maps of $r^*\Cech_{\CF}(\xi)$. The pairings agree by the reindexing compatibility in Definition~\ref{def:crossed-precosheaf-structure}.
\end{proof}

\subsection{The \v{C}ech-Loday functor}

\begin{definition}[The total \v{C}ech indexing category]
\label{def:total-cech-indexing}
Let $\SL$ be a meet-semilattice.  The category $\Cech(\SL)$ has objects
$\xi=(\fU\vert W)$, $\fU=(U_i)_{i\in I}$,
where $I$ is a finite set, $U_i,W\in \SL$, and $U_i\leq W$ for every
$i\in I$.  A morphism
$f:\upsilon=(\fV\vert W')\to\xi=(\fU\vert W)$, $\fV=(V_j)_{j\in J}$,
consists of a function $r:J\to I$ satisfying
$$
V_j\leq U_{r(j)}\quad(j\in J),
\qquad
W'\leq W.
$$
Composition is induced by composition of the underlying functions and by
transitivity of these inequalities.

Equivalently, a morphism over $r:J\to I$ is an inequality
$\upsilon\leq r^*\xi$ in $\Cech_J(\SL)$, followed by the pure reindexing
morphism $r^*\xi\to\xi$.  Thus every morphism factors canonically as
$$
\upsilon\longrightarrow r^*\xi\longrightarrow\xi.
$$
\end{definition}

\begin{remark}
The preceding definition is the Grothendieck construction of the
category-valued presheaf $I\mapsto\Cech_I(\SL)$ on the category of finite sets, with pullback of Definition~\ref{def:Cech-config-presheaf} as restriction; see Appendix~\ref{app:categorical}.  We use only the direct
description above.
\end{remark}

\begin{definition}[The category of crossed cubes of varying arity]
\label{def:total-xcube}
The category $\XCube$ has as objects crossed $I$-cubes, for all finite sets $I$.  A morphism from a crossed $J$-cube $N_\bullet$ to a
crossed $I$-cube $M_\bullet$ consists of a function $r:J\to I$ and a
morphism of crossed $J$-cubes
$N_\bullet\to r^*M_\bullet$.
Composition uses the strict identity
$(r\circ s)^*M=s^*r^*M$ of
Lemma~\ref{lem:reindexing-crossed-cubes}.

A crossed precosheaf $\CF$ therefore determines a functor
$$
\Cech_{\CF}:\Cech(\SL)\longrightarrow\XCube.
$$
It sends $\xi$ to $\Cech_{\CF}(\xi)$.  On a morphism over $r:J\to I$ it
uses fixed-fiber functoriality and Lemma~\ref{lem:reindex-cech-cube} to give
$$
\Cech_{\CF}(\upsilon)\longrightarrow r^*\Cech_{\CF}(\xi).
$$
\end{definition}

\begin{theorem}[The total Loday functor]
\label{thm:total-loday-functor}
There is a functor
$$
\BL^{\mathrm{tot}}:\XCube\longrightarrow\Ho(\SSet)
$$
whose restriction to each $\XCube_I$ is the fixed-arity Loday functor
followed by passage to the pointed homotopy category.
For $v\in I$ and the inclusion $\delta:I\setminus\{v\}\hookrightarrow I$,
it sends the coface morphism
$$
(\delta,\id):\partial_v^0M\longrightarrow M
$$
to the homotopy class of the add-map $a_v(M)$ of
Definition~\ref{def:add-map}.
\end{theorem}

\begin{proof}
For a crossed $I$-cube $M_\bullet$ put
$D_I(M):=\Diag N^I\Gamma_I(M_\bullet)$,
so that $\BL M_\bullet=\bW D_I(M)$. For $I=\varnothing$, $D_I(M)$ is
the constant simplicial group $M_\varnothing$, and we use the canonical
isomorphism $\bW D_\varnothing(M)\cong BM_\varnothing$ in arity zero.
Write $T_n^I=T_{(n,\ldots,n)}$ and let
$u^M_{\bfk}:M_{A(\bfk)}\to D_I(M)_n$, $\bfk\in T_n^I$, be
the orthant steps of Definition~\ref{def:orthant-step} in multidegree
$(n,\ldots,n)$; for $I=\varnothing$, $u^M_{()}$ is the identity of
$M_\varnothing$. By Lemma~\ref{lem:grid-normal-form}, $D_I(M)_n$ is
presented by the subgroups $u^M_{\bfk}(M_{A(\bfk)})$ with the
relations \eqref{eq:grid-presentation}, and morphisms
$f_\bullet:M_\bullet\to M'_\bullet$ act by
$u^M_{\bfk}(a)\mapsto u^{M'}_{\bfk}(f_{A(\bfk)}a)$. A
simplicial operator $\alpha:[m]\to[n]$ of the diagonal acts by $\alpha$
in every direction, so Lemma~\ref{lem:grid-operators} gives
\begin{equation}\label{eq:total-loday-simplicial-operators}
\alpha^*u_{\bfk}^M(a)=
\begin{cases}
1,&\lambda_\alpha(k_i)=\infty\text{ for some }i,\\
u_{\lambda_\alpha\bfk}^M(\mu_{A,A'}a),&\text{otherwise},
\end{cases}
\end{equation}
where $(\lambda_\alpha\bfk)_i=\lambda_\alpha(k_i)$, $A=A(\bfk)$
and $A'=A(\lambda_\alpha\bfk)$.

\emph{Reindexing maps.} For $r:J\to I$, define
$$
r_*:T_n^J\longrightarrow T_n^I,\qquad
(r_*\bfk)_i=\max_{r(j)=i}k_j,\qquad \max\varnothing=0.
$$
Then $A(r_*\bfk)=r(A(\bfk))$ and
$r_*(\bfk\vee\bfell)=r_*\bfk\vee r_*\bfell$.
Together with $h^{r^*M}_{A,B}=h^M_{r(A),r(B)}$
(Definition~\ref{def:reindexing-crossed-cubes}), the presentation of
Lemma~\ref{lem:grid-normal-form}(2) therefore gives a homomorphism
\begin{equation}\label{eq:total-loday-reindex-generators}
R_{r,M,n}:D_J(r^*M)_n\longrightarrow D_I(M)_n,
\qquad
u_{\bfk}^{r^*M}(a)\longmapsto u_{r_*\bfk}^M(a).
\end{equation}
Indeed, this rule preserves the internal group relations and sends
each relation \eqref{eq:grid-presentation} for $r^*M$ to the
corresponding one for $M$.

We check compatibility with all simplicial operators at once. The
function $\lambda_\alpha$ is nondecreasing and fixes $0$, so it
preserves finite maxima, including the empty maximum. Consequently
$$
\lambda_\alpha(r_*\bfk)=r_*(\lambda_\alpha\bfk),
$$
with maxima also taken in the ordered set containing $\infty$.
Moreover, the definition of reindexing gives, for $A'\subseteq A\subseteq J$,
$$
\mu^{r^*M}_{A,A'}=\mu^M_{r(A),r(A')}:
M_{r(A)}\longrightarrow M_{r(A')},
$$
because deleting repeated labels applies an original structure map
exactly when the last occurrence of the label is deleted. If an
infinite coordinate occurs, both orders of applying $R_{r,M}$ and
$\alpha^*$ give $1$. Otherwise
\eqref{eq:total-loday-simplicial-operators} shows that both give
$$
u_{\lambda_\alpha(r_*\bfk)}^M
   \bigl(\mu^M_{r(A),r(A')}a\bigr).
$$
Thus $R_{r,M}$ is a homomorphism of simplicial groups.

\emph{Naturality and composition.} For $f_\bullet:M_\bullet\to M'_\bullet$
in $\XCube_I$, the map $D_I(f)$ sends $u_{\bfk}^M(a)$ to
$u_{\bfk}^{M'}(f_{A(\bfk)}a)$. Hence, on generators,
\begin{equation}\label{eq:total-loday-reindex-naturality}
D_I(f)R_{r,M}=R_{r,M'}D_J(r^*f).
\end{equation}
For composable $K\xrightarrow{q}J\xrightarrow{r}I$,
$$
((r\circ q)_*\bfk)_i
=\max_{r(j)=i}\ \max_{q(x)=j}k_x
=(r_*(q_*\bfk))_i .
$$
The coefficient $a$ is unchanged, so this proves
\begin{equation}\label{eq:total-loday-reindex-composition}
R_{\id_I,M}=\id,\qquad
R_{r\circ q,M}=R_{r,M}R_{q,r^*M}.
\end{equation}
In particular, all relations among cofaces, codegeneracies, and
bijections hold: two composites representing the same indexing
function give the same homomorphism.

\emph{The functor.} Put $\vartheta_{r,M}=[\bW R_{r,M}]$, using the
canonical arity-zero comparison specified above, and define
$$
\BL^{\mathrm{tot}}(M)=\BL M,\qquad
\BL^{\mathrm{tot}}(r,\phi)=\vartheta_{r,M}[\BL(\phi)]
$$
for $\phi:N\to r^*M$. For composable morphisms
$(q,\psi):L\to N$ and $(r,\phi):N\to M$, naturality and
\eqref{eq:total-loday-reindex-composition} give
$$
\begin{aligned}
\BL^{\mathrm{tot}}(r,\phi)\BL^{\mathrm{tot}}(q,\psi)
&=\vartheta_{r,M}[\BL(\phi)]\vartheta_{q,N}[\BL(\psi)]\\
&=\vartheta_{r,M}\vartheta_{q,r^*M}
       [\BL(q^*\phi\circ\psi)]\\
&=\BL^{\mathrm{tot}}(r\circ q,q^*\phi\circ\psi).
\end{aligned}
$$
The identity law and agreement with the fixed-arity functors follow
from $R_{\id,M}=\id$. All maps preserve the identity basepoints,
so the target is $\Ho(\SSet)$.

\emph{Cofaces.} For the coface inclusion $\delta:I\setminus\{v\}\hookrightarrow I$,
the formula for $\delta_*$ inserts the coordinate $k_v=0$. On every
generator this fills the $v$-direction with identity arrows, so
$R_{\delta,M}$ is the homomorphism $e_{\scrC_v(M)}$ of
Definition~\ref{def:add-map}, under the identifications of
Lemma~\ref{lem:Gv-identifications}. Hence $\bW R_{\delta,M}$ is the
add-map $a_v(M)$. This proves the required coface compatibility.
\end{proof}

\begin{remark}\label{rem:reindexing-geometric}
The formula for $r_*$ has a simple meaning in the grid picture of
Remark~\ref{rem:grid-picture}. The function $r:J\to I$ induces
$r^\#:[n]^I\to[n]^J$, $t\mapsto(t_{r(j)})_{j\in J}$, and the preimage
under $r^\#$ of the orthant with corner $\bfk$ in $[n]^J$ is the
orthant $\{t:t_i\ge\max_{r(j)=i}k_j\}$ with corner $r_*\bfk$ in
$[n]^I$. Thus $R_{r,M}$ sends the orthant step of $a$ on an orthant of
the $J$-grid to the orthant step of the same $a$ on the preimage
orthant of the $I$-grid; compatibility with the simplicial operators
and with composition of index maps are then the statements that
preimages commute with the operators moving thresholds and with
composition. For the coface $\delta$, $\delta^\#$ forgets the
$v$-coordinate and the preimage of an orthant is the orthant constant
in direction $v$.
\end{remark}

\begin{definition}[The \v{C}ech-Loday classifying complex and functor]
\label{def:les-indexed-type}
For a crossed precosheaf $\CF$ and
$\xi\in\Cech(\SL)$, its \emph{\v{C}ech-Loday classifying complex} is
$$
\chB_{\CF}(\xi)
:=
\BL\!\left(\Cech_{\CF}(\xi)\right).
$$
We use the same notation for its image in $\Ho(\SSet)$.  The
\emph{\v{C}ech-Loday functor} is the composite
$$
\chB_{\CF}
:=
\BL^{\mathrm{tot}}\circ\Cech_{\CF}:
\Cech(\SL)\longrightarrow\Ho(\SSet).
$$
\end{definition}

\subsection{The one-coordinate fibration}

\begin{theorem}[One-coordinate \v{C}ech fibration]
\label{thm:one-coordinate-fibration}
Let $\SL$ be a meet-semilattice and  $\CF$ be a crossed precosheaf on
$\SL$. Let $\xi=((U_i)_{i\in I}\vert W)$ be a \v{C}ech configuration. Pick a $V\in \SL$ with $V\leq W$ and let
$$
\xi^+=((U_i)_{i\in I},V\vert W),\qquad
\xi|_V=((U_i\wedge V)_{i\in I}\vert V).
$$
The sequence 
$$
\chB_{\CF}(\xi|_V)
\longrightarrow
\chB_{\CF}(\xi)
\xrightarrow{\,a_V\,}
\chB_{\CF}(\xi^+)
$$
where $a_V$ is the add-map for the new coordinate, is a homotopy fibration sequence. This sequence is natural with respect to morphisms of $\Cech(\SL)$ corresponding to functions $r:J\sqcup\{u\}\to I\sqcup\{v\}$ such that $r(J)\subseteq I$ and $r(u)=v$. It is also natural with respect to $\CF$. 
\end{theorem}

\begin{figure}[t]
\centering
\begin{tikzpicture}[>=stealth, every node/.style={fill=white,inner sep=2pt}, font=\small]
  \coordinate (sh) at (2.0,1.35);
  \node (BTL) at ($(0,4.2)+(sh)$)  {$\CF(U_1{\wedge}U_2{\wedge}V)$};
  \node (BTR) at ($(6.4,4.2)+(sh)$){$\CF(U_2{\wedge}V)$};
  \node (BBL) at ($(0,0)+(sh)$)    {$\CF(U_1{\wedge}V)$};
  \node (BBR) at ($(6.4,0)+(sh)$)  {$\CF(V)$};
  \begin{scope}[on background layer]
    \node[fill=blue!8,rounded corners,fit=(BTL)(BTR)(BBL)(BBR),inner sep=4pt] (backface) {};
  \end{scope}
  \draw[->] (BTL) -- node[above]{$\mu_1$} (BTR);
  \draw[->] (BBL) -- node[below,pos=0.85]{$\mu_1$} (BBR);
  \draw[->] (BTL) -- node[right,pos=0.7]{$\mu_2$} (BBL);
  \draw[->] (BTR) -- node[right,pos=0.35]{$\mu_2$} (BBR);
  \node (FTL) at (0,4.2)   {$\CF(U_1{\wedge}U_2)$};
  \node (FTR) at (6.4,4.2) {$\CF(U_2)$};
  \node (FBL) at (0,0)     {$\CF(U_1)$};
  \node (FBR) at (6.4,0)   {$\CF(W)$};
  \draw[->] (FTL) -- node[above,pos=0.7]{$\mu_1$} (FTR);
  \draw[->] (FBL) -- node[below]{$\mu_1$} (FBR);
  \draw[->] (FTL) -- node[left]{$\mu_2$} (FBL);
  \draw[->] (FTR) -- node[left,pos=0.7]{$\mu_2$} (FBR);
  \draw[->] (BTL) -- node[above,sloped,pos=0.55]{$\mu_v$} (FTL);
  \draw[->] (BTR) -- node[above,sloped,pos=0.55]{$\mu_v$} (FTR);
  \draw[->] (BBL) -- node[above,sloped,pos=0.55]{$\mu_v$} (FBL);
  \draw[->] (BBR) -- node[above,sloped,pos=0.55]{$\mu_v$} (FBR);
\end{tikzpicture}
\caption{The crossed $3$-cube $\Cech_{\CF}(\xi^+)$ split in the
$v$-direction.  The shaded face is $\Cech_{\CF}(\xi|_V)$, the
front face is $\Cech_{\CF}(\xi)$, and the whole cube is
$\Cech_{\CF}(\xi^+)$.}
\label{fig:one-coordinate-split}
\end{figure}
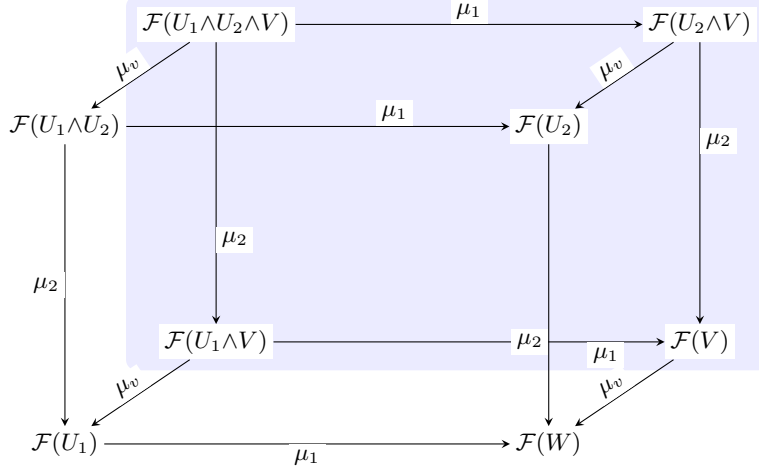

\begin{proof}
Let $M_\bullet=\Cech_{\CF}(\xi^+)$ and let $v$ denote the new
coordinate.  The face not containing $v$ is tautologically
$\Cech_{\CF}(\xi)$ (see Fig. \ref{fig:one-coordinate-split}).  To identify the other face as a crossed cube,
put
$$
\upsilon=((U_i\wedge V)_{i\in I},V\vert W),
\qquad
\zeta=((U_i\wedge V)_{i\in I},V\vert V).
$$
The fixed-fiber morphism $\upsilon\to\xi^+$ is the identity on every vertex
containing $v$. Therefore the induced morphism of crossed cubes is the identity map on every vertex group of the face containing $v$. Since this morphism preserves the $h$-pairings, the pairings on that face agree. Likewise, the fixed-fiber
morphism $\zeta\to\upsilon$ is the identity on the $v$-face of $\zeta$, so it identifies
those pairings with the pairings on the $v$-face of
$\Cech_{\CF}(\zeta)$.

Every structure map in the $v$-direction of
$\Cech_{\CF}(\zeta)$ is the identity, since the added entry equals
the ambient object $V$.  The crossed-cube compatibility axiom
$$
\mu_v h^{\zeta}_{A\cup\{v\},B\cup\{v\}}
=
h^{\zeta}_{A,B}(\mu_v,\mu_v)
$$
therefore gives
$$
h^{\zeta}_{A\cup\{v\},B\cup\{v\}}
=
h^{\zeta}_{A,B}.
$$
Thus the two opposite faces of $\Cech_{\CF}(\zeta)$ have the same
vertex groups, structure maps, and pairings.  Deleting $v$ from $\zeta$
gives $\xi|_V$.  The preceding two fixed-fiber identifications therefore
show that the face of $M_\bullet$ containing $v$ is
$\Cech_{\CF}(\xi|_V)$ as a crossed cube, not merely vertexwise. 

Theorem~\ref{thm:face-fibration-crossed-cubes}, applied to the
$v$-direction of $M_\bullet$, now gives the displayed sequence.  Under the
face identifications, the first map is induced by the $v$-direction
structure maps and the second is the add-map $a_V$.
\end{proof}

\begin{corollary}
\label{cor:ambient-object-contractible}
Let $\mathfrak W=(W_i)_{i\in I}$ be a finite family of subobjects of
$Y\in \SL$.  If $W_{i_0}=Y$ for some $i_0$, then
$\chB_{\CF}(\mathfrak W\vert Y)$
is weakly contractible.
\end{corollary}

\begin{proof}
Write $\fV=(W_i)_{i\neq i_0}$, so that $\mathfrak W=(\fV,Y)$
with the entry $Y$ in position $i_0$.  Theorem~\ref{thm:one-coordinate-fibration},
applied to the coordinate $i_0$, gives
$$
\chB_{\CF}(\fV\vert Y)
\xrightarrow{\id}
\chB_{\CF}(\fV\vert Y)
\longrightarrow
\chB_{\CF}(\fV,Y\vert Y),
$$
because restricting every member of $\fV$ to $Y$ changes nothing.
The long exact homotopy sequence, together with connectedness of Loday
classifying complexes, shows that the last term is weakly contractible.
\end{proof}

\begin{proposition}
\label{prop:dominated-entry-addition}
Let
$\xi=((U_i)_{i\in I}\vert W)$
and suppose $V\leq U_{i_0}$ for some $i_0\in I$.  If $\xi^+$ is obtained
by adjoining $V$, then the add-map
$$
\chB_{\CF}(\xi)
\longrightarrow
\chB_{\CF}(\xi^+)
$$
is a weak equivalence.  The corresponding drop-map is its inverse in
$\Ho(\SSet)$.
\end{proposition}

\begin{proof}
By Theorem~\ref{thm:one-coordinate-fibration}, the homotopy fiber is
$\chB_{\CF}(\xi|_V)$.  The family underlying $\xi|_V$ contains
$U_{i_0}\wedge V=V$, its ambient object, so
Corollary~\ref{cor:ambient-object-contractible} makes this fiber weakly
contractible.  Hence the add-map is a weak equivalence.  The drop-map is a retraction of the add-map; since the add-map is an isomorphism in the homotopy category, the drop-map is its inverse.
\end{proof}

For every $W\in \SL$, the empty configuration $(\,\vert W)$ satisfies
$$
\chB_{\CF}(\,\vert W)=B\CF(W).
$$
For $\xi=(\fU\vert W)$, the unique morphism
$(\,\vert W)\to\xi$ therefore induces a canonical map
$$
\kappa_{\CF}(\xi):
B\CF(W)\longrightarrow\chB_{\CF}(\xi).
$$
These maps are natural in $\xi$.

\subsection{Covers, refinements, and the pro-type}

At this point we assume finite joins and distributivity. This is the first
place where they are needed: finite covers are defined by joins, and common
refinements of covers are formed by pairwise meets together with
distributivity.

\begin{remark}[Pro-objects and cofiltered preorders]
Recall that a category $D$ is \emph{cofiltered} if it is nonempty, if any
two objects of $D$ admit a common predecessor, and if any two parallel
arrows in $D$ are equalized after precomposition by some arrow. A pro-object
of a category $\scrC$ is a functor $D\to\scrC$ with $D$
cofiltered.

For a preorder the parallel-arrow condition is automatic. Thus a preorder is
cofiltered exactly when it is nonempty and any two objects admit a common
predecessor. 
\end{remark}

\begin{definition}[Covers and the refinement preorder]\label{def:cover-refinement-preorder}
Let $\SL$ be a distributive lattice and fix $X\in \SL$. A finite cover of $X$ is a finite nonempty family $(U_i)_{i\in I}$ such that $\bigvee_{i\in I}U_i=X$. Let
$\Cov(X)\subset \Cech(\SL)$
be the full subcategory spanned by all finite  covers of $X$. Thus, a morphism $(\fV\vert X)\to(\fU\vert X)$ in
$\Cov(X)$ is a refinement function $r:J\to I$ such that
$V_j\leq U_{r(j)}$ for all $j\in J$.

The \emph{refinement preorder of covers of $X$}, denoted
$\Cov_{\preceq}(X)$, has the same objects as
$\Cov(X)$ and at most one morphism
$(\fV\vert X)\to(\fU\vert X)$: such a morphism exists exactly when
some refinement function $\fV\to\fU$ exists. There is a
canonical functor
$$
q_X:\Cov(X)
\longrightarrow
\Cov_{\preceq}(X)
$$
which is the identity on objects and sends every refinement function to the
corresponding order relation.
\end{definition}

\begin{lemma}[Independence of refinement functions]\label{lem:refinement-independence}
Let $\SL$ be a distributive lattice, let $\CF$ be a crossed
precosheaf on $\SL$, and fix $X\in \SL$. Suppose
$r,r':\fV\to \fU$
are two refinement functions between finite covers of $X$. Then the
two induced morphisms
$$
\chB_{\CF}(\fV\vert X)
\longrightarrow
\chB_{\CF}(\fU\vert X)
$$
in $\Ho(\SSet)$ are equal.
\end{lemma}

\begin{proof}
Write $\fU=(U_i)_{i\in I}$ and
$\fV=(V_j)_{j\in J}$. Form the meet cover
$$
\mathfrak M=(M_{(i,i')})_{(i,i')\in I\times I},
\qquad
M_{(i,i')}=U_i\wedge U_{i'},
$$
This is a cover of $X$, since its diagonal entries are the original cover.
Define refinement functions
$w:J\to I\times I$, $w(j)=(r(j),r'(j))$,
and
$p,p':I\times I\to I$, $p(i,i')=i$, $p'(i,i')=i'$.
They are legitimate refinements because
$$
V_j\leq U_{r(j)}\wedge U_{r'(j)}=M_{w(j)},
\qquad
M_{(i,i')}\leq U_i,
\qquad
M_{(i,i')}\leq U_{i'}.
$$
Also let
$\delta:I\to I\times I$, $\delta(i)=(i,i)$.
This is a refinement $\fU\to\mathfrak M$, since
$U_i=M_{\delta(i)}$.

We first show that $\chB_{\CF}(\delta)$ is an isomorphism in
$\Ho(\SSet)$. The diagonal subfamily of $\mathfrak M$ is just
$\fU$. Starting from this diagonal subfamily, add the off-diagonal
entries $U_i\wedge U_{i'}$ one at a time. Each such entry is dominated by
an already present diagonal entry, for example
$U_i\wedge U_{i'}\leq U_i$. By Proposition~\ref{prop:dominated-entry-addition},
each corresponding add-map is represented by a weak equivalence.
After all off-diagonal entries have been added, the resulting family
is $\mathfrak M$. Equivalently, the set map $\delta$ factors as a
bijection onto the diagonal block, followed by the cofaces which add the
off-diagonal entries. Thus the pure reindexing morphism over $\delta$ is
sent to an isomorphism
$$
\chB_{\CF}(\delta):
\chB_{\CF}(\fU\vert X)
\longrightarrow
\chB_{\CF}(\mathfrak M\vert X)
$$
in $\Ho(\SSet)$.

As functions, $p\circ\delta=p'\circ\delta=\id_I$, hence
the corresponding composites in $\Cov(X)$ are the identity
morphism of $(\fU\vert X)$. Functoriality of
$\chB_{\CF}:\Cech(\SL)\to\Ho(\SSet)$ gives
$$
\chB_{\CF}(p)\circ \chB_{\CF}(\delta)
=
\id,
\qquad
\chB_{\CF}(p')\circ \chB_{\CF}(\delta)
=
\id.
$$
Since $\chB_{\CF}(\delta)$ is an isomorphism, both
$\chB_{\CF}(p)$ and $\chB_{\CF}(p')$ are equal to its inverse.
Therefore $\chB_{\CF}(p)=\chB_{\CF}(p')$.

Finally, $r=p\circ w$ and $r'=p'\circ w$. Applying functoriality once
more gives
$$
\chB_{\CF}(r)
=
\chB_{\CF}(p)\circ \chB_{\CF}(w)
=
\chB_{\CF}(p')\circ \chB_{\CF}(w)
=
\chB_{\CF}(r'),
$$
as required.
\end{proof}

\begin{corollary}[The \v{C}ech-Loday pro-homotopy type]
\label{cor:nonabelianprotype}
Let $\SL$ be a distributive lattice and fix $X\in \SL$.  For every crossed
precosheaf $\CF$ on $\SL$, the restricted diagram
$$
\chB_{\CF}|_{\Cov(X)}:
\Cov(X)\longrightarrow\Ho(\SSet)
$$
descends uniquely through $q_X$ to a functor
$$
\chB_{\CF,X}:
\Cov_{\preceq}(X)\longrightarrow\Ho(\SSet).
$$
The refinement preorder is cofiltered, and this functor defines a
pro-object
$$
\chbfB_{\CF}(X)
\in
\Pro\bigl(\Ho(\SSet)\bigr).
$$
Moreover, the construction is functorial in $\CF$:
$$
\chbfB_{(-)}(X):
\mathbf{XPreCoSh}(\SL)
\longrightarrow
\Pro\bigl(\Ho(\SSet)\bigr).
$$
\end{corollary}

\begin{proof}
Lemma~\ref{lem:refinement-independence} shows that the restriction of
$\chB_{\CF}$ sends all refinement functions representing the same
order relation to the same morphism.  It therefore descends uniquely
through $q_X$.

The refinement preorder is nonempty because $(X\vert X)$ is a cover.
Given covers $\fU=(U_i)_{i\in I}$ and
$\fV=(V_j)_{j\in J}$, the pairwise-meet family
$\mathfrak W=(U_i\wedge V_j)_{(i,j)\in I\times J}$
is a common refinement.  It is again a cover because distributivity gives
$$
\bigvee_{(i,j)\in I\times J}(U_i\wedge V_j)
=
\left(\bigvee_iU_i\right)\wedge
\left(\bigvee_jV_j\right)
=X.
$$
The parallel-arrow condition is automatic for a preorder, so
$\Cov_{\preceq}(X)$ is cofiltered.

A morphism of crossed precosheaves induces a natural transformation of the
crossed \v{C}ech-cube functors and hence, after applying
$\BL^{\mathrm{tot}}$, a natural transformation of the descended
diagrams.  This gives the asserted functor on pro-objects.
\end{proof}

\begin{definition}[The \v{C}ech-Loday pro-homotopy type]
\label{def:nonabelian-cech-obstruction-diagram}
We call $\chbfB_{\CF}(X)$ the \emph{\v{C}ech-Loday
pro-homotopy type} of $X$ with coefficients in $\CF$.  The maps
$\kappa_{\CF}(\fU\vert X)$ are natural in covers and therefore define a
canonical map of pro-objects
$$
\kappa_\CF:
B\CF(X)\longrightarrow\chbfB_{\CF}(X).
$$
\end{definition}

\subsection{Basic computations}

We end this section by two useful computations for special classes of crossed precosheaves. 

\begin{proposition}[Fundamental group for coflasque crossed precosheaves]
\label{prop:pi1-coflasque}
Let $\SL$ be a meet-semilattice and $\CF$ a coflasque crossed
precosheaf on $\SL$. For a configuration
$\xi=(U_1,\ldots,U_n\vert W)$, $n\geq1$,
identify $\CF(U_i)$ with its image in $\CF(W)$ under corestriction.
Then the homomorphism
$$
(\kappa_\CF(\xi))_*:
\CF(W)=\pi_1B\CF(W)\longrightarrow\pi_1\chB_\CF(\xi)
$$
is surjective, with kernel
$$
\ker(\kappa_\CF(\xi))_*=\prod_{i=1}^n\CF(U_i).
$$
Here each $\CF(U_i)$ is normal in $\CF(W)$, so the product is a
normal subgroup independent of the order of its factors.
\end{proposition}

\begin{proof}
Write $q_\zeta:=(\kappa_\CF(\zeta))_*$ for any configuration $\zeta$.
We argue by induction on the number of entries. For an empty
configuration, $q_{(\,\vert W)}=\id_{\CF(W)}$.

Let
$\xi^-=(U_1,\ldots,U_{n-1}\vert W)$, $V=U_n$,
and put
$\xi^-|_V=(U_1\wedge V,\ldots,U_{n-1}\wedge V\vert V)$.
Theorem~\ref{thm:one-coordinate-fibration} gives a homotopy
fibration sequence
$$
\chB_\CF(\xi^-|_V)
\xrightarrow{j}
\chB_\CF(\xi^-)
\xrightarrow{a}
\chB_\CF(\xi).
$$
All three complexes are connected. Hence $a_*$ is surjective and
$\ker a_*=\im j_*$. Naturality of the augmentation gives
$$
q_\xi=a_*q_{\xi^-},
\qquad
j_*q_{\xi^-|_V}=q_{\xi^-}c_{V,W},
$$
where $c_{V,W}:\CF(V)\hookrightarrow\CF(W)$ is the corestriction.
By induction, $q_{\xi^-}$ and $q_{\xi^-|_V}$ are surjective, so
$q_\xi$ is surjective and
$$
\im j_*=q_{\xi^-}(\CF(V)).
$$
Consequently,
$$
\begin{aligned}
\ker q_\xi
&=q_{\xi^-}^{-1}\bigl(\im j_*\bigr)\\
&=q_{\xi^-}^{-1}\bigl(q_{\xi^-}(\CF(V))\bigr)\\
&=(\ker q_{\xi^-})\,\CF(V)\\
&=\prod_{i=1}^n\CF(U_i),
\end{aligned}
$$
with the empty product understood as the trivial subgroup.
Finally, normality of $\CF(U_i)$ in $\CF(W)$ follows from the
crossed-module structure on the corestriction
$\CF(U_i)\hookrightarrow\CF(W)$.
\end{proof}

Theorem \ref{thm:one-coordinate-fibration} also enables one to compute the \v{C}ech-Loday pro-homotopy type for a particular type of crossed precosheaves.

\begin{proposition}[Meet-preserving coflasque crossed epiprecosheaves]
\label{prop:meetpreserving}
Let $\SL$ be a distributive lattice and $\CF$ a coflasque crossed
precosheaf on $\SL$ which is an epiprecosheaf and preserves meets:
$$
\CF(U\wedge V)=\CF(U)\cap\CF(V)
\quad\text{inside }\CF(W)\text{ whenever }U,V\leq W,
$$
where the groups are identified with their images under corestriction.
Let $\xi=(U_1,\ldots,U_n\vert W)$ be a configuration with $n\geq1$. Then $(\kappa_\CF(\xi))_*$ induces a natural isomorphism
$$
\CF(W)\Big/\CF\Bigl(\bigvee_{i=1}^nU_i\Bigr)
\xrightarrow{\ \cong\ }\pi_1\chB_\CF(\xi),
$$
and
$$
\pi_k\chB_\CF(\xi)=0\qquad(k\geq2).
$$
For the empty configuration, $\chB_\CF(\,\vert W)=B\CF(W)$.
\end{proposition}

\begin{proof}
For $n\geq1$, the epiprecosheaf property gives
$$
\prod_{i=1}^n\CF(U_i)=\CF\Bigl(\bigvee_{i=1}^nU_i\Bigr)
\quad\text{inside }\CF(W).
$$
The asserted isomorphism therefore follows from
Proposition~\ref{prop:pi1-coflasque}. It is natural because it is
induced by the augmentation $\kappa_\CF$.

We prove the higher-homotopy vanishing by induction on $n$.
For $n=0$, the complex $B\CF(W)$ is a connected $1$-type.
For $n=1$, Theorem~\ref{thm:one-coordinate-fibration} gives
$$
B\CF(U_1)\longrightarrow B\CF(W)
\longrightarrow\chB_\CF(U_1\vert W).
$$
The first map is injective on fundamental groups by coflasqueness.
Since its source and target are $1$-types, the long exact homotopy
sequence gives vanishing of all homotopy groups of the last term
in degrees at least two.

Now let $n\geq2$ and put
$\xi^-=(U_1,\ldots,U_{n-1}\vert W)$, $V=U_n$, $S=\bigvee_{i=1}^{n-1}U_i$,
$\xi^-|_V=(U_1\wedge V,\ldots,U_{n-1}\wedge V\vert V)$.
Theorem~\ref{thm:one-coordinate-fibration} gives a homotopy
fibration sequence
$$
\chB_\CF(\xi^-|_V)
\xrightarrow{j}\chB_\CF(\xi^-)
\longrightarrow\chB_\CF(\xi).
$$
By induction, the first two terms have no homotopy groups in degrees
at least two. The long exact sequence therefore gives
$$
\pi_k\chB_\CF(\xi)=0\quad(k\geq3),\qquad
\pi_2\chB_\CF(\xi)\cong\ker(j_*).
$$
Distributivity and the epiprecosheaf property give
$$
\prod_{i=1}^{n-1}\CF(U_i\wedge V)
=\CF\Bigl(\bigvee_{i=1}^{n-1}(U_i\wedge V)\Bigr)
=\CF(S\wedge V).
$$
By naturality of the augmentation and the fundamental-group
identification above, $j_*$ is the homomorphism
$
\CF(V)/\CF(S\wedge V)\to\CF(W)/\CF(S)
$
induced by corestriction. Its kernel is
$\bigl(\CF(V)\cap\CF(S)\bigr)/\CF(S\wedge V)=1,$
since $\CF$ preserves meets. Thus $\pi_2\chB_\CF(\xi)=0$,
completing the induction.
\end{proof}

\begin{corollary}\label{cor:contractible-protype}
    Assume in addition that $(U_1,\ldots,U_n)$ is a cover of $W$. Then $\chB_{\CF}(\xi)$ is weakly contractible. Thus, for a coflasque crossed precosheaf $\CF$ which preserves meets and is an epiprecosheaf, the \v{C}ech-Loday pro-homotopy type is constant and contractible. 
\end{corollary}

\section{\v{C}ech-Loday pro-type and 't Hooft anomalies}\label{sec:tHooft}

Here we return to the discussion of Section~\ref{sec:gaugingnew} and make precise the sense in which the canonical maps $\kappa_\CF:B\CF(X)\to \chbfB_\CF(X)$ obstruct gauging. Throughout, $\SL$ is a distributive lattice with a top element $X\in \SL$, and $G$ is a group.

\subsection{The anomaly of a group action}

Let $\CF$ be a crossed precosheaf of symmetries of a concrete quantum system. 
As explained in Section \ref{sec:gaugingnew}, to gauge a $G$-action $\rho:G\to\CF(X)$ one needs to choose a gauge precosheaf.
Let us restate the definition of a gauge precosheaf in the language of crossed precosheaves.

\begin{definition}[Gauge precosheaf as a crossed precosheaf]\label{def:gauge-precosheaf}
A \emph{gauge precosheaf for $G$ over $(X,\SL)$} is a pair $(\CG,\iota)$ consisting of a crossed precosheaf $\CG$ on $\SL$ and an injective  homomorphism $\iota\colon G\to\CG(X)$ such that
\begin{enumerate}
\item $\CG$ is coflasque and its pairings are the ordinary commutators in $\CG(X)$, so that $\CG$ is a crossed precosheaf of subgroups of $\CG(X)$ in the sense of Example~\ref{ex:crossed-precosheaf-subgroups};
\item meet preservation: $\CG(U\wedge V)=\CG(U)\cap\CG(V)$ for all $U,V\in \SL$;
\item $\CG$ is an epiprecosheaf (Definition~\ref{def:epi}): $\CG(U\vee V)=\CG(U)\CG(V)$ for all $U,V\in \SL$.
\end{enumerate}
\end{definition}

Examples of gauge precosheaves in the continuum and on the lattice are given in Section \ref{sec:gaugingnew}.

By Corollary~\ref{cor:contractible-protype}, for every finite cover $\fU$ of $X$, the \v{C}ech-Loday classifying complex $\chB_{\CG}(\fU|X)$ of a gauge precosheaf is  weakly contractible.

\begin{definition}[Anomaly]\label{def:anomaly}
Let $\rho:G\to\CF(X)$ be a homomorphism. The \emph{anomaly of $(\CF,\rho)$ at a finite cover $\fU$ of $X$} is the homotopy class
$$
\alpha_{\CF,\rho}(\fU):=\kappa_{\CF}(\fU|X)\circ B\rho\colon BG\longrightarrow\chB_{\CF}(\fU|X).
$$
The \emph{anomaly of $(\CF,\rho)$} is the morphism of pro-objects
$$
\alpha_{\CF,\rho}:=\kappa_\CF\circ B\rho\in\Hom_{\Pro(\Ho(\SSet))}\bigl(BG,\chbfB_{\CF}(X)\bigr).
$$
We say that $(\CF,\rho)$ is \emph{anomaly-free} if $\alpha_{\CF,\rho}$ is trivial, i.e. if $\alpha_{\CF,\rho}(\fU)$ is null-homotopic for every finite cover $\fU$ of $X$.
\end{definition}

The anomaly depends on the theory, through $\CF$, and on the action, through $\rho$, but not on the choice of the gauge precosheaf $(\CG,\iota)$. 

The main result of this section is that anomalies are obstructions to the existence of gauging.

\begin{definition}[Gauging]\label{def:gauging}
A \emph{gauging of $(\CF,\rho)$ is a triple $(\CG,\iota,
\varrho)$}  where $(\CG,\iota)$ is a gauge precosheaf for $G$ and $\varrho\colon\CG\to\CF$ is a morphism of crossed precosheaves such that $\varrho(X)\circ\iota=\rho$.
\end{definition}

\begin{remark}\label{rem:gauging-pairings}
In Sections~\ref{sec:intro} and~\ref{sec:gaugingnew}, a gauging was
defined as a morphism of precosheaves of groups. A morphism of crossed
precosheaves (Definition~\ref{def:morphism-crossed-precosheaves-precosheaf})
is a morphism of precosheaves which in addition respects the pairings:
for $U,V\in\SL$, $a\in\CG(U)$ and $b\in\CG(V)$,
$$
\varrho(U\wedge V)\bigl(h^{\CG}_{U,V}(a,b)\bigr)
=
h^{\CF}_{U,V}\bigl(\varrho(U)a,\varrho(V)b\bigr).
$$
Since both pairings lift the commutator, the two sides have the same
image in $\CF(X)$; the condition therefore holds automatically when
$\CF$ is coflasque. 
\end{remark}

\begin{theorem}[Anomalies are obstructions to gauging]\label{thm:gauging-obstruction}
Suppose there exists a gauging  $(\CG,\iota,\varrho)$ of $(\CF,\rho)$. Then the anomaly of $(\CF,\rho)$ is trivial. 
\end{theorem}

\begin{proof}
Fix a finite cover $\fU$ of $X$ and consider the diagram
$$
\begin{tikzcd}[column sep=huge,row sep=large]
BG \arrow[r,"B\rho"] \arrow[d,"B\iota"'] \arrow[rr,bend left=22,"\alpha_{\CF,\rho}(\fU)"{pos=0.5}]
  & B\CF(X) \arrow[r,"\kappa_\CF(\fU|X)"] & \chB_\CF(\fU|X) \\
B\CG(X) \arrow[ur,dashed,"B\varrho(X)"{sloped,pos=0.55}] \arrow[r,"\kappa_\CG(\fU|X)"']
  & {\chB_\CG(\fU|X)\simeq\ast} \arrow[ur,dashed,"\chB_\varrho(\fU|X)"'{sloped,pos=0.4}] &
\end{tikzcd}
$$
in $\Ho(\SSet)$, in which the dashed arrows are supplied by the gauging. The left triangle commutes because $\varrho(X)\circ\iota=\rho$, and the square commutes by the naturality of $\kappa$ in the crossed precosheaf: $\kappa_{\CF}(\fU|X)\circ B\varrho(X)=\chB_\varrho(\fU|X)\circ\kappa_{\CG}(\fU|X)$. Hence $\alpha_{\CF,\rho}(\fU)=\chB_\varrho(\fU|X)\circ \alpha_{\CG,\iota}(\fU)$ factors through $\chB_\CG(\fU|X)$, which is weakly contractible. Therefore $\alpha_{\CF,\rho}(\fU)$ is null-homotopic for every $\fU$.
\end{proof}

\begin{remark}\label{rem:gauging-remarks}
The theorem does not assert a converse: anomaly-freeness need not imply the existence of a gauging.
\end{remark}

\subsection{Dependence on the cover}

The anomaly was defined as a compatible family over all finite covers of $X$. The components of this family are related by the refinement maps in a way that makes ``anomalous at some cover'' a well-defined notion.

\begin{proposition}[Monotonicity under refinement]\label{prop:monotone}
Let $\CF$ be a crossed precosheaf on $\SL$ and let $r\colon{\fV}\to\fU$ be a refinement of finite covers of $X$. Then
$$
\kappa_{\CF}(\fU|X)=\chB_{\CF}(r)\circ\kappa_{\CF}({\fV}|X)
\qquad\text{in }\Ho(\SSet),
$$
and consequently $\alpha_{\CF,\rho}(\fU)=\chB_{\CF}(r)\circ\alpha_{\CF,\rho}({\fV})$ for every $\rho$. In particular, if $\alpha_{\CF,\rho}(\fU)$ is not null-homotopic, then neither is $\alpha_{\CF,\rho}({\fV})$ for any refinement ${\fV}$ of $\fU$.
\end{proposition}

\begin{proof}
The first identity is the naturality of $\kappa_{\CF}$ with respect to the morphism $r\colon({\fV}|X)\to(\fU|X)$ of $\Cech(\SL)$, together with the fact that $r$ composed with $(|X)\to({\fV}|X)$ is $(|X)\to(\fU|X)$. Composing with $B\rho$ gives the second. If $H$ were a null-homotopy of $\alpha_{\CF,\rho}({\fV})$, then $\chB_{\CF}(r)\circ H$ would be one of $\alpha_{\CF,\rho}(\fU)$.
\end{proof}

Thus an anomaly detected at a cover persists under every refinement of
that cover: refinement can only detect more obstructions, never fewer.
The refinement maps need not be injective on homotopy classes, so an
obstruction detected at a fine cover may become invisible at a coarser
one.
The pro-object $\chbfB_{\CF}(X)$ retains all covers at once,
and by monotonicity the anomaly is trivial if and only if its
components at a cofinal family of covers are.

In favorable cases a single cover suffices. Recall that a pro-object is
\emph{essentially constant} if it is isomorphic in
$\Pro(\Ho(\SSet))$ to a constant pro-object $cZ$. We say that
$\chbfB_{\CF}(X)$ is \emph{represented by the cover $\fU_0$} if,
for every refinement ${\fV}$ of $\fU_0$, the map
$\chB_{\CF}({\fV}|X)\to\chB_{\CF}(\fU_0|X)$ is an
isomorphism in $\Ho(\SSet)$.

\begin{proposition}\label{prop:represented-by-cover}
If $\chbfB_{\CF}(X)$ is represented by $\fU_0$, then it is
essentially constant, isomorphic to $c\,\chB_{\CF}(\fU_0|X)$, and for
every $\rho$ the anomaly $\alpha_{\CF,\rho}$ corresponds to
$\alpha_{\CF,\rho}(\fU_0)\in[BG,\chB_{\CF}(\fU_0|X)]$ under this
isomorphism. In particular $(\CF,\rho)$ is anomaly-free if and only if
$\alpha_{\CF,\rho}(\fU_0)$ is null-homotopic.
More generally, if $\chbfB_{\CF}(X)\cong cZ$, the
isomorphism is represented by a map $f\colon\chB_{\CF}(\fU_1|X)\to Z$
for some cover $\fU_1$, and $\alpha_{\CF,\rho}$ corresponds to
$f\circ\alpha_{\CF,\rho}(\fU_1)\in[BG,Z]$.
\end{proposition}

\begin{proof}
The covers refining $\fU_0$ are cofinal in the refinement preorder,
and on them all transition maps to $\chB_{\CF}(\fU_0|X)$ are
isomorphisms; restricting a pro-object to a cofinal subcategory does not
change it up to isomorphism, and a diagram all of whose maps are
isomorphisms is isomorphic to the constant diagram on any of its values.
Under $\Hom_{\Pro}(BG,c\,\chB_{\CF}(\fU_0|X))=[BG,\chB_{\CF}(\fU_0|X)]$
the compatible family $(\alpha_{\CF,\rho}(\fU))_{\fU}$ corresponds to its
component at $\fU_0$.

A morphism from a pro-object to a constant object is represented by a
map out of some stage, 
$$
\Hom_{\Pro}(\chbfB_{\CF}(X),cY)
=\varinjlim_{\fU}[\chB_{\CF}(\fU|X),Y].
$$
Composing the representative
$f$ with the anomaly at that stage gives the image of $\alpha_{\CF,\rho}$
under the isomorphism.
\end{proof}

The simplest situation is when the category of covers has an initial object with respect to the refinement preorder, i.e. the finest cover. Then the \v{C}ech-Loday pro-type and the anomaly are represented by this cover. This is the situation for 1d quantum spin systems, see Section~\ref{sec:1d}. In higher
dimensions there is no finest cover. 

\subsection{Relative anomalies}

As discussed in Section \ref{sec:gaugingnew}, when $G$ is a Lie group, natural choices of $\CG$ are crossed precosheaves but are not gauge precosheaves in the sense of Definition \ref{def:gauge-precosheaf} because they fail either meet-preservation or epiprecosheafness. Nevertheless, we can define an anomaly of a $G$-action relative to some choice of $(\CG,\iota)$.

\begin{definition}[Relative gauging]
    Let $\CF$ be a crossed precosheaf over $\SL$ and $\rho:G\to \CF(X)$ be a homomorphism. Let $\CG$ be a crossed precosheaf over $\SL$ and $\iota:G\to\CG(X)$ be an injective homomorphism. A gauging of $(\CF,\rho)$ relative to $(\CG,\iota)$ is a morphism of crossed precosheaves $\varrho:\CG\to\CF$ such that $\varrho(X)\circ\iota=\rho$.
\end{definition}

\begin{definition}[Relative anomaly]\label{def:relative-anomaly}
Let $\CG$ be a crossed precosheaf over $\SL$ and $\iota:G\to\CG(X)$ be an injective homomorphism. Pre-composition with the anomaly $\alpha_{\CG,\iota}$ (Definition \ref{def:anomaly} specialized to $\CF=\CG$ and $\rho=\iota$)  defines a map
$$
a_{\CG,\iota}\colon\Hom_{\Pro(\Ho(\SSet))}\bigl(\chbfB_{\CG}(X),\chbfB_{\CF}(X)\bigr)\longrightarrow\Hom_{\Pro(\Ho(\SSet))}\bigl(BG,\chbfB_{\CF}(X)\bigr).
$$
We say that $(\CF,\rho)$ is \emph{anomaly-free relative to $(\CG,\iota)$} if $\alpha_{\CF,\rho}$ lies in the image of $a_{\CG,\iota}$, and we call the class of $\alpha_{\CF,\rho}$ in the quotient set
$$
\Hom_{\Pro(\Ho(\SSet))}\bigl(BG,\chbfB_{\CF}(X)\bigr)\big/\im a_{\CG,\iota}
$$
the \emph{anomaly of $(\CF,\rho)$ relative to $(\CG,\iota)$}.
\end{definition}
If $(\CF,\rho)$ is not anomaly-free relative to $(\CG,\iota)$, then no gauging of $(\CF,\rho)$ relative to $(\CG,\iota)$ exists.

\section{Applications to quantum spin systems}\label{sec:qspin}

\subsection{Quantum spin systems and quantum cellular automata}\label{sec:qca1}
Let $(X,\rho)$ be a metric space, $\rho$ being the distance function. For any $U,V\subseteq X$ we let $\dist(U,V)=\inf_{x\in U,y\in V}\rho(x,y)$ and for any $z\in X$ we let $\dist(z,U)=\dist(\{z\},U)$. For any $r\geq 0$ and any $U\subseteq X$ we let 
$$
U^r=\left\{x\in X\vert \dist(x,U)\leq r\right\}.
$$

\begin{definition}[Multiplicity functions]
    A multiplicity function on $(X,\rho)$ is a locally-finite function $q:X\ra\NN$, in the sense that for any bounded $U\subseteq X$, the set $U\cap q^{-1}(\NN\backslash \{1\})$ is finite.
\end{definition}
From now on we will suppress $\rho$ from the notation, i.e. $X$ will always mean a set $X$ with a fixed distance function. 
\begin{definition}[Quantum spin systems and their algebras of observables]
    Let $q$ be a multiplicity function on $X$. A pair $(X,q)$ is called a quantum spin system. The trivial quantum spin system over $X$ corresponds to the function $q_x=1$ for all $x\in X$. For any bounded $U\subseteq X$ we let 
    $$
    \CA(U,q)=\bigotimes_{x\in U} \Mat(\bbC^{q_x}).
    $$
    $\CA(U,q)$ is a $*$-algebra. Its group of unitary elements is denoted $\CU(U,q)$. If $U\subseteq V$ and $U,V$ are bounded, there is a canonical injective homomorphism $\CA(U,q)\to\CA(V,q)$. For an arbitrary $U\subseteq X$, we let
    $$
    \CA(U,q)=\varinjlim_{V\subseteq U, V\, {\rm bounded}}\CA(V,q).
    $$
    In particular, $\Aloc:=\CA(X,q)$ is called the $*$-algebra of local observables of the quantum spin system $(X,q)$. The group of unitary elements of $\Aloc$ will be denoted $\Uloc$.
\end{definition}

\begin{definition}[Quantum Cellular Automata]
    A Quantum Cellular Automaton (QCA) acting on a quantum spin system $(X,q)$ is a $*$-automorphism  $\alpha:\CA(X,q)\ra \CA(X,q)$ such that $\exists r\geq 0$ satisfying  $\alpha(\CA(U,q))\subset \CA(U^r,q)$ for all $U\subseteq X$. One says that $\alpha$ has spread $r$. We denote the set of QCAs by $\CQ(X,q)$. 
\end{definition}
\begin{definition}[QCA localized on a subset]
    For a $U\subseteq X$, let $\CQ(U)$ be the subset of $\CQ(X)$ defined by the following condition. $\alpha\in\CQ(U)$ iff $\exists r\geq 0$ such that $\alpha(a)=a$ whenever $a\in\CA(\{x\})$ and $\dist(x,U)>r$. If $r\geq 0$ satisfies this condition, then we say that $\alpha$ is $r$-localized on $U$. If $\alpha\in\CQ(U)$, we say that $\alpha$ is localized on $U$.
\end{definition}

From now on, we fix $(X,q)$ and no longer indicate $q$ explicitly. For example, instead of $\CA(U,q)$ we write $\CA(U)$ and instead of $\CQ(U,q)$ we write $\CQ(U)$. 

The following theorem summarizes some well-known properties of QCAs.

\begin{proposition}[Basic properties of QCAs]\label{thm:basicQCA}

    \begin{enumerate}
        \item $\CQ(X)$ is a subgroup of the group of $*$-automorphisms of $\CA(X)$.
        \item For any $U\subseteq X$, $\CQ(U)$ is a normal subgroup of $\CQ(X)$. 
        \item For any $U,V\subseteq X$ such that $U\subseteq V^r$ for some $r\geq 0$, $\CQ(U)\subseteq \CQ(V)$.
        \item If $U\subseteq X$ is bounded, any $\alpha\in\CQ(U)$ has the form $\alpha(a)=\Ad_u (a):=u a u^{-1},$ where $u$ is a unitary element of $\CA(X)$. Conversely, any $*$-automorphism of this form is a QCA $r$-localized at a point of $X$. The map $u\mapsto\Ad_u$ is a group homomorphism whose kernel is isomorphic to $U(1)$. 
    \end{enumerate}
\end{proposition}
These facts are well known; for completeness, proofs are given in Appendix~\ref{app:qca}.

The definition of $\CQ(U)$ is inherently ``fuzzy'': it
consists of QCAs localized within $U^r$ for some $r\geq 0$, rather
than QCAs supported strictly inside $U$. Moreover, by Proposition~\ref{thm:basicQCA}, if
$U\subseteq V^s$ for some $s\geq 0$, then 
$\CQ(U)\subseteq\CQ(V)$.
Thus the assignment $U\mapsto\CQ(U)$ is naturally
monotone when subsets of $X$ are ordered up to bounded thickening.
This motivates the following definition.

\begin{definition}[Fuzzy pre-order]\label{def:fuzzy-preorder}

A fuzzy pre-order on the set of all nonempty subsets of $X$ is defined as follows: $U\leq V$ iff $U\subseteq V^r$ for some $r\geq 0$. We regard this pre-ordered set as a thin category. In particular, if $U,V\subseteq X$ are non-empty and bounded, then $U\leq V$ and $V\leq U$, and thus $U$ and $V$ are isomorphic objects in the corresponding category. 
\end{definition}
Proposition~\ref{thm:basicQCA} implies that $\CQ(U)$ depends only on the isomorphism class of $U$ with respect to the fuzzy pre-order.

\subsection{Meet-subsemilattices on \texorpdfstring{$X$}{X}}\label{sec:meetsubsemi}

Recall that the definition of a crossed precosheaf requires a
meet-semilattice, namely a partially ordered set in which every finite
nonempty family has a meet. The fuzzy pre-order on subsets of $X$ is
generally not a partial order, since distinct subsets may satisfy
$U\leq V$ and $V\leq U$. One can obtain a partial order by imposing
the equivalence relation
$$
U\sim V
\quad\Longleftrightarrow\quad
U\leq V\ \text{and}\ V\leq U
$$
and passing to the quotient. Let us call this poset $P_X$
. In the following sections, by a meet-subsemilattice $\SL$ on $X$ we
mean a subset $\SL\subseteq P_X$ such that any two $U,V\in \SL$ the meet of $U,V$ {\emph {within $P_X$}} exists and is an element of $\SL$. Clearly, a meet-subsemilattice on $X$ is a meet-semilattice with respect to the fuzzy partial order. 
For general $X$, it is not clear how to choose $\SL$. When $X=\RR^d$, a suitable choice is obtained by
restricting from arbitrary subsets to closed semilinear subsets \cite{AKY}. As we recall below, after passing to the quotient poset one gets a meet-subsemilattice on $\RR^d$.

\begin{definition}[Closed semilinear sets]
A nonempty subset of $\RR^d$ is called \emph{semilinear} if it is a
finite union of subsets defined by finitely many linear equalities and
strict linear inequalities. Thus, a closed semilinear set is a finite
union of closed polyhedra.

Let $\CSets_d$ be the collection of closed semilinear subsets of
$\RR^d$. We equip $\RR^d$ with the $\ell^\infty$-metric, so that the
$r$-thickening $U^r$ of a closed semilinear set is again closed
semilinear, and we equip $\CSets_d$ with the fuzzy preorder
$$
U\leq V\quad\Longleftrightarrow\quad U\subseteq V^r\ \text{for some }r\geq0 .
$$
Write
$U\sim V$ if and only if $U\leq V$ and $V\leq U$
for the associated isomorphism relation.
\end{definition}

\begin{proposition}[\cite{AKY}]
\label{prop:cssets}
Elements of $\CSets_d$ have binary meets in $P_{\RR^d}$ up to
isomorphism. Explicitly, for any $U,V\in\CSets_d$, there exists
$R_0\geq 0$ such that, for every $R\geq R_0$,
$U^R\cap V^R$ is a greatest lower bound of $U$ and $V$ up to
$\sim$. Consequently, the quotient poset
$\CSets_d/{\sim}$
is a meet-subsemilattice, with
$$
[U]\wedge[V]=[U^R\cap V^R]
$$
for all sufficiently large $R$.
\end{proposition}

\begin{remark}
The choice of the $\ell^\infty$-metric is a convenience: thickenings
with respect to any two norms on $\RR^d$ are sandwiched between each
other, so the fuzzy preorder on subsets of $\RR^d$, the poset
$P_{\RR^d}$, and hence the meet-subsemilattice $\CSets_d/{\sim } $ do not depend on the norm. With
the $\ell^\infty$-metric, however, the representatives $U^R\cap V^R$ of
meets in Proposition~\ref{prop:cssets} are themselves closed semilinear.
\end{remark}

The quotient meet-semilattice $\CSets_d/{\sim}$ admits an
equivalent concrete description in terms of spherical semilinear subsets
of the sphere at infinity, ordered by ordinary inclusion. We now
introduce this description.

\begin{definition}[Spherical semilinear regions]
Let
$$
S^{d-1}
=
(\RR^d\setminus\{0\})/\RR_{>0}
$$
be the sphere at infinity. For $A\subseteq S^{d-1}$, let
$c(A)\subseteq\RR^d$ be the closed cone over $A$, with the
convention $c(\varnothing)=\{0\}$. A subset $A\subseteq S^{d-1}$ is a \emph{spherical polyhedron} if
$c(A)$ is a polyhedron in $\RR^d$ and $A$ is contained in
an open hemisphere. A \emph{spherical closed semilinear set} is a
finite union of spherical polyhedra.

Let $\SCSets_d$ be the poset of spherical closed semilinear sets,
ordered by inclusion. It is a lattice with intersection and union as meet and join.
\end{definition}

For $U\in\CSets_d$, define its set of directions at infinity by
$$
\widehat U
:=
\{a\in S^{d-1}\mid c(\{a\})\leq U\}.
$$

\begin{theorem}[\cite{AKY}]
Viewing the fuzzy preorder on $\CSets_d$ as a thin category, the
functors
$$
\CSets_d\longrightarrow\SCSets_d,
\qquad
U\longmapsto\widehat U,
$$
and
$$
\SCSets_d\longrightarrow\CSets_d,
\qquad
A\longmapsto c(A),
$$
define an equivalence of categories. This equivalence identifies
$\SCSets_d$ as a skeletal model of $\CSets_d$.
Equivalently, the quotient meet-semilattice of
Proposition~\ref{prop:cssets} is described by
$$
\CSets_d/{\sim}
\cong
\SCSets_d.
$$
Under this identification, bounded closed semilinear subsets of
$\RR^d$ correspond to the bottom element  
$\varnothing\in\SCSets_d$.
\end{theorem}
In what follows we use the symbol $\bott$ to denote the bottom element in the poset $\SCSets_d$ as well as the equivalence class of bounded objects in the pre-ordered set $\CSets_d$. 

\subsection{The crossed precosheaf of QCAs}\label{sec:qca2}

\begin{proposition}
\label{prop:qca-meet-preservation}
Let $\SL$ be a meet-subsemilattice on $X$, for every $U,V\in \SL$,
$$
\CQ(U\wedge V)
=
\CQ(U)\cap\CQ(V)
$$
as subgroups of $\CQ(X)$.
\end{proposition}
\begin{proof}
Put $S=U\wedge V$. Since $S\leq U$ and $S\leq V$,
Proposition~\ref{thm:basicQCA} gives
$$
\CQ(S)\subseteq\CQ(U)\cap\CQ(V).
$$

For the reverse inclusion, let
$\gamma\in\CQ(U)\cap\CQ(V)$. Choose $r,s\geq0$ such that
$\gamma$ is $r$-localized on $U$ and $s$-localized on $V$,
and set
$T=U^r\cap V^s$.
By choosing $r$ and $s$ large enough we can ensure that $T$ is nonempty.
If $x\notin T$, then either $\dist(x,U)>r$ or
$\dist(x,V)>s$. In either case, $\gamma$ fixes
$\CA(\{x\})$, so $\gamma$ is localized on $T$.
Since $T\leq U$ and
$T\leq V$, and $S$ is the meet of $U$ and $V$ in the
fuzzy preorder, one has $T\leq S$. Hence there exists $t\geq0$
such that
$T\subseteq S^t$.
If $\dist(x,S)>t$, then $x\notin T$, so $\gamma$ fixes
$\CA(\{x\})$. Therefore $\gamma$ is $t$-localized on $S$,
and hence $\gamma\in\CQ(S)$.
\end{proof}

\begin{theorem}[The crossed precosheaf of QCAs]
\label{thm:crossed-precosheaf-QCA}
Let $\SL$ be a meet-subsemilattice on $X$ with respect
to the fuzzy preorder. Define a precosheaf
$\CQ:\SL\to\Grp$, $U\longmapsto\CQ(U)$,
whose corestriction maps are the subgroup inclusions
$\CQ(U)\hookrightarrow\CQ(V)$ whenever $U\leq V$.

For every $I$-indexed \v{C}ech configuration
$\xi=((U_i)_{i\in I}\vert W)\in\Cech_I(\SL)$
and every $A,B\subseteq I$, define
$$
h^\xi_{A,B}:
\CQ(U_A^\xi)\times\CQ(U_B^\xi)
\longrightarrow
\CQ(U_{A\cup B}^\xi)
$$
by
$h^\xi_{A,B}(\alpha,\beta):=[\alpha,\beta]$.

Then $\CQ$, equipped with these commutator pairings, is a crossed
precosheaf in the sense of
Definition~\ref{def:crossed-precosheaf-structure}. Moreover, this crossed structure is canonical. 
\end{theorem}
\begin{proof}
By Proposition~\ref{thm:basicQCA}, the assignment
$U\mapsto\CQ(U)$ is order-preserving, each $\CQ(U)$ is normal in
$\CQ(X)$, and all corestriction maps are subgroup inclusions.
Thus $\CQ$ is coflasque precosheaf of subgroups of $\CQ(X)$.
Moreover, for every $U,V\in \SL$,
$$
[\CQ(U),\CQ(V)]
\subseteq
\CQ(U)\cap\CQ(V)
=
\CQ(U\wedge V).
$$
The first relation follows from normality, and the equality follows from
Proposition~\ref{prop:qca-meet-preservation}. Thus $\CQ$ satisfies the
commutator-containment condition. Therefore $\CQ$, equipped with the
ordinary group commutator pairings, is an instance of the coflasque crossed
precosheaf of subgroups described in
Example~\ref{ex:crossed-precosheaf-subgroups}. 
\end{proof}

\subsection{Finite-depth circuits}\label{sec:circuits}

Throughout this section $(X,q)$ is a fixed quantum spin system, and
$X_q := \{x \in X \mid q_x > 1\}$
is its set of \emph{occupied} sites. By local finiteness every bounded subset of $X$ contains finitely many occupied sites. Since $\CA(S) = \CA(S \cap X_q)$ for every $S \subseteq X$, it is convenient to partition $X_q$ rather than $X$.

\begin{definition}[Partitioned unitaries]\label{def:PU}
Let $R \ge 0$. A partition $\Lambda$ of $X_q$ is \emph{$R$-bounded} if every block $\lambda \in \Lambda$ is contained in a ball of radius $R$. A \emph{partitioned unitary} on $(X,q)$ is a pair
$g = \bigl(\Lambda, (u_\lambda)_{\lambda \in \Lambda}\bigr)$,
where $\Lambda$ is an $R$-bounded partition of $X_q$ for some $R \ge 0$ and each $u_\lambda$ is a unitary element of $\CA(\lambda)$. We call the $u_\lambda$ the \emph{gates} of $g$, call $\lambda$ \emph{nontrivial} if $u_\lambda \ne 1$, and write informally $g = \prod_{\lambda \in \Lambda} u_\lambda$. The set of partitioned unitaries is denoted $\Part(X)$. We put $g^{-1} := (\Lambda, (u_\lambda^{-1}))$. For $K \subseteq X$, $g$ is \emph{localized on $K$} if there is $r \ge 0$ with $\lambda \subseteq K^r$ for every nontrivial block $\lambda$ of $g$.
\end{definition}

Let $g \in \Part(X)$ with $\Lambda$ $R$-bounded. A bounded $S \subseteq X$ meets only finitely many blocks, since these lie in the bounded set $S^{2R}$. Put
$$
g_S := \prod_{\lambda \cap S \ne \emptyset} u_\lambda \in \Uloc ,
$$
a finite product of commuting unitaries. If $S \subseteq S'$ and $a \in \CA(S)$, the extra factors of $g_{S'}$ commute with $a$, so $g_{S'} a g_{S'}^{-1} = g_S a g_S^{-1}$. Hence there is a unique $*$-automorphism $\Ad_g$ of $\CA(X)$ with
$$
\Ad_g(a) = g_S\, a\, g_S^{-1} \qquad (S \text{ bounded},\ a \in \CA(S)),
$$
and $\Ad_{g^{-1}} = \Ad_g^{-1}$. Since $g_S \in \CA(S^{2R})$, $\Ad_g$ is a QCA of spread $2R$; if $g$ is $r$-localized on $K$, then $\Ad_g$ fixes $\CA(\{x\})$ whenever $\dist(x,K) > r$, so $\Ad_g \in \CQ(K)$.

\begin{definition}[Circuits]\label{def:Circ}
The \emph{group of circuits} $\Circ(X)$ is the free group on the set $\Part(X)$. For a nonempty $K \subseteq X$, $\Circ(K) \le \Circ(X)$ is the subgroup generated by the partitioned unitaries localized on $K$. The homomorphism
$\phi : \Circ(X) \to \CQ(X)$
is the unique one with $\phi(g) = \Ad_g$ for $g \in \Part(X)$. We put
$$
\CC(K) := \phi\bigl(\Circ(K)\bigr) \subseteq \CQ(K), \qquad \CC(X) := \phi\bigl(\Circ(X)\bigr),
$$
and call $\CC(K)$ the group of \emph{circuit-generated automorphisms localized on $K$}.
\end{definition}

The relation among partitioned unitaries, formal circuit words, and the
corresponding automorphisms is summarized by
$$
\begin{tikzcd}[column sep=large]
\Part(X)
  \arrow[r,hook,"\substack{\text{free}\\\text{generators}}"]
&[1.5em] \Circ(X)
  \arrow[r,two heads,"\phi"]
& \CC(X)
  \arrow[r,hook]
& \CQ(X).
\end{tikzcd}
$$
The first arrow is a map of sets into the free group on $\Part(X)$, the
middle arrow is surjective by the definition of $\CC(X)$, and the
composite $\Part(X)\to\CC(X)$ sends $g$ to $\Ad_g$.

If $K \subseteq K'^{\,s}$ for some $s \ge 0$, a partitioned unitary localized on $K$ is localized on $K'$, so $\Circ(K) \subseteq \Circ(K')$ and $\CC(K) \subseteq \CC(K')$. Both therefore depend only on the class of $K$ in the fuzzy preorder, and we write $\Circ(U)$, $\CC(U)$ for $U \in \SL$.

A \emph{word} is a finite sequence $(g_1^{\eps_1}, \dots, g_n^{\eps_n})$ with $g_i \in \Part(X)$ and $\eps_i = \pm 1$; it represents $A = g_1^{\eps_1} \cdots g_n^{\eps_n} \in \Circ(X)$, and every element of $\Circ(K)$ is represented by a word in partitioned unitaries localized on $K$. The number $n$ is the \emph{depth} of the word, the $g_i^{\eps_i}$ are its \emph{layers}, and the gates of the layers are its \emph{gates}, the gates of $g^{-1}$ being the $u_\lambda^{-1}$. We say the gates of a word lie in $S\subseteq X$ if every nontrivial block of every layer is contained in $S$. Then
$\phi(A) = \Ad_{g_1}^{\eps_1} \circ \cdots \circ \Ad_{g_n}^{\eps_n}$.

\begin{remark}\label{rem:CnormalQ}
Suppose $X_q$ is uniformly locally finite (i.e. for any $r>0$ there exists $N>0$ such that the number of points of $X_q$ in any ball of radius $r$ is upper-bounded by $N$). Then $\CC(X)$ is a normal subgroup of $\CQ(X)$ \cite{LudewigQCAcoarse}. Normality also holds for any locally-finite $q$ when $X$ is suitably constrained  \cite{JiYang}. However, the theory of circuit-generated automorphisms can be developed without this assumption.
\end{remark}

\begin{definition}[Truncation]\label{def:trunc}
Let $K \subseteq X$ be nonempty and $r \ge 0$. For $g = (\Lambda,(u_\lambda)) \in \Part(X)$, the \emph{truncation} $\tau_{K,r}(g) \in \Part(X)$ has the same partition $\Lambda$ and gates
$$
u'_\lambda := \begin{cases} u_\lambda, & \lambda \subseteq K^r, \\ 1, & \text{otherwise.} \end{cases}
$$
We extend $\tau_{K,r}$ to the endomorphism of the free group $\Circ(X)$ with $\tau_{K,r}(g) $ as above on generators; thus $\tau_{K,r}(g_1^{\eps_1} \cdots g_n^{\eps_n}) = \tau_{K,r}(g_1)^{\eps_1} \cdots \tau_{K,r}(g_n)^{\eps_n}$, and $\tau_{K,r}$ preserves commutators. Since every nontrivial block of $\tau_{K,r}(g)$ lies in $K^r$, one has $\tau_{K,r}(\Circ(X)) \subseteq \Circ(K)$.
\end{definition}

\begin{remark}\label{rem:trunc-rep}
Elements of $\SL$ are regions up to the fuzzy preorder, whereas $\tau_{K,r}$ needs an actual subset of $X$; for $K \in \SL$ we choose a representative. Two representatives $\widetilde K, \widetilde K'$ satisfy $\widetilde K \subseteq \widetilde K'^{\,s}$ and $\widetilde K' \subseteq \widetilde K^{\,t}$ for some $s,t \ge 0$, so changing the representative changes the truncation radius by a bounded amount. All uses of truncation below allow the radius to be taken arbitrarily large and depend only on the stabilized result, so we suppress the choice.
\end{remark}

\begin{definition}[The unitary of a bounded circuit]\label{def:hat}
Let $\Circ_{\mathrm b}(X) \le \Circ(X)$ be the subgroup generated by partitioned unitaries localized on a bounded set; when $\bot \in \SL$ this is $\Circ(\bot)$. Such a $g$ has finitely many nontrivial blocks, and we put
$$
\hat g := \prod_{\lambda \text{ nontrivial}} u_\lambda \in \Uloc .
$$
Let $A \mapsto \hA$ be the homomorphism $\Circ_{\mathrm b}(X) \to \Uloc$ extending $g \mapsto \hat g$; on a word, $\hA = \hat g_1^{\,\eps_1} \cdots \hat g_n^{\,\eps_n}$ in the order of the word. Since $\Ad_{\hat g} = \Ad_g$, one has $\phi(A) = \Ad_{\hA}$ for all $A \in \Circ_{\mathrm b}(X)$.
\end{definition}

Localization properties of circuits are summarized by the following elementary lemma.
\begin{lemma}[Light-cone bound]\label{lem:lightcone}
Let $R\ge0$, let $S\subseteq X$, and let
$$
W=\bigl(g_1^{\eps_1},\dots,g_d^{\eps_d}\bigr),
\qquad
W'=\bigl(g_1'^{\,\eps_1},\dots,g_d'^{\,\eps_d}\bigr)
$$
be words of the same depth $d\ge0$ with the same exponents, such that for every $j$ the partitioned unitaries $g_j=(\Lambda_j,(u_{j,\lambda}))$ and $g'_j=(\Lambda_j,(u'_{j,\lambda}))$ have the same $R$-bounded partition $\Lambda_j$. Let $A,A'\in\Circ(X)$ be the elements represented by $W$ and $W'$.
\begin{enumerate}
\item $\phi(A)\bigl(\CA(S)\bigr)\subseteq\CA(S^{2Rd})$.
\item If $u_{j,\lambda}=u'_{j,\lambda}$ for every $j$ and every block $\lambda\in\Lambda_j$ meeting $S^{2Rd}$, then $\phi(A)=\phi(A')$ on $\CA(S)$. In particular, if no nontrivial block of any layer of $W$ meets $S^{2Rd}$, then $\phi(A)$ is the identity on $\CA(S)$.
\item Let $K\subseteq X$ be nonempty and $r\ge0$. If every nontrivial block of every layer of $W$ which meets $S^{2Rd}$ is contained in $K^r$, then
$$
\phi(A)=\phi\bigl(\tau_{K,r}A\bigr)\qquad\text{on }\CA(S).
$$
\item If $S$ and $K$ are bounded and $K$ is nonempty, then $A_{K,r}:=\tau_{K,r}A$ lies in $\Circ_{\mathrm b}(X)$ for every $r\ge0$, and there is $r_1\ge0$ such that for all $r\ge r_1$
$$
\phi(A)=\phi(A_{K,r})=\Ad_{\hA_{K,r}}\qquad\text{on }\CA(S).
$$
\end{enumerate}
\end{lemma}

\begin{proof}
First, consider single-layer words. Let $g=(\Lambda,(u_\lambda))$ be a partitioned unitary with $\Lambda$ $R$-bounded, let $T\subseteq X$, and let $a\in\CA(T)$. Choose a bounded $T'\subseteq T$ with $a\in\CA(T')$. Since $(g^{-1})_{T'}=g_{T'}^{-1}$, we have
$$
\Ad_g^{\pm1}(a)=g_{T'}^{\pm1}\,a\,g_{T'}^{\mp1},
\qquad
g_{T'}=\prod_{\lambda\cap T'\ne\emptyset}u_\lambda\in\CA\bigl(T'^{\,2R}\bigr).
$$
Hence $\Ad_g^{\pm1}(a)\in\CA(T^{2R})$, and $\Ad_g^{\pm1}(a)$ is determined by $a$ and by the gates of $g$ on the blocks meeting $T$. This proves (1) and (2) for $d=1$. The general case easily follows by induction on $d$. 

(3) The element $\tau_{K,r}A$ is represented by the word $W'$ obtained by truncating every layer of $W$; it has the same exponents and the same partitions as $W$. Let $\lambda$ be a block of the $j$-th layer meeting $S^{2Rd}$. If $\lambda$ is trivial, the gates of $g_j$ and of $\tau_{K,r}(g_j)$ on $\lambda$ are both equal to $1$. If $\lambda$ is nontrivial, then $\lambda\subseteq K^r$ by hypothesis, so its gate is retained by $\tau_{K,r}$. The claim now follows from (2).

(4) Every block meeting $S^{2Rd}$ is contained in $S^{2R(d+1)}$. This set is bounded, hence there is $r_1\ge0$ with $S^{2R(d+1)}\subseteq K^{r}$ for all $r\ge r_1$. For such $r$ the hypothesis of (3) holds, which gives the first equality. Every nontrivial block of every layer of the truncated word lies in the bounded set $K^r$, so $A_{K,r}\in\Circ_{\mathrm b}(X)$, and the second equality, which holds on all of $\CA(X)$, is part of Definition~\ref{def:hat}.
\end{proof}

The first consequence of this lemma is commutator locality for the coflasque precosheaf $U\mapsto \CC(U)$.
\begin{proposition}\label{prop:comm-rep}
Let $U, V \subset X$ be non-empty subsets such that $K:=U\wedge V$ exists in $P_X$, let $\alpha \in \CC(U)$, $\beta \in \CC(V)$, and let $A \in \Circ(U)$, $B \in \Circ(V)$ satisfy $\phi(A) = \alpha$, $\phi(B) = \beta$. Then there is $r_0 \ge 0$ such that for all $r \ge r_0$,
$$
[\alpha,\beta] = \phi\bigl(\tau_{K,r}([A,B])\bigr) = \bigl[\phi(\tau_{K,r}A),\, \phi(\tau_{K,r}B)\bigr].
$$
In particular $[\CC(U),\CC(V)] \subseteq \CC(U \wedge V)$.
\end{proposition}
\begin{proof}
Fix a representative of $K$ in $X$, denoted by the same letter. Choose words for $A$ and $B$ of depths $n$ and $m$, in partitioned unitaries localized on $U$ and on $V$ respectively; choose $a, b \ge 0$ such that the gates of $A$ lie in $U^a$ and those of $B$ in $V^b$, and $R \ge 0$ such that all layers are $R$-bounded.

Put
$$
T := U^{a + 2R(m+1)} \cap V^{b + 2R(n+1)} .
$$
We may choose $a,b$ large enough for $T$ to be non-empty. Then $T \le U$ and $T \le V$ in the fuzzy preorder on nonempty subsets of $X$, hence $T \le K$, and we choose $r_0$ with $T \subseteq K^{r_0}$. Fix $r \ge r_0$ and set
$$
\tilde A := \tau_{K,r}(A), \quad \tilde B := \tau_{K,r}(B), \quad \talpha := \phi(\tilde A), \quad \tilde\beta := \phi(\tilde B),
$$
where $\tilde A$, $\tilde B$ are represented by the words obtained by truncating each layer of $A,B$.

First, $\alpha=\talpha$ on $\CA(V^b)$ and $\beta=\tilde\beta$ on
$\CA(U^a)$. Indeed, let $\lambda$ be a nontrivial block of a layer of $A$
meeting $(V^b)^{2Rn}$. Then $\lambda\subseteq U^a$ and
$\lambda\subseteq V^{b+2R(n+1)}$, so $\lambda\subseteq T\subseteq K^r$.
Lemma~\ref{lem:lightcone}(3) with $S=V^b$ gives the first claim; the second is symmetric.

Second, $\gamma := \talpha^{-1}\alpha$ commutes with $\beta$, and $\delta := \tilde\beta^{-1}\beta$ commutes with $\talpha$. Indeed, we proved above that $\gamma$ is a QCA acting trivially on $\CA(V^b)$, which contains every gate of $B$. Let $g$ be a layer of $B$ and $a \in \CA(X)$. Choose a bounded $S$ with $a, \gamma^{-1}(a) \in \CA(S)$. Then
$$
\gamma \Ad_g \gamma^{-1}(a) = \gamma\bigl(g_S\,\gamma^{-1}(a)\,g_S^{-1}\bigr) = \gamma(g_S)\,a\,\gamma(g_S)^{-1} = g_S\, a\, g_S^{-1} = \Ad_g(a),
$$
since $\gamma$ fixes each factor of $g_S$. Hence $\gamma$ commutes with each $\Ad_g^{\pm 1}$ and so with $\beta$. The gates of $\tilde A$ are among those of $A$, hence lie in $\CA(U^a)$, on which $\delta$ acts trivially; the same computation shows that $\delta$ commutes with $\talpha$.
This implies
$$
[\alpha,\beta] = \talpha\gamma\beta\gamma^{-1}\talpha^{-1}\beta^{-1} = \talpha\beta\talpha^{-1}\beta^{-1},
$$
and therefore
$$
\talpha\beta\talpha^{-1}\beta^{-1} = \talpha\tilde\beta\,\delta\talpha^{-1}\delta^{-1}\,\tilde\beta^{-1} = [\talpha,\tilde\beta].
$$
Finally $[\talpha,\tilde\beta] = \phi([\tilde A,\tilde B]) = \phi(\tau_{K,r}([A,B]))$ because $\phi$ and $\tau_{K,r}$ are homomorphisms, and $\tau_{K,r}([A,B]) \in \Circ(K)$ by Definition~\ref{def:trunc}.
\end{proof}

\subsection{The crossed precosheaf \texorpdfstring{$\CC$}{C}}

\begin{theorem}[The crossed precosheaf of circuits]
\label{thm:crossed-precosheaf-circuits}
Let $\SL$ be a meet-subsemilattice on $X$. Define a precosheaf
$\CC:\SL\to\Grp$, $U\longmapsto\CC(U)$,
where $\CC(U)$ is the group of circuit-generated automorphisms
localized on $U$, and the corestriction maps are the subgroup
inclusions
$\CC(U)\hookrightarrow\CC(V)$ whenever $U\leq V$.

For every $I$-indexed \v{C}ech configuration
$
\xi=((U_i)_{i\in I}\vert W)\in\Cech_I(\SL)
$
and every $A,B\subseteq I$, define the commutator pairing
$$
h^\xi_{A,B}:
\CC(U_A^\xi)\times\CC(U_B^\xi)
\longrightarrow
\CC(U_{A\cup B}^\xi)
$$
by 
$h^\xi_{A,B}(\alpha,\beta):=[\alpha,\beta]$. Then $\CC$, equipped with these commutator pairings, is a crossed
precosheaf in the sense of
Definition~\ref{def:crossed-precosheaf-structure}. 
\end{theorem}

\begin{proof}
The assignment $U\mapsto\CC(U)$ is
order-preserving, so $\CC(U)$ is a coflasque precosheaf of subgroups of $\CC(X)$.
By Proposition~\ref{prop:comm-rep},
$[\CC(U),\CC(V)]\subseteq \CC(U\wedge V)
$
for all $U,V\in \SL$. Therefore
$\CC$, equipped with the ordinary group commutator pairings, is an
instance of the coflasque crossed precosheaf of subgroups described in
Example~\ref{ex:crossed-precosheaf-subgroups}.
\end{proof}

The fuzzy preorder on nonempty subsets of $X$ has binary joins: for nonempty $U, V \subseteq X$ one has $(U \cup V)^r = U^r \cup V^r$, so $U \cup V$ is a least upper bound of $U$ and $V$. We say that a meet-subsemilattice $\SL$ on $X$ is \emph{closed under joins} if for $U, V \in \SL$ the class of $U \cup V$ lies in $\SL$; it is then a sublattice of $P_X$. For $X = \RR^d$ the meet-subsemilattice $\SCSets_d$ is closed under joins, since $c(A \cup B) = c(A) \cup c(B)$ for spherical closed semilinear sets $A, B$.

\begin{proposition}\label{prop:C-epi}
Let $U, V \subseteq X$ be nonempty. Then
$$
\CC(U \cup V) = \CC(U)\,\CC(V).
$$
Consequently, if $\SL$ is a meet-subsemilattice on $X$ closed under joins, the crossed precosheaf $\CC$ of Theorem~\ref{thm:crossed-precosheaf-circuits} is a coflasque crossed epiprecosheaf on $\SL$ in the sense of Definition~\ref{def:epi}.
\end{proposition}

\begin{proof}
Fix representatives of $U$ and $V$. Since $U, V \le U \cup V$, both $\CC(U)$ and $\CC(V)$ are subgroups of $\CC(U \cup V)$. Since $U$ is the meet of $U$ and $U\cup V$ and $V$ is the meet of $V$ and $U\cup V$, Proposition \ref{prop:comm-rep} implies that $\CC(U),\CC(V)$ are normal subgroups of $\CC(U\cup V)$. Therefore $\CC(U)\CC(V)$ is a subgroup of $\CC(U \cup V)$ as well. 

It remains to prove that every generator of $\CC(U \cup V)$ lies in $\CC(U)\CC(V)$.
Let $g = (\Lambda, (u_\lambda)) \in \Part(X)$ be localized on $U \cup V$, with $\Lambda$ $R$-bounded, and choose $a \ge 0$ such that every nontrivial block of $g$ lies in $(U \cup V)^a = U^a \cup V^a$. Define two partitioned unitaries with the same partition $\Lambda$,
$$
g^U := \bigl(\Lambda, (u^U_\lambda)\bigr), \qquad u^U_\lambda := \begin{cases} u_\lambda, & \lambda \cap U^a \ne \emptyset, \\ 1, & \text{otherwise,} \end{cases}
\qquad
g^V := \bigl(\Lambda, (u^V_\lambda)\bigr), \qquad u^V_\lambda := \begin{cases} u_\lambda, & \lambda \cap U^a = \emptyset, \\ 1, & \text{otherwise.} \end{cases}
$$
Every nontrivial block of $g^U$ meets $U^a$ and hence lies in $U^{a + 2R}$, so $g^U$ is localized on $U$. Every nontrivial block of $g^V$ lies in $U^a \cup V^a$ but does not meet $U^a$, so it lies in $V^a$, and $g^V$ is localized on $V$. The nontrivial blocks of $g^U$ and $g^V$ are disjoint, so for every bounded $S \subseteq X$ the finite products satisfy $g_S = g^U_S\, g^V_S$ with commuting factors, and therefore
$$
\Ad_g = \Ad_{g^U} \circ \Ad_{g^V} \in \CC(U)\,\CC(V).
$$
\end{proof}

In general, it is not true that $\CC$ is meet-preserving. However, the following weaker result is often useful.
\begin{lemma}\label{lem:CC-and-meets}
Let $U,V\in\SL$ be such that $\CC(U\wedge V)=\CQ(U\wedge V)$. Then $\CC(U\wedge V)=\CC(U)\cap \CC(V).$
\end{lemma}
\begin{proof}
    Let $\alpha\in\CC(U)\cap\CC(V)$. Then $\alpha\in\CQ(U)\cap\CQ(V)=\CQ(U\wedge V)=\CC(U\wedge V)$. Hence $\CC(U)\cap\CC(V)\subseteq \CC(U\wedge V)$. The reverse inclusion is immediate. 
\end{proof}

\subsection{The crossed precosheaf \texorpdfstring{$\CE$}{E}}\label{sec:E}

Throughout this subsection $\SL$ is a meet-subsemilattice on $X$ with $\bott \in \SL$. We assume $X_q \ne \emptyset$; otherwise $\CA(X) = \bbC$, $\Uloc = U(1)$, and everything below is trivial. We fix a bounded representative of $\bott$; then $\bott^r$ is a bounded set for every $r$, and every bounded subset of $X$ lies in $\bott^r$ for $r$ large. For $A \in \Circ(X)$ we write
$$
A_r := \tau_{\bott, r}(A) \in \Circ(\bott), \qquad \hA_r \in \Uloc,
$$
using Definitions~\ref{def:trunc} and~\ref{def:hat}.

For $u \in \Uloc$ choose a bounded $S \subseteq X$ with $u \in \CA(S)$ and $S \cap X_q \ne \emptyset$, and let $g_u \in \Part(X)$ be the partitioned unitary whose partition consists of the single block $\lambda_u := S \cap X_q$ together with the singletons of $X_q \setminus \lambda_u$, with gate $u$ on $\lambda_u$ and $1$ elsewhere. Then $g_u$ is localized on $\bott$, $\hat g_u = u$, and $\phi(g_u) = \Ad_u$. In particular $\Ad_u \in \CC(\bott) \subseteq \CC(V)$ for every $V \in \SL$, and conversely $\CC(\bott) = \Ad(\Uloc)$, since a circuit localized on $\bott$ has finitely many nontrivial gates.

\begin{definition}[The precosheaf $\CE$]\label{def:E}
Define
$$
\CE(U) := \begin{cases} \CC(U), & U \ne \bott, \\ \Uloc, & U = \bott, \end{cases}
$$
with corestriction maps, for $U \le V$,
$$
\mu^{\CE}_{U,V} := \begin{cases} \text{the inclusion } \CC(U) \hookrightarrow \CC(V), & U \ne \bott, \\ \id_{\Uloc}, & U = V = \bott, \\ \Ad : \Uloc \to \CC(V), & U = \bott \ne V. \end{cases}
$$
These compose correctly, so $\CE : \SL \to \Grp$ is a precosheaf.
\end{definition}

\begin{definition}[The projection $\pi : \CE \to \CC$]\label{def:pi}
Let $\pi_U := \id$ for $U \ne \bott$ and $\pi_{\bott} := \Ad : \Uloc \to \CC(\bott)$. This is a morphism of precosheaves. For $x \in \CE(U)$ we write $\bx := \pi_U(x) \in \CC(U)$; thus $\bx = x$ if $U \ne \bott$ and $\bx = \Ad_x$ if $U = \bott$. The map $x \mapsto \bx$ is a homomorphism, and $\overline{\mu^{\CE}_{U,V}(x)} = \bx$ for all $U \le V$.
\end{definition}

\begin{definition}[Commutator pairing \cite{KapustinXu}]\label{def:eta}
Let $U, V \in \SL$ with $U \wedge V = \bott$, let $\alpha \in \CC(U)$, $\beta \in \CC(V)$, and choose $A \in \Circ(U)$, $B \in \Circ(V)$ with $\phi(A) = \alpha$, $\phi(B) = \beta$. The \emph{commutator pairing} is
$$
\eta(\alpha,\beta) := \lim_{r \to \infty} [\hA_r, \hB_r] = \lim_{r \to \infty} \alpha(\hB_r)\, \hB_r^{-1} = \lim_{r \to \infty} \hA_r\, \beta(\hA_r^{-1}) \;\in \Uloc .
$$
\end{definition}

\begin{proposition}[Properties of the commutator pairing \cite{KapustinXu}]\label{prop:eta}
In the situation of Definition~\ref{def:eta}, we have
\begin{enumerate}
\item \emph{Stabilization.} There is $r_1 \ge 0$ such that for all $r \ge r_1$ the three expressions $[\hA_r, \hB_r]$, $\alpha(\hB_r)\hB_r^{-1}$, and $\hA_r \beta(\hA_r^{-1})$ coincide and are independent of $r$. The common value depends only on the automorphisms $\alpha$ and $\beta$, not on $U$, $V$, or the chosen representatives.
\item \emph{Lift of the commutator.} $\Ad_{\eta(\alpha,\beta)} = [\alpha,\beta]$.
\item \emph{Crossed multiplicativity.} For $\alpha, \alpha' \in \CC(U)$ and $\beta, \beta' \in \CC(V)$,
$$
\eta(\alpha\alpha', \beta) = \alpha\bigl(\eta(\alpha',\beta)\bigr)\, \eta(\alpha,\beta), \qquad
\eta(\alpha, \beta\beta') = \eta(\alpha,\beta)\, \beta\bigl(\eta(\alpha,\beta')\bigr).
$$
\item \emph{Equivariance.} For every $\gamma \in \CC(X)$, $\eta(\gamma\alpha\gamma^{-1}, \gamma\beta\gamma^{-1}) = \gamma\bigl(\eta(\alpha,\beta)\bigr)$.
\end{enumerate}
\end{proposition}
\begin{proof}
\emph{1.}
Fix representatives $U,V$ and circuit representatives $A,B$ of
$\alpha,\beta$ whose nontrivial gates are contained in $U,V$,
respectively.
By Lemma~\ref{lem:lightcone}(3), the restriction of $\alpha$ to
$\CA(V)$ depends only on gates of $A$ contained in $U\cap V^{2Rd}$.
Since $U\wedge V=\bott$, this set is bounded, so all these gates
are retained in $A_r$ for sufficiently large $r$.
Applying the same argument to $B$, we obtain, for all sufficiently
large $r,s$,
$$
\Ad_{\hA_r}\big|_{\CA(V)}
=\alpha\big|_{\CA(V)},
\qquad
\Ad_{\hB_s}\big|_{\CA(U)}
=\beta\big|_{\CA(U)}.
$$
Since $\hA_r\in\CA(U)$ and $\hB_s\in\CA(V)$, it follows that
$$
[\hA_r,\hB_s]
=\alpha(\hB_s)\hB_s^{-1}
=\hA_r\,\beta(\hA_r^{-1}).
$$
The middle expression is independent of $r$, and the last is
independent of $s$. Thus the commutator is constant for all
sufficiently large $r,s$. Taking $r=s$ proves the asserted equality
and stabilization.

We next prove independence of the circuit representatives and
localization regions. Suppose also that
$\alpha\in\CC(U')$ and $\beta\in\CC(V')$, where
$U',V'\in\SL$ and $U'\wedge V'=\bott$.
Fix representatives $U',V'$ and circuit representatives $A',B'$
of $\alpha,\beta$ whose nontrivial gates are contained in $U',V'$,
respectively. The set
$$
F:=(U\cap U')\cap(V\cup V')
\subseteq (U\cap V)\cup(U'\cap V')
$$
is bounded.
Since $\alpha$ is localized on $U\cap U'$, it preserves
$\CA(U\cap U')$.
Choose a bounded $K\subseteq U\cap U'$ such that $\CA(K)$ contains
both $\CA(F)$ and $\alpha(\CA(F))$. The inclusion and the restriction
of $\alpha$ are two unital $*$-embeddings of the same full matrix
algebra into $\CA(K)$, so they are conjugate by a unitary
$w\in\CA(K)$. Thus $\Ad_w$ and $\alpha$ agree on $\CA(F)$,
and both act trivially outside $U\cap U'$. Consequently,
$$
\Ad_w\big|_{\CA(V\cup V')}
=\alpha\big|_{\CA(V\cup V')}.
$$
For sufficiently large $r$, we therefore have
$$
\begin{aligned}
[\hA_r,\hB_r]
&=\alpha(\hB_r)\hB_r^{-1}
=w\hB_r w^{-1}\hB_r^{-1},\\
[\hA'_r,\hB'_r]
&=\alpha(\hB'_r)(\hB'_r)^{-1}
=w\hB'_r w^{-1}(\hB'_r)^{-1}.
\end{aligned}
$$
Since $w\in\CA(K)$ and $K$ is bounded,
Lemma~\ref{lem:lightcone}(3) implies that, for sufficiently large $r$,
the truncations $B_r$ and $B'_r$ retain all gates needed to compute
$\beta$ on $\CA(K)$. Hence
$$
\Ad_{\hB_r}(w^{-1})
=\beta(w^{-1})
=\Ad_{\hB'_r}(w^{-1}),
$$
and consequently
$$
[\hA_r,\hB_r]
=w\,\beta(w^{-1})
=[\hA'_r,\hB'_r].
$$

\emph{2.}
Choose circuit representatives $A,B$ of $\alpha,\beta$.
For sufficiently large $r$, stabilization gives
$$
\eta(\alpha,\beta)=[\hA_r,\hB_r].
$$
Since $\Ad$ is a group homomorphism and
$\Ad_{\hA_r}=\phi(A_r)$, we obtain
$$
\Ad_{\eta(\alpha,\beta)}
=[\Ad_{\hA_r},\Ad_{\hB_r}]
=[\phi(A_r),\phi(B_r)]
=[\alpha,\beta],
$$
where the last equality follows from
Proposition~\ref{prop:comm-rep}.

\emph{3.}
Choose circuit representatives $A,A'\in\Circ(U)$ of
$\alpha,\alpha'$ and $B\in\Circ(V)$ of $\beta$.
Truncation and the hat map preserve products. For sufficiently
large $r$, we have
$$
\begin{aligned}
\eta(\alpha\alpha',\beta)
&=[\hA_r\hA'_r,\hB_r]\\
&=\hA_r[\hA'_r,\hB_r]\hA_r^{-1}
  [\hA_r,\hB_r]\\
&=\hA_r\,\eta(\alpha',\beta)\,\hA_r^{-1}
  \eta(\alpha,\beta)\\
&=\alpha\bigl(\eta(\alpha',\beta)\bigr)\,\eta(\alpha,\beta).
\end{aligned}
$$
The first and third equalities follow from stabilization, the second
from the commutator identity, and the last from
Lemma~\ref{lem:lightcone}(3), since $\eta(\alpha',\beta)$ is a local
unitary. The second identity follows similarly.

\emph{4.}
Write $\widetilde\alpha=\gamma\alpha\gamma^{-1}$ and
$\widetilde\beta=\gamma\beta\gamma^{-1}$. By normality,
$\widetilde\alpha\in\CC(U)$ and $\widetilde\beta\in\CC(V)$.
Fix representatives $U,V$ and circuit representatives $A,\widetilde B$
of $\alpha,\widetilde\beta$ whose nontrivial gates are contained in
$U,V$, respectively. Let $\ell$ bound the propagation of
$\gamma$ and $\gamma^{-1}$.

As in item~(1), Lemma~\ref{lem:lightcone}(3) gives, for sufficiently
large $r,s$,
$$
\Ad_{\hA_r}\big|_{\CA(V^\ell)}
=\alpha\big|_{\CA(V^\ell)},
\qquad
\Ad_{\widehat{\widetilde B_s}}\big|_{\CA(U^\ell)}
=\widetilde\beta\big|_{\CA(U^\ell)}.
$$
The first equality implies that $\Ad_{\gamma(\hA_r)}$ agrees
with $\widetilde\alpha$ on $\CA(V)$.
Since $\widehat{\widetilde B_s}\in\CA(V)$ and
$\gamma(\hA_r)\in\CA(U^\ell)$, these equalities and stabilization
give
$$\eta(\widetilde\alpha,\widetilde\beta)
=\widetilde\alpha(\widehat{\widetilde B_s})
  \widehat{\widetilde B_s}^{-1}
=[\gamma(\hA_r),\widehat{\widetilde B_s}]
=\gamma(\hA_r)\,
  \widetilde\beta\bigl(\gamma(\hA_r^{-1})\bigr)
=\gamma\bigl(\hA_r\,\beta(\hA_r^{-1})\bigr)
=\gamma\bigl(\eta(\alpha,\beta)\bigr).$$
\end{proof}

We will also need the following elementary consequences.
\begin{lemma}\label{lem:eta-more}
Let $U \wedge V = \bott$, $\alpha \in \CC(U)$, $\beta \in \CC(V)$, $u, v \in \Uloc$.
\begin{enumerate}
\item \emph{Skew-symmetry.} $\eta(\beta,\alpha) = \eta(\alpha,\beta)^{-1}$.
\item \emph{Normalization.} $\eta(1,\beta) = \eta(\alpha,1) = 1$.
\item \emph{Inner arguments.}
$$
\eta(\Ad_u, \beta) = u\,\beta(u^{-1}), \qquad \eta(\alpha, \Ad_v) = \alpha(v)\,v^{-1}, \qquad \eta(\Ad_u, \Ad_v) = [u,v].
$$
In particular these expressions depend on $u$, $v$ only through $\Ad_u$, $\Ad_v$.
\end{enumerate}
\end{lemma}
\begin{proof}
(1) By Proposition~\ref{prop:eta}(1), for $r$ large $\eta(\alpha,\beta) = [\hA_r, \hB_r]$ and $\eta(\beta,\alpha) = [\hB_r, \hA_r] = [\hA_r, \hB_r]^{-1}$.
(2) Take the empty word as representative of $1$.
(3) Since $\eta$ depends only on the automorphisms, we may use $g_u$ as a representative of $\Ad_u \in \CC(\bott)$; then $(g_u)_r = g_u$ and $\hat{(g_u)}_r = u$ for $r$ large, so the third expression in Definition~\ref{def:eta} gives $\eta(\Ad_u,\beta) = u\beta(u^{-1})$. The second formula follows from the second expression in the same way, and the third from either.
\end{proof}

Now we are ready to define the crossed structure on $\CE$.

\begin{theorem}[The unitary-lifted crossed precosheaf]\label{thm:E}
For every $I$-indexed \v{C}ech configuration $\xi = ((U_i)_{i \in I} \mid W)$ and all $A, B \subseteq I$, define
$$
h^{\CE,\xi}_{A,B} : \CE(U^\xi_A) \times \CE(U^\xi_B) \longrightarrow \CE(U^\xi_{A \cup B}), \qquad
h^{\CE,\xi}_{A,B}(x,y) := \begin{cases} [x,y], & U^\xi_{A\cup B} \ne \bott, \\ \eta(\bx, \by), & U^\xi_{A \cup B} = \bott. \end{cases}
$$
Then $\CE$, equipped with these pairings, is a crossed precosheaf on $\SL$ in the sense of Definition~\ref{def:crossed-precosheaf-structure}.
\end{theorem}

Recall that $U^\xi_{A\cup B}=U^\xi_A\wedge U^\xi_B$.  If this region is
not $\bott$, the pairing is the ordinary commutator and is well defined by
Proposition~\ref{prop:comm-rep}; if it is $\bott$, the pairing is the unitary
lift $\eta(\bx,\by)$.  Thus all displayed pairings are well defined.

\begin{proof}
Away from $\bott$, all crossed-cube identities are inherited from the
crossed precosheaf $\CC$.  At $\bott$, compatibility with the structure
maps, crossed multiplicativity, and equivariance follow from
Proposition~\ref{prop:eta} and Lemma~\ref{lem:eta-more}; the only substantial
remaining case, the Hall-Witt identity, reduces after truncation to the
ordinary Hall-Witt identity in $\Uloc$.  The detailed verification of the
crossed-cube, fixed-fiber, and reindexing axioms is given in
Appendix~\ref{app:qca}.
\end{proof}

The projection $\pi:\CE\to\CC$ of Definition~\ref{def:pi} is in fact a
surjective morphism of crossed precosheaves.  Indeed, preservation of the
pairings at a $\bott$-intersection is exactly the identity
$\Ad_{\eta(\alpha,\beta)}=[\alpha,\beta]$ from
Proposition~\ref{prop:eta}(2); elsewhere it is immediate.  Its kernel is the
skyscraper precosheaf of abelian groups
$$
\mathcal Z_{\bott}(U)=
\begin{cases}
U(1),&U=\bott,\\
1,&U\ne\bott,
\end{cases}
$$
with trivial pairings.

\begin{remark}\label{rem:E-actions}
Let $A \subseteq B \subseteq I$, so that $U^\xi_B \le U^\xi_A$, and let $x \in \CE(U^\xi_A)$, $y \in \CE(U^\xi_B)$. The crossed-cube action ${}^x y := h^{\CE,\xi}_{A,B}(x,y)\,y$ is as follows.
\begin{itemize}
\item If $U^\xi_B \ne \bott$: ${}^x y = [x,y]\,y = xyx^{-1}$.
\item If $U^\xi_B = \bott \ne U^\xi_A$, so $y = u \in \Uloc$: ${}^x u = \eta(x, \Ad_u)\,u = x(u)u^{-1}u = x(u)$ by Lemma~\ref{lem:eta-more}(3).
\item If $U^\xi_A = U^\xi_B = \bott$, so $x = u$, $y = v \in \Uloc$: ${}^u v = \eta(\Ad_u,\Ad_v)\,v = [u,v]\,v = uvu^{-1} = \Ad_u(v)$.
\end{itemize}
In all cases: if $y \in \Uloc$ then ${}^x y = \bx(y)$, and always $\overline{{}^x y} = \bx\,\by\,\bx^{-1}$. Thus circuit-generated automorphisms act on each other by conjugation and on local unitaries by evaluation, and local unitaries act by inner conjugation.
\end{remark}

\subsection{Anomalies for quantum spin systems}

Let $(X,q)$ be a quantum spin system, and let $\SL$ be a meet-subsemilattice on $X$ with respect to the fuzzy preorder (Section~\ref{sec:meetsubsemi}) which is closed under joins and is a distributive lattice; for $X=\RR^d$ one may take $\SL=\SCSets_d$. Let $G$ be any group. By Lemma~\ref{lem:fuzzification} and the discussion following it, the
fuzzified precosheaf $\CGos$ of on-site gauge symmetries of
Section~\ref{sec:gaugingnew}, with the diagonal embedding
$\iota\colon G\to\CGos(X)$, is a gauge precosheaf for $G$ over
$(X,\SL)$ in the sense of Definition~\ref{def:gauge-precosheaf}.

\begin{corollary}\label{cor:onesite-anomaly-free}
    Let $\rho$ be a homomorphism from $G$ to $\CQ(X)$. If $(\CQ,\rho)$ admits a gauging with respect to $(\CGos,\iota)$, then the anomaly $\alpha_{\CQ,\rho}:BG\ra \chB_\CQ(\fU|X)$ is null-homotopic for any cover $\fU$. The same is true if one replaces $\CQ$ with $\CC$ or $\CE$.
\end{corollary}

Call an action $\rho_0:G\to\CC(X)$ \emph{projectively on-site} if
$$
\rho_0(g)=\prod_{x\in X_q}\Ad_{u_x(g)},
$$
where each $g\mapsto u_x(g)\in\CU(\{x\})$ is a projective unitary
representation.  It is \emph{on-site} if all these representations are
linear.

\begin{corollary}[On-site actions up to conjugacy]
\label{cor:onsite-up-to-conjugacy}
Let $\rho:G\to\CQ(X)$ be an action.
\begin{enumerate}
\item If $\rho$ is $\CC(X)$-conjugate to a projectively on-site action,
then $\alpha_{\CC,\rho}$ is trivial.
\item If $\rho$ is $\CQ(X)$-conjugate to a projectively on-site action,
then $\alpha_{\CQ,\rho}$ is trivial.
\item If $\rho$ takes values in $\CC(X)=\CE(X)$ and is
$\CC(X)$-conjugate to an on-site action, then
$\alpha_{\CE,\rho}$ is trivial.
\end{enumerate}
\end{corollary}

\begin{proof}
A projectively on-site action has a gauging into $\CC$ given by
$$
\varrho(U)(f)=
\prod_{x\in X_q}\Ad_{u_x(f(x))},
\qquad f\in\CGos(U),
$$
and hence, after composing with $\CC\hookrightarrow\CQ$, a gauging into
$\CQ$.  If the on-site representations are linear, this gauging lifts to
$\CE$ by setting
$$
\varrho(\bott)(f)=\prod_{x\in X_q}u_x(f(x)).
$$
Conjugating these morphisms by an element of $\CC(X)$, or by an element of
$\CQ(X)$ in the $\CQ$-case, gives gaugings of the conjugate actions.
The result follows from Corollary~\ref{cor:onesite-anomaly-free}.
\end{proof}

\section{Computations for quantum spin systems
}\label{sec:computations}
In this section, we compute the \v{C}ech-Loday homotopy types and anomalies of quantum spin systems in one and two spatial dimensions. Section~\ref{sec:1d} treats the 1d case. Section~\ref{sec:2d} computes the anomaly for a particular cover of the 2d plane. Section~\ref{sec:2d-cofinal} shows that the 2d pro-homotopy type is essentially constant and isomorphic to the type computed in Section~\ref{sec:2d}.

\subsection{One spatial dimension}\label{sec:1d}
Take $X=\RR$. Its sphere at infinity is the two-point set
$\{L,R\}$,
and hence
$\SCSets_1=\{\bott,L,R,X\}$. The cover $\fU=(L,R)$ is the initial object in the category $\Cov_\preceq(X)$, thus for any crossed precosheaf $\CF$ the \v{C}ech-Loday pro-homotopy type $\chbfB_{\CF}(X)$ is isomorphic to the constant pro-homotopy type $\chB_{\CF}(L,R\vert X)$. This property is special to the 1d case. 

Below we compute this homotopy type for the crossed precosheaves $\CC,\CE,$ and $\CQ$. We will see that in the first two cases the answer does not depend on the quantum spin system being considered (i.e. on the function $q$), while in the third case it does.

We start with the following observation.
\begin{lemma}
    For $X=\RR$ the crossed precosheaf $\CC$ is meet-preserving. 
\end{lemma}
\begin{proof}
    We need to check that $\CC(U\wedge  V)=\CC(U)\cap\CC(V)$ for all $U,V\in \SCSets_1$. The only nontrivial case is $U=L$ and $V=R$. The result follows from Lemma \ref{lem:CC-and-meets} and $\CQ(\bott)=\im\Ad$ where $\Ad:\Uloc\to \CQ(X)$ is the inner action of $\Uloc$ on $\CA(X)$. 
\end{proof}
Combining this with Proposition \ref{prop:C-epi} we get
\begin{corollary}
    The Cech-Loday homotopy type $\chB_\CC(L,R|X)$ is contractible.
\end{corollary}

Next, let us compute $\chB_\CE(L,R|X)$. 
Here $\CE$ agrees with $\CC$ except at $\bott$, where its value is $\Uloc$. The non-identity corestrictions from $\bott$ are given by $\Ad$, and the other non-identity corestrictions are the subgroup inclusions $\CC(R)\hookrightarrow\CC(X)$ and $\CC(L)\hookrightarrow\CC(X)$.
Whenever $U\wedge V\neq\bott$, the crossed-precosheaf pairing
is the ordinary commutator. Since $R\wedge L=\bott$, the
pairing between $\alpha\in\CC(R)$ and $\beta\in\CC(L)$ is instead
the lift
$$
h_{R,L}(\alpha,\beta)
=
\eta(\alpha,\beta)\in\Uloc,
\qquad
\Ad_{\eta(\alpha,\beta)}=[\alpha,\beta].
$$
For $u,v\in\Uloc$, the pairings involving $\bott$ are
$$
h_{\bott,R}(u,\alpha)=u\alpha(u^{-1}),
\qquad
h_{R,\bott}(\alpha,u)=\alpha(u)u^{-1},
\qquad
h_{\bott,\bott}(u,v)=[u,v],
$$
with the analogous formulas for $\CC(L)$.

By Theorem~\ref{thm:E}, these
corestriction maps and commutator pairings make $\CE$ into a crossed
precosheaf on $\SCSets_1$.

Consider the one-element \v{C}ech configuration
$\xi=(\bott\vert R)$.
The associated crossed $1$-cube $\Cech_{\CE}(\xi)$ is the crossed module
$$
\partial=\Ad:\Uloc\longrightarrow\CC(R).
$$
The action of $\CC(R)$ on $\Uloc$ is given by evaluation. The homotopy groups of the corresponding 2-type are $\pi_1=\CC(R)/{\im\Ad}$ and $\pi_2=U(1)$, with $\pi_1$ acting trivially on $\pi_2$ (see Example \ref{ex:loday-crossed-module}). 

To every crossed module with homotopy groups $\pi_1$ and $\pi_2$ one can associate an element $\frl\in H^3(B\pi_1,\pi_2)$ \cite{EilenbergMacLane1947}. It is the Postnikov class of the classifying space of the crossed module \cite{Noohi}. For the crossed module $\Cech_{\CE}(\bott|R)$ the class $\frl$ is the universal Else-Nayak anomaly class \cite{KapustinXu}. More precisely, let $G$ be a group and $\rho:G\ra \CC(X)$ be a homomorphism. Since $\CC(X)=\CC(R)\CC(L)$ and $\CC(R)\cap\CC(L)=\im\Ad$, there is a canonical truncation homomorphism $t:\CC(X)\ra \CC(R)/{\im\Ad}$. The Else-Nayak anomaly class for the $G$-action $\rho$ is $(t\circ\rho)^*\frl\in H^3(BG,U(1))$ \cite{KapustinXu}. 

We will now show that the same anomaly class  can be obtained from the \v{C}ech-Loday homotopy type $\chB_\CE(L,R\vert X)$. 

\begin{theorem}[The anomaly of 1d spin systems]\label{thm:1d-anomaly}
There is a weak equivalence
$\chB_{\CE}(L,R\vert X)\simeq K(U(1),3)$.
which identifies $\pi_3(\chB_{\CE}(L,R\vert X))$ with $U(1)$ via the connecting homomorphism  specified in the proof. Let
$u\in H^3(\chB_{\CE}(L,R\vert X);U(1))$
be the corresponding fundamental class.
The canonical map
$\kappa:B\CC(X)\to\chB_{\CE}(L,R\vert X)$ then defines
$\frl':=\kappa^*u\in H^3(B\CC(X);U(1))$.
This class is given by 
$\frl'=\pm t^*\frl$, where $t:\CC(X)\to \CC(R)/\im{\Ad}$ is the truncation homomorphism. Thus for any group $G$ and any $G$-action $\rho:G\to \CC(X)$ we have $(t\circ\rho)^*\frl=\pm\rho^*\frl'$. That is, the Else-Nayak anomaly is the obstruction for the map $BG\to B\CC(X)\to\chB_{\CE}(L,R\vert X)$ to be null-homotopic.
\end{theorem}

\begin{proof}
    The cover $(L,R)$ of $X$ can be built by starting with the empty cover and adding $L$ and then $R$. By Theorem~\ref{thm:one-coordinate-fibration}, the first stage gives a fibration sequence
    $$
    B\CC(L)\to B\CC(X)\xrightarrow{a_L} \chB_{\CE}(L\vert X).
    $$
    The left map is induced by the injection $\CC(L)\to\CC(X)$, therefore the long exact sequence implies $\chB_{\CE}(L\vert X)\simeq B(\CC(X)/\CC(L))\simeq B(\CC(R)/{\im\Ad})$ and that under this identification $a_L=B t$. 
    
    By the same theorem, the second stage gives a fibration sequence
    $$
    \chB_{\CE}(\bott\vert R)\to \chB_{\CE}(L\vert X)\xrightarrow{a_R} \chB_{\CE}(L,R\vert X).
    $$
    The fiber inclusion induces a map of fundamental groups $\CC(R)/\im, \Ad\to\CC(X)/\CC(L)$ which is a composition of inclusion $\CC(R)\to\CC(X)$ and quotient $\CC(X)\to\CC(X)/\CC(L)$. This map is injective because $\CC(R)\cap\CC(L)=\im\,\Ad$ and surjective because $\CC(X)=\CC(R)\CC(L)$.
    Since $\chB_{\CE}(\bott\vert R)$ is 2-truncated, $\chB_{\CE}(L\vert X)$ is 1-truncated, and the fiber inclusion here induces an isomorphism of fundamental groups, the leftmost map is a model for the first Postnikov truncation. Thus $\chB_{\CE}(L,R\vert X)\simeq K(U(1),3)$ and $a_R$ is the classifying map of the Postnikov truncation. If $u\in H^3(\chB_{\CE}(L,R\vert X),U(1))$ is a fundamental class (well-defined up to a sign convention), then $\frl=\pm a_R^*u$.  

    On the other hand, the canonical map $B\CC(X)\to \chB_{\CE}(L,R\vert X)$ is the composite of $a_L$ and $a_R$. Putting this together, we get the desired result.  
\end{proof}

The homotopy fiber $Y$ of the canonical map $B\CE(X)\to K(U(1),3)$ is a 2-type with $\pi_1(Y)=\CC(X)$, $\pi_2(Y)=U(1)$, and the Postnikov class $\pm t^*\frl$, where the sign depends on the normalization convention for the Postnikov class. If we define the higher symmetry homotopy type to be $Y$, the group of 0-form symmetries is $G_0=\pi_1(Y)=\CC(X)$ and the group of 1-form symmetries is $G_1=\pi_2(Y)=U(1)$, in agreement with expectations.

Finally, let us compute $\chB_\CQ(L,R|X)$. For definiteness, we will assume that $q(x)=N$ if $x\in\ZZ\subset\RR$ and $q(x)=1$ otherwise, so that the quantum spin system has a spatial symmetry which acts by translations. It is known that for such a system $\CC(L)=\CQ(L)$, $\CC(R)=\CQ(R)$, while $\CQ(X)/\CC(X)\simeq\ZZ^P$, where $P$ is the number of distinct prime factors in $N$ \cite{GNVW}. The isomorphism $\CQ(X)/\CC(X)\simeq\ZZ^P$ is known as the GNVW index. 

The crossed square corresponding to the \v{C}ech configuration $(L,R|X)$ is
$$
\begin{tikzcd}[column sep=large,row sep=large]
\CQ(\bott) \arrow[r,hook] \arrow[d,hook]
& \CQ(R) \arrow[d,hook]
\\
\CQ(L) \arrow[r,hook]
& \CQ(X),
\end{tikzcd}
$$
where all arrows are subgroup inclusions. Since $\CQ$ is coflasque and meet-preserving, Example \ref{ex:loday-crossed-square} implies that the corresponding \v{C}ech-Loday type is a 1-type whose fundamental group is $\CQ(X)/(\CQ(L)\CQ(R))=\CQ(X)/\CC(X)\simeq\ZZ^P$. The canonical map $B\CQ(X)\to \chB_\CQ(L,R|X)$ is induced by the GNVW homomorphism $\CQ(X)\to \ZZ^P$.

\subsection{Two spatial dimensions}\label{sec:2d}

Let $X=\RR^2$, and identify its sphere at infinity with $S^1$. Choose
complementary closed semicircles $H,D\subset S^1$. Choose a point $M$ in the
interior of $D$, and write
$$
D=D_L\vee D_R,
\qquad
M=D_L\wedge D_R,
$$
where $D_L$ and $D_R$ are the two closed arcs into which $M$ divides $D$.
Put
$$
L:=H\wedge D_L,
\qquad
R:=H\wedge D_R,
\qquad
\bdH:=H\wedge D=L\vee R.
$$
Thus $L,R,M$ are points of the sphere at infinity, corresponding to rays in
$\RR^2$, while $H,D,D_L,D_R$ correspond to half-planes and sectors. As
before, we use the same symbols for the corresponding conical regions in
$X$. We also have $H\wedge D_L\wedge D_R=\bott$.
The three-element family
$\fU=(D_L,D_R,H)$
is a cover of $X$, and no proper subfamily covers $X$.

We first record the localization identities needed below.

\begin{lemma}
\label{lem:two-dimensional-localization}
For the rays $J=L,R,M$ one has $\CQ(J)=\CC(J)$. Moreover,
$$
\CC(D_L)\cap\CC(D_R)=\CC(M),
\qquad
\CC(L)\cap\CC(R)=\CC(\bott)=\im\Ad,
$$
and
$$
\CC(D_L)\CC(D_R)=\CC(D),
\qquad
\CC(H)\CC(D)=\CC(X),
\qquad
\CC(L)\CC(R)=\CC(\bdH).
$$
Put
$$
K:=\CC(H)\cap\CC(D).
$$
Then
$$
\CC(\bdH)\subseteq K\subseteq\CQ(\bdH),
\qquad
[\CC(H),K]\subseteq\CC(\bdH).
$$
In particular, $K/\CC(\bdH)$ is a central subgroup of
$\CC(H)/\CC(\bdH)$, and hence is abelian.
\end{lemma}

\begin{proof}
A QCA localized on a ray has trivial GNVW index and is therefore
generated by a circuit localized on that ray \cite{GNVW}. Thus
$\CQ(J)=\CC(J)$ for $J=L,R,M$. Applying
Lemma~\ref{lem:CC-and-meets} to $D_L,D_R$ gives the first intersection
identity. Applying it to $L,R$, using
$\CQ(\bott)=\CC(\bott)=\im\Ad$, gives the second. The three product
identities follow from Proposition~\ref{prop:C-epi}.

Monotonicity gives $\CC(\bdH)\subseteq K$, while
Proposition~\ref{prop:qca-meet-preservation} gives
$$
K\subseteq\CQ(H)\cap\CQ(D)=\CQ(\bdH).
$$
The groups $\CC(H)$ and $\CC(D)$ are normal in $\CC(X)$, by
commutator locality, so their intersection $K$ is normal as well.
For $a\in\CC(H)$ and $k\in K$, regard $k$ as an element of $\CC(D)$.
Commutator locality for $\CC$ then gives
$$
[a,k]\in[\CC(H),\CC(D)]
\subseteq\CC(H\wedge D)=\CC(\bdH).
$$
This proves centrality of $K/\CC(\bdH)$ in
$\CC(H)/\CC(\bdH)$; taking $a\in K$ also gives abelianness
directly.
\end{proof}

Set
$$
\Pi:=\CC(H)/\CC(\bdH),
\qquad
\bPi:=\CC(H)/K,
\qquad
\CI:=K/\CC(\bdH).
$$
By Lemma~\ref{lem:two-dimensional-localization}, these groups fit into
a central extension
\begin{equation}
\label{eq:two-dimensional-central-extension}
1\longrightarrow\CI
\longrightarrow\Pi
\xrightarrow{\,q_K\,}\bPi
\longrightarrow1.
\end{equation}
The inclusion $K\hookrightarrow\CQ(\bdH)$ induces an injection
$$
\CI\hookrightarrow
\CQ(\bdH)/\CC(\bdH)
\cong\im(\operatorname{ind}_{\mathrm{GNVW}})
\subseteq\mathbb Q_{>0}^{\times}.
$$
Thus $\CI$ is the group of boundary indices realized by automorphisms
which admit circuit implementations in both half-planes. In general
it need not be the full boundary index group.

We build the cover $\fU$ by adjoining $D_L$, then $D_R$, and finally $H$:
$$
\xi_0=(\,\vert X),
\qquad
\xi_1=(D_L\vert X),
\qquad
\xi_2=(D_L,D_R\vert X),
\qquad
\xi_3=(D_L,D_R,H\vert X).
$$
Only $\xi_3$ is a cover; the preceding configurations are intermediate
stages in its construction. Put
$$
B_D:=\chB_{\CE}(D_L,D_R\vert X),
\qquad
B_{\fU}:=\chB_{\CE}(D_L,D_R,H\vert X).
$$

\begin{proposition}
\label{prop:two-dimensional-half-plane-stage}
There is a weak equivalence
$$
B_D\simeq B\bPi.
$$
Under this equivalence, the canonical map
$B\CC(X)\to B_D$
is induced by the homomorphism
$$
\overline t_H:
\CC(X)\longrightarrow \CC(X)/\CC(D)
\xrightarrow{\ \simeq\ }
\bPi.
$$
Equivalently, if $\alpha=\alpha_H\alpha_D$ with
$\alpha_H\in\CC(H)$ and $\alpha_D\in\CC(D)$, then
$\overline t_H(\alpha)=\alpha_H K$. Two such factorizations give
$H$-factors differing by an element of $K$, so this is truncation to
$H$ modulo precisely that ambiguity.
\end{proposition}

\begin{proof}
Adjoining $D_L$ to the empty configuration and applying
Theorem~\ref{thm:one-coordinate-fibration} gives a fibration sequence
$$
B\CC(D_L)\longrightarrow B\CC(X)
\longrightarrow\chB_{\CE}(D_L\vert X).
$$
Since $\CC(D_L)\trianglelefteq\CC(X)$, the last term is weakly equivalent to
$B(\CC(X)/\CC(D_L))$.

Adjoining $D_R$ gives a fibration sequence
$$
\chB_{\CE}(M\vert D_R)
\longrightarrow
\chB_{\CE}(D_L\vert X)
\longrightarrow
B_D.
$$
The first term is the classifying complex of the injective crossed module
$\CC(M)\hookrightarrow\CC(D_R)$, and hence is weakly equivalent to
$B(\CC(D_R)/\CC(M))$. Under the preceding identification, the map on
fundamental groups is
$$
\CC(D_R)/\CC(M)
\longrightarrow
\CC(X)/\CC(D_L).
$$
It is injective by Lemma~\ref{lem:two-dimensional-localization}. The long
exact homotopy sequence therefore gives
$$
B_D
\simeq
B\bigl(\CC(X)/(\CC(D_L)\CC(D_R))\bigr)
=
B\bigl(\CC(X)/\CC(D)\bigr).
$$
Finally, Lemma~\ref{lem:two-dimensional-localization} and the second
isomorphism theorem give
$$
\CC(X)/\CC(D)
\cong
\CC(H)/(\CC(H)\cap\CC(D))
=
\CC(H)/K
=
\bPi.
$$
Tracing the two add-maps shows that the canonical map from $B\CC(X)$ is the
classifying map of the displayed quotient homomorphism $\overline t_H$.
\end{proof}

The last member of the cover is $H$. Restricting the first two members to
$H$ gives
$$
(D_L,D_R\vert X)|_H
=
(D_L\wedge H,D_R\wedge H\vert H)
=
(L,R\vert H).
$$

\begin{proposition}
\label{prop:kapustin-xu-crossed-square}
Put
$\CK_H:=\chB_{\CE}(L,R\vert H)$.
The underlying crossed square is the crossed square of \cite{KapustinXu}:
$$
\begin{tikzcd}[column sep=large,row sep=large]
\Uloc \arrow[r,"\Ad"] \arrow[d,"\Ad"']
& \CC(R) \arrow[d,hook]
\\
\CC(L) \arrow[r,hook]
& \CC(H),
\end{tikzcd}
$$
with commutator lifting
$\eta:\CC(L)\times\CC(R)\to\Uloc$. Its Loday classifying complex is a connected $3$-type
with
$$
\pi_j(\CK_H)\cong
\begin{cases}
\Pi, & j=1,\\
U(1), & j=3,\\
1, & j\neq1,3.
\end{cases}
$$
Consequently, there is a homotopy fibration sequence
\begin{equation}
\label{eq:kapustin-xu-postnikov-fibration}
\CK_H
\longrightarrow
B\Pi
\xrightarrow{\,\mathfrak t_H\,}
K(U(1),4),
\end{equation}
where $\mathfrak t_H$ represents the Postnikov class
$[\mathfrak t_H]\in H^4(\Pi;U(1))$.
\end{proposition}

\begin{proof}
The Moore complex of the crossed square (see Example \ref{ex:loday-crossed-square}) is
$$
\Uloc
\xrightarrow{\,u\mapsto(\Ad_u^{-1},\Ad_u)\,}
\CC(L)\rtimes\CC(R)
\xrightarrow{\,(\alpha,\beta)\mapsto\alpha\beta\,}
\CC(H).
$$
Its degree-zero homology is
$$
\CC(H)/(\CC(L)\CC(R))
=
\CC(H)/\CC(\bdH)
=
\Pi.
$$
Its middle homology vanishes because by Lemma \ref{lem:CC-and-meets} we have 
$$
\CC(L)\cap\CC(R)=\CC(\bott)
=
\im\Ad.
$$
Its degree-two homology is $\ker\Ad=U(1)$. This gives the displayed
homotopy groups. The action of $\Pi$ on $U(1)$ is trivial because every QCA
fixes scalar operators. The Postnikov decomposition of this $3$-type is
therefore \eqref{eq:kapustin-xu-postnikov-fibration}. By construction,
$\mathfrak t_H$ is the universal anomaly class of
\cite{KapustinXu}, up to the choice of normalization for the Postnikov
class.
\end{proof}

\begin{proposition}
\label{prop:two-dimensional-homotopy-types}
There is a homotopy fibration sequence
\begin{equation}
\label{eq:three-cone-fibration}
\CK_H
\longrightarrow
B\bPi
\xrightarrow{\,a_H\,}
B_{\fU},
\end{equation}
where the middle term is identified using
Proposition~\ref{prop:two-dimensional-half-plane-stage}. The map on
fundamental groups induced by the fiber inclusion is the quotient
$q_K:\Pi\to\bPi$. Moreover,
$$
\pi_j(B_{\fU})\cong
\begin{cases}
\CI, & j=2,\\
U(1), & j=4,\\
1, & j\neq2,4.
\end{cases}
$$
In particular, $B_{\fU}$ is a simply connected $4$-type.
\end{proposition}

\begin{proof}
The fibration sequence is Theorem~\ref{thm:one-coordinate-fibration},
applied when $H$ is adjoined to $(D_L,D_R\vert X)$. Its fiber is
$$
\chB_{\CE}(D_L\wedge H,D_R\wedge H\vert H)
=
\chB_{\CE}(L,R\vert H)
=
\CK_H.
$$
Under the identification
$B_D\simeq B\bPi$, the map on fundamental groups is induced
by $\CC(H)\hookrightarrow\CC(X)$ followed by quotienting by $\CC(D)$. By
Lemma~\ref{lem:two-dimensional-localization}, this is precisely
$$
q_K:\CC(H)/\CC(\bdH)
\longrightarrow
\CC(H)/K.
$$
Since $q_K$ is surjective with kernel $\CI$, the long exact
homotopy sequence gives $\pi_1(B_{\fU})=1$ and
$\pi_2(B_{\fU})\cong\CI$. It also gives
$\pi_3(B_{\fU})=0$ and $\pi_4(B_{\fU})\cong U(1)$, with all higher
groups zero. These identifications use
\eqref{eq:three-cone-fibration}, Proposition~\ref{prop:kapustin-xu-crossed-square},
and the exact sequence \eqref{eq:two-dimensional-central-extension}.
\end{proof}

Let
$$
p:B_{\fU}\longrightarrow K(\CI,2)
$$
be the second Postnikov truncation, normalized to induce the
identification $\pi_2(B_{\fU})\cong\CI$ just obtained. Its homotopy
fiber is $K(U(1),4)$. Put
$$
e:=p\circ a_H:B\bPi\longrightarrow K(\CI,2),
\qquad
F:=\hofib(e).
$$
There is a natural map
$\hofib(a_H)\to F$
induced by $p$. Both fibers are connected. Since $p$ induces an
isomorphism on $\pi_2$ and $B\bPi$ is a $1$-type, comparison
of the long exact sequences identifies the short exact sequence
$$
1\longrightarrow\CI\longrightarrow\pi_1(F)
\longrightarrow\bPi\longrightarrow1
$$
with \eqref{eq:two-dimensional-central-extension}. In particular, the
map from $\hofib(a_H)\simeq\CK_H$ induces an isomorphism
$\Pi\cong\pi_1(F)$. The same long exact sequence shows that $F$ is a
$1$-type. Hence $F\simeq B\Pi$ over $B\bPi$, and $e$
classifies the central extension
\eqref{eq:two-dimensional-central-extension}.

\begin{lemma}
\label{lem:two-dimensional-three-by-three}
With the preceding identifications and compatible choices of the
$U(1)$ coefficient identification, there is a homotopy-commutative
diagram
$$
\begin{tikzcd}[column sep=large,row sep=large]
\CK_H
  \arrow[r]
  \arrow[d,"\simeq"']
& B\Pi
  \arrow[r,"\mathfrak t_H"]
  \arrow[d,"Bq_K"]
& K(U(1),4)
  \arrow[d]
\\
\hofib(a_H)
  \arrow[r]
  \arrow[d]
& B\bPi
  \arrow[r,"a_H"]
  \arrow[d,"e"]
& B_{\fU}
  \arrow[d,"p"]
\\
*
  \arrow[r]
& K(\CI,2)
  \arrow[r,equal]
& K(\CI,2),
\end{tikzcd}
$$
in which all rows and columns are homotopy fibration sequences.
\end{lemma}

\begin{proof}
For the composable maps $a_H,p$, the standard fiber-sequence
comparison (the Verdier lemma, \cite[Lemma~1.2.7]{MayPonto}) gives a
homotopy fibration sequence
$$
\hofib(a_H)
\longrightarrow\hofib(p\circ a_H)
\longrightarrow\hofib(p)
$$
and the corresponding $3\times3$ diagram. The preceding paragraph
identifies its middle fiber with $B\Pi$ and its map to
$B\bPi$ with $Bq_K$. Its upper-left term is $\CK_H$, and the
map $\CK_H\to B\Pi$ induces the identity on $\pi_1=\Pi$; it is
therefore the first Postnikov truncation. The resulting map
$B\Pi\to K(U(1),4)$ represents the Postnikov class of $\CK_H$.
After choosing the coefficient identification compatibly with
\eqref{eq:kapustin-xu-postnikov-fibration}, this map is
$\mathfrak t_H$. This proves the asserted diagram, including the
identification of its top row.
\end{proof}

Let
$\alpha:BG\longrightarrow B_{\fU}$
be a pointed map. Since
$$
K(U(1),4)\longrightarrow
B_{\fU}
\xrightarrow{\,p\,}
K(\CI,2)
$$
is the Postnikov fibration of $B_{\fU}$, the problem of
null-homotoping $\alpha$ naturally separates into two stages.

First, any null-homotopy of $\alpha$ induces a null-homotopy of
$p\circ\alpha$. Hence the class
$$
o_2(\alpha)
:=
[p\circ\alpha]
\in
[BG,K(\CI,2)]
\cong
H^2(G;\CI)
$$
is an obstruction to null-homotoping $\alpha$. It is called the
\emph{primary obstruction}, because $K(\CI,2)$ is the first
nontrivial Postnikov stage of $B_{\fU}$.

Suppose that $o_2(\alpha)=0$, and choose a pointed null-homotopy
$\nu:p\circ\alpha\simeq *$.
By the definition of the homotopy fiber, the pair $(\alpha,\nu)$ determines
a pointed map
$$
\alpha^{(4)}_{\nu}:
BG\longrightarrow
\hofib(p)
\simeq
K(U(1),4)
$$
whose composite with the fiber inclusion
$K(U(1),4)\to B_{\fU}$ is homotopic to $\alpha$. The class
$$
o_4(\alpha,\nu)
:=
[\alpha^{(4)}_{\nu}]
\in
[BG,K(U(1),4)]
\cong
H^4(G;U(1))
$$
is the \emph{secondary obstruction} associated with the chosen
trivialization $\nu$ of the primary obstruction. It vanishes precisely
when $\alpha$ admits a null-homotopy compatible with $\nu$.

The secondary obstruction is therefore defined only after the primary
obstruction has been trivialized, and it may depend on the choice of that
trivialization. Consequently, $\alpha$ is null-homotopic if and only if $o_2(\alpha)=0$ and there exists a null-homotopy $\nu$ of
$p\circ\alpha$ for which $o_4(\alpha,\nu)=0$.

We now apply this general description to
$$
\alpha
=
\alpha_{\CE,\rho}(\fU)
=
a_H\circ B\overline\rho.
$$

\begin{theorem}[The two-dimensional anomaly]
\label{thm:2d-anomaly}
Let $G$ be a group and let
$\rho:G\to\CC(X)=\CE(X)$
be an action by circuit-generated automorphisms. Put
$\overline\rho:=\overline t_H\circ\rho:G\to\bPi$.
Then the anomaly of $(\CE,\rho)$ at the cover $\fU=(D_L,D_R,H)$ is
represented by
\begin{equation}
\label{eq:two-dimensional-anomaly-map}
\alpha_{\CE,\rho}(\fU):
BG
\xrightarrow{\,B\overline\rho\,}
B\bPi
\xrightarrow{\,a_H\,}
B_{\fU}.
\end{equation}
Its primary obstruction is
$$
o_2(\rho)
:=
\bigl[e\circ B\overline\rho\bigr]
\in
H^2(G;\CI).
$$
It vanishes precisely when $B\overline\rho$ lifts through
$Bq_K:B\Pi\to B\bPi$, equivalently when one can choose a homomorphism
$$
\widetilde\rho:G\longrightarrow\Pi,
\qquad
q_K\circ\widetilde\rho=\overline\rho.
$$
After such a lift has been chosen, the secondary obstruction is
$$
o_4(\rho,\widetilde\rho)
:=
\bigl[\mathfrak t_H\circ B\widetilde\rho\bigr]
\in
H^4(G;U(1)).
$$
The anomaly \eqref{eq:two-dimensional-anomaly-map} is null-homotopic if and
only if $o_2(\rho)=0$ and there exists a lift $\widetilde\rho$ for which
$o_4(\rho,\widetilde\rho)=0$.

The secondary class is the pullback of the Postnikov class of the
crossed square of \cite{KapustinXu}. The primary class takes values
in $\CI$, the subgroup of boundary indices realizable in both
half-planes. Its image under the coefficient inclusion into the
full boundary GNVW-index group is the corresponding index-valued
factorization obstruction.
\end{theorem}

\begin{proof}
By Proposition~\ref{prop:two-dimensional-half-plane-stage}, the composite
of the first two add-maps in the construction of $\fU$ is
$B\overline t_H$. The final add-map is $a_H$, proving
\eqref{eq:two-dimensional-anomaly-map}.

By Lemma~\ref{lem:two-dimensional-three-by-three}, the composite
$p\circ\alpha_{\CE,\rho}(\fU)$ is
$e\circ B\overline\rho$. Since $e$ classifies the central extension
\eqref{eq:two-dimensional-central-extension}, its null-homotopy is exactly
the primary obstruction to lifting $B\overline\rho$ to $B\Pi$.

A choice of lift $B\widetilde\rho$ together with its homotopy over
$B\bPi$ specifies a trivialization of the primary
obstruction. By Lemma~\ref{lem:two-dimensional-three-by-three}, a
null-homotopy of $\alpha_{\CE,\rho}(\fU)$ compatible with this
trivialization is equivalent to lifting $B\widetilde\rho$ through
the homotopy fiber of
$\mathfrak t_H:B\Pi\to K(U(1),4)$. By
Proposition~\ref{prop:kapustin-xu-crossed-square}, that homotopy fiber is
$\CK_H\simeq\hofib(a_H)$. The obstruction to this
second lift is precisely
$[\mathfrak t_H\circ B\widetilde\rho]\in H^4(G;U(1))$. This proves all the
assertions.
\end{proof}

\begin{remark}[The uniform square lattice]
\label{rem:uniform-square-lattice-index}
Suppose that $q$ is a constant $N\geq1$ on $\ZZ^2$ and is $1$
elsewhere. Then every QCA localized on $\bdH$ can be realized
by a circuit localized in either $H$ or $D$. Indeed, the GNVW
classification \cite{GNVW} reduces this assertion to realizing shifts
of prime-dimensional tensor factors along the boundary: after such
shifts are removed, the remaining boundary QCA has index $1$ and
belongs to $\CC(\bdH)$. Each prime-dimensional shift is the
boundary action of a half-plane exchange circuit built from SWAP
gates, using the corresponding tensor factor at every lattice site;
reflection and inversion give a realization in the other half-plane.
This is the standard half-plane realization discussed in
\cite[Section~2]{KapustinXu}. Consequently,
$$
K=\CQ(\bdH),
\qquad
\CI=\CQ(\bdH)/\CC(\bdH),
$$
so $\CI$ is the full GNVW-index group for this spin system.
More explicitly, if $P_N$ is the set of prime divisors of $N$, then
$$
\CI\cong
\left\{\prod_{p\in P_N}p^{n_p}:n_p\in\ZZ\right\}
\cong\ZZ^{|P_N|}.
$$
In this case the primary obstruction above is the full boundary
GNVW obstruction, and the two-stage description agrees with the
usual two-dimensional anomaly construction of \cite{KapustinXu}.
\end{remark}

\subsection{Cofinal covers in two spatial dimensions}
\label{sec:2d-cofinal}
Recall that $\SCSets_2$ consists of finite unions of closed arcs and points of $S^1$, with $\bott=\varnothing$. Under the identification with conical regions of $X=\RR^2$, a finite cover of $X$ corresponds to a finite family of such subsets whose union is $S^1$. We now single out a useful family of covers.

\begin{definition}[Fan covers]
\label{def:planar-fans}
A \emph{fan cover} is a cover $F=(A_1,\ldots,A_m)$ obtained by cutting
$S^1$ at $m\geq3$ distinct points and taking the consecutive closed arcs.
Let $\Fan(X)$ be the full subpreorder of
$\Cov_{\preceq}(X)$ consisting of these covers
(Definition~\ref{def:cover-refinement-preorder}). Up to relabeling,
refinements in this family are precisely subdivisions of arcs.
\end{definition}

We now show that refinements between fan covers induce weak equivalences of their \v{C}ech-Loday classifying complexes and that $\Fan(X)$ is cofinal among finite covers of $X$.

\begin{proposition}
\label{prop:planar-fan-refinements}
Every refinement between covers in $\Fan(X)$ induces a
weak equivalence of their \v{C}ech-Loday classifying complexes.
\end{proposition}

\begin{proof}
It suffices to subdivide one member $V$ of a fan cover
$(U_1,\ldots,U_k,V)$, where $k\geq2$, into consecutive closed arcs
$V_1,V_2$. Thus $V_1\vee V_2=V$ and $V_1\wedge V_2$ is a single
interior point of $V$.
Since $V_1,V_2\leq V$, Proposition~\ref{prop:dominated-entry-addition}
gives weak equivalences
\begin{equation}\label{eq:twoaddmaps}
\begin{aligned}
\chB_{\CE}(U_1,\ldots,U_k,V\vert X)
&\xrightarrow[\simeq]{\,+V_1\,}
\chB_{\CE}(U_1,\ldots,U_k,V_1,V\vert X)\\
&\xrightarrow[\simeq]{\,+V_2\,}
\chB_{\CE}(U_1,\ldots,U_k,V_1,V_2,V\vert X).
\end{aligned}
\end{equation}
To compare with the subdivided cover, adjoin $V$ to
$(U_1,\ldots,U_k,V_1,V_2)$. Theorem~\ref{thm:one-coordinate-fibration}
gives the homotopy fibration sequence
\begin{multline*}
\chB_{\CE}(U_1\wedge V,\ldots,U_k\wedge V,V_1,V_2\vert V)
\longrightarrow
\chB_{\CE}(U_1,\ldots,U_k,V_1,V_2\vert X)
\xrightarrow{\ a_V\ }
\chB_{\CE}(U_1,\ldots,U_k,V_1,V_2,V\vert X).
\end{multline*}
Because the original fan has at least three arcs,
$$
U_i\cap V=
\begin{cases}
\{p\}, & \text{where $p$ is an endpoint of $V$ and
                 $\{p\}\leq V_1$ or $\{p\}\leq V_2$},\\
\bott, & \text{with $\bott\leq V_1$ and $\bott\leq V_2$}.
\end{cases}
$$
Thus every entry $U_i\cap V$ is dominated by one of the retained
entries $V_1,V_2$ and can be dropped by
Proposition~\ref{prop:dominated-entry-addition}. Consequently, the
fiber is weakly equivalent to $\chB_{\CE}(V_1,V_2\vert V)$.

All regions in this reduced configuration are nonbottom. By
Definition~\ref{def:E} and Theorem~\ref{thm:E}, its crossed square is
$$
\begin{tikzcd}
\CC(V_1\wedge V_2) \arrow[r,hook] \arrow[d,hook]
  & \CC(V_2) \arrow[d,hook] \\
\CC(V_1) \arrow[r,hook] & \CC(V).
\end{tikzcd}
$$
The meet $V_1\wedge V_2$ is a single point of $S^1$, whose associated
cone in $X$ is a half-ray. Thus the GNVW index of any QCA in $\CQ(V_1\wedge V_2)$ is trivial and therefore $
\CQ(V_1\wedge V_2)=\CC(V_1\wedge V_2)$. By Lemma \ref{lem:CC-and-meets}, we have 
$\CC(V_1)\cap\CC(V_2)=\CC(V_1\wedge V_2)$.
The restriction of $\CC$ to the distributive sublattice
$\{V_1\wedge V_2,V_1,V_2,V\}$ is therefore a meet-preserving crossed
precosheaf. This restricted precosheaf is coflasque, since its
corestriction maps are inclusions, and is an epiprecosheaf by
Proposition~\ref{prop:C-epi}, since the sublattice inherits its joins
from the ambient lattice. As $(V_1,V_2)$ covers $V$, we may apply Corollary~\ref{cor:contractible-protype} to this restricted precosheaf.
The associated crossed square consequently has weakly contractible
classifying complex:
$$
\chB_{\CE}(V_1,V_2\vert V)
=\chB_{\CC}(V_1,V_2\vert V)\simeq *.
$$ 
Thus $a_V$ is a weak equivalence. 

The subdivision refinement
$r:(U_1,\ldots,U_k,V_1,V_2)
\to (U_1,\ldots,U_k,V)$
is the composite $r=d\circ a_V$ where $d:(U_1,\ldots,U_k,V_1,V_2,V)\to (U_1,\ldots,U_k,V)$ is a drop-map. The latter retracts the composite of two weak equivalences of (\ref{eq:twoaddmaps}) and
therefore also induces a weak equivalence. Hence $r$ induces a weak equivalence.

Up to relabeling, every refinement in $\Fan(X)$ can be
obtained by finitely many steps, each dividing one arc into two.
The assertion follows by composition, with
Lemma~\ref{lem:refinement-independence} ensuring independence of the
refinement function.
\end{proof}

\begin{proposition}
\label{prop:planar-fans-cofinal}
The preorder $\Fan(X)$ is cofiltered and cofinal in
$\Cov_{\preceq}(X)$: every finite cover of $X$ admits a
refinement in $\Fan(X)$.
\end{proposition}

\begin{proof}
Given a cover $(U_1,\ldots,U_n\vert X)$, collect, for each $U_i$,
the endpoints of its closed-arc components and its isolated points.
Subdivide $S^1$ at all these points, adding further distinct cut
points if necessary to obtain at least three, and denote the
resulting consecutive closed arcs by $V_1,\ldots,V_m$.
Each $V_j$ satisfies 
$V_j\leq U_i$ for some $U_i$. Thus $(V_1,\ldots,V_m\vert X)$ is a refinement
of the original cover and belongs to $\Fan(X)$.

To see cofilteredness, let two fan covers have endpoint sets
$\{a_1,\ldots,a_r\}$ and $\{b_1,\ldots,b_s\}$.
The consecutive closed arcs obtained by cutting at
$\{a_1,\ldots,a_r\}\cup\{b_1,\ldots,b_s\}$
form a fan cover refining both. Since $\Fan(X)$ is
nonempty, it is therefore cofiltered.
\end{proof}

\begin{theorem}[Essential constancy in two dimensions]
\label{thm:planar-protype-constant}
Let $F_3=(D_L,D_R,H)$ be the cover of Section~\ref{sec:2d}. Then
$$
\chbfB_{\CE}(X)
\cong c\,\chB_{\CE}(F_3\vert X)
\qquad\text{in }\Pro\bigl(\Ho(\SSet)\bigr),
$$
where $c$ denotes the constant pro-object. This isomorphism identifies
$\kappa_{\CE}$ with $\kappa_{\CE}(F_3\vert X)$.
\end{theorem}

\begin{proof}
By Proposition~\ref{prop:planar-fans-cofinal}, $\Fan(X)$
is cofiltered, so the restriction of the \v{C}ech-Loday functor to
$\Fan(X)$ defines a pro-object. Since this family is
cofinal in $\Cov_{\preceq}(X)$, the resulting pro-object
is isomorphic to $\chbfB_{\CE}(X)$.

By Proposition~\ref{prop:planar-fan-refinements}, every transition map
in this restricted diagram is an isomorphism in $\Ho(\SSet)$.
A cofiltered diagram of isomorphisms is isomorphic to the constant
diagram at any one of its values. Taking the value at
$F_3\in\Fan(X)$ therefore gives
$$
\chbfB_{\CE}(X)
\cong c\,\chB_{\CE}(F_3\vert X).
$$
The identification of $\kappa_{\CE}$ with
$\kappa_{\CE}(F_3\vert X)$ follows from naturality.
\end{proof}

\begin{remark}
The choice of three arcs in $F_3$ does not affect the resulting
homotopy type. Indeed, since $\Fan(X)$ is cofiltered,
any two three-arc covers $F_3$ and $\widetilde F_3$ admit a common
refinement $F\in\Fan(X)$. By
Proposition~\ref{prop:planar-fan-refinements}, both refinement maps
induce weak equivalences:
$$
\chB_{\CE}(F_3\vert X)
\xleftarrow{\ \simeq\ }
\chB_{\CE}(F\vert X)
\xrightarrow{\ \simeq\ }
\chB_{\CE}(\widetilde F_3\vert X).
$$
By naturality, these equivalences also identify the corresponding
canonical maps $\kappa_{\CE}$.
\end{remark}

\begin{corollary}[Universality of the two-dimensional anomaly]
\label{cor:universal-planar-anomaly}
For every action $\rho:G\to\CC(X)=\CE(X)$, the anomaly
$\alpha_{\CE,\rho}(F_3)$ is null-homotopic if and only if
$\alpha_{\CE,\rho}(\fU)$ is null-homotopic for every finite cover
$\fU$ of $X$. Thus the obstructions of Theorem~\ref{thm:2d-anomaly}
give a universal criterion for anomaly-freeness in the sense of
Definition~\ref{def:anomaly}.
\end{corollary}

\begin{proof}
Choose a fan $F$ refining both $F_3$ and $\fU$. By
Proposition~\ref{prop:monotone}, its anomaly maps to those at both
covers. Since $\chB_{\CE}(F\vert X)\to
\chB_{\CE}(F_3\vert X)$ is an equivalence, null-homotopy at $F_3$
implies null-homotopy at $F$, and hence at $\fU$. The converse follows
by taking $\fU=F_3$.
\end{proof}

\begin{remark}
    In both one and two spatial dimensions, the \v{C}ech-Loday pro-homotopy type $\chbfB_{\CE}(X)$ is essentially constant. It is natural to ask whether this remains true in higher dimensions. The proof of Proposition~\ref{prop:planar-fan-refinements}, however, uses
$$
\CC(V_1)\cap\CC(V_2)=\CC(V_1\wedge V_2),
$$
which we establish there using the fact that $V_1\wedge V_2$ is a single point of $S^1$. In dimensions $d\geq 3$, intersections arising from analogous subdivisions will not be a single point, so the same argument does not apply. A higher-dimensional result may require a suitable notion of stabilization for QCAs and circuits.
\end{remark}

\appendix
\section{Categorical formulation of crossed precosheaves}
\label{app:categorical}

This appendix supplies the proof of Lemma~\ref{lem:reindexing-crossed-cubes} and the categorical reformulation of crossed precosheaves mentioned in Remark~\ref{rem:categorical}. Throughout, $\Fin$ denotes the category of finite sets and all functions, and $\Cech_\bullet(\SL)$ denotes the $\Cat$-valued presheaf $I\mapsto\Cech_I(\SL)$ on $\Fin$, with pullback along functions as restriction. The reformulation is not used in the body of the paper.

\begin{proof}[Proof of Lemma~\ref{lem:reindexing-crossed-cubes}]
For $A\subseteq J$ and $j\in J$ we have
$r(A\setminus\{j\})=r(A)\setminus\{r(j)\}$ if $j\in A$ and
$r(j)\notin r(A\setminus\{j\})$, and $r(A\setminus\{j\})=r(A)$ otherwise.
Hence, in the notation for composite structure maps introduced before
Lemma~\ref{lem:pairing-restriction}, the structure maps of
Definition~\ref{def:reindexing-crossed-cubes} are
\begin{equation}\label{eq:reindexed-structure-maps}
\mu^{r^*M}_{j,A}=\mu^M_{r(A),\,r(A\setminus\{j\})}.
\end{equation}
We first verify the identities of Definition~\ref{def:crossed-n-cube}
for $r^*M_\bullet$.

\textbf{Commutativity of the cube.}
By \eqref{eq:reindexed-structure-maps}, for $j,k\in J$ both composites
$M_{r(A)}\to M_{r(A\setminus\{j,k\})}$ are equal to
$\mu^M_{r(A),\,r(A\setminus\{j,k\})}$. More generally, for
$A'\subseteq A\subseteq J$ the composite structure map of $r^*M_\bullet$
from the vertex $A$ to the vertex $A'$ is $\mu^M_{r(A),r(A')}$.

\textbf{Compatibility of the pairings with the structure maps.}
Since $h^{r^*M}_{A,B}=h^M_{r(A),r(B)}$, $r(A)\cup r(B)=r(A\cup B)$ and
$r(A\setminus\{j\})\cup r(B\setminus\{j\})=r((A\cup B)\setminus\{j\})$,
the required identity
$$
\mu^{r^*M}_{j,A\cup B}\,h^{r^*M}_{A,B}(a,b)
=h^{r^*M}_{A\setminus\{j\},\,B\setminus\{j\}}
\bigl(\mu^{r^*M}_{j,A}a,\;\mu^{r^*M}_{j,B}b\bigr)
$$
is Lemma~\ref{lem:pairing-restriction} for $M_\bullet$ and the subsets
$r(A\setminus\{j\})\subseteq r(A)$ and $r(B\setminus\{j\})\subseteq r(B)$.

\textbf{Independence of a common coordinate.}
Let $j\in A\cap B$. Then $r(A\setminus\{j\})\cup r(B)=r(A\cup B)$, so
Lemma~\ref{lem:pairing-restriction} for the subsets
$r(A\setminus\{j\})\subseteq r(A)$ and $r(B)\subseteq r(B)$ gives
$h^{r^*M}_{A,B}(a,b)=h^{r^*M}_{A\setminus\{j\},B}\bigl(\mu^{r^*M}_{j,A}a,b\bigr)$.
The identity
$h^{r^*M}_{A,B}(a,b)=h^{r^*M}_{A,B\setminus\{j\}}\bigl(a,\mu^{r^*M}_{j,B}b\bigr)$
follows in the same way.

The remaining identities (the diagonal commutator identity,
skew-symmetry, normalization, crossed multiplicativity, the
Hall-Witt identity, and equivariance of the pairings) do not involve
the structure maps. They are therefore exactly the corresponding
identities in $M_\bullet$, with $A,B,C$ replaced by
$r(A),r(B),r(C)$. Hence $r^*M_\bullet$ is a crossed $J$-cube.

Next, let $f:M_\bullet\to N_\bullet$ be a morphism of crossed
$I$-cubes. The maps $(r^*f)_A=f_{r(A)}$ are compatible with the
structure maps by \eqref{eq:reindexed-structure-maps}, because $f$
commutes with the composite structure maps of $M_\bullet$ and
$N_\bullet$, and with the pairings because
$h^{r^*M}_{A,B}=h^M_{r(A),r(B)}$. Hence $r^*$ defines a functor
$\XCube_I\to\XCube_J$.

Finally, we verify functoriality in $r$. Let
$K\xrightarrow{q}J\xrightarrow{r}I$ be maps in $\Fin$. For every
$A\subseteq K$, both $(r\circ q)^*M$ and $q^*r^*M$ have vertex
$M_{r(q(A))}$. By \eqref{eq:reindexed-structure-maps}, applied first
to $q$ and then to $r$, the structure map of $q^*r^*M$ in the direction
$k\in K$ at the vertex $A$ is
$$
\mu^{r^*M}_{q(A),\,q(A\setminus\{k\})}
=\mu^M_{r(q(A)),\,r(q(A\setminus\{k\}))},
$$
which is the structure map of $(r\circ q)^*M$. The
pairings and reindexed morphisms agree immediately from
$r(q(A))=(r\circ q)(A)$. Therefore
$$
(r\circ q)^*=q^*r^*,
\qquad
\id_I^*=\id
$$
strictly.
\end{proof}

\begin{definition}[The presheaf of crossed cubes]\label{def:Xcube-presheaf}
    Let $\XCube_\bullet$ be a functor $\Fin^{op}\to\Cat$ whose value on a finite set $I$ is the category $\XCube_I$ of $I$-labeled crossed cubes, and whose value on $r:J\to I$ is given by the reindexing functor $r^*:\XCube_I\to\XCube_J$. Equivalently, $\XCube_\bullet$ is a presheaf on $\Fin$ valued in $\Cat$.
\end{definition}

\begin{definition}[Strict morphisms of $\Cat$-valued presheaves on $\Fin$]
\label{def:strict-cat-presheaf-morphism}
Let
$\scrA_\bullet,\scrB_\bullet: \Fin^{\mathrm{op}}\to \Cat$
be $\Cat$-valued presheaves. A \emph{strict morphism of
$\Cat$-valued presheaves}
$\Phi:\scrA_\bullet\to\scrB_\bullet$ means an ordinary natural
transformation in the functor category
$\Fun(\Fin^{\mathrm{op}},\Cat)$. Thus it consists of
functors
$\Phi_I:\scrA_I\to\scrB_I$, $I\in\Fin$,
such that, for every map $r:J\to I$ in $\Fin$, the square
$$
\begin{tikzcd}
\scrA_I \arrow[r,"\Phi_I"] \arrow[d,"r^*"']
&
\scrB_I \arrow[d,"r^*"]
\\
\scrA_J \arrow[r,"\Phi_J"']
&
\scrB_J
\end{tikzcd}
$$
commutes strictly. Equivalently, the two composite functors have the same
values on objects and morphisms.
\end{definition}

We can now give a more concise definition of a crossed precosheaf.

\begin{definition}[Crossed precosheaf: categorical formulation]
\label{def:crossed-precosheaf-categorical}
Let $\SL$ be a meet-semilattice. A \emph{crossed precosheaf} on $\SL$
is a strict morphism of $\Cat$-valued presheaves
$$
\CF_\bullet:
\check{\mathbf C}_\bullet(\SL)
\longrightarrow
\XCube_\bullet
$$
satisfying the following normalization condition.

For every $I\in\Fin$ and every $I$-indexed \v{C}ech configuration
$\xi=((U_i)_{i\in I}\vert W)\in\check{\mathbf C}_I(\SL)$,
the crossed $I$-cube
$\CF_I(\xi)\in\XCube_I$
has $A$-vertex group
$$
\bigl(\CF_I(\xi)\bigr)_A
=
\CF_{\varnothing}(U_A^\xi),
\qquad 
U_A^\xi=
\begin{cases}
W, & A=\varnothing,\\[2mm]
\displaystyle\bigwedge_{i\in A}U_i, & A\neq\varnothing,
\end{cases}
$$
under the canonical identifications
$\check{\mathbf C}_{\varnothing}(\SL)=\SL$, $\XCube_{\varnothing}=\Grp$.
For $i\in A$, its structure map is the corestriction map of
$\CF_{\varnothing}$ associated to
$U_A^\xi\leq U_{A\setminus\{i\}}^\xi$.
Moreover, for $\xi\leq\xi'$, the $A$-component of the induced
morphism
$\CF_I(\xi)\to\CF_I(\xi')$
is the corestriction map of $\CF_{\varnothing}$ associated to
$U_A^\xi\leq U_A^{\xi'}$.
\end{definition}

\begin{definition}[Morphism of crossed precosheaves: categorical formulation]
\label{def:morphism-crossed-precosheaves-categorical}
Let $\SL$ be a meet-semilattice, and let $\CF$ and $\CG$ be crossed
precosheaves on $\SL$. Thus, for every finite set $I$, they
determine functors
$\CF_I,\CG_I: \Cech_I(\SL) \to \XCube_I$.

A \emph{morphism of crossed precosheaves} $\psi:\CF\longrightarrow\CG$ is a family of natural transformations
$\psi_I:\CF_I\Longrightarrow\CG_I$, $I\in\Fin$,
satisfying the following conditions.

\begin{enumerate}
\item
\textbf{Normalization.}
The empty-index component
$\psi_{\varnothing}: \CF_{\varnothing} \Longrightarrow \CG_{\varnothing}$
is a natural transformation between the underlying precosheaves. For
every $\xi\in\Cech_I(\SL)$ and every $A\subseteq I$, the map at the
$A$-vertex of the crossed-cube morphism
$\psi_I(\xi): \CF_I(\xi) \to \CG_I(\xi)$
is
$$
\bigl(\psi_I(\xi)\bigr)_A
=
\bigl(\psi_{\varnothing}\bigr)_{U_A^\xi}:
\CF_{\varnothing}(U_A^\xi)
\longrightarrow
\CG_{\varnothing}(U_A^\xi).
$$

\item
\textbf{Reindexing compatibility.}
For every map $r:J\to I$ in $\Fin$ and every
$\xi\in\Cech_I(\SL)$,
$$
\psi_J(r^*\xi)
=
r^*\bigl(\psi_I(\xi)\bigr).
$$
this is an equality of morphisms in $
\Hom_{\XCube_J}
\left(
\CF_J(r^*\xi),
\CG_J(r^*\xi)
\right).
$
\end{enumerate}
\end{definition}

The formulations of Definitions~\ref{def:crossed-precosheaf-categorical} and~\ref{def:crossed-precosheaf-structure} are equivalent: the reindexing compatibility of Definition~\ref{def:crossed-precosheaf-structure} is precisely the strictness of the square in Definition~\ref{def:strict-cat-presheaf-morphism}, and the normalization condition identifies the vertex groups. The same comparison identifies Definitions~\ref{def:morphism-crossed-precosheaves-categorical} and~\ref{def:morphism-crossed-precosheaves-precosheaf}.

\section{Small gauge transformations form an epiprecosheaf}
\label{app:path-epi}

We sketch a proof of the claim made at the end of Section~\ref{sec:crossed-precosheaves}. Notation is that of Example~\ref{ex:support-functions} and Remark~\ref{rem:Fpath}.

\begin{proposition}\label{prop:path-epi}
Let $X$ be a paracompact manifold and let $G$ be a Lie group. As in Remark \ref{rem:Fpath}, let
$\CHpath(U)$
denote the path component of the identity in the topological group
$\CH(U)$, where $\CH(U)$ has the compact-open topology. Then
$\CHpath$ is a coflasque crossed epiprecosheaf. In
particular, if $W=U_1\cup\cdots\cup U_m$ is a finite open cover, then
$$
\CHpath(W)
=
\CHpath(U_1)\cdots \CHpath(U_m).
$$
\end{proposition}

\begin{proof}[Sketch of proof]
By Remark~\ref{rem:Fpath}, $\CHpath$ is a coflasque crossed
precosheaf. In particular, each $\CHpath(U_i)$ is contained in
the group $\CHpath(W)$, so it remains to show that every
$f\in\CHpath(W)$ is a product $f_1\cdots f_m$ with
$f_i\in\CHpath(U_i)$.
Since $X$ is locally compact Hausdorff, a path from the identity to $f$
in $\CH(W)$ is the same as a continuous map
$\Phi:X\times[0,1]\to G$ such that $\Phi(x,0)=e$, $\Phi(x,1)=f(x)$, and
$\supp\Phi(-,t)\subseteq W$ for every $t$.

First, one modifies $\Phi$, without changing its endpoints, so that it is
equal to $e$ on $N_0\times[0,1]$ for some open neighborhood $N_0$ of
$X\setminus W$. To this end, choose a neighborhood $V$ of $e$ in $G$ with
a contraction $c_s:V\to V$, $s\in[0,1]$, such that $c_0\equiv e$,
$c_1=\id$ and $c_s(e)=e$. By compactness of $[0,1]$ there is an open
neighborhood $O$ of $X\setminus W$ with $\Phi(O\times[0,1])\subseteq V$.
Using normality of $X$, choose open sets $N_0,N$ with
$X\setminus W\subseteq N_0$, $\overline{N_0}\subseteq N$ and
$\overline N\subseteq O\setminus\supp f$, and a continuous
function $\chi:X\to[0,1]$ which vanishes on $\overline{N_0}$ and is equal
to $1$ on $X\setminus N$. Then replace $\Phi(x,t)$ by
$c_{\chi(x)}\bigl(\Phi(x,t)\bigr)$ for $x\in\overline N$.

Second, choose a partition of unity $\lambda_0,\lambda_1,\ldots,\lambda_m$
subordinate to the open cover $N_0,U_1,\ldots,U_m$ of $X$, put
$s_i:=\lambda_1+\cdots+\lambda_i$ for $0\leq i\leq m$, and define
$$
\Phi_i(x,t)
:=
\Phi\bigl(x,t\,s_{i-1}(x)\bigr)^{-1}\,
\Phi\bigl(x,t\,s_i(x)\bigr),
\qquad
f_i:=\Phi_i(-,1),
\qquad 1\leq i\leq m.
$$
Since $\Phi_i(x,t)=e$ whenever $\lambda_i(x)=0$, we have
$\supp\Phi_i(-,t)\subseteq\supp\lambda_i\subseteq U_i$,
so $\Phi_i$ is a path in $\CH(U_i)$ from the identity to $f_i$, and
$f_i\in\CHpath(U_i)$. The product telescopes:
$f_1(x)\cdots f_m(x)=\Phi(x,s_m(x))$. This is equal to $f(x)$, because
$s_m(x)=1$ for $x\notin N_0$, while $\Phi(x,t)=e=f(x)$ for $x\in N_0$.
Hence $f=f_1\cdots f_m$.
\end{proof}

\section{Proof of the grid normal form lemma}
\label{app:internal-categories}

\begin{proof}[Proof of Lemma \ref{lem:grid-normal-form}.]
Throughout, write $\Gamma=\Gamma_I(M_\bullet)$. In the  case
$I=\varnothing$ all three statements are trivialities.

(1) We induct on $|I|$, the case $I=\varnothing$ being trivial. Let
$i$ be the first element of $I$ in the chosen order, so that the
lexicographic order on $T_{\bfm}$ compares the $i$-coordinate
first, and put $I'=I\setminus\{i\}$ and $\bfm=(m_i,\bfm')$
with $\bfm'\in\NN^{I'}$. Taking the nerve in direction $i$ last
(Lemma~\ref{lem:nerve-limits}), $(N^I\Gamma)_{\bfm}$ is the group
of $m_i$-strings of composable arrows of the internal category
$\scrC$ in groups obtained by evaluating the internal category
$N^{I'}\Gamma$ in $I'$-simplicial groups in multidegree $\bfm'$.
By Lemma~\ref{lem:faces}, its object group and source kernel are
$$
C_0=(N^{I'}\Gamma_{I'}\partial^0_iM_\bullet)_{\bfm'},
\qquad
K=(N^{I'}\Gamma_{I'}\partial^1_iM_\bullet)_{\bfm'},
$$
with $\iota^{\partial^0_iM}_A(a)$ and $\iota^{\partial^1_iM}_A(b)$
corresponding to $\iota^M_A(a)$ and $\iota^M_{A\cup\{i\}}(b)$; the
identifications are isomorphisms of $\cat^{I'}$-groups, so they commute
with the operators of the directions in $I'$.

For $x\in C_0$ and $k\in K$, $H_0^*x$ is the string of $m_i$ identity
arrows at $x$, and $H_j^*k$ is the string with $k$ at step $j$,
identity arrows at $\sigma k=1$ before it and at $\tau k$ after it.
Hence the $j$-th arrow of $H_{m_i}^*(k_{m_i})\cdots H_1^*(k_1)\,H_0^*(x)$
is $k_j\,\varepsilon(\tau k_{j-1})\cdots\varepsilon(\tau k_1)\,\varepsilon(x)
=k_j\,\varepsilon(x_{j-1})$ in the notation of
Lemma~\ref{lem:source-kernel-normal-form}, i.e.\ this product is
$\Theta_{m_i}(k_1,\ldots,k_{m_i};x)$. That lemma therefore says that
every element of $(N^I\Gamma)_{\bfm}$ is uniquely
$H_{m_i}^*(c_{m_i})\cdots H_1^*(c_1)\,H_0^*(x)$ with $x\in C_0$ and
$c_j\in K$. By the induction hypothesis for $\partial^0_iM_\bullet$
and $\partial^1_iM_\bullet$, with the order induced from $I$, $x$ and
each $c_j$ are uniquely products
$\prod_{\bfk'}u^{\partial^0_iM}_{\bfk'}(a_{\bfk'})$ and
$\prod_{\bfk'}u^{\partial^1_iM}_{\bfk'}(b_{j,\bfk'})$ over
$T_{\bfm'}$ in decreasing order. Writing $(j,\bfk')$ for the
element of $T_{\bfm}$ with $i$-coordinate $j$ and remaining
coordinates $\bfk'$, and using that operators in distinct
directions commute,
$$
H_0^*\,u^{\partial^0_iM}_{\bfk'}(a)
=\Bigl(\prod_{j'\in I'}H^*_{k'_{j'}}\Bigr)H_0^*\,\iota^M_{A(\bfk')}(a)
=u^M_{(0,\bfk')}(a),
\qquad
H_j^*\,u^{\partial^1_iM}_{\bfk'}(b)
=\Bigl(\prod_{j'\in I'}H^*_{k'_{j'}}\Bigr)H_j^*\,\iota^M_{A(\bfk')\cup\{i\}}(b)
=u^M_{(j,\bfk')}(b),
$$
since $A((0,\bfk'))=A(\bfk')$ and
$A((j,\bfk'))=A(\bfk')\cup\{i\}$ for $j>0$. Substituting the
expansions of $x$ and the $c_j$ and using that the $H_j^*$ are
homomorphisms gives an expression \eqref{eq:grid-normal-form}, since
the decreasing lexicographic order on $T_{\bfm}$ is: first by
$j$, then by $\bfk'$. It is unique because
$x,c_1,\ldots,c_{m_i}$ are determined by the element and their
expansions are unique.

(2) We first prove \eqref{eq:grid-presentation} by restricting to the
top-dimensional cells of the grid. For $1\le j\le m$ let
$e_j:[1]\to[m]$ be the inclusion of the edge $[j-1,j]$, and for $m=0$
let $e_1:[1]\to[0]$ be the unique map. For an internal category
$\scrC$ in groups, $e_j^*:(N\scrC)_m\to C_1$ is the projection
onto the $j$-th arrow of a string if $m\ge1$, and is $\varepsilon$ if
$m=0$; in both cases $(e_j^*)_{1\le j\le\max(m,1)}$ is an injective
homomorphism into $C_1^{\max(m,1)}$. Taking the nerves one direction
at a time, it follows that the homomorphism
$$
(N^I\Gamma)_{\bfm}\longrightarrow\prod_{\mathbf j}\Gamma,
\qquad
g\longmapsto\bigl(e^*_{\mathbf j}g\bigr)_{\mathbf j},
\qquad
e^*_{\mathbf j}=\prod_{i\in I}e_{j_i}^*,
\quad
\mathbf j\in\prod_{i\in I}\{1,\ldots,\max(m_i,1)\},
$$
where $e^*_{j_i}$ acts in direction $i$ and $\Gamma=(N^I\Gamma)_{(1,\ldots,1)}$,
is injective. So it suffices to check \eqref{eq:grid-presentation}
after applying each $e^*_{\mathbf j}$.

Let $\bfk\in T_{\bfm}$, $a\in M_A$ with $A=A(\bfk)$, and
put $A'=\{i\in A:j_i=k_i\}$. In direction $i$,
$e_{j_i}^*H_{k_i}^*=(H_{k_i}\circ e_{j_i})^*$, and $H_k\circ e_j$ is
$c_0$ if $j<k$, the identity if $j=k$, and $H_0$ if $j>k$ (in
particular if $k=0$); on $\Gamma$ the operators of direction $i$
induced by $c_0$ and $H_0$ are $\varepsilon\sigma=s_i$ and
$\varepsilon\tau=t_i$. Since $j_i<k_i$ forces $i\in A$, and
$s_i\iota_A(a)=1$ for $i\in A$, $t_i\iota_A(a)=\iota_{A\setminus\{i\}}(\mu_ia)$
for $i\in A$ and $t_i\iota_A(a)=\iota_A(a)$ for $i\notin A$
(Definition~\ref{def:catrecover}), we get
\begin{equation}\label{eq:cell-of-orthant-step}
e^*_{\mathbf j}u_{\bfk}(a)=
\begin{cases}
1,& j_i<k_i\text{ for some }i\in A,\\[2pt]
\iota_{A'}\bigl(\mu_{A,A'}a\bigr),&\text{otherwise}.
\end{cases}
\end{equation}
Now let also $\bfell\in T_{\bfm}$, $b\in M_B$ with
$B=A(\bfell)$, and $B'=\{i\in B:j_i=\ell_i\}$. If $j_i<k_i$ for
some $i\in A$, then $e^*_{\mathbf j}$ kills $u_{\bfk}(a)$, and
also $u_{\bfk\vee\bfell}(h_{A,B}(a,b))$ because
$j_i<k_i\le(\bfk\vee\bfell)_i$ with $i\in A\cup B$; so both
sides of \eqref{eq:grid-presentation} restrict to $1$, and likewise if
$j_i<\ell_i$ for some $i\in B$. Otherwise
$\{i\in A\cup B:j_i=\max(k_i,\ell_i)\}=A'\cup B'$, because
$j_i=\max(k_i,\ell_i)$ holds iff $j_i=k_i>0$ or $j_i=\ell_i>0$. By
\eqref{eq:cell-of-orthant-step} and the defining relations of
$\Gamma$, the two sides of \eqref{eq:grid-presentation} restrict to
$\iota_{A'\cup B'}\bigl(h_{A',B'}(\mu_{A,A'}a,\mu_{B,B'}b)\bigr)$ and
$\iota_{A'\cup B'}\bigl(\mu_{A\cup B,A'\cup B'}h_{A,B}(a,b)\bigr)$. These agree by Lemma~\ref{lem:pairing-restriction}.

For the presentation, let $P$ be the quotient of
$\ast_{\bfk}M_{A(\bfk)}$ by the normal subgroup generated by
the elements displayed in (2). By \eqref{eq:grid-presentation}, the
$u_{\bfk}$ induce a homomorphism $\phi:P\to(N^I\Gamma)_{\bfm}$,
surjective by (1). For $\bfell>\bfk$ the relations give
$j_{\bfk}(a)\,j_{\bfell}(b)\,j_{\bfk}(a)^{-1}
=j_{\bfk\vee\bfell}(h(a,b))\,j_{\bfell}(b)$ with
$\bfk\vee\bfell\ge\bfell$, so the subgroup
generated by the $j_{\bfell}(M_{A(\bfell)})$ with
$\bfell>\bfk$ is normalized by $j_{\bfk}(M_{A(\bfk)})$.
By descending induction on $\bfk$, every element of $P$ is a
product $\prod_{\bfk}j_{\bfk}(a_{\bfk})$ in decreasing
order, and by the uniqueness in (1) such a product lies in $\ker\phi$
only if all $a_{\bfk}=1$. Hence $\phi$ is an isomorphism.
\end{proof}

\section{Proof of Theorem~\ref{thm:crossed-module-fibration}}
\label{app:crossed-module-fibration}

Let us outline our strategy for proving Theorem \ref{thm:crossed-module-fibration}. Recall that $p_{\scrC}=\bW e_{\scrC}$, where
$e_{\scrC}:C_0\to\Diag N\scrC$ sends $g$ to the string of identity
arrows at $g$. The sequence of simplicial groups
$$
K\xrightarrow{\ \partial\ }C_0\xrightarrow{\ e_{\scrC}\ }
\Diag N\scrC
$$
is not short exact, so the theorem does
not follow from the fact that $\bW$ turns short exact sequences into
fiber sequences. Instead, we use the classical path-fibration argument,
carried out inside internal categories in simplicial groups. There are
three steps.

First, the homomorphism $e_{\scrC}$ is not of the form $\Diag N(-)$ as it
stands, because its source $C_0$ is a simplicial group rather than an
internal category. We therefore regard $C_0$ as an internal category
$\Disc(C_0)$ with only identity arrows. Then
$\BDelta\Disc(C_0)\cong\BDelta C_0$, and $e_{\scrC}$
becomes $\Diag N$ of the internal functor
$S_{\scrC}:\Disc(C_0)\to\scrC$ which is the identity
on objects and $\varepsilon$ on arrows
(Definition~\ref{def:disc} and Lemma~\ref{lem:disc-classifying}). This puts all maps in the theorem in the
image of $\BDelta$, so that they can be manipulated by means of internal
functors and internal natural transformations.

Second, the map $p_{\scrC}$ need not be a Kan fibration, so we replace its domain
by an equivalent one. We introduce the path category
$\operatorname{Path}(\scrC)$, whose objects are the arrows of
$\scrC$. It comes with three internal functors
$$
\begin{tikzcd}[column sep=large]
\Disc(C_0)
  \arrow[r,shift left=1ex,"\idt"]
&
\operatorname{Path}(\scrC)
  \arrow[l,shift left=1ex,"\src"]
  \arrow[r,"\tgt"]
&
\scrC ,
\end{tikzcd}
$$
which send an arrow to its source and its target, and an object to its
identity arrow. They satisfy
$\src\circ\idt=\id$ and
$\tgt\circ\idt=S_{\scrC}$. We show:
\begin{enumerate}
\item[(a)] there is an internal natural transformation
$\idt\circ\src\Rightarrow\id$; hence, after applying
$\BDelta$, the functor $\src$ becomes a simplicial homotopy
equivalence $r$, and $p_{\scrC}\circ r$ becomes homotopic to
$f:=\BDelta\tgt$;
\item[(b)] $\tgt$ is surjective on objects and on arrows, with kernel
$\Disc(\ker\tau)\cong\Disc(K)$; hence $f$ is a Kan
fibration with fiber $\BDelta K$, and the restriction of $r$ to this
fiber is $\BDelta\partial$.
\end{enumerate}

Thus we get a diagram of simplicial sets
\begin{equation}\label{eq:diagramD}
\begin{tikzcd}[column sep=large]
\BDelta K
  \arrow[r,"\iota_K"]
  \arrow[dr,"\BDelta\partial"']
&
\BDelta\operatorname{Path}(\scrC)
  \arrow[r,"f"]
  \arrow[d,"r","\sim"']
&
\BDelta\scrC
\\
&
\BDelta C_0
  \arrow[ur,"p_{\scrC}"']
&
\end{tikzcd}
\end{equation}
in which the top row is a Kan fibration with its fiber, the left triangle
commutes, and the right triangle commutes up to a pointed homotopy. Third, a
general recognition lemma
(Lemma~\ref{lem:recognition-fibration-sequence}) says that in this
situation the lower composite is a homotopy fibration sequence.

Apart from the recognition lemma and the elementary
Lemma~\ref{lem:disc-classifying}, we need two further lemmas. The first
converts internal natural transformations into simplicial homotopies and is
used in~(a). The second converts short exact sequences of internal
categories into Kan fibrations and is used in~(b).

\begin{definition}
\label{def:disc}
Let $M$ be a simplicial group. The discrete internal category on $M$
is the internal category $\Disc(M)$ in simplicial groups with
$$
C_0=C_1=M,
\qquad
\sigma=\tau=\varepsilon=\id_M ,
$$
and with composition the identity of $C_1\times_{C_0}C_1\cong M$. A
homomorphism of simplicial groups $\phi:M\to M'$ induces the internal
functor
$\Disc(\phi):\Disc(M)\to\Disc(M')$
equal to $\phi$ on objects and on arrows. This defines a functor
$\Disc$ from simplicial groups to internal categories in
simplicial groups.

For an internal category $\scrC$ in simplicial groups, the identity-arrow functor is the internal functor
$$
S_{\scrC}:\Disc(C_0)\longrightarrow\scrC,
\qquad
(S_{\scrC})_0=\id_{C_0},
\qquad
(S_{\scrC})_1=\varepsilon .
$$
\end{definition}

In each simplicial degree $q$, $\Disc(M)_q$ is the
$\cat^1$-group $(M_q,\,s=t=\id)$, which corresponds to the crossed module
$1\to M_q$. The elements of $M_q$ are its objects, and its only
arrows are identities. The functor $\Disc$ is
left adjoint to $\scrC\mapsto C_0$, and $S_{\scrC}$ is the counit
of this adjunction; in particular, $S_{\scrC}$ is natural in
$\scrC$.

\begin{lemma}
\label{lem:disc-classifying}
For every simplicial group $M$ there is an isomorphism of
simplicial groups $\theta_M:\Diag N\Disc(M)\cong M$, natural in $M$.
Consequently there is an isomorphism of simplicial sets
$$
c_M:={\bW}\theta_M:\BDelta\Disc(M)
=\bW\bigl(\Diag N\Disc(M)\bigr)
\xrightarrow{\ \cong\ }
\bW M=\BDelta M ,
$$
natural in $M$. For every internal category $\scrC$ in simplicial groups, we have $\Diag N S_\scrC=e_\scrC\circ \theta_{C_0}$. Consequently
\begin{equation}
\label{eq:p-via-disc}
p_{\scrC}=\bW e_\scrC=\BDelta S_{\scrC}\circ c_{C_0}^{-1}.
\end{equation}
\end{lemma}

\begin{proof} 
A string of $r$ composable arrows in $\Disc(M)_q$ has the
form $(m,\dots,m)$ with $m\in M_q$, so
$N_r\Disc(M)_q\cong M_q$. Under this identification all faces
and degeneracies in the nerve direction are identity maps. Thus
$N\Disc(M)$ is the bisimplicial group which is constant in the
nerve direction and equal to $M$ in the other direction, and its diagonal
is canonically isomorphic to $M$. This gives $\theta_M$, and it is clearly natural in $M$. Applying
$\bW$ gives $c_M$ and its naturality.

For the second statement, note that in nerve degree $r$, the functor $S_{\scrC}$ sends $(g,\dots,g)$
to $(\varepsilon g,\dots,\varepsilon g)$. Hence, under the isomorphism $\theta_{C_0}$, $\Diag N S_{\scrC}$ sends $g\in C_{0,q}$ to the string of $q$
identity arrows at $g$, which is $e_{\scrC}(g)$. Applying $\bW$
gives $\BDelta S_{\scrC}=p_{\scrC}\circ c_{C_0}$.
\end{proof}

\begin{lemma}
\label{lem:internal-natural-transformation-homotopy}
Let $\scrC$ and $\scrD$ be internal categories in simplicial
groups, let
$F,G:\scrC\to\scrD$
be internal functors, and let
$\eta:F\Longrightarrow G$
be an internal natural transformation. Then $\eta$ induces a pointed
simplicial homotopy
$$
H_\eta:
\BDelta\scrC\times\Delta^1
\longrightarrow
\BDelta\scrD
$$
from $\BDelta F$ to $\BDelta G$.
\end{lemma}

\begin{proof}
Let $U$ denote the forgetful functor from simplicial groups to simplicial
sets. Fix an outer simplicial degree $q$ and denote by $C_0,C_1$ and $D_0,D_1$ the groups of objects and arrows in the internal categories in $\Grp$ $\scrC_q$ and ${\scrD}_q$. We also suppress the outer degree in the structure maps of these internal categories and in 
$F_0,F_1,G_0,G_1$, and $\eta$. For a string
$$
x_0\xrightarrow{a_1}x_1\xrightarrow{a_2}\cdots
\xrightarrow{a_r}x_r
$$
in $(N\scrC_q)_r$ and an $r$-simplex
$\theta:[r]\to[1]$ of $\Delta^1$, consider a map
$$
\lambda_{q,r,\theta}:(N\scrC_q)_r
\longrightarrow(N\scrD_q)_r
$$
which sends the vertex $x_i$ to 
$$
y_i=
\begin{cases}
F_0(x_i),&\theta(i)=0,\\
G_0(x_i),&\theta(i)=1,
\end{cases}
$$
and the arrow $a_i$ to 
$$
b_i=
\begin{cases}
F_1(a_i),&\theta(i-1)=\theta(i)=0,\\[2mm]
G_1(a_i)\circ\eta(x_{i-1}),
&\theta(i-1)=0,\ \theta(i)=1,\\[2mm]
G_1(a_i),&\theta(i-1)=\theta(i)=1.
\end{cases}
$$
For $r=0$, the image is just the vertex $y_0$. 

First, one can check that for every $q$ the maps $\lambda_{q,r,\theta}$ define a simplicial map  $N\scrC_q\times \Delta^1\to N{\scrD}_q$. This uses the functoriality of $F$ and $G$ and the identity 
$$
G_1(a_i)\circ\eta(x_{i-1})
=\eta(x_i)\circ F_1(a_i)
$$
which follows from the naturality of $\eta$. 

Second, the maps $\lambda_{q,r,\theta}$ are compatible with the outer simplicial operators because $F$, $G$, and $\eta$ are simplicial.

Third, we check that each $\lambda_{q,r,\theta}$ is a group homomorphism. Since
$(N\scrD_q)_r$ is a subgroup of $D_1^{\,r}$, it suffices to show that
for each $i$ the map $(a_1,\ldots,a_r)\mapsto b_i$ is a homomorphism from
$(N\scrC_q)_r$ to $D_1$. If $\theta(i-1)=\theta(i)$, this map is the
projection to the $i$-th arrow followed by $F_1$ or $G_1$. In the mixed case
$\theta(i-1)=0$, $\theta(i)=1$ (which occurs for at most one $i$), recall
that $x_{i-1}=\sigma(a_i)$ and consider
$$
\psi:C_1\longrightarrow D_1\times D_1,
\qquad
\psi(a)=\bigl(G_1(a),\,\eta(\sigma a)\bigr).
$$
It is a homomorphism because $G_1$, $\sigma$, and the component map
$\eta:C_0\to D_1$ of the internal natural transformation are homomorphisms.
Its image lies in the subgroup
$D_1\times_{D_0}D_1\subseteq D_1\times D_1$ of composable pairs, since
$\sigma G_1(a)=G_0(\sigma a)=\tau\,\eta(\sigma a)$. By the definition of an
internal category in groups, internal composition $\mu_{\scrD}$ is a
homomorphism from $D_1\times_{D_0}D_1$ to $D_1$. Hence
$b_i=\mu_{\scrD}\bigl(\psi(a_i)\bigr)$ depends homomorphically on the
string. For $r=0$, the claim follows from the homomorphism property of $F_0$
and $G_0$.

Put $A:=\Diag N\scrC$ and $B:=\Diag N\scrD$. Taking the
full diagonal gives a simplicial homotopy
$$
\Lambda_\eta:UA\times\Delta^1\longrightarrow UB,
\qquad
(\Lambda_\eta)_n(a,\theta)=\lambda_{n,n,\theta}(a),
$$
from $U\Diag NF$ to $U\Diag NG$. A \emph{loop homotopy} between
homomorphisms of simplicial groups means a simplicial homotopy of their
underlying maps which is multiplicative in the group variable
\cite[Remarks~27.5]{MaySimplicialObjects}. Here this condition is
$$
(\Lambda_\eta)_n(aa',\theta)
=(\Lambda_\eta)_n(a,\theta)(\Lambda_\eta)_n(a',\theta)
$$
for all $n\geq0$, $a,a'\in A_n$, and
$\theta\in(\Delta^1)_n$. It holds because each
$\lambda_{n,n,\theta}$ is a homomorphism. In particular,
$(\Lambda_\eta)_n(1,\theta)=1$.

By the same reference, $\Lambda_\eta$ induces a simplicial homotopy
from $\bW(\Diag NF)=\BDelta F$ to
$\bW(\Diag NG)=\BDelta G$. In our codiagonal model,
this induced homotopy can be described explicitly. Write an $n$-simplex
of $\bW A=\nabla(B_\bullet A)$ as
$$
x=(x_0,\ldots,x_n),\qquad
x_p=(a_{p,1},\ldots,a_{p,p})\in A_{n-p}^{\,p}.
$$
For $\theta:[n]\to[1]$, let
$$
\theta^{(p)}:[n-p]\longrightarrow[1],
\qquad \theta^{(p)}(j)=\theta(p+j).
$$
Define
$$
\bigl(H_{\eta,n}(x,\theta)\bigr)_p
=
\bigl((\Lambda_\eta)_{n-p}(a_{p,j},\theta^{(p)})\bigr)_{j=1}^{p},
\qquad 0\leq p\leq n,
$$
with the empty tuple for $p=0$. The matching conditions are preserved:
$\fd_0\theta^{(p)}=\theta^{(p+1)}$ for $p<n$, and the homomorphisms
$(\Lambda_\eta)_{n-p}(-,\theta^{(p)})$ commute with the bar operators.
The simplicial identities for $\Lambda_\eta$ and restriction of
$\theta$ to these final segments likewise give compatibility with the
codiagonal faces and degeneracies. Thus these formulas define the simplicial map $H_\eta$.
At the constant simplices $\theta=0$ and $\theta=1$, the formulas give
$\BDelta F$ and $\BDelta G$, respectively. Since every homomorphism
in the construction preserves the identity, $H_\eta$ fixes the
basepoint.
\end{proof}

\begin{lemma}
\label{lem:internal-ses-fibration}
Let
$$
1\longrightarrow\scrA
\xrightarrow{\alpha}\scrB
\xrightarrow{\beta}\scrC
\longrightarrow1
$$
be a sequence of internal categories in simplicial groups, short exact
on both object and arrow groups in every simplicial degree. Then
$\BDelta\beta$ is a Kan fibration, and $\BDelta\alpha$ identifies
$\BDelta\scrA$ canonically with its fiber over the basepoint.
Consequently, the square
$$
\begin{tikzcd}
\BDelta\scrA \arrow[r,"\BDelta\alpha"] \arrow[d]
& \BDelta\scrB \arrow[d,"\BDelta\beta"]\\
* \arrow[r] & \BDelta\scrC
\end{tikzcd}
$$
is homotopy cartesian, naturally with respect to morphisms of short
exact sequences. In particular, the top and right maps form a homotopy
fibration sequence with the constant nullhomotopy of their composite.
\end{lemma}

\begin{proof}
For an internal category $\mathscr{X}$, put
$K_{\mathscr{X}}:=\ker\sigma_{\mathscr{X}}$. The induced sequence
$$
1\longrightarrow K_{\scrA}
\longrightarrow K_{\scrB}
\longrightarrow K_{\scrC}
\longrightarrow1
$$
is short exact. Only surjectivity requires verification. Fix a
simplicial degree, let $c\in K_{\scrC}$, and choose
$b\in\scrB_1$ with $\beta_1(b)=c$. Then
$\beta_0(\sigma_{\scrB}b)=1$, so
$$
\widetilde b:=b\varepsilon_{\scrB}(\sigma_{\scrB}b)^{-1}
$$
still maps to $c$ and belongs to $K_{\scrB}$.

The natural source-kernel normal form of
Lemma~\ref{lem:source-kernel-normal-form}, applied in each outer
simplicial degree, now gives surjectivity at every nerve level.
The nerve and diagonal preserve kernels by
Lemmas~\ref{lem:nerve-limits} and~\ref{lem:levelwise-diag}.
Hence there is a short exact sequence of simplicial groups
$$
1\longrightarrow\Diag N\scrA
\longrightarrow\Diag N\scrB
\longrightarrow\Diag N\scrC
\longrightarrow1.
$$
By \cite[Theorem~1.1]{FranzFibrations}, applying $\bW$ gives a
simplicial fiber bundle with fiber
$\bW\Diag N\scrA=\BDelta\scrA$. This fiber is Kan
\cite[Lemma~21.3]{MaySimplicialObjects}, so the bundle projection is a
Kan fibration \cite[Proposition~18.4]{MaySimplicialObjects}.
The canonical fiber identification is induced by $\BDelta\alpha$:
in the codiagonal formula for $\bW$, mapping to the basepoint means
that every group entry lies in the kernel. Thus the displayed square
is a pullback along a Kan fibration, and hence is homotopy cartesian.
Naturality follows from the functoriality of the nerve, diagonal, and
$\bW$ constructions.
\end{proof}

\begin{lemma}
\label{lem:recognition-fibration-sequence}
Let $Y$ and $Z$ be pointed Kan complexes and let $p:Y\to Z$ be a pointed
map. Suppose given
\begin{itemize}
\item a pointed Kan fibration $f:X\to Z$, together with a pointed
simplicial set $A$ and a pointed map $\iota:A\to X$ which is an isomorphism
of $A$ onto the fiber $f^{-1}(*)$;
\item a pointed weak equivalence $r:X\to Y$;
\item a pointed simplicial homotopy $L:X\times\Delta^1\to Z$ with
$L(-,0)=p\circ r$ and $L(-,1)=f$.
\end{itemize}
Then
$$
A\xrightarrow{\ r\iota\ }Y\xrightarrow{\ p\ }Z
$$
is a homotopy fibration sequence in the sense of
Definition~\ref{def:homotopy-fibration-sequence}, with respect to the
pointed nullhomotopy
$$
H:=L\circ(\iota\times\id_{\Delta^1}):A\times\Delta^1\longrightarrow Z .
$$
\end{lemma}

\begin{proof}
The simplicial set $A$ is a Kan complex, being isomorphic to a fiber of a
Kan fibration. Since $L(-,0)\circ\iota=p\circ r\iota$ and
$L(-,1)\circ\iota=f\iota=*$, the map $H$ is a pointed nullhomotopy of
$p\circ r\iota$. We must show that
$(r\iota,H^\#):A\to\hofib_*(p)$ is a weak equivalence.

Let $Z^{\Delta^1}$ be the simplicial path space, with endpoint maps
$\ev_0,\ev_1:Z^{\Delta^1}\to Z$, and put
$$
\widetilde Y
:=Y\times_{p,Z,\ev_0}Z^{\Delta^1}.
$$
It is pointed by the basepoint of $Y$ together with the constant path at the
basepoint of $Z$. Let
$$
\rho:\widetilde Y\longrightarrow Y,
\qquad
\widetilde p:\widetilde Y\longrightarrow Z
$$
be given by $\rho(y,\gamma)=y$ and $\widetilde p(y,\gamma)=\gamma(1)$.

Since $Z$ is Kan, $\ev_0$ is a trivial Kan fibration, so its
pullback $\rho$ is a trivial Kan fibration. The endpoint map
$$
(\ev_0,\ev_1):
Z^{\Delta^1}\longrightarrow Z\times Z
$$
is a Kan fibration, and its pullback along $p\times\id_Z$ is
$(\rho,\widetilde p):\widetilde Y\to Y\times Z$. Since $Y$ is Kan, the
projection $Y\times Z\to Z$ is also a Kan fibration. Hence $\widetilde p$ is
a Kan fibration; see \cite[Ch.~I, \S5]{GoerssJardine}. Its fiber over the
basepoint is
$$
\widetilde p^{-1}(*)
= Y\times_{Z\times Z}Z^{\Delta^1}
=\hofib_*(p),
$$
where the map $Y\to Z\times Z$ is $(p,*)$.

Let $L^\#:X\to Z^{\Delta^1}$ be adjoint to $L$. Because
$\ev_0L^\#=p r$, the pair
$$
\widetilde r:=(r,L^\#):X\longrightarrow\widetilde Y
$$
is well defined, pointed, and satisfies
$$
\rho\,\widetilde r=r,
\qquad
\widetilde p\,\widetilde r=\ev_1L^\#=f.
$$
As $\rho$ and $r$ are weak equivalences, so is $\widetilde r$. Thus
$\widetilde r$ is a weak equivalence between the Kan fibrations $f$ and
$\widetilde p$ over the same base $Z$, and therefore induces a weak
equivalence of their fibers over the basepoint
\cite[\S13.3]{Hirschhorn}. Precomposing with the isomorphism
$\iota:A\to f^{-1}(*)$, we obtain the weak equivalence
\[
A\longrightarrow\hofib_*(p).
\qedhere
\]
\end{proof}

We are finally ready to prove Theorem \ref{thm:crossed-module-fibration}.

\begin{proof}[Proof of Theorem~\ref{thm:crossed-module-fibration}]
Throughout, $K=\ker\sigma$ and $\partial=\tau|_K$, and we use the
notation $\Disc$, $S_{\scrC}$ of
Definition~\ref{def:disc} and $c_M$ of Lemma~\ref{lem:disc-classifying}. By \eqref{eq:p-via-disc},
$p_{\scrC}=\BDelta S_{\scrC}\circ c_{C_0}^{-1}$. We put
$$
Y:=\BDelta C_0,
\qquad
Z:=\BDelta\scrC,
\qquad
p:=p_{\scrC}:Y\longrightarrow Z .
$$
Both are Kan complexes, being values of $\bW$.

Define an internal category $\scrP=\operatorname{Path}(\scrC)$ in
simplicial groups by
$$
\scrP_0=C_1,
\qquad
\scrP_1=C_1\times_{C_0}C_1
=\{(a,b)\in C_1\times C_1\mid\sigma a=\sigma b\}.
$$
Here $(a,b)$ is an arrow from $b$ to $a$, the identity at $a$ is $(a,a)$,
and $(c,a)\circ(a,b)=(c,b)$. All structure maps are evidently homomorphisms
of simplicial groups. Thus an object of $\scrP$ is an arrow of
$\scrC$, and there is a unique arrow between two objects precisely when
they have the same source in $\scrC$.

By Proposition~\ref{prop:internal-groupoid}, categorical inversion
$a\mapsto a^{\circ-1}$ is a homomorphism of simplicial groups
$C_1\to C_1$. Define internal functors
$$
\tgt:\scrP\longrightarrow\scrC,
\qquad
\src:\scrP\longrightarrow\Disc(C_0),
\qquad
\idt:\Disc(C_0)\longrightarrow\scrP
$$
by
$$
\begin{aligned}
\tgt_0(a)&=\tau a,
& \tgt_1(a,b)&=a\circ b^{\circ-1},\\
\src_0(a)&=\sigma a,
& \src_1(a,b)&=\sigma a=\sigma b,\\
\idt_0(g)&=\varepsilon(g),
& \idt_1(g)&=(\varepsilon(g),\varepsilon(g)).
\end{aligned}
$$
The composite $a\circ b^{\circ-1}$ is defined because
$\tau(b^{\circ-1})=\sigma b=\sigma a$, and it is an arrow from $\tau b$ to
$\tau a$. The map $\tgt_1$ is a homomorphism because
$(a,b)\mapsto(a,b^{\circ-1})$ is a homomorphism into the group of composable
pairs and internal composition is a homomorphism on that group. It preserves
identities, since $a\circ a^{\circ-1}=\varepsilon(\tau a)$, and composition,
since
$$
\tgt_1(c,a)\circ\tgt_1(a,b)
=(c\circ a^{\circ-1})\circ(a\circ b^{\circ-1})
=c\circ b^{\circ-1}
=\tgt_1(c,b).
$$
The internal-functor identities for $\src$ and $\idt$ are
immediate. Directly from the formulas,
\begin{equation}
\label{eq:path-category-identities}
\src\circ\idt=\id_{\Disc(C_0)},
\qquad
\tgt\circ\idt=S_{\scrC}.
\end{equation}
For the second identity, note that $\tau\varepsilon(g)=g$ and
$\varepsilon(g)\circ\varepsilon(g)^{\circ-1}=\varepsilon(g)$.

Put
$$
X:=\BDelta\scrP,
\qquad
f:=\BDelta\tgt:X\longrightarrow Z,
\qquad
r:=c_{C_0}\circ\BDelta\src:X\longrightarrow Y,
\qquad
s:=\BDelta\idt\circ c_{C_0}^{-1}:Y\longrightarrow X .
$$
All three maps are pointed, and \eqref{eq:path-category-identities} gives
\begin{equation}
\label{eq:path-space-identities}
r s=\id_Y,
\qquad
f s=p .
\end{equation}

Next, we show that $r$ is a weak equivalence, and $p r\simeq f$.
Define
$$
\kappa:\scrP_0=C_1\longrightarrow\scrP_1,
\qquad
\kappa(a)=(a,\varepsilon\sigma a).
$$
This is a homomorphism of simplicial groups, and $\kappa(a)$ is an arrow
from $\varepsilon\sigma a=(\idt\circ\src)_0(a)$ to $a$.
For an arrow $(a,b):b\to a$ of $\scrP$ we have
$$
(a,b)\circ\kappa(b)
=(a,b)\circ(b,\varepsilon\sigma b)
=(a,\varepsilon\sigma b),
\qquad
\kappa(a)\circ(\idt\circ\src)_1(a,b)
=(a,\varepsilon\sigma a)\circ(\varepsilon\sigma a,\varepsilon\sigma a)
=(a,\varepsilon\sigma a),
$$
and these agree because $\sigma a=\sigma b$. Hence $\kappa$ is an internal
natural transformation
$\kappa:\idt\circ\src\Longrightarrow\id_{\scrP}$.
By Lemma~\ref{lem:internal-natural-transformation-homotopy}, it induces a
pointed simplicial homotopy
$$
H_\kappa:X\times\Delta^1\longrightarrow X,
\qquad
H_\kappa(-,0)=\BDelta(\idt\circ\src)=s r,
\qquad
H_\kappa(-,1)=\id_X .
$$
Together with $r s=\id_Y$, this shows that $r$ is a simplicial homotopy
equivalence, and in particular a weak equivalence. Moreover,
$$
L:=f\circ H_\kappa:X\times\Delta^1\longrightarrow Z
$$
is a pointed simplicial homotopy with
$$
L(-,0)=f s r=p r,
\qquad
L(-,1)=f,
$$
where we used \eqref{eq:path-space-identities}.

So far, we have constructed the right triangle in the diagram \ref{eq:diagramD} and showed that it is homotopy commutative. Let us show that $f$ is a Kan fibration with fiber $\BDelta K$.
There is a short exact sequence of internal categories in simplicial groups
$$
1\longrightarrow\Disc(\ker\tau)
\longrightarrow\scrP
\xrightarrow{\ \tgt\ }\scrC
\longrightarrow1,
$$
in which the first functor is the inclusion $\ker\tau\subseteq C_1$ on
objects and $a\mapsto(a,a)$ on arrows. Indeed, $\tgt_0=\tau$ is
split surjective with kernel $\ker\tau$. The map $\tgt_1$ is
surjective because $\tgt_1(g,\varepsilon\sigma g)=g$. Finally,
$$
\tgt_1(a,b)=1
\quad\Longleftrightarrow\quad
a=b\in\ker\tau .
$$
For the forward implication, recall that $1\in C_1$ is the identity arrow at
$1\in C_0$; applying source and target gives $\tau a=\tau b=1$, and
composing $a\circ b^{\circ-1}=1$ with $b$ gives $a=b$. The converse holds
because $a\circ a^{\circ-1}=\varepsilon(\tau a)$.

By Lemma~\ref{lem:internal-ses-fibration}, $f$ is a Kan fibration, and
$\BDelta$ of the inclusion identifies
$\BDelta\Disc(\ker\tau)$ with the fiber $f^{-1}(*)$.
We now identify $\ker\tau$ with $K$. Categorical inversion exchanges
source and target, hence restricts to an isomorphism of simplicial groups
$$
j:K\xrightarrow{\ \cong\ }\ker\tau,
\qquad
j(k)=k^{\circ-1}=k^{-1}\varepsilon(\partial k),
$$
and
\begin{equation}
\label{eq:sigma-j}
\sigma j=\partial .
\end{equation}
Let $J:\Disc(K)\to\scrP$ be the internal functor with
object map $j$ and arrow map $k\mapsto(j(k),j(k))$. It is the composite of
$\Disc(j)$ with the inclusion of
$\Disc(\ker\tau)$, so
$$
\iota_K:=\BDelta J\circ c_K^{-1}:\BDelta K\longrightarrow X
$$
is a pointed map which is an isomorphism of $\BDelta K$ onto
$f^{-1}(*)$. 

Finally, we prove the commutativity of the left triangle in diagram \ref{eq:diagramD}. We need to compute $r\iota_K$. By \eqref{eq:sigma-j}, the composite
$\src\circ J:\Disc(K)\to\Disc(C_0)$ is
$k\mapsto\sigma j(k)=\partial k$ both on objects and on arrows, that is,
$\src\circ J=\Disc(\partial)$.
Hence, by functoriality of $\BDelta$,
$$
r\,\iota_K
=c_{C_0}\circ\BDelta\src\circ\BDelta J\circ c_K^{-1}
=c_{C_0}\circ\BDelta\Disc(\partial)\circ c_K^{-1}.
$$
By the naturality of $c$ with respect to
$\partial:K\to C_0$, we have
$c_{C_0}\circ\BDelta\Disc(\partial)=\BDelta\partial\circ c_K$,
and therefore $
r\,\iota_K=\BDelta\partial $. 

We now established all the claimed properties of the diagram \ref{eq:diagramD}. 
The data $(f,\iota_K,r,L)$ satisfy the hypotheses of
Lemma~\ref{lem:recognition-fibration-sequence} with $A=\BDelta K$.
Since $r\iota_K=\BDelta\partial$, the lemma shows that
$$
\BDelta K\xrightarrow{\BDelta\partial}\BDelta C_0
\xrightarrow{p_{\scrC}}\BDelta\scrC
$$
is a homotopy fibration sequence, with respect to the pointed nullhomotopy
$H=L\circ(\iota_K\times\id_{\Delta^1})$.

It remains to check naturality. 
For an internal functor $T=(T_0,T_1):\scrC\to\scrC'$, the map
$T_1$ sends $K$ to $K'$ because $\sigma'T_1=T_0\sigma$, and its restriction
$T_K:K\to K'$ satisfies
$$
T_0\,\partial=\partial'\,T_K,
\qquad
(\Diag NT)\circ e_{\scrC}=e_{\scrC'}\circ T_0 .
$$
The first identity follows from $\tau'T_1=T_0\tau$, and the second from
$T_1\varepsilon=\varepsilon'T_0$. Applying $\bW$ proves naturality.
\end{proof}

\section{Model structures on crossed cubes}
\label{app:model-structures-crossed-cubes}

In this appendix we show that the Loday classifying functor is part of a
richer structure.  For every $n\geq 1$, the category $\XCube_n$ of
crossed $n$-cubes carries a natural model structure, and its homotopy
theory agrees with that of pointed connected $(n+1)$-types.  Thus one
obtains more than a functor from crossed cubes to simplicial sets: the
category of crossed cubes itself supports a homotopy theory.  The natural
framework for this statement is the language of model categories; for
background we refer to \cite{Hirschhorn}.
The discussion in this appendix follows Adrom \cite{Adrom}.

\subsection{The transferred model structure}

Fix $n\geq 1$.  The model structure is obtained from the following
sequence of adjunctions:
$$
\SSet
\underset{U}{\overset{F}{\rightleftarrows}}
\sGrp
\underset{\Codiag}{\overset{\sigma^\ast}{
  \scalebox{2}[1]{$\rightleftarrows$}}}
\mathbf{s}^{n}\Grp
\underset{N^n}{\overset{\tau}{\rightleftarrows}}
\cat^n\Grp .
$$
When $n=1$, the middle adjunction is understood to be the identity
adjunction.  Here $F$ is the levelwise reduced  free-group functor and $U$ is
the forgetful functor.  The standard model structure on simplicial groups
is transferred across $F\dashv U$: a map of simplicial groups is a weak
equivalence or a fibration precisely when its underlying map of simplicial
sets is so. The functor $N^n$ is the multinerve, and its left adjoint $\tau$ is the fundamental $n$-fold categorical-group functor.  The functor
$\sigma^*$ is induced by the ordinal-sum functor.  Explicitly,
$$
  (\sigma^*G)_{k_1,\ldots,k_n}
  =G_{k_1+\cdots+k_n+n-1}.
$$
Its right adjoint is a right Kan extension.  We denote this right adjoint
by $\Codiag$, since for $n=2$ it is the usual
Artin-Mazur codiagonal $\nabla$ and for general $n$ it is its iterated analog.

Transfer the weak equivalences and fibrations to $\cat^n\Grp$ by
declaring that $f$ is a weak equivalence, respectively a fibration, when
$U\Codiag N^n(f)$
is a weak equivalence, respectively a fibration, of simplicial sets.
Cofibrations are then defined by the left lifting property with respect to
trivial fibrations.

\begin{theorem}[Model structures on categorical groups and crossed cubes]
\label{thm:model-structures-xcube}
The transferred classes above define a cofibrantly generated model
structure on $\cat^n\Grp$.  Under the Ellis-Steiner equivalence
$$
  \Phi_n:\cat^n\Grp
  \rightleftarrows
  \XCube_n:\Gamma_n
$$
this model structure transfers to a cofibrantly generated model structure
on $\XCube_n$.
\end{theorem}
\begin{proof}
The model structure on simplicial groups is transferred from $\SSet$
along $F\dashv U$.  Adrom applies the model-structure transfer theorem
to the composite adjunction
$$
  \tau\sigma^*:\sGrp
  \rightleftarrows
  \cat^n\Grp:\Codiag N^n
$$
and verifies the required acyclicity condition; see
\cite[Chapter~II, Theorem~1.2, and Chapter~III, Section~2]{Adrom}.
This gives the first assertion.

For the second assertion, transport the three distinguished classes across
the Ellis-Steiner equivalence $\Phi_n\dashv\Gamma_n$.  Model-category axioms and cofibrant generation are
invariant under an equivalence of categories.  They therefore give the
stated model structure on $\XCube_n$.
\end{proof}

\subsection{Comparison with simplicial sets}

Now that we have a model structure on $\XCube_n$ for each $n$, it is natural to ask how it compares with the standard model structure on $\SSet$ through the Loday classifying functor $\BL$. 
First we note that $\bW$ has a natural left adjoint $G$ known as the Kan-Loop functor. 
$$
  G:\SSet_0
  \rightleftarrows
  \sGrp:\bW .
$$

Let $\SSet_0^n$ denote the model category of reduced simplicial sets
representing pointed connected $(n+1)$-types.

\begin{theorem}[Adrom]
\label{thm:adrom-quillen-equivalence}
There is a Quillen equivalence
$$
\SSet_0^n
\underset{\bW\,\Codiag N^n\Gamma_n}
{\overset{\Phi_n\tau\sigma^\ast G}{
  \scalebox{5}[1]{$\rightleftarrows$}}}
\XCube_n .
$$
\end{theorem}

\begin{proof}
Adrom constructs $\SSet_0^n$ as a left Bousfield localization and proves
a Quillen equivalence between $\cat^n\Grp$ and $\SSet_0^n$
see \cite[Chapter~III, Theorem~6.1]{Adrom}. Composing this adjunction with the Ellis-Steiner equivalence
gives the stated left and right adjoints and hence the claimed Quillen
equivalence.
\end{proof}
\begin{remark}
    Theorem~\ref{thm:adrom-quillen-equivalence}, together with Lemma~\ref{lem:diagonal-codiagonal} below, gives another proof of Corollary~\ref{cor:loday-connected-truncated}.
\end{remark}

The right adjoint in Theorem~\ref{thm:adrom-quillen-equivalence} is not
literally the Loday classifying functor. For a crossed $n$-cube $M$,
the latter is
$$
  \BL
  =\bW\Diag N^n\Gamma_n,
$$
whereas Adrom's right adjoint uses the codiagonal.  The following comparison
shows that the distinction disappears after passage to homotopy theory.

\begin{lemma}[Diagonal-codiagonal comparison]
\label{lem:diagonal-codiagonal}
For every $n$-simplicial set $Z$, the diagonal and codiagonal are
naturally weakly equivalent:
$$
  \Diag Z\simeq\Codiag Z.
$$
The same conclusion holds for $n$-simplicial groups. 
\end{lemma}

\begin{proof}
This is \cite[Chapter~III, Lemma~2.7]{Adrom}. 
\end{proof}

Applying Lemma~\ref{lem:diagonal-codiagonal} to the multinerve of a crossed
cube, and then applying $\bW$, gives a natural weak equivalence
\begin{equation}
\label{eq:loday-codiagonal-comparison}
  \BL
  =\bW\,\Diag N^n\Gamma_n
  \simeq
  \bW\,\Codiag N^n\Gamma_n.
\end{equation}
Thus the derived Loday classifying functor agrees with the derived right
adjoint in Theorem~\ref{thm:adrom-quillen-equivalence}.

\begin{proposition}
\label{prop:derived-mapping-spaces-xcube}
For crossed $n$-cubes $X$ and $Y$, there is a natural weak equivalence
$$
  \Map^h_{\XCube_n}(X,Y)
  \simeq
  \Map^h_{\SSet_0^n}\bigl(\BL X,\BL Y\bigr),
$$
where the Loday classifying functor and both mapping spaces are understood
derivatively.
\end{proposition}

\begin{proof}
Write
$B_{\mathrm{co}} :=\bW\,\Codiag N^n\Gamma_n$,
for the right adjoint in
Theorem~\ref{thm:adrom-quillen-equivalence}. The right adjoint of a
Quillen equivalence induces weak equivalences on derived mapping spaces;
see \cite[ Proposition~17.4.16]{Hirschhorn}. Hence there is a natural weak equivalence
$$
  \Map^h_{\XCube_n}(X,Y)
  \simeq
  \Map^h_{\SSet_0^n}
  \bigl(B_{\mathrm{co}}X,B_{\mathrm{co}}Y\bigr).
$$
By \eqref{eq:loday-codiagonal-comparison}, there are natural weak
equivalences
$$
  B_{\mathrm{co}}X\simeq \BL X,
  \qquad
  B_{\mathrm{co}}Y\simeq \BL Y.
$$
Derived mapping spaces are invariant under weak equivalences in each
variable \cite[Theorem~17.6.3]{Hirschhorn}; therefore
$$
  \Map^h_{\SSet_0^n}
  \bigl(B_{\mathrm{co}}X,B_{\mathrm{co}}Y\bigr)
  \simeq
  \Map^h_{\SSet_0^n}
  \bigl(\BL X,\BL Y\bigr).
$$
Combining these two weak equivalences proves the assertion.
\end{proof}

Proposition~\ref{prop:derived-mapping-spaces-xcube} shows that all the
homotopical information carried by the Loday classifying spaces at fixed
arity $n$, including derived morphisms and their higher homotopies, may
be recovered within the model category of crossed $n$-cubes. Here we give examples of how this works in $n=1$ and $n=2$

\begin{remark}
\label{rem:model-structure-for-12}
For crossed modules ($n=1$) and crossed squares $n=2$, homotopy groups can be computed neatly. See \ref{ex:loday-crossed-module} and \ref{ex:loday-crossed-square}. In higher arities one should not expect comparably simple formulas in general.  Nevertheless, the model structure shows that every homotopy invariant of the corresponding Loday classifying spaces can, at least in principle, be computed entirely within $\XCube_n$.
\end{remark}

\section{Proofs for quantum cellular automata and circuits}
\label{app:qca}

This appendix contains the proofs of Proposition~\ref{thm:basicQCA} and Theorem~\ref{thm:E}. These are, respectively, the basic properties of localized QCAs and the verification of the crossed-precosheaf structure on $\CE$. Notation is that of Section~\ref{sec:qspin}.

\begin{proof}[Proof of Proposition~\ref{thm:basicQCA}]
We first note the following easy fact. Let $S\subseteq X$, and let $a\in \CA(X)$. If $[a,b]=0$
for every $x\notin S$ and every $b\in \CA(\{x\})$, then $a\in \CA(S)$. This follows from local finiteness of $q$ and the fact that in a finite tensor product of full matrix algebras, the commutant of one tensor factor is the tensor product of the
remaining factors.

We prove the four assertions in order.

\emph{1.} The identity automorphism is a QCA of spread zero. Suppose
that $\alpha$ has spread $r$ and $\beta$ has spread $s$. Then, for
every $U\subseteq X$,
$$
(\alpha\circ\beta)(\CA(U))
\subseteq
\alpha(\CA(U^s))
\subseteq
\CA((U^s)^r)
\subseteq
\CA(U^{r+s}).
$$
Thus $\alpha\circ\beta$ is a QCA of spread $r+s$.

It remains to prove that $\alpha^{-1}$ is a QCA. We claim that it
has spread $r$. Let $a\in \CA(U)$, let $x\in X$ satisfy
$\dist(x,U)>r,$
and let $b\in \CA(\{x\})$. Since $\alpha$ has spread $r$, $\alpha(b)\in \CA(\{x\}^r)$.
The sets $U$ and $\{x\}^r$ are disjoint, so $
[a,\alpha(b)]=0,$ and therefore
$$
[\alpha^{-1}(a),b]
=
\alpha^{-1}\bigl([a,\alpha(b)]\bigr)
=
0.
$$
This holds for every $x\notin U^r$ and every
$b\in \CA(\{x\})$, therefore $
\alpha^{-1}(a)\in \CA(U^r).$
Hence
$\alpha^{-1}(\CA(U))\subseteq \CA(U^r)$
for every $U\subseteq X$, and $\alpha^{-1}$ has spread $r$.
Therefore $\CQ(X)$ is a subgroup of the group of
$*$-automorphisms of $\CA(X)$.

\emph{2.} We first show that $\CQ(U)$ is a subgroup. Suppose that
$\alpha$ is $r$-localized on $U$ and $\beta$ is $s$-localized on
$U$. If $\dist(x,U)>\max\{r,s\},$
then both $\alpha$ and $\beta$ act identically on $\CA(\{x\})$.
Consequently, $\alpha\circ\beta$ is
$\max\{r,s\}$-localized on $U$. Moreover, if $\alpha$ acts
identically on $\CA(\{x\})$, then so does $\alpha^{-1}$. Thus
$\alpha^{-1}$ is $r$-localized on $U$, and $\CQ(U)$ is a subgroup.

Now let $\alpha\in \CQ(U)$ be $r$-localized on $U$, and let
$\gamma\in \CQ(X)$ have spread $\ell$. By the first part,
$\gamma^{-1}$ also has spread $\ell$. Suppose that $\dist(x,U)>r+\ell$
and let $a\in \CA(\{x\})$. Then
$\gamma^{-1}(a)\in \CA(\{x\}^{\ell})$.
Every $y\in\{x\}^{\ell}$ satisfies
$$
\dist(y,U)
\geq
\dist(x,U)-\dist(x,y)
>
r.
$$
Hence $\alpha$ acts identically on every one-site algebra
$\CA(\{y\})$ with $y\in\{x\}^{\ell}$, and therefore acts identically
on the whole algebra $\CA(\{x\}^{\ell})$. It follows that
$$
(\gamma\alpha\gamma^{-1})(a)
=
\gamma\bigl(\alpha(\gamma^{-1}(a))\bigr)
=
\gamma(\gamma^{-1}(a))
=
a.
$$
Thus $\gamma\alpha\gamma^{-1}$ is $(r+\ell)$-localized on $U$.
Therefore $
\gamma \CQ(U)\gamma^{-1}\subseteq \CQ(U),$
and $\CQ(U)$ is a normal subgroup of $\CQ(X)$.

\emph{3.} Suppose that $U\subseteq V^s$
for some $s\geq 0$, and let $\alpha\in \CQ(U)$ be $r$-localized on
$U$. Then 
$U^r\subseteq V^{s+r}.$
Consequently, if
$\dist(x,V)>s+r,$
then $\dist(x,U)>r$, and $\alpha$ acts identically
on $\CA(\{x\})$. Thus $\alpha$ is $(s+r)$-localized on $V$, so 
$\CQ(U)\subseteq \CQ(V).$

\emph{4.} Let $U$ be bounded and let $\alpha\in \CQ(U)$ be $r$-localized on $U$. Put $B=U^r$, then $B$ is bounded and by local finiteness of $q$ the $*$-algebra $\CA(B)$ is a full matrix algebra.
The automorphism $\alpha$ fixes every one-site algebra outside $B$. We claim that it preserves $\CA(B)$. Indeed, let $a\in \CA(B)$, let
$x\notin B$, and let $b\in \CA(\{x\})$. Since $\alpha(b)=b$ and
$[a,b]=0$, one has
$$
[\alpha(a),b]
=
[\alpha(a),\alpha(b)]
=
\alpha([a,b])
=
0.
$$
Since $x\notin B$ and $b\in \CA(\{x\})$ were arbitrary, 
$\alpha(a)\in \CA(B)$. Thus $\alpha$ restricts to a $*$-endomorphism of the full matrix algebra $\CA(B)$. It is actually a $*$-automorphism, since exactly the same argument applies to $\alpha^{-1}$. Every such $*$-automorphism is inner, hence  there is
a unitary $u\in \CA(B)$ such that
$\alpha(a)=uau^{-1}$ for every $a\in\CA(B)$.

Both $\alpha$ and $\Ad_u$ act identically on every
one-site algebra $\CA(\{x\})$ with $x\notin B$. Since $\CA(X)$ is
generated by $\CA(B)$ together with these one-site algebras, it follows
that $\alpha=\Ad_u$ on all of $\CA(X)$. Conversely, it is easy to check that for any unitary $u\in \CA(X)$ the $*$-automorphism $\Ad_u$ is a QCA.

Finally, $u\mapsto\Ad_u$ is a group homomorphism, and its kernel consists of unitary central elements of $\CA(X)$. This group is  isomorphic to $U(1)$.
\end{proof}

In the remainder of this appendix we verify that $\CE$ is a crossed precosheaf (Theorem \ref{thm:E}).

\begin{lemma}\label{lem:nested}
Let $Y_1, Y_2, Y_3 \in \SL$ with $Y_1 \wedge Y_2 \wedge Y_3 = \bott$, let $\alpha \in \CC(Y_1)$, $\beta \in \CC(Y_2)$, $\gamma \in \CC(Y_3)$, and let $P \in \Circ(Y_1)$, $Q \in \Circ(Y_2)$, $R \in \Circ(Y_3)$ be circuits with $\phi(P) = \alpha$, $\phi(Q) = \beta$, $\phi(R) = \gamma$. Then $[\alpha^{-1},\beta] \in \CC(Y_1 \wedge Y_2)$, so that $\eta([\alpha^{-1},\beta],\gamma)$ is defined, and for all sufficiently large $r$,
$$
\alpha\Bigl(\eta\bigl([\alpha^{-1},\beta],\gamma\bigr)\Bigr) = \hP_r\, \bigl[[\hP_r^{-1}, \hat Q_r], \hR_r\bigr]\, \hP_r^{-1}.
$$
\end{lemma}

\begin{proof}
Put $\delta := [\alpha^{-1},\beta]$. By Proposition~\ref{prop:comm-rep}, $\delta \in \CC(Y_1 \wedge Y_2)$, and $(Y_1 \wedge Y_2) \wedge Y_3 = \bott$, so $v := \eta(\delta,\gamma) \in \Uloc$ is defined. Put
$$
w_r := [\hP_r^{-1}, \hat Q_r], \qquad \delta_r := \phi\bigl([P_r^{-1}, Q_r]\bigr) = \Ad_{w_r},
$$
where the last equality holds by Definition~\ref{def:hat}. By Proposition~\ref{prop:eta}(1), applied to the second expression in Definition~\ref{def:eta} with $R$ representing $\gamma$, we have $v = \delta(\hR_r)\hR_r^{-1}$ for $r$ large. We claim that
$$
\delta(\hR_r) = \delta_r(\hR_r) \qquad \text{for $r$ large.}
$$
Then $v = w_r \hR_r w_r^{-1} \hR_r^{-1} = [w_r, \hR_r]$ for $r$ large.

To prove the claim, fix representatives of $Y_1$, $Y_2$, $Y_3$ in $X$, choose words for $P$, $Q$, $R$ of depths $n$, $m$, $\ell$ with $\varrho$-bounded layers, and choose $s_1, s_2, s_3 \ge 0$ such that their gates lie in $Y_1^{s_1}$, $Y_2^{s_2}$, $Y_3^{s_3}$. Let $W$ be the resulting word for $[P^{-1},Q]$, of depth $d := 2(n+m)$, so that $\delta = \phi(W)$. Truncating its layers gives a word $W_r$ for $[P_r^{-1},Q_r]$, so that $\delta_r = \phi(W_r)$. Put $\rho := 2\varrho d$.

Let $x \in X$ with $x \notin Y_1^{s_1+\rho}$. Then no nontrivial block of a layer of $P$ meets $\{x\}^\rho$. Let $W'$ be the word obtained from $W$ by replacing all gates of the layers coming from $P^{\pm1}$ by $1$. Then $\phi(W') = \beta\beta^{-1} = \id$, and $W$ and $W'$ have the same gates on all blocks meeting $\{x\}^\rho$, so $\delta$ is trivial on $\CA(\{x\})$ by Lemma~\ref{lem:lightcone}(2). The same argument applies to $W_r$, whose nontrivial blocks are among those of $W$, and, with the roles of $P$ and $Q$ exchanged, to points $x \notin Y_2^{s_2+\rho}$. Hence $\delta$ and $\delta_r$ are trivial on $\CA(\{x\})$ unless
$$
x \in T := Y_1^{s_1 + \rho} \cap Y_2^{s_2 + \rho} .
$$
Now let $\lambda$ be a nontrivial block of a layer of $R_r$, so $\lambda \subseteq Y_3^{s_3}$. If $\lambda$ does not meet $T$, then $\delta$ and $\delta_r$ both fix $\CA(\lambda)$. If $\lambda$ meets $T$, then $\lambda \subseteq S_0 := T^{2\varrho} \cap Y_3^{s_3}$. The set $S_0$ is nonempty and satisfies $S_0 \le Y_1$, $S_0 \le Y_2$, $S_0 \le Y_3$ in the fuzzy preorder, hence $S_0 \le Y_1 \wedge Y_2 \wedge Y_3 = \bott$; that is, $S_0$ is bounded. Since $S_0$ does not depend on $r$, Lemma~\ref{lem:lightcone}(4), applied to $W$ with $S = S_0$ and $K = \bott$, gives $\delta = \delta_r$ on $\CA(S_0)$ for $r$ large. Thus $\delta(u_\lambda) = \delta_r(u_\lambda)$ for every gate $u_\lambda$ of $R_r$, and therefore $\delta(\hR_r) = \delta_r(\hR_r)$, proving the claim.

Finally, choose a bounded $S \subseteq X$ with $v \in \CA(S)$. By Lemma~\ref{lem:lightcone}(4), applied to a word for $P$ with $K = \bott$, we have $\alpha = \Ad_{\hP_r}$ on $\CA(S)$ for $r$ large. Hence
$$
\alpha(v) = \hP_r\, v\, \hP_r^{-1} = \hP_r\, \bigl[[\hP_r^{-1}, \hat Q_r], \hR_r\bigr]\, \hP_r^{-1}
$$
for all sufficiently large $r$.
\end{proof}

\begin{proposition}\label{prop:E-cube}
Let $\xi = ((U_i)_{i\in I} \mid W)$ be an $I$-indexed \v{C}ech configuration. Put $M_A := \CE(U^\xi_A)$, let $\mu_i : M_A \to M_{A \setminus \{i\}}$ be the corestriction maps of $\CE$, and let $h_{A,B} := h^{\CE,\xi}_{A,B}$. Then $(M_A, \mu_i, h_{A,B})$ is a crossed $I$-cube of groups.
\end{proposition}

\begin{proof}
We verify the conditions of Definition~\ref{def:crossed-n-cube}. Whenever all regions occurring in an identity are different from $\bott$, the groups, structure maps, and pairings are those of the crossed precosheaf $\CC$, and the identity holds by Theorem~\ref{thm:crossed-precosheaf-circuits}. We treat the remaining cases, using throughout that $x \mapsto \bx$ is a homomorphism compatible with corestrictions (Definition~\ref{def:pi}) and the description of the actions in Remark~\ref{rem:E-actions}.

\textbf{Commutativity of the cube.} $\mu_i\mu_j$ and $\mu_j\mu_i$ are both the corestriction $\CE(U^\xi_A) \to \CE(U^\xi_{A\setminus\{i,j\}})$.

\textbf{Compatibility of the pairings with the structure maps.}
We must show
$$
\mu_i\bigl(h_{A,B}(x,y)\bigr)
=
h_{A\setminus\{i\},B\setminus\{i\}}(\mu_i x,\mu_i y).
$$
Put $K:=U^\xi_{A\cup B}$ and
$K':=U^\xi_{(A\cup B)\setminus\{i\}}$, so $K\leq K'$. If
$K\ne\bott$, both sides are the ordinary commutator. If
$K=K'=\bott$, then $\mu_i$ is the identity on the target $\Uloc$, and
both sides equal $\eta(\bx,\by)$ because
$\overline{\mu_i x}=\bx$. If $K=\bott\ne K'$, the left side is
$\Ad_{\eta(\bx,\by)}=[\bx,\by]$ by
Proposition~\ref{prop:eta}(2), and the right side is
$[\mu_i x,\mu_i y]=[\bx,\by]$.

\textbf{Independence of a common coordinate.} If $i \in A \cap B$, then $U^\xi_{(A\setminus\{i\})\cup B} = U^\xi_{A \cup (B \setminus \{i\})} = U^\xi_{A\cup B}$, so the three pairings are in the same case, and $\overline{\mu_i x} = \bx$, $\overline{\mu_i y} = \by$.

\textbf{Diagonal commutator identity.} If $U^\xi_A = \bott$, then $h_{A,A}(x,x') = \eta(\Ad_x, \Ad_{x'}) = [x,x']$ by Lemma~\ref{lem:eta-more}(3).

\textbf{Skew-symmetry and normalization.} If $U^\xi_{A\cup B} = \bott$, these are Lemma~\ref{lem:eta-more}(1),(2).

\textbf{Crossed multiplicativity.} If $U^\xi_{A \cup B} = \bott$, then by Remark~\ref{rem:E-actions} the action of $x$ on $\Uloc$ is evaluation by $\bx$, and Proposition~\ref{prop:eta}(3) gives
$$
h_{A,B}(xx',y) = \eta(\bx\bx', \by) = \bx\bigl(\eta(\bx',\by)\bigr)\,\eta(\bx,\by) = {}^x h_{A,B}(x',y)\; h_{A,B}(x,y),
$$
and similarly in the second variable.

\textbf{Equivariance of the pairings.} Let $A \subseteq B$, $A \subseteq C$ and $U^\xi_{B\cup C} = \bott$. By Remark~\ref{rem:E-actions}, $\overline{{}^a b} = \ba\bar b\ba^{-1}$ and $\overline{{}^a c} = \ba\bar c\ba^{-1}$, while ${}^a$ acts on $\Uloc$ as $\ba$. Proposition~\ref{prop:eta}(4) with $\gamma = \ba$ gives
\[
h_{B,C}({}^a b, {}^a c) = \eta(\ba\bar b\ba^{-1}, \ba\bar c\ba^{-1}) = \ba\bigl(\eta(\bar b,\bar c)\bigr) = {}^a h_{B,C}(b,c). \qedhere
\]
\end{proof}

\begin{proposition}\label{prop:E-fixedfiber}
Let $\xi = ((V_i)_{i\in I} \mid W) \le \xi' = ((U_i)_{i \in I} \mid W')$ in $\Cech_I(\SL)$, and let $\rho_A : \CE(U^\xi_A) \to \CE(U^{\xi'}_A)$ be the corestriction maps. Then $(\rho_A)_{A \subseteq I}$ is a morphism of crossed $I$-cubes $\Cech_{\CE}(\xi) \to \Cech_{\CE}(\xi')$.
\end{proposition}

\begin{proof}
Compatibility with the structure maps holds because all maps involved are corestrictions of $\CE$. For the pairings, put $K := U^\xi_{A\cup B} \le K' := U^{\xi'}_{A \cup B}$ and note $\overline{\rho_A(x)} = \bx$. If $K \ne \bott$, both pairings are commutators and the $\rho$'s are inclusions. If $K = K' = \bott$, then $\rho_{A\cup B} = \id_{\Uloc}$ and both sides are $\eta(\bx,\by)$. If $K = \bott \ne K'$, then $\rho_{A\cup B} = \Ad$ and $\rho_{A \cup B}(\eta(\bx,\by)) = [\bx,\by] = [\rho_A(x), \rho_B(y)]$ by Proposition~\ref{prop:eta}(2).
\end{proof}

Reindexing compatibility holds because $h^{\CE,\xi}_{A,B}$ depends on $(\xi,A,B)$ only through the regions $U^\xi_A$ and $U^\xi_B$, and $U^{r^*\xi}_A=U^\xi_{r(A)}$ for every $r:J\to I$ and $A\subseteq J$. This completes the proof of Theorem~\ref{thm:E}.

\bibliographystyle{unsrturl}
\bibliography{part1_references}

\end{document}